\documentstyle[12pt,epsfig,wrapfig,amss]{report}
\newcommand{\bild} [2]{
                       \refstepcounter{figure}
                       {\begin{center}
     \parbox{#1}{\footnotesize\baselineskip=12pt Fig.~\thefigure. #2}
                       \end{center}}
                       }
\newcommand{\scaption}[1]{\caption{\protect{\footnotesize  #1}}}

\newcommand{\av}[1]{\mbox{$ \langle #1 \rangle $}}
\newcommand{\ol}[1]{\mbox{$\overline{#1}$}}
\newcommand{\order}[1]{\mbox{${\cal O}(#1)$}}
\newcommand{\ncs}{\mbox{$N_{\rm CS}~$}}
\newcommand{\ein}{\mbox{$E_{\rm in}~$}}
\newcommand{\eout}{\mbox{$E_{\rm out}~$}}
\newcommand{\einp}{\mbox{$E_{\rm in}'~$}}
\newcommand{\eoutp}{\mbox{$E_{\rm out}'~$}}

\newcommand{\qprimesq}{\mbox{$Q'^2~$}}
\newcommand{\xprime}{\mbox{$x'~$}}

\newcommand{\qprimex}{\mbox{$Q'$}}
\newcommand{\qprimesqx}{\mbox{$Q'^2$}}
\newcommand{\xprimex}{\mbox{$x'$}}

\newcommand{\flim}{\mbox{$f_{\rm lim}~$}}
\newcommand{\finst}{\mbox{$f_{I}~$}}

\newcommand{\shat}{\mbox{$\hat{s}$}}
\newcommand{\xpom}{\mbox{$x_{I\!P}$}}
\newcommand{\pom}{\mbox{$I\!P$}}

\newcommand{\ftwoc}{\mbox{$F_2^{c\overline{c}}$}}
\newcommand{\sigc}{\mbox{$\sigma_{c\overline{c}}$}}
\newcommand{\bbbar}{\mbox{$b\overline{b}$}}
\newcommand{\ccbar}{\mbox{$c\overline{c}$}}
\newcommand{\qqbar}{\mbox{$q\overline{q}$}}
\newcommand{\ppbar}{\mbox{$p\overline{p}$}}
\newcommand{\degr}{\mbox{$^\circ$}}
\newcommand{\kjet}{\mbox{$k_{T\rm{jet}}$}}
\newcommand{\xjet}{\mbox{$x_{\rm{jet}}$}}
\newcommand{\xgluon}{\mbox{$x_{g/p}$}}
\newcommand{\xgobs}{\mbox{$x_g^{\rm obs}$}}
\newcommand{\xgp}{\mbox{$x_{g/p}$}}
\newcommand{\xqp}{\mbox{$x_{q/p}$}}
\newcommand{\ejet}{\mbox{$E_{\rm{jet}}$}}
\newcommand{\thjet}{\mbox{$\theta_{\rm{jet}}$}}
\newcommand{\thlab}{\mbox{$\theta_{\rm{lab}}$}}
\newcommand{\ptjet}{\mbox{$p_{T\rm{jet}}$}}
\newcommand{\W}{\mbox{$W~$}}
\newcommand{\Q}{\mbox{$Q~$}}
\newcommand{\xb}{\mbox{$x~$}}  

\newcommand{\sigtotal}{\mbox{$\sigma_{\rm tot}~$}}
\newcommand{\sigtot}{\mbox{$\sigma_{\rm tot}^{\gamma^* p}$}}
\newcommand{\sigt}{\mbox{$\sigma_T~$}}
\newcommand{\sigl}{\mbox{$\sigma_L~$}}
\newcommand{\xf}{\mbox{$x_F~$}}  
\newcommand{\xp}{\mbox{$x_p~$}}  %
\newcommand{\xe}{\mbox{$x_E~$}}  %

\newcommand{\Qsq}{\mbox{$Q^2~$}}
\newcommand{\Qsqx}{\mbox{$Q^2$}}
\newcommand{\knot}{\mbox{$K^0~$}}
\newcommand{\ls}{\mbox{$\lambda_s$}}
\newcommand{\mt}{\mbox{$m_T~$}}
\newcommand{\mur}{\mbox{$\mu_R$}}

\newcommand{\et}{\mbox{$E_T~$}}
\newcommand{\kt}{\mbox{$k_T~$}}
\newcommand{\pt}{\mbox{$p_T~$}}
\newcommand{\pz}{\mbox{$p_z~$}}
\newcommand{\ftwo}{\mbox{$F_2~$}}
\newcommand{\fl}{\mbox{$F_L~$}}

\newcommand{\ethad}{\mbox{$E_T^{\rm had}~$}}
\newcommand{\etpar}{\mbox{$E_T^{\rm par}~$}}
\newcommand{\ptintr}{\mbox{$p_T^{\rm intr}~$}}
\newcommand{\ptfrag}{\mbox{$p_T^{\rm frag}~$}}
\newcommand{\ptrad}{\mbox{$p_T^{\rm rad}~$}}
\newcommand{\pzmax}{\mbox{$p_z^{\rm max}~$}}

\newcommand{\pmax}{\mbox{$p^{\rm max}~$}}
\newcommand{\nmax}{\mbox{$n_{\rm max}~$}}
\newcommand{\emax}{\mbox{$E^{\rm max}~$}}
\newcommand{\pzmin}{\mbox{$p_z^{\rm min}~$}}

\newcommand{\as}{\mbox{$\alpha_s~$}}
\newcommand{\asx}{\mbox{$\alpha_s$}}
\newcommand{\ycut}{\mbox{$y_{\rm cut}~$}}
\newcommand{\kcut}{\mbox{$k_{\rm cut}~$}}

\newcommand{\mrsap}{\mbox{${\rm MRSA}^\prime~$}}

\newcommand{\lmsfive}{\mbox{$\Lambda^{(5)}_{\rm \ol{MS}}~$}}
\newcommand{\lambdams}{\mbox{$\Lambda_{\rm \ol{MS}}~$}}
\newcommand{\lambdaqcd}{\mbox{$\Lambda_{\rm QCD}~$}}
\newcommand{\leff}{\mbox{$\Lambda_{\rm eff}~$}}
\newcommand{\enel}{\mbox{$E_e~$}}
\newcommand{\thel}{\mbox{$\theta_e~$}}
\newcommand{\gstar}{\mbox{$\gamma^\ast~$}}
\newcommand{\dstar}{\mbox{$D^\star~$}}
\newcommand{\zmin}{\mbox{$z_{\rm min}~$}}
\newcommand{\ymin}{\mbox{$y_{\rm min}~$}}
\newcommand{\ymax}{\mbox{$y_{\rm max}~$}}
\newcommand{\dif}{\mbox{\rm d}}
\newcommand{\dd}{{\rm d}}
\newcommand{\sys}{{\rm sys}}
\newcommand{\stat}{{\rm stat}}
\newcommand{\rec}{{\rm rec}}

\newcommand{\fm}{\mbox{\rm ~fm}}
\newcommand{\TeV}{\mbox{\rm ~TeV~}}
\newcommand{\TeVx}{\mbox{\rm TeV}}
\newcommand{\GeV}{\mbox{\rm ~GeV~}}
\newcommand{\GeVx}{\rm GeV}

\newcommand{\GeVsq}{\mbox{${\rm ~GeV}^2~$}}
\newcommand{\GeVsqx}{\mbox{${\rm GeV}^2$}}
\newcommand{\nb}{\mbox{${\rm ~nb}~$}}

\newcommand{\pbx}{\mbox{${\rm pb}$}}

\newcommand{\pbinv}{\mbox{${\rm ~pb^{-1}}~$}}

\newcommand{\mmm}{\mbox{$\cdot 10^{-3}$}}
\newcommand{\mmmm}{\mbox{$\cdot 10^{-4}$}}

\newcommand{\nn}{\mbox{$1\cdot 10^{-2}$}}

\newcommand{\nnnn}{\mbox{$1\cdot 10^{-4}$}}

\newcommand{\epem}{\mbox{$e^+e^-$}}
\newcommand{\ep}{\mbox{$ep~$}}
\def\herafuture{Proc. Workshop
                    on ``Future Physics at HERA'', Hamburg 1995-1996,
                    eds. A. De Roeck, G. Ingelman and R. Klanner}
\def\heraphysics{Proc. Workshop on
                     ``Physics at HERA'', Hamburg 1991, eds.
                     W. Buchm\"uller and G. Ingelman}
\def\ringberg{to appear in Proc.
              Workshop ``New Trends in HERA Physics'',
              Schlo\ss\ Ringberg, Tegernsee 1997}
\def\madrid{to appear in Proc.
            Madrid Workshop on ``Low $x$ Physics'',
            Miraflores de la Sierra 1997}
\def\rome{Proc. Workshop DIS96 on
               ``Deep Inelastic Scattering and Related Phenomena'',
               Rome 1996, eds. G. D'Agostini and A. Nigro}
\def\paris{Proc. Workshop DIS95 on ``Deep Inelastic Scattering and QCD'',
           Paris 1995, eds. JF. Laporte and Y. Sirois}
\def\marseille{Proc. EPS-HEP93, Marseille 1993,
               eds. J. Carr and M. Perrottet}
\def\chicago{to appear in Proc.
               Workshop DIS97, Chicago 1997, eds. J. Repond and D. Krakauer}
\def\jerusalem{to appear in Proc. EPS-HEP97 Conference, Jersualem 1997}
\def\zuoz{Proc. ``Hadronic Aspects of Collider Physics'', Zuoz 1994}
\def\hamburg{to appear in Proc. Lepton-Photon Conf., Hamburg 1997}
\def\warsaw{Proc. ICHEP 96, Warsaw 1996}
\def\protvino{Proc. XIXth Workshop on ``High Energy Physics
              and Field Theory'', Protvino 1996,
              eds. V. A. Petrov, A. P. Samokhin
              and R. N. Rogalyov}
\def\blois{Proc. 6th Rencontre de Blois,
           ``The Heart of the Matter'', Blois 1994, eds. J.-F. Mathiot and
            J. Tran Thanh Van}

\begin{document}
%
%
%
\begin{titlepage}

\noindent
{\tt MPI-PhE/97-33}                  \\
{\tt hep-ph/9712505}                  \\
{\tt December 1997}                  \\

\begin{center}

\vspace*{1cm}

\begin{Large}

{\bf  QCD and the Hadronic Final
State in Deep Inelastic Scattering at HERA
   }\\[1.5cm]

\vspace*{1.cm}
Michael Kuhlen \\
\end{Large}

\vspace*{0.5cm}
  Max-Planck-Institut f\"ur Physik \\
  Werner-Heisenberg-Institut        \\
  F\"ohringer Ring 6                 \\
  D-80805 M\"unchen                  \\
  Germany       \\
  E-mail: kuhlen@desy.de
\end{center}

\vspace*{0.5cm}

\begin{center}
{\bf Abstract}
\end{center}
\begin{footnotesize}

\noindent
The measurements of the hadronic final state in deep inelastic
scattering at HERA are reviewed and discussed
in the context of QCD.
Covered are the general event properties in terms of energy flows,
charged particle production, and charm and strangeness production.
Quark fragmentation properties are studied in the Breit frame.
Event shape measurements
allow ``power corrections'' to be applied and the strong
coupling \as to be extracted. Other \as measurements are based
on dijet rates. Jet rates as well as charm production have been
used to determine the gluon density in the proton.
Indications have been found in the hadronic final state
for unconventional, non-DGLAP evolution at small $x$,
which could be explained
with BFKL evolution.
Signatures for QCD instanton effects are discussed
and first search results are presented.
\end{footnotesize}

\vspace{0.5cm}
\begin{center}
{\sl Habilitationsschrift,
submitted to Universit\"at Hamburg in December 1997}
\end{center}

\begin{abstract}
The measurements of the hadronic final state in deep inelastic
scattering at HERA are reviewed and discussed
in the context of QCD.
Covered are the general event properties in terms of energy flows,
charged particle production, and charm and strangeness production.
Quark fragmentation properties are studied in the Breit frame.
Event shape measurements
allow ``power corrections'' to be applied and the strong
coupling \as to be extracted. Other \as measurements are based
on dijet rates. Jet rates as well as charm production have been
used to determine the gluon density in the proton.
Indications have been found in the hadronic final state
for unconventional, non-DGLAP evolution at small $x$,
which could be explained
with BFKL evolution.
Signatures for QCD instanton effects are discussed
and first search results are presented.
\end{abstract}

\end{titlepage}

\tableofcontents

\chapter{Introduction \label{ch:intro}}                          
  \section{Overview  \label{sn:view}}     
\subsubsection{Motivation}

HERA\footnote{``Hadron-Elektron-Ring-Anlage''}
is the world's first electron\footnote{
HERA
can operate with either electrons or positrons. In the following,
the generic name electron is used
for electrons as well as for positrons.}-proton collider.
The large centre of mass (CM) energy of 300 GeV allows to explore
new regimes:
new particles with masses up to 300 GeV can be produced,
the structure of the proton can be studied with a resolving power
varying over 5 orders of magnitude
down to dimensions of $10^{-18}$~m,
and partons with very small fractional proton momenta
(Bjorken $x$ down to $10^{-6}$) become experimentally accessible.
The hadronic final state
which emerges when the proton breaks up has an invariant mass
up to 300 GeV.
It provides a laboratory to study quantum chromodynamics (QCD) under
varying experimental conditions, which can be controlled by measuring
the scattered electron.
The hadronic final state carries
information on the structure of the proton.
It is made use of to study
the dynamics of the proton's constituents, complementary
to structure function measurements.

In contrast to ``clean'' \epem~ interactions,
the initial state in $ep$ collisions
contains already a strongly interacting particle.
That makes the physics more complicated, but leads also to
interesting effects that cannot be studied in \epem~
collisions.
At HERA, well established fields are being studied, exploiting
the tunable kinematic conditions.
These comprise jet physics and comparisons with perturbative QCD
to extract the strong coupling $\alpha_s$ and the density of gluons
in the proton, or for example the measurement of
fragmentation functions.
Other topics have only blossomed with the advent of HERA.
To name a few, a class of events with large
rapidity gaps and small $x$ physics
allow longstanding problems in QCD to be addressed, which
are connected with scattering cross sections
at high energies and confinement.
There are also exotic effects like instanton induced reactions,
which, if discovered at HERA, would alter significantly our view
of particle physics.

\subsubsection{Purpose}

This work provides a review of the hadronic final state
measurements at HERA in deep inelastic scattering (DIS).
The emphasis is on experimental results, because in many cases at HERA,
experiment is driving theory.
Many measurements are being performed without any theoretical prediction,
and often they have not yet found
an unambiguous theoretical explanation.
Nevertheless, the results are
discussed in the context of the theory, where possible.

The review of the experimental situation is complete up to the fall of 1997.
It can therefore be consulted for quick access to the HERA data.
Basic concepts are explained in a partly pedagogical fashion
to serve physicists from outside the HERA community and
newcomers to HERA physics.

\subsubsection{Contents}

In chapter \ref{ch:intro}
the HERA machine and the H1 and ZEUS experiments
are introduced, and the kinematic variables are discussed.
In chapter \ref{ch:theo}
the theoretical framework of deep inelastic scattering is set up with
evolution equations and a special section on the interest
in small $x$ physics.
The hadronic final state should not be discussed without knowledge of
the inclusive $ep$ cross section and the proton
structure functions (chapter \ref{ch:sf}).
In chapter \ref{ch:models} we start with simple models for hadron
production. They are subsequently being refined in chapter \ref{ch:models}
and serve as the basis for the
discussion of the data.

Measurements of basic event properties are presented in chapter
\ref{ch:properties}: energy flows, charged particle spectra, charm
and strangeness contents, Bose-Einstein correlations.
The fragmentation of the scattered quark is studied in chapter
\ref{ch:qfrag} and compared with quark fragmentation in \epem~ annihilation
and QCD calculations.
Measurements of event shape variables allow a new
view on hadronziation properties with ``power corrections'',
offering a potentially powerful
tool for measurements of the strong coupling $\alpha_s$.
Jet production has been compared to perturbative QCD predictions to measure
\as and the gluon density in the proton
(chapter \ref{ch:jets}).
By measuring jet rates,
regions of phase space have been
identified where the measured jet rates are not well understood yet,
and where the underlying physics is
possibly departing from the
conventional picture of deep inelastic scattering.

Also the energy flow measurements at small $x$ have not yet found
an unambiguous theoretical interpretation.
In chapter \ref{ch:lowx} on low $x$ physics dedicated searches for
``footprints'' of new QCD effects (BFKL) are discussed:
energy flows, high \pt particles and ``forward jets''.
Chapter \ref{ch:inst} deals with
the possibility to discover QCD instantons at HERA,
which would have far reaching consequences for our understanding
of field theories and for cosmology.

Some related topics and neighbouring fields could only be touched upon.
The interested reader is referred to other reviews on
structure functions \cite{rev:levy,rev:badelek},
rapidity gaps \cite{rev:diffr},
photoproduction \cite{rev:erdmann}
and on
hadron production at fixed target experiments \cite{rev:schmitz}.

Throughout, unless stated otherwise,
the data shown have been corrected for detector effects
and QED radiation, and the errors comprise statistical
and systematic errors added in quadrature.

  \section{The HERA Machine  \label{sn:machine}}  
In the HERA machine,
electrons and and protons are accelerated and stored in two separate
rings. The circumference of the machine is 6.3 km.
The magnets of the proton ring are superconducting, the magnets
of the electron ring are conventional.
The final beam energies are $E_e=27.5\GeV$ for electrons and
$E_p=820\GeV$ for protons with a
collision centre of mass energy of
$\sqrt{s}=\sqrt{4 E_e E_p}=300~\GeVx$. Early data were collected
with $E_e=26.7~\GeVx$.

The beams are collided head-on in two interaction regions occupied by
the experiments H1 and ZEUS.
There are 220 bunch positions in the beam, of which typically 190 are filled
with a few $10^{10}$ particles per bunch. The time between
bunch crossings is 96 ns. The longitudinal bunch length is about
60 cm, leading to an approximately Gaussian distribution of
interaction points along the beam line with width 10 cm. The
transverse beam size is 300 $\mu$m
horizontally by 70 $\mu$m vertically.

In 1997 the average peak luminosity was
$8.4 \cdot 10^{30} {\rm cm}^{-2}{\rm s}^{-1}$ with
average beam currents
at the beginning of a fill of 77 mA for protons and 36 mA for electrons.
The total integrated luminosity for the 1997 run was 35 \pbinv.
Most analyses in this review are based on data
from the runs 1992-1994,
corresponding to an integrated luminosity of \order{1-2}\pbinv.

  \section{The H1 and ZEUS Detectors  \label{sn:det}}  

\begin{figure}[tbh]
   \centering
   \epsfig{file=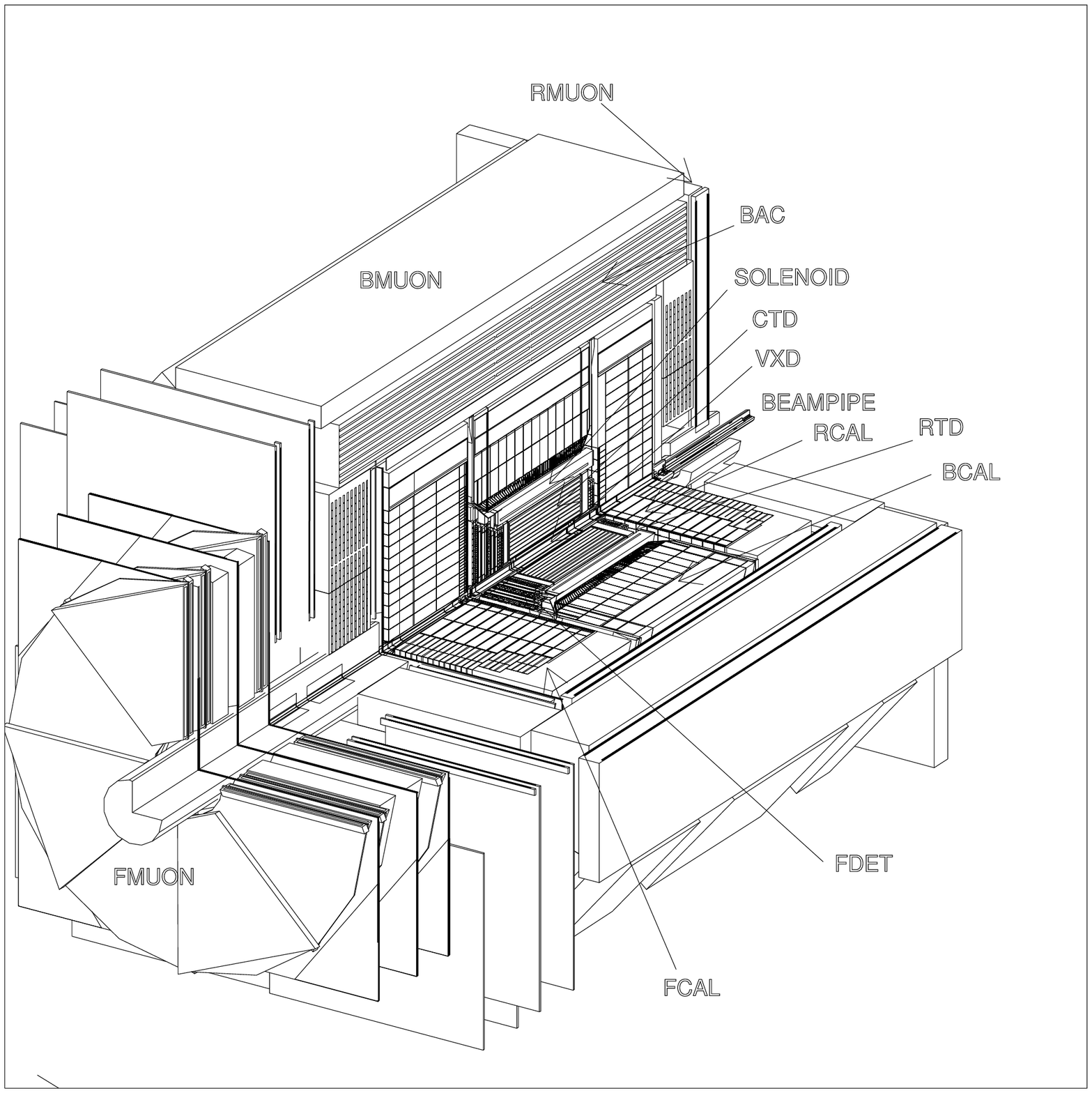,
   width=11cm,%
   bbllx=16pt,bblly=141pt,bburx=579pt,bbury=706pt,clip=}
   \scaption{
              The ZEUS detector. Protons come towards
              the observer. Shown are central, forward, backward  and
              vertex tracking detectors (CTD, FDET, RTD, VXD), the uranium
              calorimeter (BCAL, RCAL, FCAL), the muon system
              (RMUON, BMUON, FMUON) and the backing calorimeter (BAC).
              The dimension of the whole detector
              is roughly $10 \times 10 \times 18 {\rm ~m}^3$. }
   \label{detz}
\end{figure}

\begin{figure}[tbh]
   \centering
   \epsfig{file=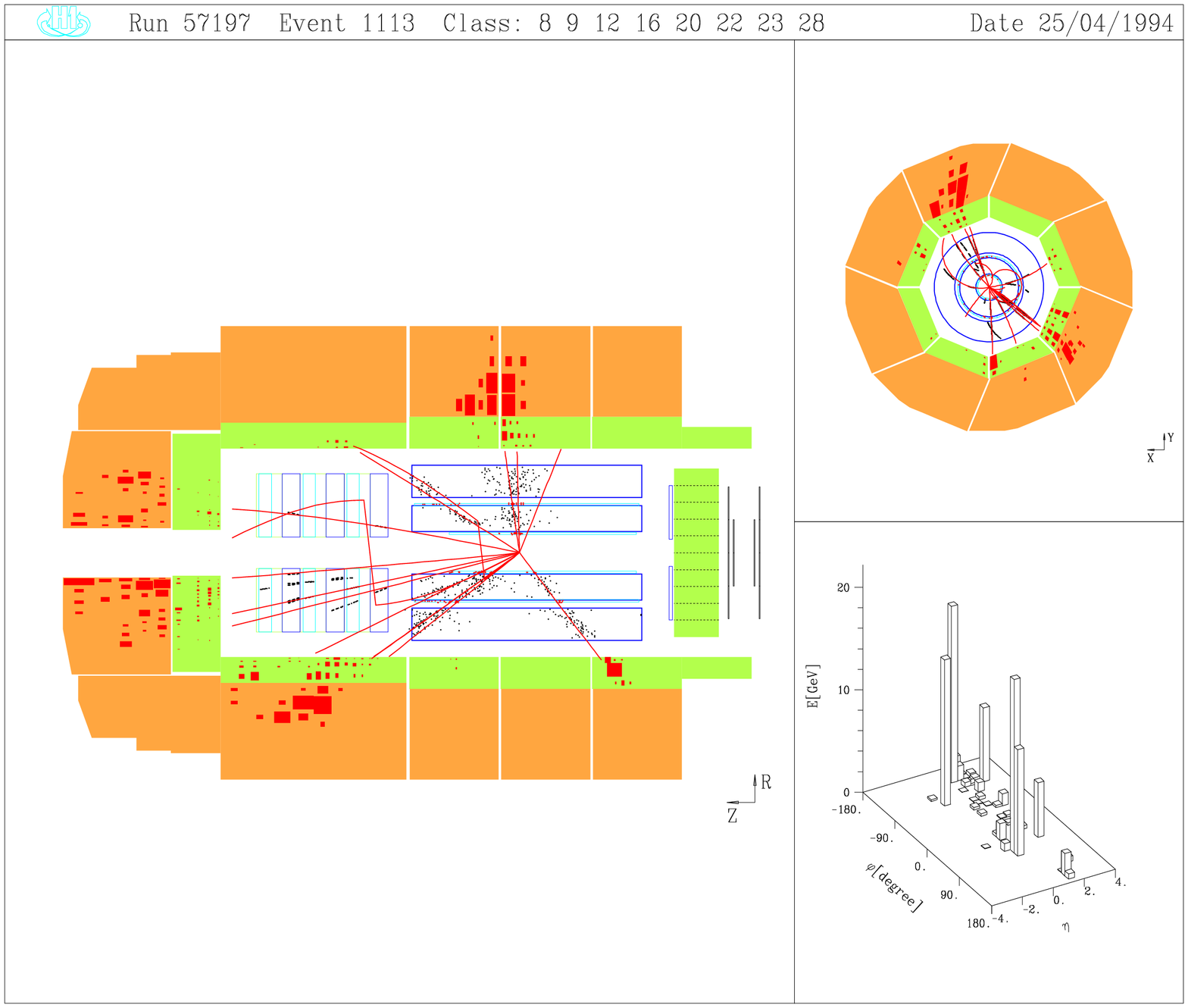,
   width=7cm,%
   bbllx=109pt,bblly=323pt,bburx=401pt,bbury=746pt,angle=90,clip=}
\begin{picture}(0,0) \put(-120,60){{\bf EMC}} \end{picture}
\begin{picture}(0,0) \put(-120,40){{\bf HAC}} \end{picture}
\begin{picture}(0,0) \put(-70,60){{\large $e^\prime$}} \end{picture}
\begin{picture}(0,0) \put(-100,120){{\bf CT}} \end{picture}
\begin{picture}(0,0) \put(-210,120){{\bf FT}} \end{picture}
\begin{picture}(0,0) \put(-60,80){{\bf BEMC}} \end{picture}
\begin{picture}(0,0) \put(-350,110){{\bf remnant}} \end{picture}
\begin{picture}(0,0) \put(-140,210){{\bf jet}} \end{picture}
\begin{picture}(0,0) \put(-270,15){{\bf jet}} \end{picture}
   \scaption{
                            DIS event
              with the scattered electron $e^\prime$ and
              two well separated jets detected in the H1 detector.
              The proton remnant leaves mostly
              undetected in the $+z$ direction.
              Shown are the central and forward tracking chambers (CT,FT),
              the electromagnetic and hadronic sections of the liquid
              argon calorimeter (EMC, HAC), and the backward electromagnetic
              calorimeter (BEMC).
}
   \label{deth}
\end{figure}

The H1 \cite{h1:nim1,h1:nim2}
and ZEUS detectors \cite{z:det} serve to detect the scattered electron
in $ep$ collisions and to measure the emerging hadrons. The
individual detector components are mounted concentrically around
the beam line. Due to the asymmetric beam energies, the hadronic
system is boosted into the proton direction
($+z$).
Therefore the detectors
are also asymmetric with respect to the interaction point, with
enhanced instrumentation for hadrons in the $+z$ (forward) direction.
The acceptances and resolutions of the main detector components
for the analyses presented here are given in table \ref{tab:det}.
Fig. \ref{detz} shows a drawing of the ZEUS detector, and
fig. \ref{deth} an event display from H1.

\begin{footnotesize}
\begin{table}[tbh]
\begin{center}
\begin{tabular}{|l||c|c||c|c|}
\hline
   & \multicolumn{2}{c||}{H1} & \multicolumn{2}{c|}{ZEUS} \\
\hline
\hline
tracking   &  $\theta$ acceptance & resol. $\sigma_{p_T}/p_T$ &
            $\theta$ acceptance & resol. $\sigma_{p_T}/p_T$    \\
\hline
forward & $ 7\degr - 25\degr$ & $0.02 \cdot p_T/\sin\theta $ &
          $ 7.5\degr - 28\degr$ &    \\
central & $20\degr -160\degr$ & $0.009 \cdot p_T $ &
          $15\degr -164\degr$ & $0.005 \cdot p_T $  \\
\hline
\hline
calorimetry  &  $\theta$ acceptance & resol. $\sigma_{E}/E$ &
                $\theta$ acceptance & resol. $\sigma_{E}/E$    \\
  \hline
electromagnetic &   $4\degr-153 \degr$ (LAr)     & $0.11/\sqrt{E} $   &
     $2.2\degr-176.5\degr$ (U)    & $0.18/\sqrt{E} $   \\
 & $155.5\degr-174.5\degr$ (BEMC) &
        $0.1/\sqrt{E} $ & & \\
 & $151\degr-177.5\degr$ (SPACAL) &
        $0.075/\sqrt{E} $ & & \\
\hline
hadronic & $0.7\degr-3.3\degr$ (PLUG)   & $\approx 1.5/\sqrt{E}$ & & \\
 & $4\degr-153\degr$ (LAr)      & $0.5/\sqrt{E} $ &
   $2.2\degr-176.5\degr$ (U)& $0.35/\sqrt{E} $ \\
 & $153\degr-178\degr$ (SPACAL) & $\approx 0.3/\sqrt{E}$ & & \\
\hline
\end{tabular}
\end{center}
\scaption{Acceptances and resolutions of the main detector
          components from H1 and ZEUS.
    In the resolution formulae, energy $E$ and transverse momentum $p_T$ are
    to be taken in GeV. Additional constant contributions to
    the resolutions of \order{1-3\%} are not shown.
    }
\label{tab:det}
\end{table}
\end{footnotesize}

Closest to the beam line are wire chambers for measuring charged particle
trajectories. The particles' momenta are determined from their track curvature
in a longitudinal magnetic field provided by a superconducting coil.

Electromagnetic and hadronic showers are measured in calorimeters
surrounding the inner tracking devices. H1 emphasizes electron
detection with a finely segmented lead (inner electromagnetic part)
and steel (outer hadronic part)
liquid argon calorimeter (LAr) with
good energy resolution for electrons, supplemented by a dedicated
electromagnetic backward calorimeter.
From 1992-1994 an electromagnetic lead/scintillator sandwich calorimeter was
installed in the backward region (BEMC), and from 1995 onwards a
lead/scintillating fibre calorimeter (SPACAL). A copper calorimeter
with silicon readout (PLUG) covers part of the forward beam hole.
Compensation for the
different response to hadronic and
electromagnetic showers in the LAr is
done offline by a software weighting technique.
ZEUS achieves
better hadronic energy resolution with a self-compensating
uranium/scintillator calorimeter (U), and compromises on electromagnetic
energy resolution.

The calorimeters are surrounded by chambers and absorber
plates for measuring shower leakage and for muon detection.
Further specialized detectors are
installed very close to the beam line to detect particles that
are scattered under small angles. Silicon detectors that have
already been installed as vertex detectors or are being planned
have not yet been used in physics analyses.

  \section{Kinematics \label{sn:kine}}      

\subsubsection{Definition of kinematic variables}

\begin{figure}[htb]
   \centering
   \begin{picture}(1,1) \put(0.,40.){QPM} \end{picture}
   \epsfig{file=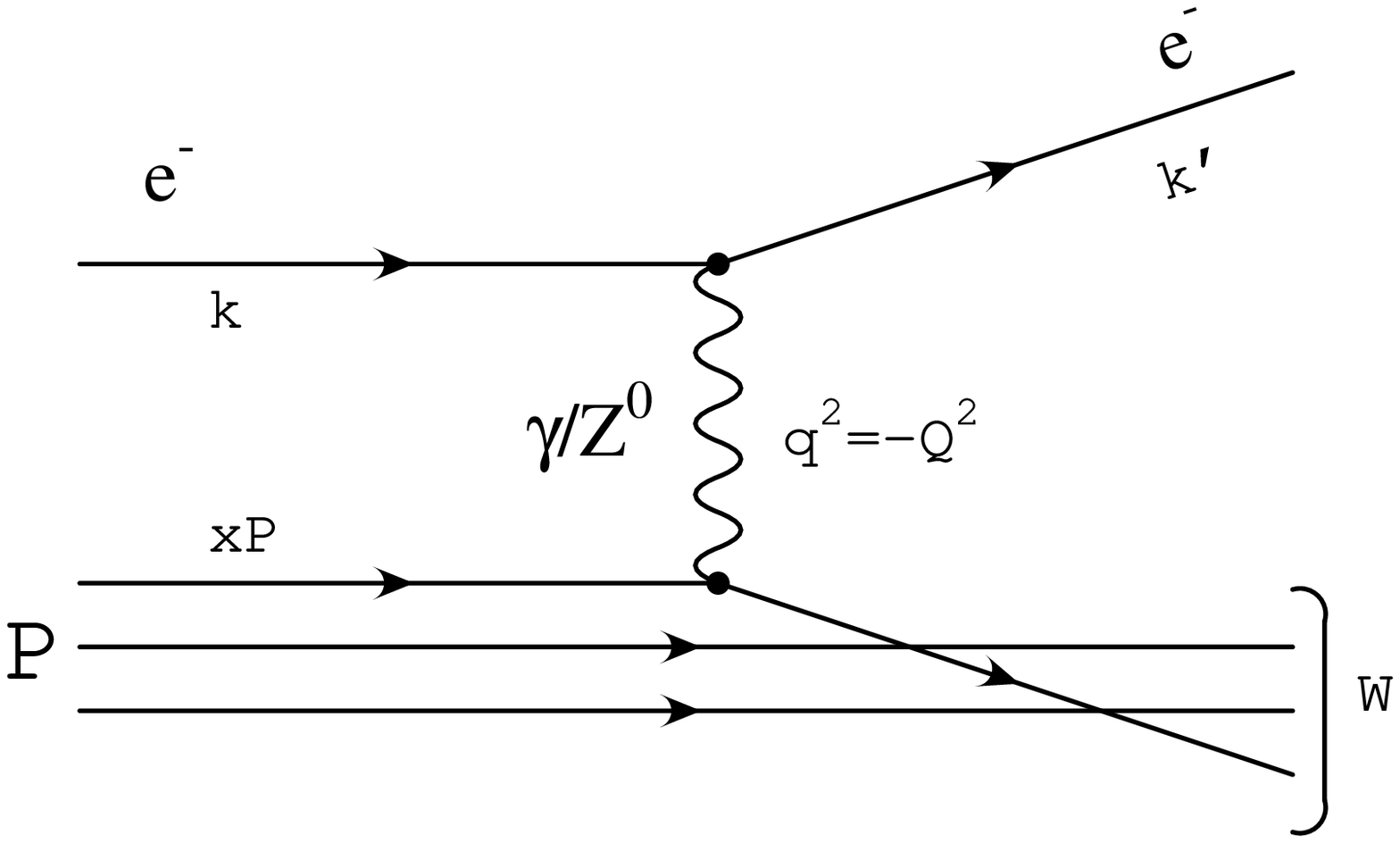,
    width=7cm,
    bbllx=50pt,bblly=483pt,bburx=522pt,bbury=771pt,clip=}
%
%
   \scaption{
              Basic diagram for DIS in
              $O(\alpha_s^0)$ (quark parton model - QPM).}
   \label{kin}
\end{figure}

The kinematics of the basic $ep$ scattering process in Fig.~\ref{kin}
can be characterized by any set of two
Lorentz-invariants out of $Q^2,x,y$ and $W$,
which are built from the 4-momentum
transfer $q=k-k'$ mediated by the virtual boson and from the
4-momentum $P$ of the incoming proton. The $ep$ invariant mass squared
is $s=(e+P)^2$.
These kinematic variables are then:
\begin{equation}
  Q^2 := - q^2,
\end{equation}
   which gives the transverse resolving power
   of the probe with wavelength $\lambda =1/Q$
   (we set $\hbar = c = 1$);
\begin{equation}
   x:=\frac{Q^2}{2Pq},
\end{equation}
   the Bjorken scaling variable ($0\leq x \leq1$), which can be
   interpreted as the momentum fraction of the proton which is
   carried by the struck quark (in a frame where the proton is fast,
   and assuming the quark-parton model to be a good approximation);
\begin{equation}
   y:=\frac{P q}{Pk}= \frac{Q^2}{x(s-m_p^2)},
\end{equation}
   the transferred energy fraction from the electron
   to the proton in the proton rest frame ($0\leq y \leq 1$); and
\begin{equation}
   W^2=\Qsq \frac{1-x}{x} + m_p^2 \approx sy-\Qsq,
\end{equation}
   the invariant mass squared of the outgoing hadronic
   system $H$.
   The invariant
\begin{equation}
   \nu := \frac{Pq}{m_p},
\end{equation}
   is rarely used at HERA.
   In the proton rest frame it gives the energy transfer from the
   lepton to the proton.

\begin{figure}[htb]
   \centering
   \epsfig{file=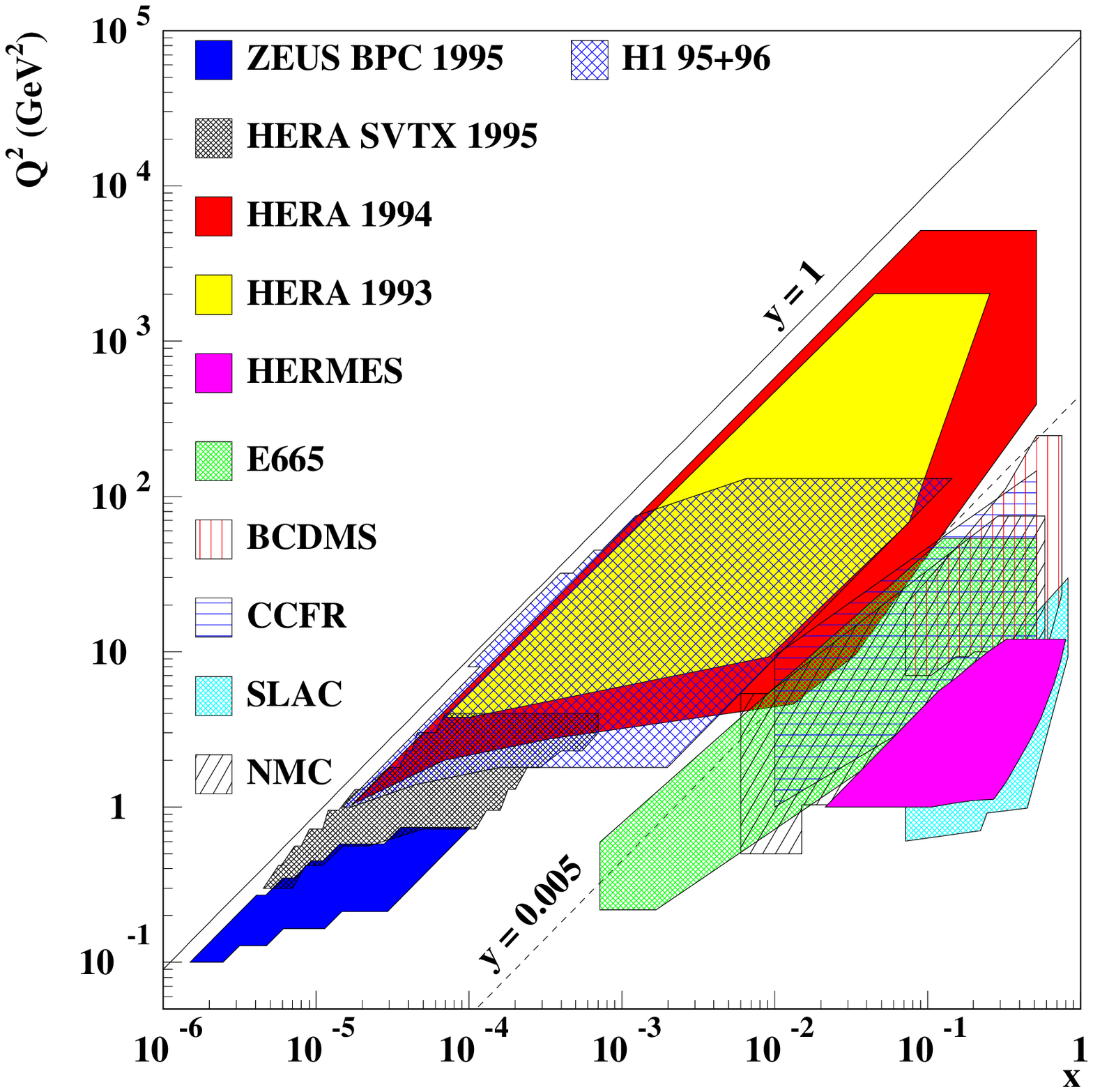,
   width=12cm}
   \scaption{Coverage in the kinematic plane $(x,\Qsq)$
             of various DIS experiments. The kinematic boundary for
             HERA is given by the line $y=1$. There exist also HERA
             data in the region $10^4 \GeVsq < \Qsq < 10^5 \GeVsq$.
              }
   \label{hera_kine}
\end{figure}

The large CM energy gives access to
kinematic regions both at very small $x$
and at large \Qsq (Fig. \ref{hera_kine}). The HERA data
cover roughly
$\Qsq = 0.2 - 10^4 \GeVsq$,
$x=10^{-5} - 10^{-1}$ and
$W=40-300 \GeV$.

\subsubsection{Experimental reconstruction of the event kinematics}

The kinematics can be determined either from the electron alone, or from
the measured hadronic system alone, or from a combination of both,
permitting important systematic cross checks. The hadronic measurement
relies mostly upon the calorimeters. At large $y$ the precision
of the hadron method can
be improved by momentum measurements in the trackers.

\begin{description}

\item[The electron method] The kinematic variables are calculated from
the energy $E_e'$ and angle \thel of the scattered electron (measured
with respect to the proton direction):
\begin{equation}
 y_e = 1 - \frac{E_e'}{\enel} \sin^2 \frac{\thel}{2}
 \hspace{1cm}
 Q^2_e = 4 E_e' \enel \cos^2 \frac{\thel}{2}
  = \frac{E_e'^2 \sin^2 \thel}{1-y_e} = \frac{p_{Te}^2}{1-y_e}.
\label{eq:kinel}
\end{equation}

\item[The hadron method]
The kinematics is measured entirely with the hadronic system:
\begin{equation}
 y_h = \frac{E_H - p_{zH}}{2\enel}
 \hspace{1cm}
 Q^2_h = \frac{p_{xH}^2+p_{yH}^2}{1-y_h}.
\label{eq:kinh}
\end{equation}
Here $E_H,p_{xH},p_{yH}$ and $p_{zH}$ denote the 4-vector components of
the hadronic system $H$, which are calculated as the
4-momentum sum over all
final state hadrons $h$. Jacquet and Blondel \cite{hera:jb} have shown
that the contribution from hadrons lost in the beam pipe
is insignificant.

\item[The Sigma method \cite{h1:bassler}]
Here the denominator of $y_h$ is replaced with
$\sum_{i} (E_i - p_{zi})$,
where
$i$ runs over all final state particles, {\it including} the scattered
electron.
This expression equals $2\enel$ due
to energy momentum conservation.
In case the incident electron had radiated off
photons which escape detection, the sum yields the
true electron energy which goes into the $ep$ interaction.
This method relies on both electron and hadron measurements.
With
\begin{equation}
\Sigma = \sum_{h} (E_h - p_{zh}),
\end{equation}
where the sum runs over all
hadronic final state particles, the kinematic variables can
be written as
\begin{equation}
y_\Sigma=\frac{\Sigma}{\Sigma+E_e'(1-\cos\thel)}
\hspace{1cm}
\Q^2_\Sigma  = \frac{E_e'^2 \sin^2 \thel}{1-y_\Sigma}.
\end{equation}
The denominator $\Sigma+E_e'(1-\cos\thel)$ is twice the energy
of the ``true'' incident electron, after QED radiation from the
incoming electron beam.

\item[The double angle method]
We define the angle $\gamma$ by
\begin{equation}
\cos \gamma = \frac{p_{TH}^2-(E_H - p_{zH})^2}
                   {p_{TH}^2+(E_H - p_{zH})^2}.
\end{equation}
In the simple quark parton model $\gamma$ would be the
angle of the scattered (massless) quark. The kinematic
variables can be calculated from $\gamma$ regardless of
its interpretation:
\begin{equation}
Q^2_{\rm DA} = 4 E_e^2
  \frac{\sin\gamma(1+\cos\gamma)}
       {\sin\gamma+\sin\thel-\sin(\thel+\gamma)}
\end{equation}
and
\begin{equation}
x_{\rm DA} = \left(\frac{\enel}{E_p}\right)
  \frac{\sin\gamma + \sin\thel + \sin(\thel+\gamma) }
       {\sin\gamma + \sin\thel - \sin(\thel+\gamma) }
\end{equation}
\end{description}

For most of the phase space the electron method is superior.
At small $y$ the hadron method has a better resolution than the
electron method. The ``mixed method'' uses \Qsq reconstructed from
the electron method and $y$ reconstructed with the hadron method.
Also
the double angle method and the sigma method use information from both
the electron and the hadronic system,
thus interpolating between the pure electron and hadron methods.
The sigma method has the advantage that it corrects for initial state
radiation.

\chapter{Theoretical framework \label{ch:theo} }              
  \section{Deep Inelastic Scattering  \label{sn:dis}}   
The fundamental measurement in DIS concerns the cross section
for $ep\rightarrow e'H$ as a function of the kinematic variables
(any pair of two independent ones).
The quark parton model (QPM) offers a physical picture:
the scattering takes place via a virtual photon which is
radiated off the scattering electron, and which couples
to a pointlike constituent inside the proton, that is
a quark or antiquark.
The cross section is then proportional to the quark density inside
the proton.

The differential cross section
$ep\rightarrow e'H$ can be expressed in terms of
two\footnote{For the sake of simplicity, $Z$
exchange (a 1\% correction
for $\Qsq\approx 1000 \GeVsq$) has been neglected, and the structure
function $F_3$ thus been omitted.}
independent
structure functions $F_1(x,Q^2)$ and $F_2(x,Q^2)$:
\begin{equation}
\begin{array}{ll}
\frac{\dd^2 \sigma}{\dd x \dd Q^2}
 &  = \frac{4\pi \alpha^2}{xQ^4} \left[ (1-y) F_2 + y^2 x F_1 \right] \\
 &  = \frac{4\pi \alpha^2}{xQ^4}
            \left[ 1 - y + \frac{y^2}{2}\frac{1}{1+R} \right] F_2     \\
 & = \frac{4\pi \alpha^2}{xQ^4} \left[ \left(1-y+\frac{y^2}{2}\right)
                             \cdot F_2
                          -  \frac{y^2}{2} \cdot F_L \right].
\end{array}
\label{eq:dsig}
\end{equation}
$\alpha$ is the electromagnetic coupling constant.
Here we have expressed the cross section also in terms of
the longitudinal structure function $F_L(x,Q^2)$ and the ratio $R$,
defined as
\begin{equation}
F_L := F_2 - 2x F_1 \hspace{2cm}  R := \frac{F_L}{F_2-F_L}
                                     = \frac{F_2-2xF_1}{2xF_1}
                                     = \frac{\sigl}{\sigt}.
\end{equation}
$R$ can be interpreted as the ratio of
the cross sections
\sigt and \sigl for the absorption of transversely
and longitudinally polarized virtual
photons
on protons, with
$\sigtot = \sigl+\sigt$.
The structure function \ftwo can be
expressed\footnote{We use the Hand convention \cite{th:hand}
for the definition of the virtual photon flux.} in terms of
\sigt and \sigl,
\begin{equation}
\ftwo = \frac{Q^2(1-x)}{4\pi^2\alpha}
\frac{Q^2}{Q^2+4 m_p^2 x^2} \cdot \sigtot
\approx \frac{\Qsq}{4\pi^2 \alpha}(\sigl+\sigt),
\end{equation}
where the small \xb approximation has been applied.
Similarly,
\begin{equation}
F_L=\frac{Q^2 (1-x)}{4\pi^2\alpha} \cdot \sigl
\approx \frac{Q^2}{4\pi^2\alpha} \cdot \sigl.
\end{equation}

In the ``DIS'' scheme \ftwo can be written
in terms
of the quark and antiquark densities, $q_i$ and $\ol{q}_i$,
and their couplings to the photon, i.e. their charges $e_{q_i}$:
\begin{equation}
F_2(x,Q^2) = x \sum_{i} e_{q_i}^2 \left[q_i(x,Q^2) + \ol{q}_i(x,Q^2)\right],
\label{eq:ftwodis}
\end{equation}
where the sum runs over all quark
flavours\footnote{
Equation \ref{eq:ftwodis} represents the ``leading twist''
(called twist 2) contribution
to the  structure function $F_2$,
when expanded in powers of $Q^2$ \cite{rev:badelek},
\begin{equation}
   F_2(x,Q^2) = \sum_{n=0}^{\infty} C_n(x,Q^2)/(Q^2)^n.
\end{equation}
The coefficients $C_n(x,Q^2)$ are varying logarithmically with $Q^2$.
The ``higher twist'' terms ($n>0$, called twist 4, 6 etc.)
arise from interactions of
the struck parton with the remnant and are suppressed by $(1/Q^2)^n$.
We shall not pursue higher twist effects any further, but
note that they may not be negligible at small $x$ for $Q^2$ as large as a few
\GeVsq \cite{rev:badelek,lowx:bartels_chicago}.}.
In other schemes (for example the ``$\ol{MS}$'' scheme) the relation between
\ftwo and the parton densities
(eq. \ref{eq:ftwodis}) holds only in leading order perturbation theory.
The longitudinal structure
function $F_L$ vanishes in zero'th order $\alpha_s$,
and will be discussed in section \ref{sn:fl}.

In the simple quark parton model the proton
consists just of 3 valence quarks. Their distribution functions in
fractional proton momentum $x$, $xq(x)$,  would peak at
$x\approx 1/3$ and tend
towards zero
for $x\rightarrow 0,1$.
In a static model of the proton, they would not depend on \Qsqx.
It follows that \ftwo should not depend on $Q^2$, just on $x$
(Bjorken scaling).

When QCD is ``turned on'' the
quarks may radiate (and absorb)
gluons, which in turn may split into quark --
antiquark pairs or gluon pairs.
More and more of these fluctuations can be resolved
with increasingly shorter wavelength of the photonic probe,
$\lambda = 1/Q$.
With \Qsq increasing, we have a depletion of quarks
at large $x$, and a corresponding accumulation
at lower $x$.
In addition, ``sea quarks'' from $g\rightarrow q\ol{q}$
splittings populate small $x$. In fact, at small \xb it is
the gluon content with distribution function $g(x,Q^2)$
that governs the proton and gives rise
to the DIS cross section via the creation of \qqbar~ pairs.

  \section{Evolution Equations  \label{sn:evoleq}}    

\begin{figure}[tbh]
 \centering
\begin{picture}(0,0) \put(0,0){{\bf a)}} \end{picture}
   \hspace{0.5cm}
 \epsfig{%
   file=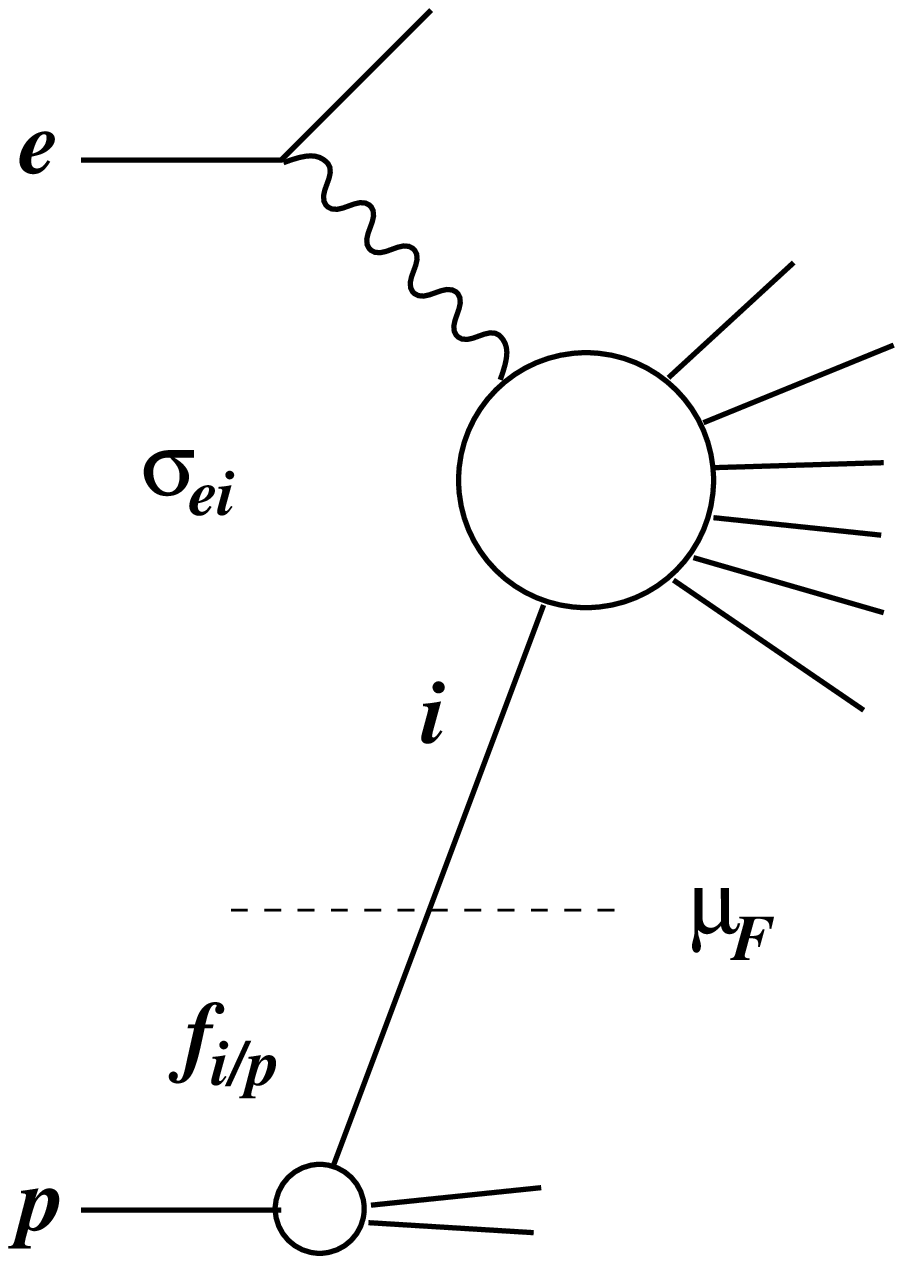,
   width=4cm,bbllx=164pt,bblly=200pt,bburx=471pt,bbury=600,clip=}
   \hspace{2cm}
\begin{picture}(0,0) \put(0,0){{\bf b)}} \end{picture}
   \epsfig{file=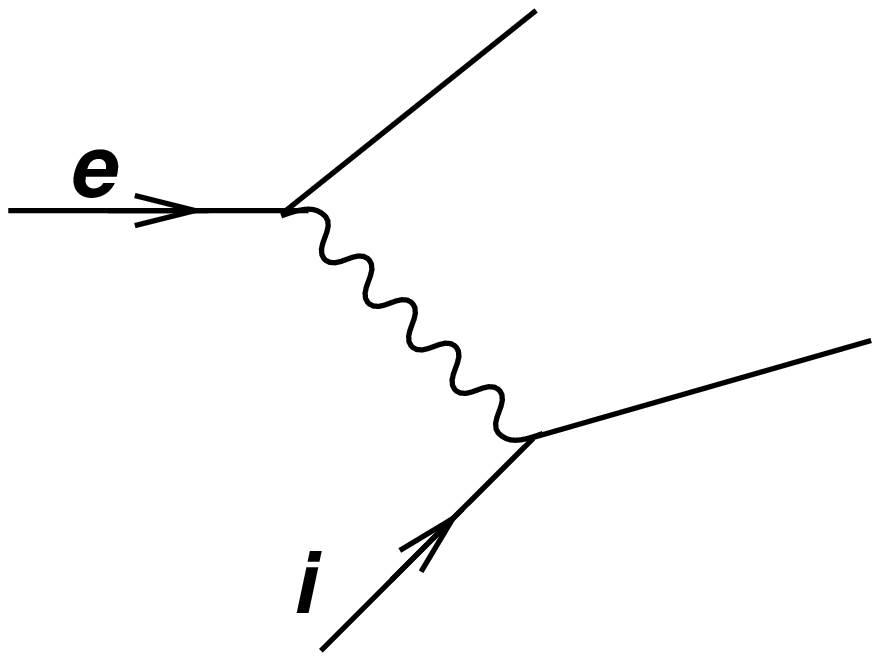,width=4cm}
\scaption{
   {\bf a)} Deep inelastic $ep$ scattering. The $ep$ cross section
   is factorized into electron - parton cross sections $\sigma_{ei}$
   and parton densities $f_{i/p}$ with the factorization scale $\mu_F$:
   $\sigma_{ep} = \sum_i [f_{i/p}(\mu_F^2) \otimes \sigma_{ei}(\mu_F^2)]$.
    {\bf b)} The lowest order diagram (Born graph) contributing
    to $\sigma_{ei}$ in  {\bf a)}.
   }
   \label{factorization}
\end{figure}

It has not yet been possible to calculate the structure of the
hadrons
from first principles, involving the building blocks
of hadronic matter, the quarks and gluons, and their mutual
interactions as given by QCD.
Therefore also the lepton-nucleon scattering cross section
cannot be calculated from first principles.
Due to the factorization theorem of QCD one can however
split the problem.
The cross section can be calculated by folding initial
parton distribution functions $f_{i/p}$, giving the density
of partons $i$ in the proton $p$,
with a perturbatively
calculable lepton-parton scattering cross section. Symbolically
\begin{equation}
    \sigma_{ep} =  \sum_i \left[ f_{i/p} \otimes \sigma_{ei} \right]
\end{equation}
for $ep$ scattering (see fig.\ref{factorization}a).
The initial parton distributions cannot be calculated.
They have to be determined experimentally.
They are however universal in the sense that once they have
been measured in one reaction,
they can be used for calculations of other processes.
Rather than employing the
rigorous operator product expansion (OPE) technique
for the evolution of the structure functions (see e.g. \cite{books:roberts}),
we shall use in this section the more intuitive picture of Feynman
diagram summation.

The expansion parameter for the perturbation series is the
strong coupling $\alpha_s$.
The coupling is scale dependent according to
the ``renormalization group equation''
(see \cite{rev:pdg} for a concise
summary).
The two-loop expression (next-to-leading order = NLO) for the
``running'' coupling \as as a function of the
renormalization scale \mur~ is
\begin{equation}
 \alpha_s(\mur^2) =
 \frac{4\pi}{\beta_0 \ln(\mur^2/\Lambda^2)}
 \left[
    1 - \frac{2\beta_1}{\beta_0^2} \frac{\ln [ \ln(\mur^2/\Lambda^2)]}
                {\ln(\mur^2/\Lambda^2)}
 \right]
\label{eq:asrun}
\end{equation}
where
\begin{equation}
\beta_0 = 11-\frac{2}{3} n_f \hspace{2cm} \beta_1=51-\frac{19}{3}n_f.
\end{equation}
The renormalization scale \mur~ is set by the length scale ($1/\mur$)
over which the interaction takes place, given for example
by the virtuality of the probing photon \Qsq in DIS, or
by the \pt of a parton.
By eq.~\ref{eq:asrun} the QCD scale parameter $\Lambda$ is
introduced. In contrast to $\alpha_s$,
its definition depends on the number of active flavours $n_f$ and
on the renormalization scheme
(for example the ``minimal subtraction scheme'' $\ol{MS}$).
The strong coupling decreases with increasing scale -- at short
distances partons become asymptotically free.
\as grows beyond all bounds for small scales
($\mur\rightarrow\Lambda$)
or large distances, when perturbation theory breaks down and
confinement sets in. The size of a hadron $\approx$ 1~fm provides an
estimate when that happens. Therefore
$\Lambda = \order{1/{\rm fm}} \approx 0.2$ GeV.
The current world average for \as at the scale set by the $Z$ mass is
$\alpha_s(m_Z^2) = 0.118\pm0.003$ or
$\lmsfive=0.209^{+0.039}_{-0.033} \GeV$
for 5 flavours \cite{rev:pdg}.

\begin{figure}[tbp]
 \centering
\begin{tabular}{cccc}


 &
 \epsfig{%
   file=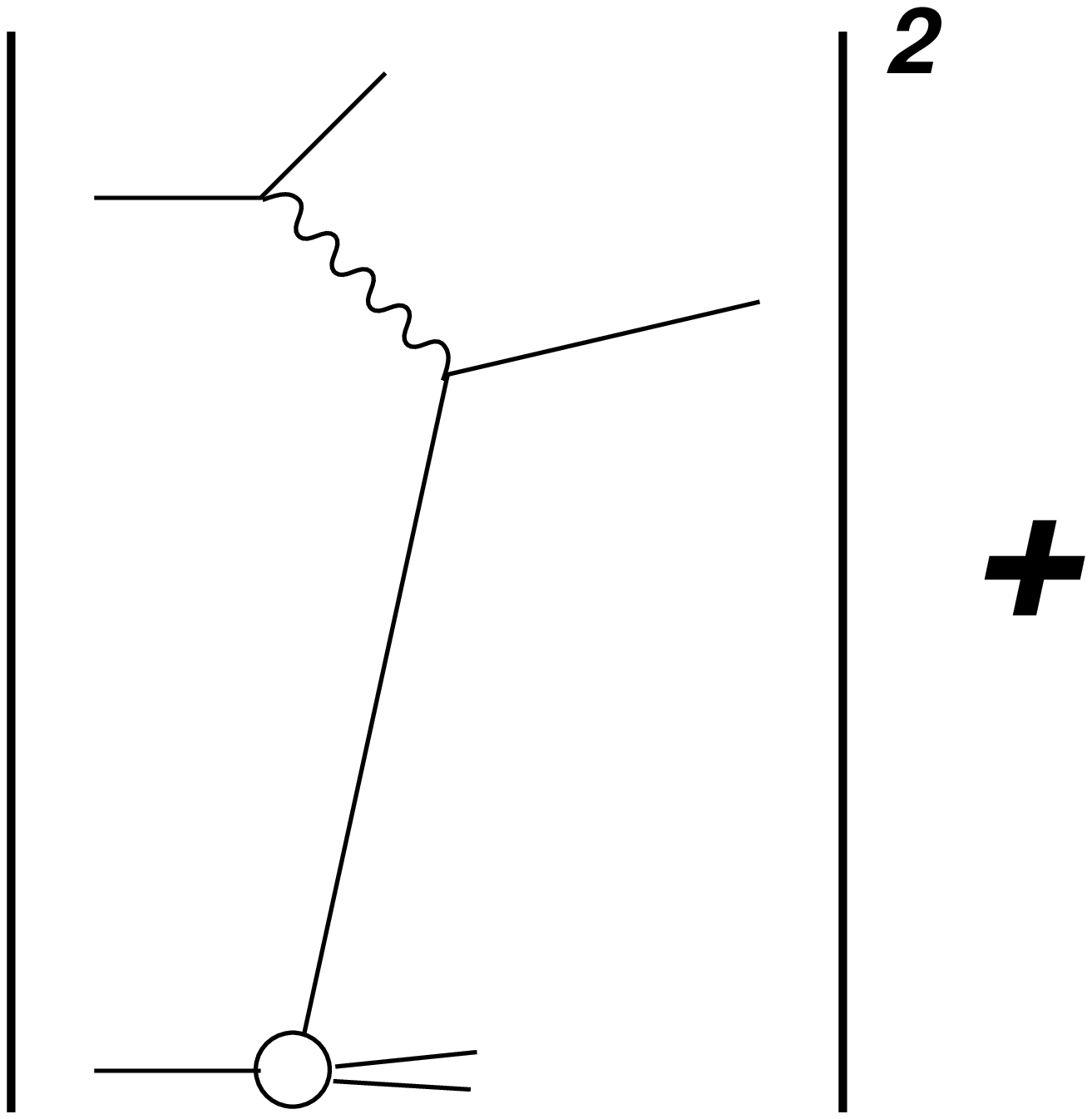,
   width=3cm,bbllx=70pt,bblly=200pt,bburx=570pt,bbury=600,clip=}
 &
 \epsfig{%
   file=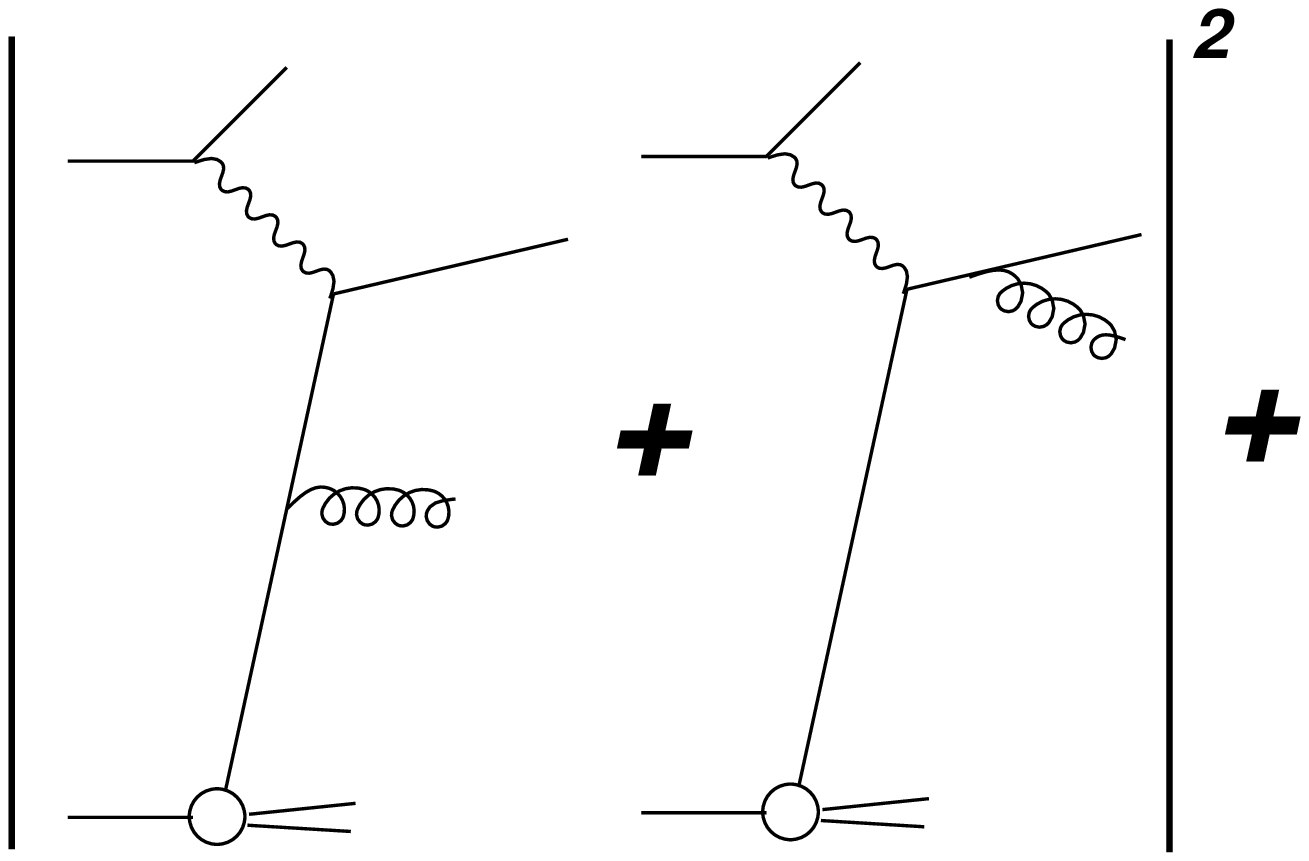,
   width=5cm,bbllx=30pt,bblly=265pt,bburx=516pt,bbury=531,clip=}
 &
 \epsfig{%
   file=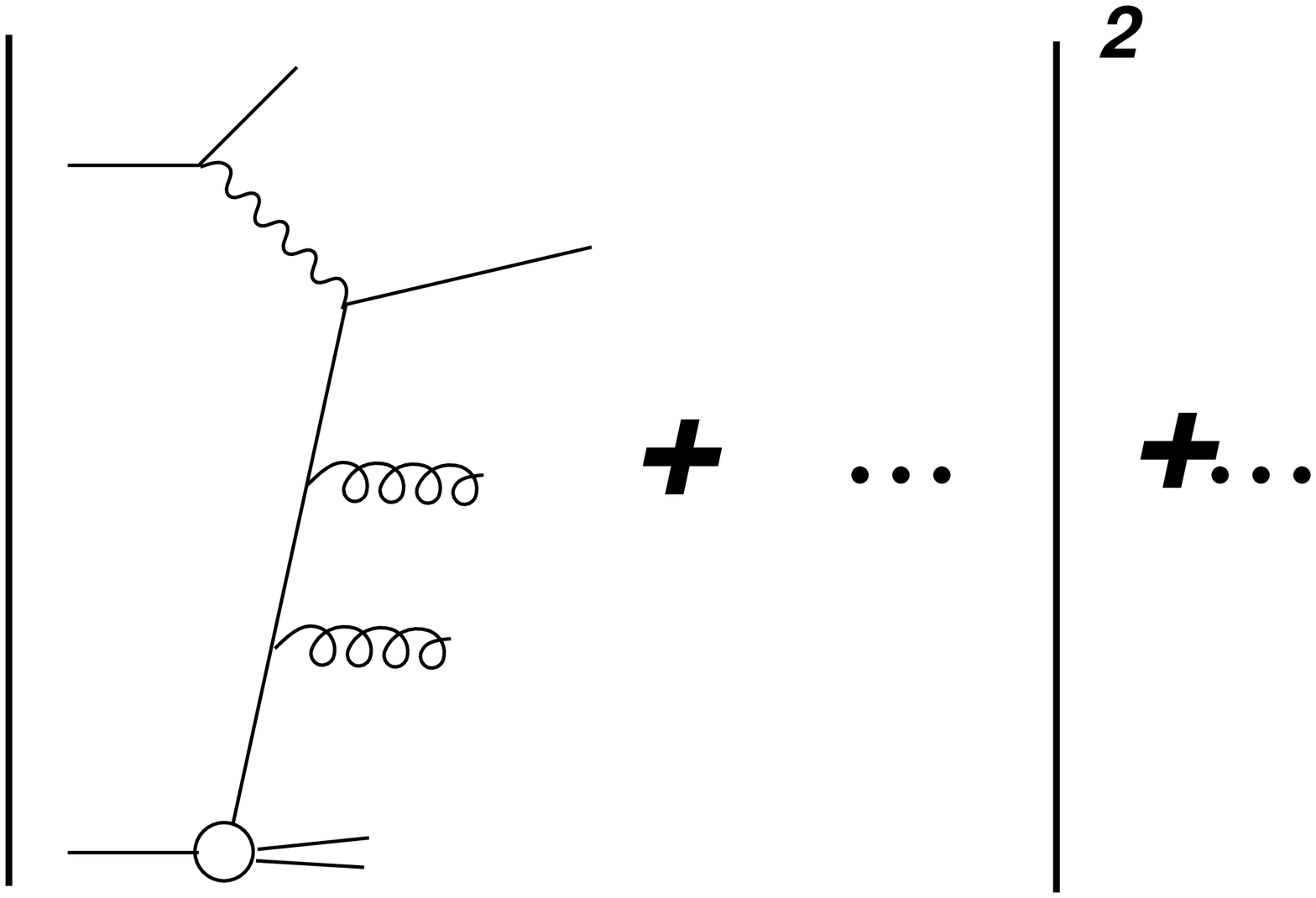,
   width=4cm,bbllx=8pt,bblly=200pt,bburx=593pt,bbury=600,clip=} \\

 &
 \epsfig{%
   file=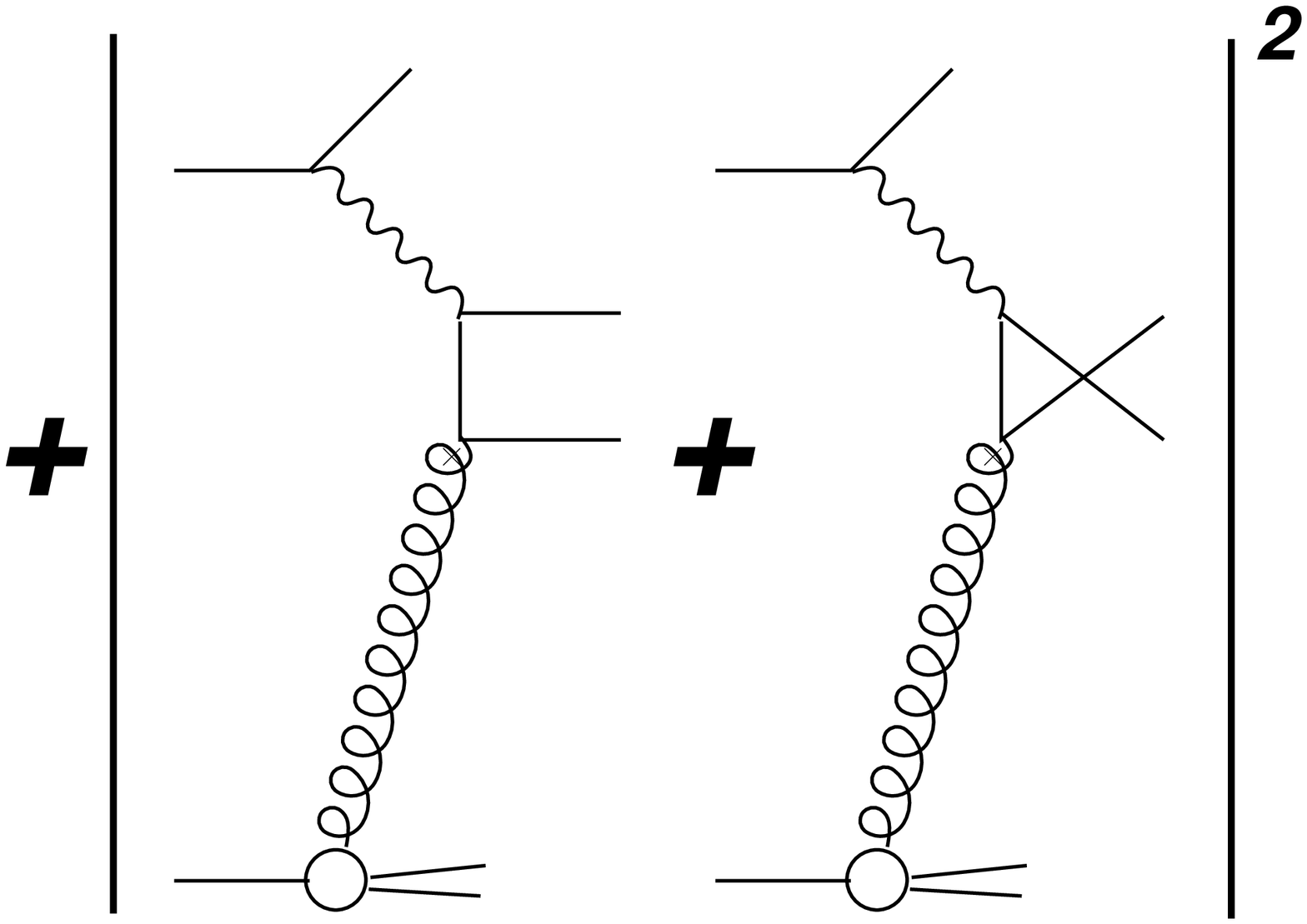,
   width=4cm,bbllx=30pt,bblly=200pt,bburx=593pt,bbury=600,clip=}
 &
 \epsfig{%
   file=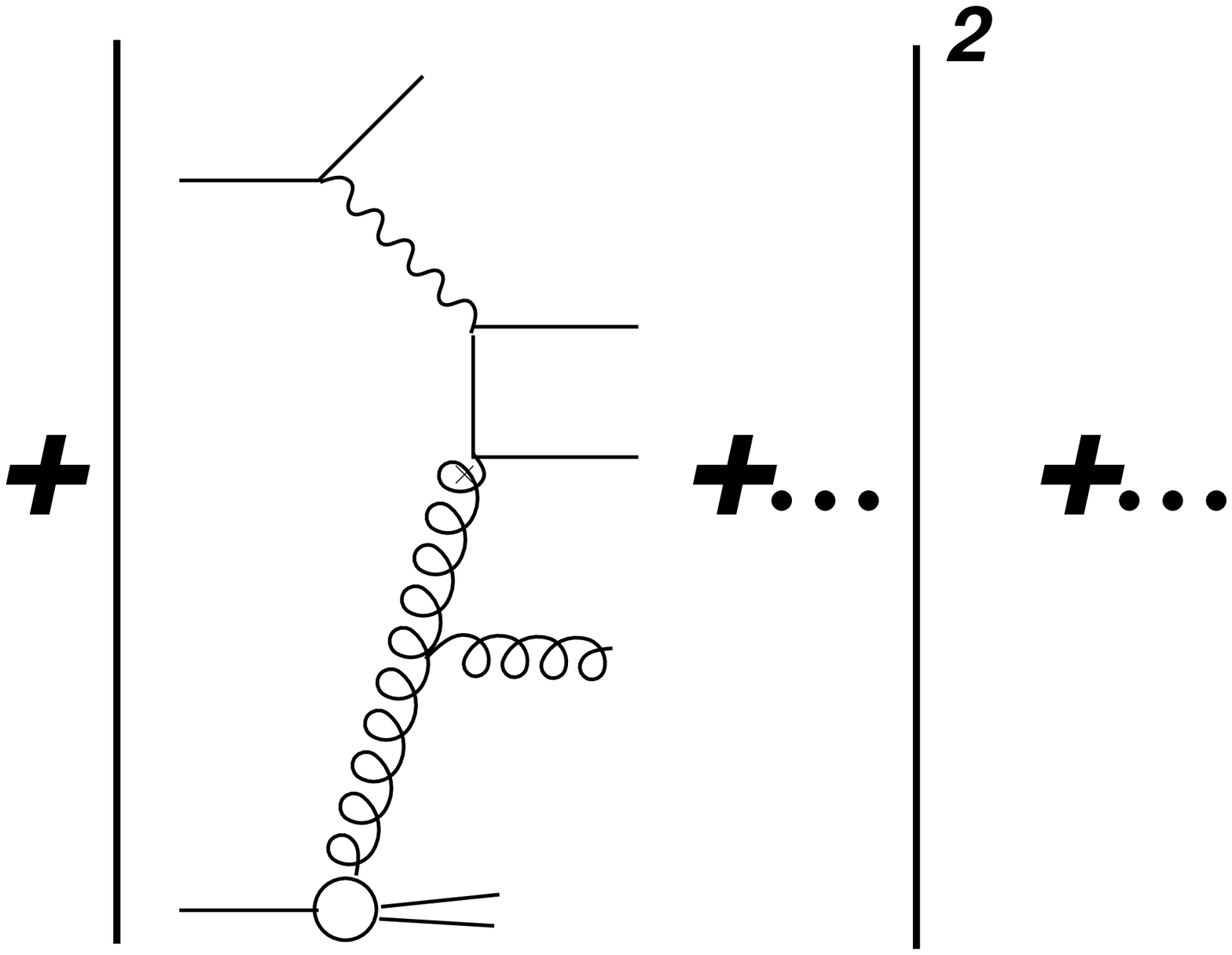,
   width=4cm,bbllx=30pt,bblly=200pt,bburx=581pt,bbury=600,clip=}
 &

\end{tabular}
 \scaption{Diagrams contributing to the DIS cross section $\sigma_{ep}$.}
 \label{sumdiag}
\end{figure}

The calculation of the cross section $\sigma_{ei}$ is a formidable
task.
It turns out that there is no fast convergence of the
perturbation series (fig.~\ref{sumdiag}).
Many diagrams contribute and have to be summed up.
One encounters two types of divergencies.
Divergencies due to the radiation of
soft quanta with small momenta $k\rightarrow 0$ are exactly cancelled
by virtual corrections to graphs where
that radiation is absent (``no emission'').
Divergencies due to
collinear radiation
(so called collinear or mass singularities for $\kt \rightarrow 0$)
can be absorbed (factorized off) into the
``bare'' parton distribution functions.
Thereby they are redefined and depend now on the (mass) factorization
scale $\mu_F$, and so does the electron-parton scattering cross section
with the singularities removed:
\begin{equation}
   \sigma_{ep} = \sum_i [f_{i/p}(\mu_F^2) \otimes \sigma_{ei}(\mu_F^2)]
\end{equation}
(see fig.~\ref{factorization}a).
The choice of the factorization scale is arbitrary.
The physical cross section $\sigma_{ep}$ is of course independent of $\mu_F$.
Often one chooses $\mu_F^2=Q^2$, because then $\sigma_{ei}$ reduces
to the Born graph
(here $ei\rightarrow ei$ by $\gamma$ exchange, fig.~\ref{factorization}b).
In the parton picture $f_{i/p}(Q^2)$ is then
interpreted as the parton density in the proton as seen
by a photon with virtuality (resolving power) $Q^2$.

Since $\sigma_{ei}(\mu_F^2)$ can in principle
be calculated perturbatively for any scale, one can also
calculate the change of the redefined parton distribution
function with a change of scale. These are the evolution
equations.
Once a parton distribution function (or structure function)
is known at one scale, it can be calculated for any other scale.
For the derivation of the evolution equations, one has to perform the
perturbative calculation of $\sigma_{ei}$, taking into account all
contributing graphs (fig. \ref{sumdiag}).
In order to carry out the calculation in practice,
one applies certain approximations, thus restricting the phase space
for radiation.
Such approximations are then valid in regions of $x$ and $Q^2$ where
the selected contributions are the dominant ones. In the following,
evolution equations will be discussed which differ in their
approximations, and therefore in their $x,Q^2$ regions of validity.
Always, in order to allow perturbation theory to be valid,
$\alpha_s(Q^2)\ll 1$ is required.

\begin{figure}[htb]
   \centering
   \epsfig{file=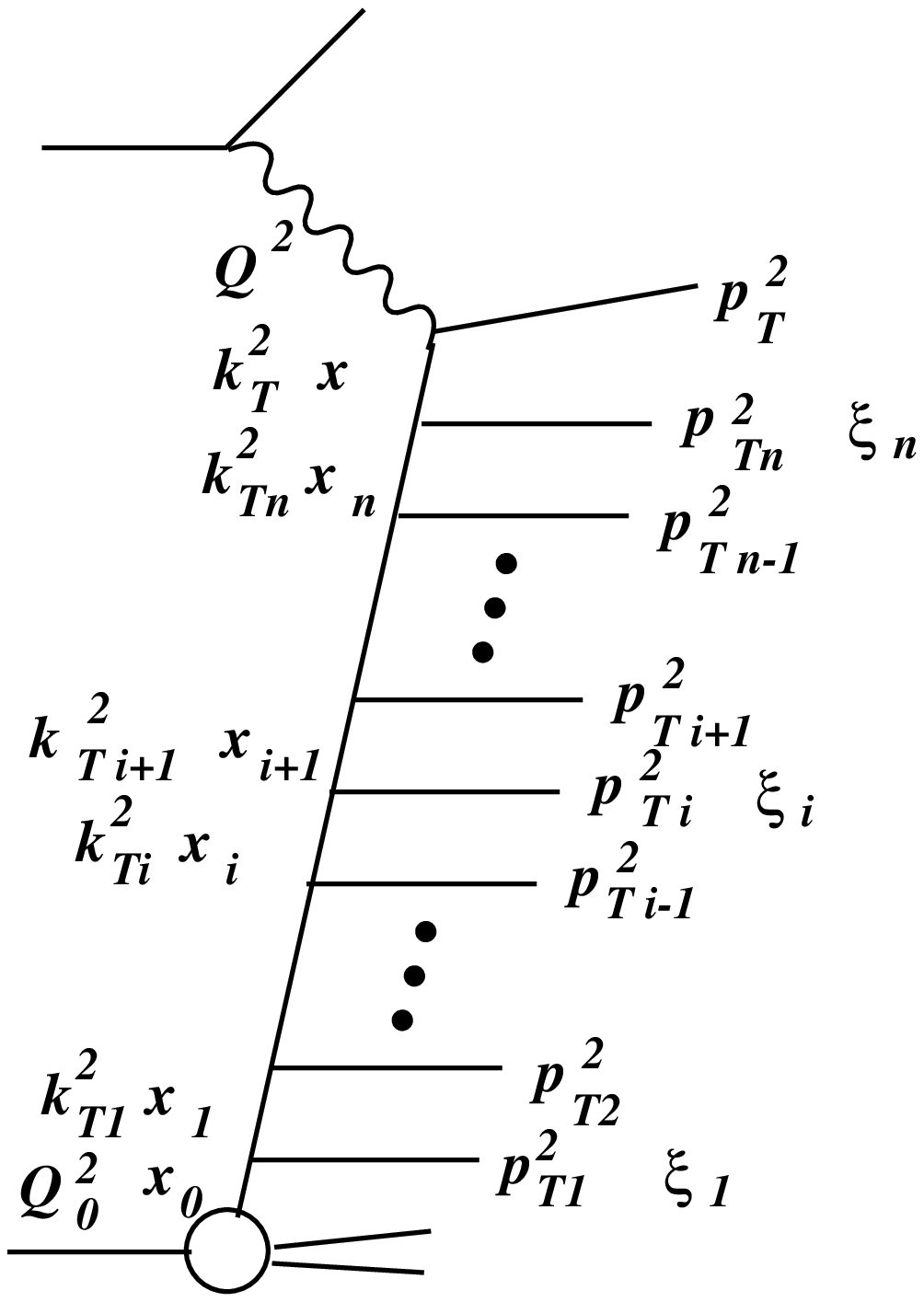,width=6cm,
           bbllx=159pt,bblly=188pt,bburx=460pt,bbury=609,clip=}
   \scaption{The notations for a ladder diagram with $n$ rungs.}
   \label{ladderlabel}
\end{figure}

In a ``physical'' gauge, in which only the physical
transverse gluon polarization states need to be taken into account,
the individual contributions of the perturbation series
can be represented by
so-called ladder diagrams (see fig.~\ref{ladderlabel}).
We work in a frame that moves parallel to the proton, and where the
proton is fast.
The transverse momenta of the emitted quanta are denoted with $p_{Ti}$.
Similarly, the transverse momenta carried by the quanta that constitute
the side rails of the ladder are  $k_{Ti}$.
The longitudinal components are given in fractions of the proton energy
$E_i/E_p$ and are labelled with $\xi_i$ for the emitted quanta and
with $x_i$ for the internal quanta. Energy-momentum conservation
requires $x_i=x_{i+1} + \xi_i$, and therefore $x_i > x_{i+1}$.

  \section{The DGLAP Equations \label{sn:dglap}}      

In the approximation leading to the
DGLAP
(Dokshitzer-Gribov-Lipatov-Altarelli-Parisi) equations \cite{th:dglap}
all ladder diagrams are summed up, in which the
transverse momenta along the side rails of the ladder are
``strongly ordered'',
$Q_0^2 \ll ... k_{Ti}^2 \ll k_{Ti+1}^2 \ll ...\ll Q^2$.
This condition implies
strong ordering also for
the emitted quanta,
$p_{Ti}\ll p_{Ti+1}$.

Where can we expect such an approximation to be valid? A good pedagogical
discussion
can be found in \cite{lowx:miraflores}.
We shall sketch the main arguments.
The evaluation of a ladder diagram with $n$ rungs requires integrations
over the internal momenta exchanged between rungs of the form
\begin{equation}
 \alpha_s \cdot \int
 \frac {\dd k_{Ti}^2}{k_{Ti}^2} .... \int \frac{\dd x_i}{x_i} ....,
\end{equation}
where the dots represent functions which depend on the actual nature of
the emitted quanta and their dynamics.
With strong $k_T$ ordering, the nested integration over all
$n$ rungs in the ladder can be carried out. The result is an expression
$\propto (\alpha_s \ln \frac{Q^2}{Q_0^2})^n$.
We see that the $k_T$ integration yields large logarithms when
the \kt 's are strongly ordered.
They compensate
the smallness of $\alpha_s$.
Clearly, since \as decreases only logarithmically with \Qsq and
is compensated by a logarithmically growing term in \Qsq,
in a perturbative expansion
all graphs with rungs up to $n=\infty$ need to be summed up.
(Often the expression ``re-summation'' is used, because one re-arranges
the perturbation series such that the largest terms come first).
This is called a leading log approximation (here in $\ln (Q^2/Q_0^2)$), since
each power $n$ in \as is accompanied by the same
(maximal) power of  $\ln (Q^2/Q_0^2)$. Subleading terms
would be $\propto \alpha_s^n (\ln \frac{Q^2}{Q_0^2})^{n-1}$.

We expect this to be a good approximation when $Q^2$ is large,
but $x$ is not too small in order not to produce also large logarithms,
\begin{equation}
  \alpha_s(Q^2) \ln \frac{1}{x} \ll
  \alpha_s(Q^2) \ln \frac{Q^2}{Q_0^2} \lesssim 1.
\end{equation}

In this approximation, the evolution equations for the quark density
$q_i$ for flavour $i$ and the gluon density $g$ are
\begin{eqnarray}
  \frac{\dd q_i(x,Q^2)}{\dd \ln Q^2} & = &
  \frac{\alpha_s}{2\pi}\int_x^1 \frac{\dd z}{z}
  \left[q_i(z,Q^2) P_{qq}\left(\frac{x}{z}\right) +
        g(z,Q^2) P_{qg}\left(\frac{x}{z}\right)\right]
   \nonumber \\
  \frac{\dd g(x,Q^2)}{\dd \ln Q^2} & = &
  \frac{\alpha_s}{2\pi}\int_x^1 \frac{\dd z}{z}
  \left[\sum_i q_i(z,Q^2) P_{gq}\left(\frac{x}{z}\right)  +
         g(z,Q^2) P_{gg}\left(\frac{x}{z}\right)\right]
   \label{eq:dglap}
\end{eqnarray}
These are the famous
DGLAP
equations \cite{th:dglap}, describing the scaling violations
of the structure functions.
They involve the calculable Altarelli-Parisi splitting functions
$P_{ij}(\zeta)$.
$\frac{\alpha_s}{2\pi} P_{ij}(\zeta)$ gives
the probability per unit of $\ln \frac{Q^2}{Q_0^2}$ for parton branchings
$q\rightarrow qg$, $g\rightarrow gg$ and  $g\rightarrow q\ol{q}$,
where the daughter parton $i$
carries a fraction $1-\zeta$ of the mother's ($j$) momentum.
The splitting functions in LO are given by
\begin{equation}
\begin{array}{l}
P_{qq}(\zeta) = \frac{4}{3} \frac{1+\zeta^2}{1-\zeta}  = P_{gq}(1-\zeta) \\
P_{gq}(\zeta) = \frac{4}{3} \frac{1+(1-\zeta)^2}{\zeta} =
                 P_{qq}(1-\zeta) \\
P_{qg}(\zeta) = \frac{1}{2} (\zeta^2 + (1-\zeta)^2) = P_{qg}(1-\zeta) \\
P_{gg}(\zeta) =
       6 (\frac{\zeta}{1-\zeta} + \frac{1-\zeta}{\zeta} + \zeta(1-\zeta))
              = P_{gg}(1-\zeta).
\label{eq:splitting}
\end{array}
\end{equation}
The singularities for soft emissions $\zeta \rightarrow 1$
are cancelled by virtual corrections to the ``no-emission'' graphs.
This is physical, because arbitrary soft emissions cannot be distinguished
from the ``no-emission''
case\footnote{Technically the singularities
can be regularized by the following prescription.
Replace $1/(1-\zeta)$ with $1/(1-\zeta)_+$, where the ``+ - prescription''
defines how integrals are to be carried out:
\begin{equation}
\int_0^1 \dd \zeta \frac{f(\zeta)}{(1-\zeta)}_+
     := \int_0^1 \dd \zeta \frac{f(\zeta)-f(1)}{1-\zeta}
\end{equation}
for any function $f(z)$. A term $2 \delta(1-\zeta)$ has to be added
to $P_{qq}$, and a term $(11/2 - n_f/3)\delta(1-\zeta)$ to $P_{gg}$.
}.
The coupled integro-differential equations for the quark and gluon
densities (eq. \ref{eq:dglap})
can be solved,
allowing to calculate them
for any value of $Q^2$ and $x>x_0$,
once they are known at a particular value $Q_0^2$ for $x>x_0$.

A special case for which the DGLAP equations can be solved analytically
(see for example \cite{books:roberts})
occurs when in addition to the above conditions also strong ordering in $x$
is required, $x \ll ...x_{i+1} \ll x_i \ll ... x_0$.
The large logarithmic terms arising from the integration are then of the
form $\propto (\alpha_s(Q^2) \ln \frac{Q^2}{Q_0^2} \ln \frac{1}{x} )^n$, which
need to be resummed. This is the double leading log approximation (DLL). It
is expected to hold when the DLL terms dominate over the others,
\begin{equation}
    \left.
    \begin{array}{l}
       \alpha_s(Q^2)\ln \frac{Q^2}{Q_0^2} \\
       \alpha_s(Q^2) \ln \frac{1}{x}
    \end{array}
    \right\}
     \ll \alpha_s(Q^2) \ln \frac{Q^2}{Q_0^2} \ln \frac{1}{x} \lesssim 1.
\end{equation}
This is the case for large $Q^2$ and small $x$.
At small $x$ the parton content of the proton is expected to be dominated by
gluons, because $P_{ij}(\zeta=x/z)$ is largest when gluons are being produced
($i=g$). When quarks are neglected, and $P_{gg} = \frac{6}{\zeta}$ is approximated,
the DGLAP equations can be solved to yield the DLL solution \cite{th:dll}
\begin{equation}
  x g(x,Q^2) \approx x g(x,Q_0^2) \exp\sqrt{\frac{144}{25}
 \ln \left[\frac{\ln (Q^2/\Lambda^2)}{\ln (Q_0^2/\Lambda^2)}\right]
 \ln (1/x) },
\label{eq:dll}
\end{equation}
provided the gluon density is not too singular at small $x$
(needs to be quantified).
At small \xb a fast rise of the gluon density
with decreasing \xb is predicted.
That is, $xg$ increases faster than
$(\ln \frac{1}{x})^\lambda$,
but slower than
$(\frac{1}{x})^{\lambda^\prime}$
for any powers $\lambda,\lambda^\prime>0$.
Apart from these shape restrictions the
actual rate of the growth is not
predicted,
it depends on the
``evolution length'' from
$Q_0^2$ to $Q^2$.

  \section{The BFKL Equation \label{sn:bfkl}}          

When $x$ is small, but $Q^2$ not large enough to reach the DLL
regime, the DGLAP approximations cease to be valid.
For the limit $1/x$ large and \Qsq finite and fixed
the BFKL (Balitsky-Fadin-Kuraev-Lipatov) \cite{th:bfkl} equation
has been derived.
It takes into account diagrams in which the $x_i$ are
strongly ordered,
$x_0 \ll ... x_i \ll x_{i+1} \ll ...\ll x$.
No ordering on $k_{Ti}$ is imposed.
Large logarithms
$\propto (\alpha_s \ln \frac{1}{x})^n$ are thus generated that
need to be resummed, leading to the
leading log approximation in $\ln \frac{1}{x}$.
The region of
validity is
\begin{equation}
  \alpha_s(Q^2) \ln \frac{Q^2}{Q_0^2} \ll
  \alpha_s(Q^2) \ln \frac{1}{x}            \lesssim 1.
\end{equation}

The BFKL equation is expressed in terms of the ``unintegrated''
gluon density $f(x,k_T^2)$, which is related to the usual
gluon density by
\begin{equation}
  xg(x,Q^2) = \int_0^{Q^2} \frac{\dd k_T^2}{k_T^2} f(x,k_T^2).
\label{eq:unintegrated}
\end{equation}
The BFKL equation is an evolution equation in $x$.
It is formulated for gluons which dominate at small $x$,
\begin{equation}
   \frac{\partial f(x,k_T^2)}{\partial \ln (1/x)}
   = \frac{3\alpha_s}{\pi} k_T^2
   \int_0^\infty \frac{ \dif k_T^{\prime 2} }{k_T^{\prime 2}}
   \left[
      \frac{ f(x,k_T^{\prime 2}) - f(x,k_T^{2}) }{|k_T^{\prime 2}-k_T^2| }
    + \frac{ f(x,k_T^2) }{\sqrt{4 k_T^{\prime 4} + k_T^4}}
   \right]
   = K \otimes f,
  \label{eq:bfkl}
\end{equation}
where $K$ is the BFKL kernel.
$f(x,k_T^2)$ can be calculated
for any (small) $x$, once it is known at some $x_0$ for all $k_T^2$.
We note in passing that if one requires strong $k_T$ ordering
($k_T^2\gg k_T^{\prime 2}$)
when solving eq. \ref{eq:bfkl},
the DLL result is retained \cite{lowx:miraflores}.

For fixed \as the equation can be solved analytically.
The result is (in the saddle point approximation)
\begin{eqnarray}
   \frac{f(x,k_T^2)}{\sqrt{k_T^2}}
              & \propto & \frac{(x/x_0)^{-\lambda}}
                           { \sqrt{2\pi\lambda^{\prime\prime} \ln(x_0/x)} }
                       \exp\left[
                           \frac{ -\ln^2(k_T^2/\ol{k}_T^2)}
                                {2\lambda^{\prime\prime} \ln(x_0/x) }
                           \right]   \nonumber  \\
              &  \propto & \left(\frac{x}{x_0}\right)^{-\lambda}
\end{eqnarray}
with $\lambda = \frac{n_c \alpha_s}{\pi} \cdot 4 \ln 2 \approx 0.5$
(for $n_c=3$ colours and $\alpha_s=0.19$), and
$\lambda^{\prime\prime}=\frac{n_c \alpha_s}{\pi} \cdot 28 \zeta(3)$
(the Riemann $\zeta$ function gives $\zeta(3)=1.202$).
$\ol{k}_T^2$ specifies the starting point for the evolution.
Therefore the gluon density is expected to rise like
a power of $1/x$ for decreasing $x$, $xg(x,Q^2) \propto x^{-\lambda}$,
faster than the DLL result eq. \ref{eq:dll}.
However, the running of \as and higher order corrections
decrease the value of $\lambda\approx 0.5$
\cite{th:ciafaloni}.

\begin{figure}[tbh]
   \centering
\begin{picture}(0,0) \put(0,0){{\bf a)}} \end{picture}
   \epsfig{file=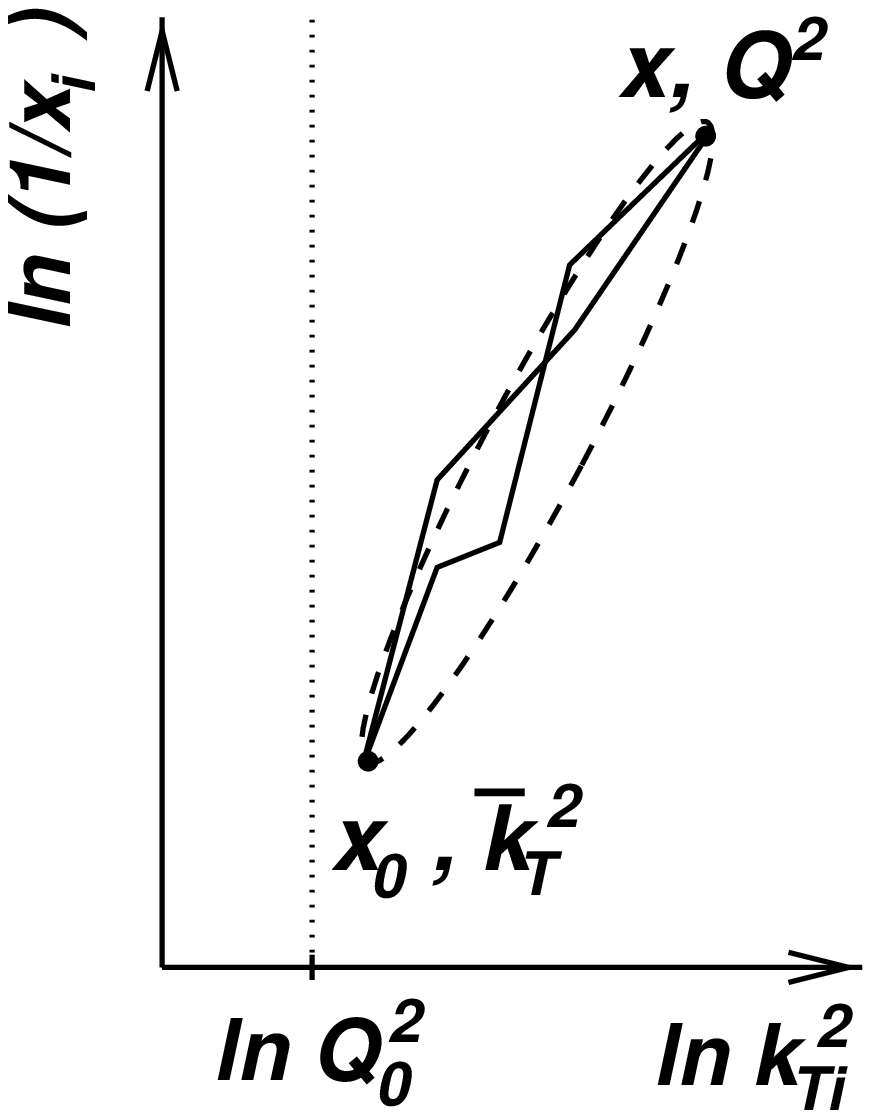,%
    width=5cm,bbllx=181pt,bblly=236pt,bburx=450pt,bbury=574pt,clip=}
   \hspace{2cm}
\begin{picture}(0,0) \put(0,0){{\bf b)}} \end{picture}
   \epsfig{file=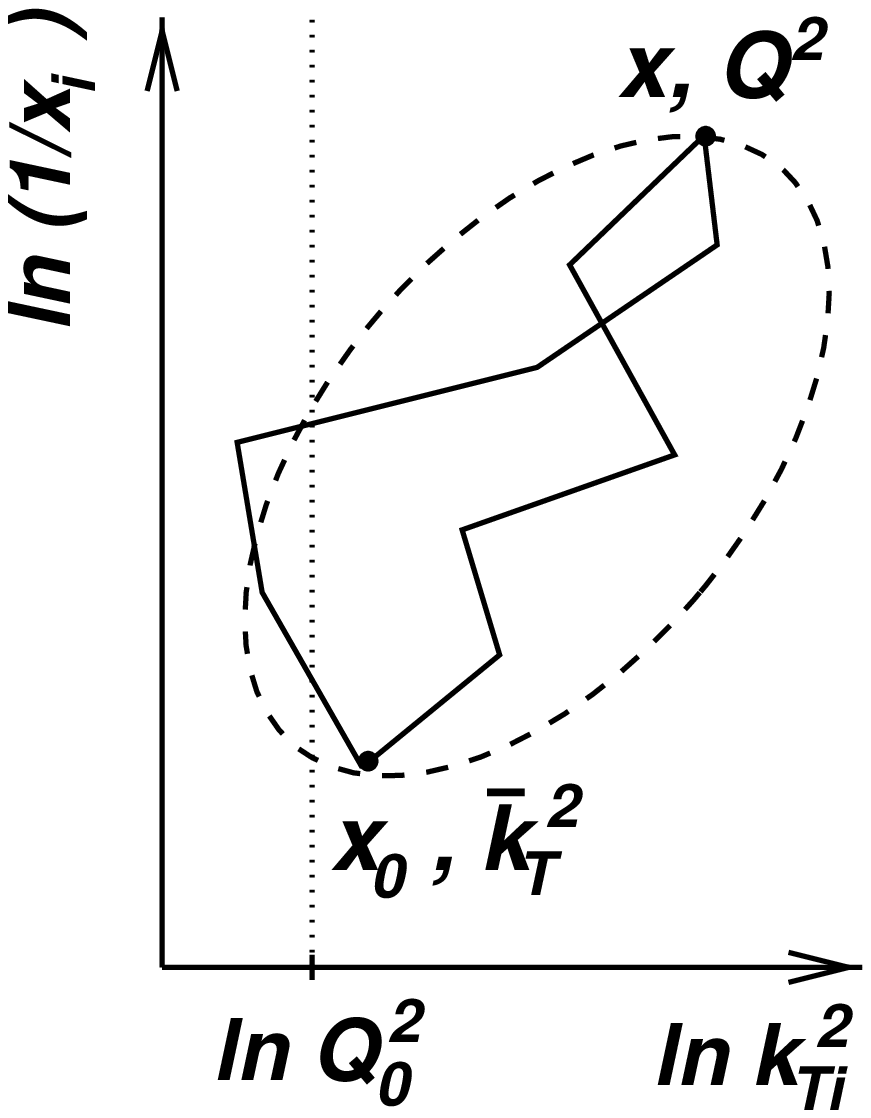,%
    width=5cm,bbllx=181pt,bblly=236pt,bburx=450pt,bbury=574pt,clip=}
   \scaption{Possible evolution paths for the parton cascade
             {\bf a)} for DGLAP evolution with strong \kt ordering and
             {\bf b)} for BFKL evolution without \kt ordering.
             Shown are the $x_i$ and $k_{Ti}$ in the ladder for
             fixed start ($x_0,\ol{k}_T^2$) and end ($x,Q^2$) points.
             The infrared region $k_{Ti}^2<Q_0^2$ is
             to the left of the dotted line.
             The dashed ``cigars'' represent the $\pm 1 \sigma$
             contours for the $(x_i,k_{Ti})$ taken from a large
             ensemble of evolution paths.
             One of the BFKL paths shown ``diffuses'' into the
             infrared region $k_{Ti}^2<Q_0^2$.}
   \label{diffusion}
\end{figure}

Another characteristic prediction of the BFKL equation is
``$k_T$ diffusion'', in contrast to $k_T$ ordering for DGLAP
(see fig. \ref{diffusion}).
The $k_T$ distribution function $f(x,k_T^2)/\sqrt{k_T^2}$ is Gaussian
in $\ln k_T^2$ with a width that increases with the
BFKL ``evolution length''
$\sqrt{\ln (x_0/x)}$.
An individual evolution path will follow a kind of random walk
in $k_T^2$. An ensemble of evolution paths exhibits a diffusion
pattern according to the Gaussian in $\ln k_T^2$.

$k_T$ diffusion poses a difficulty for the application of the
BFKL equation, because $k_T$ may diffuse into the infrared region
($k_T<Q_0$) where perturbation theory cannot be applied.
One therefore usually
introduces a lower cut-off $k_{T0}$ for the $k_T$ integration, and
studies the dependence of the result on that cut-off.
Due to \kt diffusion BFKL
looses much of its predictive power when applied
to the structure function $F_2$. The inclusive structure function
$F_2$ is probably not a good place to identify BFKL effects
unambiguously.
It is however possible
to study special final state configurations
where diffusion into the infrared region can be minimized by fixing
both start and end point of the evolution far enough above the
infrared region. The search for signs of BFKL evolution in the
hadronic final state is presented in chapter \ref{ch:lowx}.

The CCFM equation \cite{th:ccfm} developed in recent years
unifies the BFKL and DGLAP approaches \cite{lowx:martinrome}
and takes into account
coherence effects by angular ordering.
The CCFM approach leads to a reduction of the exponent $\lambda$
and a reduction of the \kt diffusion \cite{lowx:ringsalam,th:bfklldc}.
The Linked Dipole Chain model \cite{th:ldc} provides an implementation
of the CCFM equation which is suited for final state predictions.

  \section{The Interest in Small $x$
                                       \label{sn:interest}}
\subsubsection{Orthogonal evolution equations}

\begin{figure}[tbh]
   \centering
   \epsfig{file=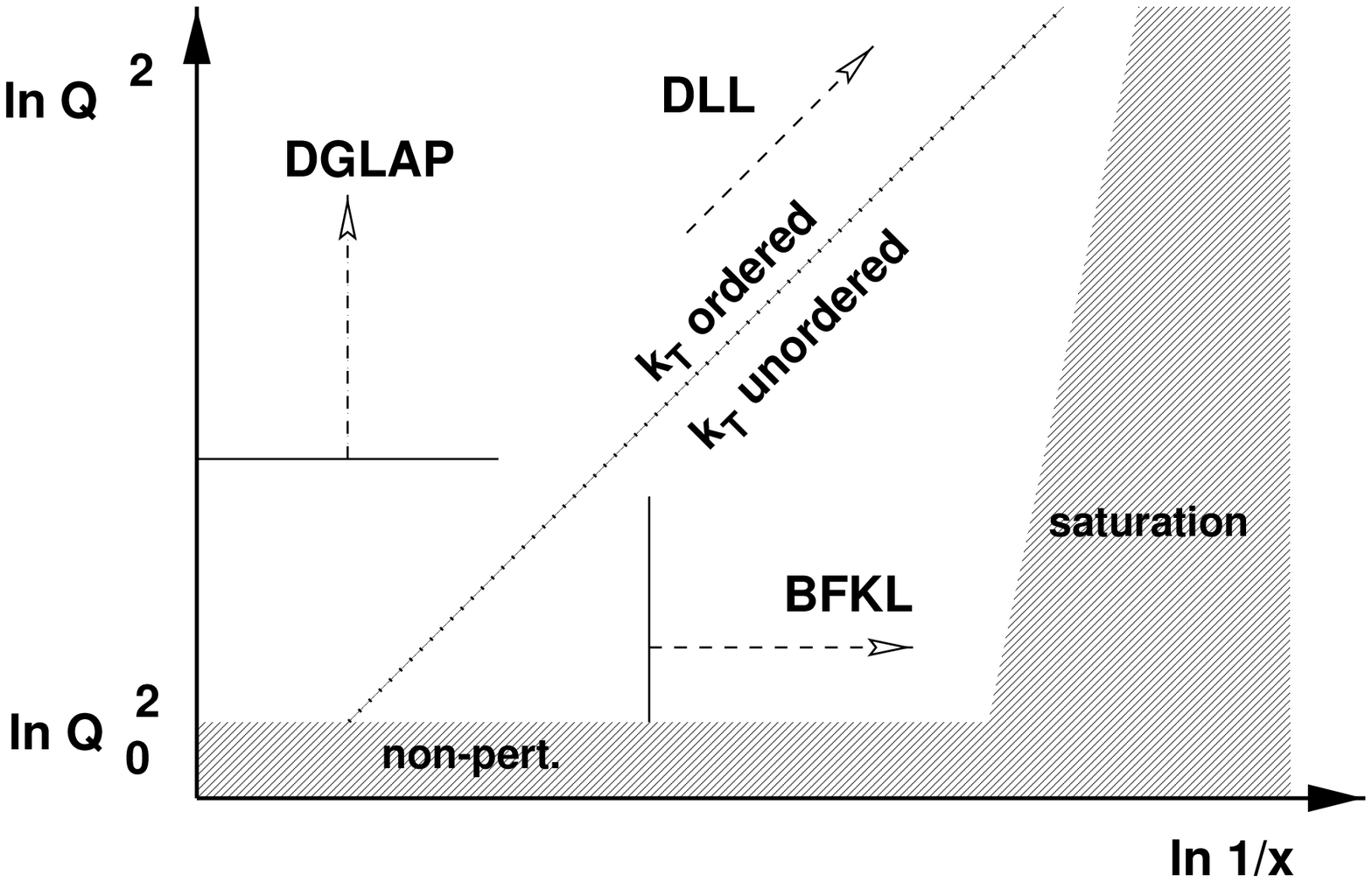,%
    width=10cm}
   \scaption{Schematic map of the kinematic plane.
            Indicated are the actions and regions of
            validity of the DGLAP, DLL and BFKL evolutions.
            The non-perturbative region ($Q^2<Q_0^2$)
            and the saturation limit are given by the shaded areas.}
   \label{bfklmap}
\end{figure}

In fig.~\ref{bfklmap} the regions of validity of
the different evolution equations are sketched. They
are not predicted precisely by the theory, but have to be
explored experimentally.

With DGLAP evolution a parton density $p(x,Q_0^2)$ known for
$x \in [x_0,1]$
can be evolved to any value of $Q^2$ for $x \in [x_0,1]$.
The behaviour
for $x<x_0$ cannot be predicted with DGLAP.
Similarly,
with BFKL evolution a parton density $p(x_0,k_T^2)$ known
for $0<k_T^2<\infty$ can be evolved to any value of $x$.
The new feature, orthogonal to the DGLAP evolution,
is that the low $x$ behaviour is predicted by the theory.
In principle DGLAP and BFKL evolution together could be used for
a M\"unchhausen trick (bootstrapping), to predict the structure of
the proton for all $x$ as long as $Q^2>Q_0^2$, above a cut-off to
avoid the non-perturbative region. The problem with $k_T$ diffusion
into the non-perturbative regime however poses a severe obstacle
for that goal to be reached.

Another motivation is connected with the cross section for
hadron-hadron scattering at high energies. Before we can make
the point, we need to make an excursion to Regge theory
(see for example \cite{rev:landshoff} as a review related to HERA physics;
\cite{books:collinsmartin} and \cite{books:forshawross}
as a modern language textbook introduction;
\cite{books:collins} for an in depth discussion of Regge theory).

\subsubsection{Regge theory}

\begin{figure}[htb]
   \centering
\begin{picture}(0,0) \put(0,0){{\bf a)}} \end{picture}
  \hspace{0.4cm}
   \epsfig{file=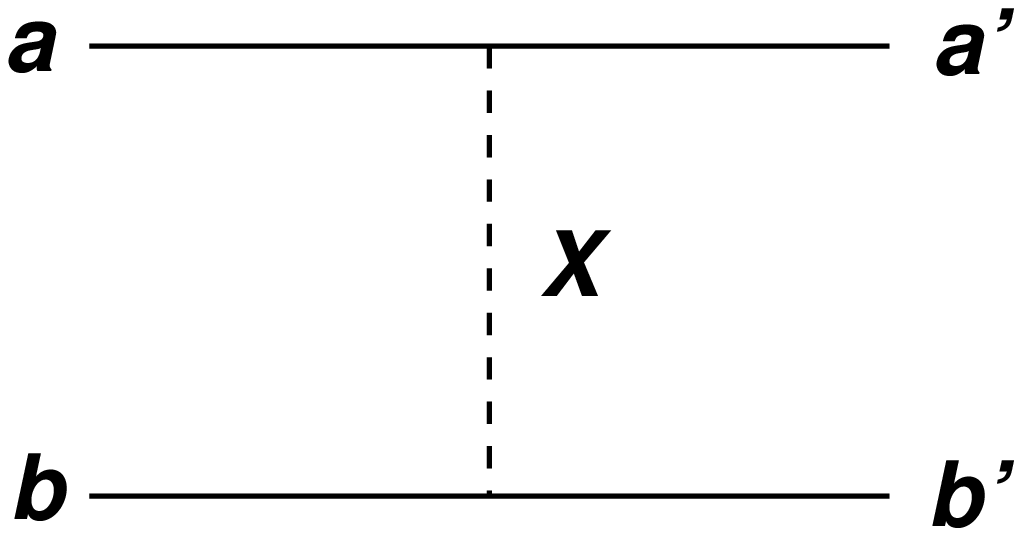,%
    width=4cm,bbllx=152pt,bblly=309pt,bburx=468pt,bbury=484pt,clip=}
   \hspace{2cm}
\begin{picture}(0,0) \put(0,0){{\bf b)}} \end{picture}
   \epsfig{file=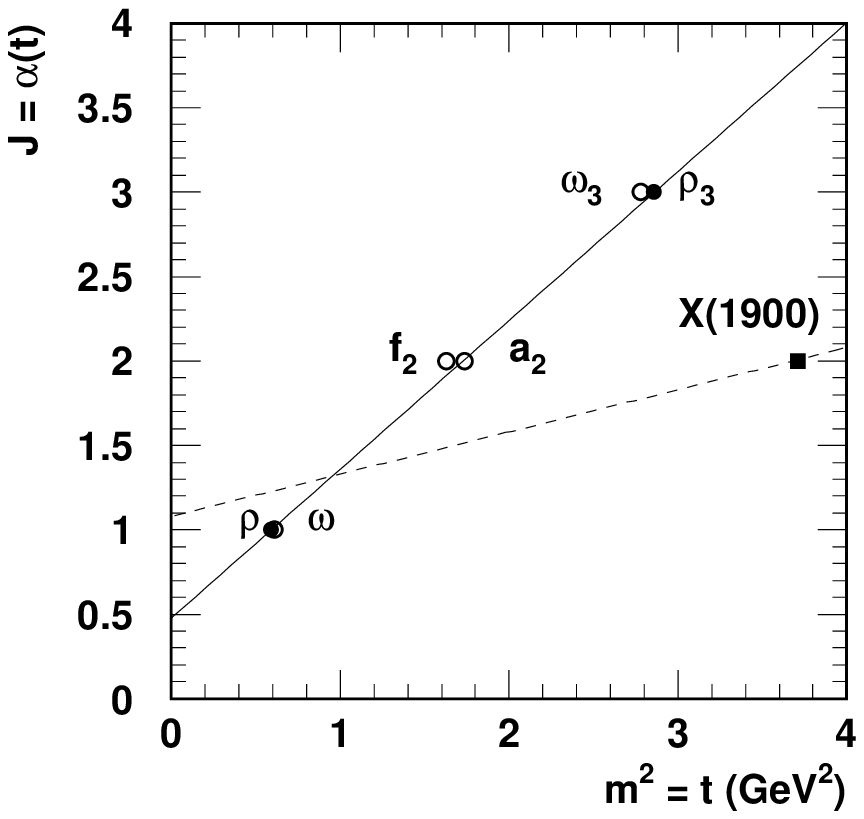,%
    width=6cm,bbllx=30pt,bblly=409pt,bburx=278pt,bbury=653pt,clip=}
   \scaption{
   {\bf a)} The elastic scattering of hadrons $a$ and $b$, mediated
   by the exchange of a quantum $X$. The 4-vectors before and after
   the scattering are $a,b$ and $a^\prime,b^\prime$.
   {\bf b)} Regge trajectories. The trajectories for the
     $\rho$, $\omega$, $f_2$ and the $a_2$ resonances almost coincide;
     they are represented here with a solid line.
     The indicated resonances are the $\rho(770)$, the
     $\omega(782)$, the $f_2(1270)$, the $a_2(1320)$, the
     $\rho_3(1690)$ and the $\omega_3(1670)$.
     The Pomeron
     trajectory (dashed line) is shown together with the
     $I(J^{PC})=0(2^{++})$ state $X(1900)$ observed by
     WA91 \cite{o:wa91x1900}.}
   \label{mytraj}
\end{figure}

Consider the elastic
scattering of hadrons $a$ and $b$, $ab\rightarrow ab$
(fig.~\ref{mytraj}a).
Their 4-momenta are denoted with $a,b$ for the initial and
$a^\prime,b^\prime$ for the final state.
The cross section can be expressed as a function
of the Mandelstam variables $s=(a+b)^2$ and $t = (b-b^\prime)^2$.
It is
the squared sum over the scattering amplitudes due to the quanta $X$
(conventionally mesons) that can
be exchanged,
\begin{equation}
   \frac{\dd \sigma^{{\rm el.}}}{\dd t} =
   \frac{1}{16 \pi s^2}
   \left| \sum_X {\cal M}_X^{{\rm el.}}(s,t)
   \right|^2.
\end{equation}
In Regge theory, where the mesons are connected via a so-called
Regge trajectory (explained below),
the sum yields
\begin{equation}
  \sum_X {\cal M}_X^{{\rm el.}}(s,t) = {\cal M}_{I\!R} (s,t)
  \propto \beta (t) \xi(\alpha(t)) s^{\alpha(t)}.
\end{equation}
$\beta(t)$ is an unknown real function.
The complex phase is given by
\begin{equation}
   \xi(\alpha(t)) = e^{-i\pi\alpha(t)/2}
\end{equation}
for exchanged particles with $C$ parity $+1$, and receives
an extra factor $i$ for negative $C$ parity.
The Regge trajectory $\alpha(t)$ gives the relationship
between the mass $m$ and the spin $J$ of the exchanged mesons,
$J = \alpha(m^2)$ (fig. \ref{mytraj}b).

Empirically, Regge trajectories can be parametrized as straight lines
with
\begin{equation}
  \alpha(t) = \alpha_0 + \alpha^\prime t.
\end{equation}
$\alpha_0$ is called the intercept (with the ordinate),
and $\alpha^\prime$ the slope of
the trajectory.
As an example, a Regge trajectory for mesons is shown in
fig.~\ref{mytraj}b.
Due to the confinement problem, meson trajectories
(hadron masses) could not yet
be calculated from first principles in QCD.

\subsubsection{The total cross section}

Starting from the elastic cross section (in principle one
has to sum over all Regge trajectories whose resonances can
be exchanged in the reaction)
\begin{equation}
   \frac{\dd \sigma^{{\rm el.}}}{\dd t}
   \propto (\beta (t))^2 \cdot s^{2\alpha(t)-2},
\end{equation}
we can use the optical theorem, relating the total cross section
to the forward scattering amplitude
\begin{equation}
    \sigma_{{\rm tot}} = \frac{1}{s} {\rm Im} {\cal M}_{{\rm el}}(s,t=0),
\end{equation}
to predict the behaviour of the total hadron-hadron scattering cross section
\begin{equation}
    \sigma_{{\rm tot}} \propto s^{\alpha_0 - 1}.
\end{equation}

\begin{figure}[htb]
   \centering
   \epsfig{file=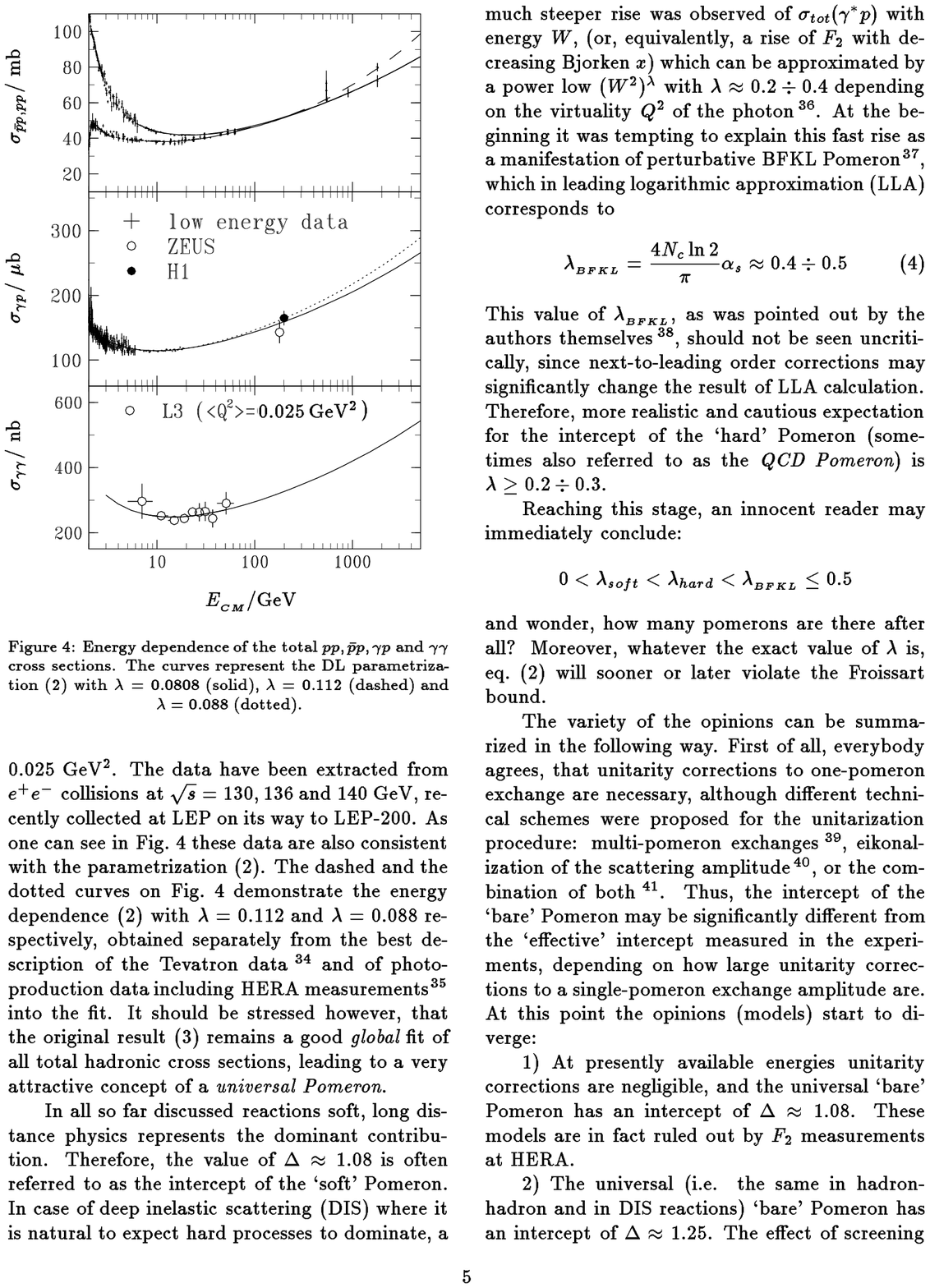,%
    width=7cm,bbllx=76pt,bblly=445pt,bburx=285pt,bbury=749pt,clip=}
   \scaption{The total cross section for $pp$ (or $p\ol{p}$),
   $\gamma p$ and $\gamma \gamma$ interactions as a function of
   the total CM energy $\sqrt{s}=E_{\rm CM}$ \cite{rev:levonian}.
   ZEUS and H1 measure $\sigma_{\gamma p}$ in $ep$ scattering with
   $Q^2 \approx 0$ (photoproduction). The curves represent the
   DL parametrizations with $\alpha_{I\!P}$ =1.0808 (solid),
   =1.112 (dashed) and =1.088 (dotted).
   }
   \label{levonian}
\end{figure}

The total cross sections for $pp$, $p\ol{p}$, $\gamma p$ and $\gamma\gamma$
reactions are plotted as a function of the CM energy $\sqrt{s}$ in
fig.~\ref{levonian}.
Their behaviour is surprisingly similar (and also for other
hadron-hadron scattering cross sections
like $\pi p$, $Kp$ etc. that are not shown in fig.~\ref{levonian}
\cite{rev:pdg}).
They fall
at small CM energy
$\sqrt{s}\lesssim 10$ GeV, and rise towards large energy.
All these cross section can be parametrized with the
universal ansatz \cite{th:dola1}
\begin{equation}
   \sigma_{{\rm tot}}  = A s^{\alpha_{I\!M}(0)-1} +
                         B s^{\alpha_{I\!P}(0)-1}.
\end{equation}
$A$ and $B$ are process dependent constants, whereas
$\alpha_{I\!M} (0) = 0.5322 \pm 0.0059$
and $\alpha_{I\!P}(0) = 1.0790 \pm 0.0011$ \cite{rev:pdg} are
universal, process independent constants.

\subsubsection{The Pomeron}

The fall off at small energies
is readily interpreted as due to meson exchange, whose Regge trajectories
have an intercept $\alpha_{I\!M} (0)<1$ (compare
fig.~\ref{mytraj}b;
here only trajectories that dominate high energy scattering are shown;
other meson trajectories have smaller intercepts
and therefore do not contribute
much to high energy scattering).
Correspondingly, the rise at
high energy is attributed to an exchange described by a Regge trajectory
with intercept $\alpha_{I\!P}(0)>1$.
There exists however no established set of
particles with such a Regge trajectory.
Nevertheless, the Regge ansatz gives a successful parametrization of
the scattering process. Therefore a hypothetical object, the Pomeron
\pom ~is postulated, which has the quantum numbers of the vacuum
(electrically and colour neutral, isospin 0 and $C$ parity $+1$), and whose
exchange is described by the
Pomeron trajectory $\alpha_{I\!P}(t) \approx 1.08 + 0.25 \cdot t$,
see fig~\ref{mytraj}b.
The object is suspected to be of gluonic nature, perhaps a glue ball.
A possible glue ball candidate with $J^{PC}=2^{++}$ in fact would
fall on the Pomeron trajectory, see fig.~\ref{mytraj}b.
Because $\alpha_{I\!P}(0)>1$, physical
states with $J=0,1$ belonging to the Pomeron trajectory cannot exist.
An up-to-date textbook on QCD and the Pomeron is \cite{books:forshawross}.

Deep inelastic scattering at small $x$ can be viewed as
virtual photon - proton scattering at high energy
$\sqrt{s_{\gamma^\ast p}} = W \approx \sqrt{Q^2/x}$.
We had connected the structure function
$F_2$ with the total cross section for $\gamma^\ast p$ scattering,
\begin{equation}
  F_2 = \frac{\Qsq}{4\pi^2 \alpha} \sigtot.
\end{equation}
If the total cross section behaviour
$\sigtotal \propto s^{0.08}$ found for hadron-hadron scattering
and real photon-hadron scattering continues to hold for
virtual photon-hadron scattering, one expects \ftwo to
rise as $\ftwo \propto (1/x)^{0.08}$ with decreasing $x$.

We note that the power growth of the
total cross section $\sigtotal \propto s^\lambda$ will eventually
violate the Froissart bound \cite{th:froissart} for hadron-hadron
scattering,
\begin{equation}
   \sigtotal \leq \frac{\pi}{m_\pi^2} \left( \ln \frac{s}{s_0} \right)^2,
\end{equation}
where $s_0$ is an unknown constant. The power growth of \sigtotal~ must
be dampened by some mechanism at large energies.

\subsubsection{The small $x$ behaviour of $F_2$}

At small $x$, $F_2$ will be determined by the dominant gluon
content of the proton, because the quarks to which the photon
couples are pair created by the gluons.
The BFKL prediction for
small $x$ was
\begin{equation}
  xg(x) \propto \left(\frac{1}{x}\right)^\lambda \Rightarrow
  \sigtot \propto \ftwo \propto  xg(x)
  \propto \left(\frac{1}{x}\right)^\lambda
  \propto s^\lambda.
\end{equation}
The (LO) BFKL expectation for the small $x$ behaviour of \ftwo is
$\ftwo \propto (1/x)^{0.5}$.
This power growth is faster than the growth expected from
eq. \ref{eq:dll}, the DLL approximation.

\begin{figure}[tbh]
   \centering
   \epsfig{file=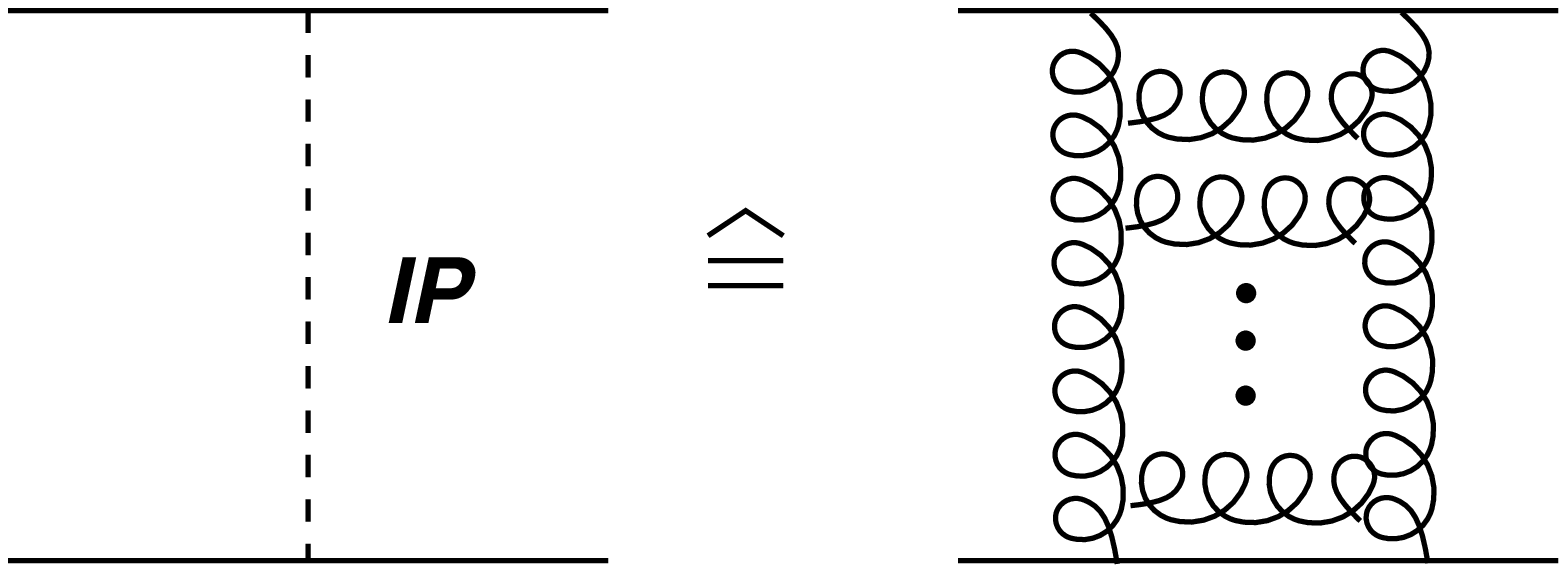,%
    width=6cm,bbllx=71pt,bblly=296pt,bburx=543pt,bbury=483pt,clip=}
   \scaption{The exchange of a (BFKL) Pomeron in the Regge language
    is equivalent to a sum over graphs with gluon ladders between
    the interacting particles in perturbative QCD.}
   \label{bfklpom}
\end{figure}

This situation is very interesting.
The experience from total cross sections at high energies would suggest that
\ftwo should rise $\propto (1/x)^{0.08}$ at small $x$, a behaviour
long known and parametrized with the soft Pomeron,
but whose origin is not understood from QCD.
On the other hand, in the BFKL approximation, QCD does make
a prediction for small $x$, which is
different from the past experience: \ftwo should rise much faster,
$\propto (1/x)^{0.5}$. The DLL expectation is in between.
The slow rise is often said to be due to the ``soft Pomeron"
or the ``non-perturbative" Pomeron,
or the ``Donnachie-Landshoff" Pomeron.
The fast rise would
be attributed to the ``hard", or ``perturbative", or ``Lipatov", or
``BFKL" Pomeron, if one still wants to use the Regge language
(see fig.~\ref{bfklpom}).
Under which conditions will we see which behaviour? How about the
transition region? Do HERA data extend into a kinematic regime where
the steep rise of \ftwo predicted by BFKL can be seen? And if a steep rise
is to be seen, is it really to be attributed to BFKL dynamics?
Some of the answers will be given by the HERA data presented
in the next chapter \ref{ch:sf} on structure function measurements.
It will turn out however that \ftwo is too inclusive a quantity
to resolve the question of BFKL dynamics. The search for specific
signatures of BFKL evolution in the hadronic final state is presented
in chapter \ref{ch:lowx}.

Ultimately the hope is that the BFKL equation offers a way to approach
the confinement problem of QCD.
It allows to make predictions
for \ftwo at small $x$, thus for the structure of hadrons at small $x$,
something which otherwise has to be assumed as non-perturbative input
for the parton densities. It also makes a prediction for the total
$\gamma^* p$ cross section at high energies, where previously we had
just a prediction based upon a parametrization of ``soft" phyics, the
``soft" Pomeron. That the two predictions do not coincide makes it all
the more interesting.

  \section{Hadronic Final States \label{sn:finst}}     

So far we have developed the theory for the total inclusive
cross section, loosely speaking a sum over everything that
can happen inside the proton. When the proton is being probed
by the virtual photon, one out of
all the possible virtual fluctuations in the proton is selected by the
measurement process. The remnants of the fluctuations materialize
in the hadronic final state and become observable when
the proton wave function is projected onto a specific state.
Also the scattered quark and radiation thereoff contribute to
the hadronic final state. Matters are even more complicated:
the
distinction between initial state
fluctuations and final state radiation may be practical and justifiable
in many cases, but is quantum mechanically not rigorous.

QCD predictions for the hadronic final state are in general
much more difficult and less rigorous
than for the inclusive cross section.
On the other hand the hadronic final state can provide much more
detailed information on the QCD processes in $ep$ scattering than
just the total cross section.

\subsubsection{Different observables and approximations}

It depends very much on the final state
observable which technique and
approximation is appropriate for a theoretical description.
For example, for a global event property
like the amount of energy on average emitted in a certain
solid angle, it may be a good approximation to interpret the
parton evolution equations in a
probabilistic way\footnote{Strictly speaking the evolution
equations have been derived for the inclusive cross section.
Cancellations between different diagrams are necessary to
ensure that the total cross section stays finite.
It is therefore not a priori clear to what extent the evolution
equations can be used for hadronic final state predictions.
}.
The
evolution equations define a cascade of parton emissions
(a parton shower) with specified
emission probabilities. One has to
sum up all emission cross sections, weighted with
the energy.
Techncially the calculation could be done analytically, numerically,
or with a ``Monte Carlo'' program.
Of course such a calculation
will only be valid where the
approximations are valid that went into the
evolution equations.

For less global quantities, one may find that one has selected
a piece of the phase space which is unimportant for the
total inclusive cross section, and which might have been
rightfully neglected in the evolution equations. For example, for
the process $\gamma^\ast g \rightarrow \qqbar$
with high transverse momenta of the quarks, the
DGLAP approximation
with strong \pt ordering will not be sufficient to describe
the cross section accurately.
In that case one will instead
use the exact, fixed order QCD matrix element for the process
$\gamma^\ast g \rightarrow \qqbar$ to cover the full phase space,
and fold it with the
probability to find a gluon in the proton.
In practice the matrix element will be calculated only to a few
orders in perturbation theory, in leading (LO) or next-to-leading
order (NLO)\footnote{Calculations in next-to next-to-leading order (NNLO)
are not yet available for DIS.}.
In turn, this approximation will be insufficient for our
first example, where many parton emissions contribute. This case
is hardly covered by the fixed order matrix element
(compare section \ref{sn:flow});
in NLO there are
at most 3 partons in the final state.
The actual implementations of the different perturbative QCD approximations
are presented in section \ref{sn:meps}.

A special r\^{ole} play ``infrared safe'' observables.
An infrared safe observable does not change its value when
an object in the final state (parton, particle, energy cluster, ...)
is split collinearly into smaller units,
or when a soft object is added.
Examples for infrared safe observables are collective observables like
energy flows and certain event shape and jet observables.
A counter example is the total multiplicity.
Infrared safety ensures that perturbative
predictions can be made
without the need to introduce a cut-off parameter
against infrared divergencies.
This property is also
experimentally advantageous, as it does not
require to know the particle composition of a certain
energy deposition in the detector that is used in the analysis.
For example, calorimeter clusters usually
do not have a one-to-one correspondence to incident particles.

\subsubsection{Event generators}

The calculational techniques differ.
Some specific final state observables can be calculated
analytically (rarely),
or numerically with a program.
More ambitious are event generators, ``Monte Carlo'' programs
(see section \ref{sn:gen}).
They attempt to model event by event the complete final state.
Ideally, an ensemble of Monte Carlo generated events would correspond
in all aspects to an ensemble of events collected in nature.
An intrinsic difficulty with this approach is that the
probabilistic generators
have to use probabilities, not probability amplitudes.
Quantum mechanical interference effects therefore
have to be implemented ``by hand'', for example by enforcing
an angular ordering for parton emissions (see section \ref{sn:meps}).

\subsubsection{Hadronization}

One of the biggest problems of QCD is to make contact with
the real world of hadrons.
Perturbative QCD makes predictions for parton
cross sections. Observable are hadrons though.
The transition from partons to hadrons (``hadronization'')
cannot be treated
perturbatively, because it happens at a scale where the
strong coupling constant becomes large.
Non-perturbative
QCD can be considered as one of the last frontiers, where new
insight into nature can be gained, or as a nuisance,
because it hides the beauty and simplicity
of the partonic world from
direct observation.

In any case, much has been learnt about hadronization
from the study of hadron production.
This has led to
some rather successful
phenomenological models to describe
hadronization (see section \ref{sn:hadron}).
Being models, they are not
uniquely defined by theory.
Rather, they depend to some extent on
ad hoc assumptions and parameters which
are being adapted to
observation.
At least the hadronization models are universal and can be used
for different kinds of reactions.
Probably
the most sophisticated model is based on a colour string
(a flux tube) that connects coloured partons.
When stretched by the separating partons,
the string breaks,
pulling new \qqbar~ pairs from the vacuum.
When no more energy is left in the string,
colour neutral hadrons are formed.


\chapter{Inclusive Cross Sections \label{ch:sf}}      
 \section{The Structure Function \mbox{\ftwo}
                                         \label{sn:ftwo}}

ZEUS \cite{z:f2old,z:f2,z:f2bpc,z:f2svd}
and H1 \cite{h1:f2pap,h1:f2,h1:f2of94,h1:f2svtx95,h1:f2jlowx,h1:fl}
have measured
\ftwo in a completely new kinematic domain
compared with fixed target experiments, most notably towards much
smaller values of $x$, and towards much larger \Qsqx.
For kinematic reasons the cross section falls $\propto (1/Q)^4$
(see eq. \ref{eq:dsig}).
Here we shall concentrate on the region of \Qsq less than a
few thousand \GeVsqx, because very little
(though not uninteresting \cite{h1:highq2,z:highq2,h1:highq2update})
data exist for higher \Qsqx, especially as the hadronic final state is
concerned.

\begin{figure}[p]
   \centering
   \epsfig{file=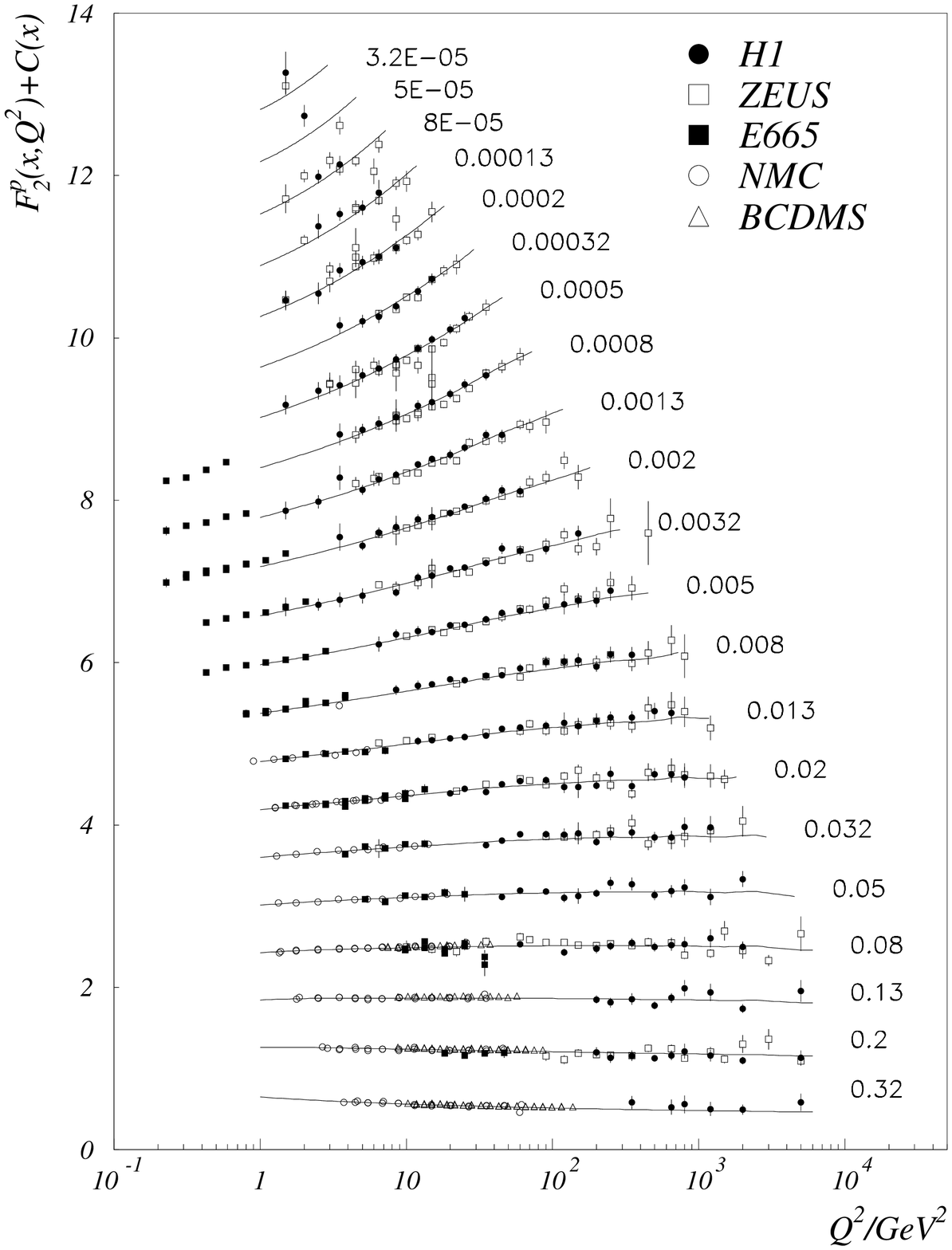,
   width=14cm,%
   bbllx=28pt,bblly=47pt,bburx=555pt,bbury=761pt}
   \scaption{The structure functions $F_2(x,Q^2)$ as a function
             of \Qsq for different \xb bins.
             For better visibility, a function $C(x)=0.6(i-0.4)$,
             with $i$ the $x$ bin number ($i=1$ for $x=0.32$), is added
             to $F_2$ in the plot. Shown are the
             HERA \cite{z:f2,h1:f2of94}
             and fixed target data \cite{o:e665,o:nmc,o:bcdms}, as well as a
             NLO QCD fit \cite{h1:f2of94}.}
   \label{sfq2}
\end{figure}

\begin{figure}[htb]
   \centering
   \epsfig{file=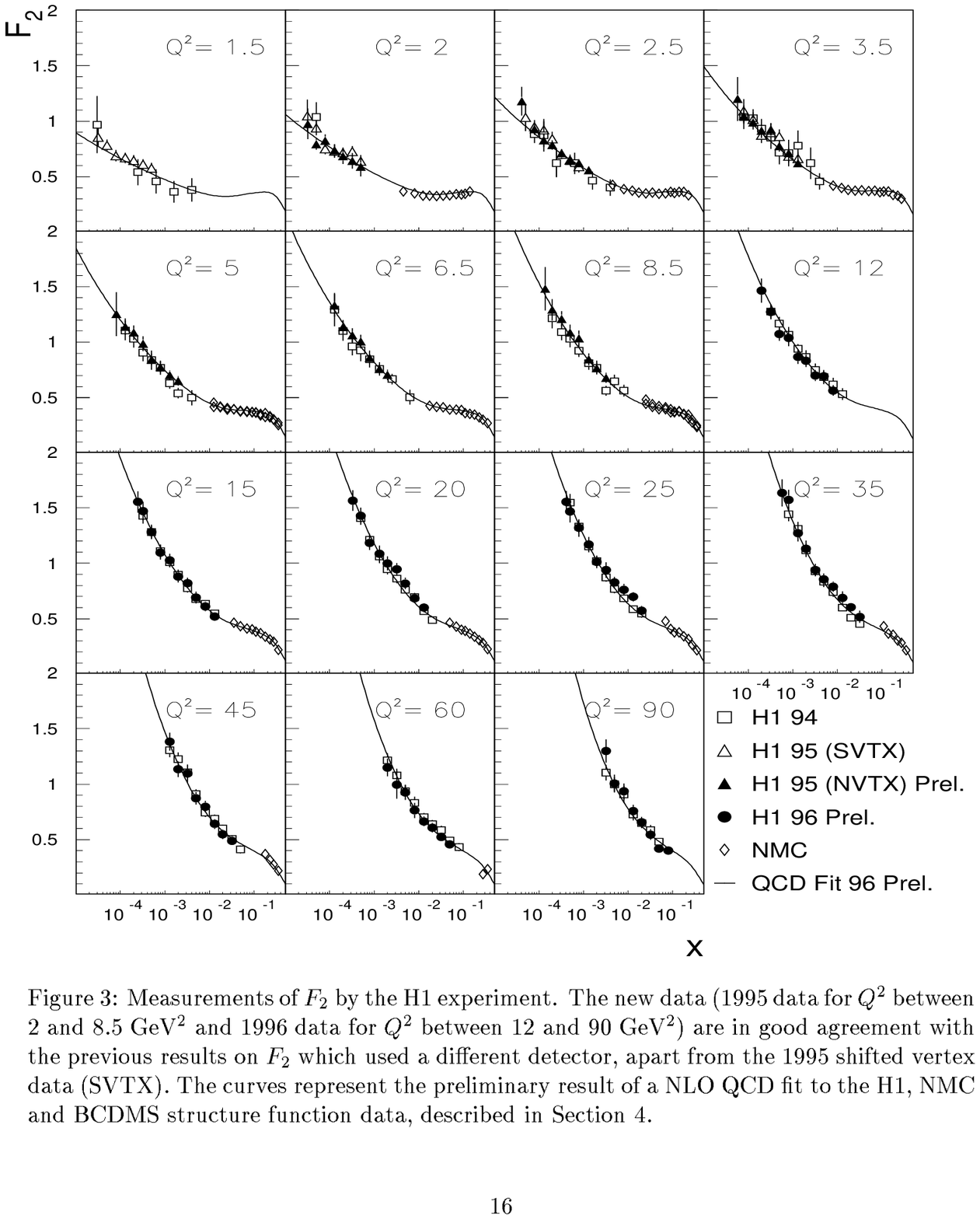,
   width=13cm,%
   bbllx=68pt,bblly=210pt,bburx=509pt,bbury=661pt,clip=}
%
   \scaption{The structure functions $F_2(x,Q^2)$ as a function
             of \xb for different \Qsq bins (\Qsq values are
             given in \GeVsqx).
             Shown are data
             from H1 \cite{h1:f2jlowx} and NMC \cite{o:nmc}, together
             with a NLO DGLAP QCD fit to the data with  $Q^2>1.5\GeVsq$
             from H1, NMC and BCDMS \cite{o:bcdms}. The starting point
             for the evolution was $Q_0^2=1~\GeVsqx$,
             and $\alpha_s(m_Z^2)=0.118$ was set.}
   \label{sfx}
\end{figure}

In Fig.~\ref{sfq2} \ftwo is shown as a function of \Qsq for fixed \xb
values. At $x\approx 0.1$ scaling is observed -- \ftwo does not depend
on \Qsq. At $x>0.1$ \ftwo decreases with \Qsq due to parton splittings,
the products of which are found then at smaller $x$. Therefore
\ftwo increases with \Qsq for $x<0.1$ in accord with the
qualitative discussion of section \ref{sn:dis}.
When plotted as a function of \xb for fixed \Qsq
(fig.~\ref{sfx}),
\ftwo exhibits a steep rise towards small \xb, which flattens at smaller
$Q^2$.
The increase signals
growing parton densities with decreasing
$x$.

From a DGLAP evolution of pre-HERA data this
sharp rise could not be predicted a priori,
because input distributions at small $x$ were not available.
(Assuming however valence-like parton distributions at a small scale
$Q_0^2\approx 0.3~\GeVsqx$, a sharp rise at larger \Qsq
was predicted from DGLAP evolution
\cite{th:grv}.)
It was known though that asymptotically for $\Qsq \rightarrow \infty$
the small $x$ behaviour is given by the DLL formula eq. \ref{eq:dll}.
Are the HERA data still consistent
with DGLAP evolution,
or is there a need for other effects, for example
BFKL,
which one may expect at very small $x$?
It turns out that the data with $\Qsq\gtrsim 1\GeVsq$
can be fit perfectly well
with parton densities which obey the next-to-leading-order (NLO)
DGLAP evolution equations (see Fig.~\ref{sfx})
\cite{z:f2,h1:f2of94,h1:f2jlowx}.
Standard QCD evolution appears to
work over many orders of magnitude in both $x$ and $Q^2$!

When fitting $F_2 \propto (1/x)^\lambda$, H1 finds that
the exponent $\lambda$ increases from $\lambda\approx 0.15$ to
 $\lambda\approx 0.40$ between $Q^2=1\GeVsq$ and  $Q^2=1000\GeVsq$
\cite{h1:f2svtx95}.
\ftwo rises faster than expected from
the soft Pomeron model ($\lambda\approx 0.08$), but less fast
than expected from the LO BFKL equations
($\lambda=0.5$ in LO; $\lambda$ is expected to decrease in NLO).
In fact, both the \xb and the \Qsq dependence of \ftwo can
be attributed predominantly
to the DLL formula eq. \ref{eq:dll},
which can be displayed nicely with
a suitable variable transformation
(``double asymptotic scaling'' \cite{th:das}).
A unified BFKL and DGLAP description of the \ftwo data using
the unintegrated gluon distribution (eq.~\ref{eq:unintegrated})
 is also possible, with
significant contributions from the $\ln 1/x$ resummation
\cite{lowx:unified}.
The structure function data are thus compatible with pure DGLAP
evolution, but cannot exclude significant contributions from
BFKL evolution \cite{lowx:martinrome}.
The structure function data are probably too
inclusive to resolve the question of non-DGLAP evolution.
One
has to resort to less inclusive measurements on the hadronic
final state, which will be discussed in chapter \ref{ch:lowx}.

The \ftwo data can also be
described by evolving flat or valence-like
input quark and gluon distributions from
a very low scale, $Q_0^2=0.35 \GeVsq$
up in \Qsq as was shown by Gl\"uck, Reya and Vogt (GRV) \cite{th:grv}.
The steep rise with decreasing \xb is achieved by the
long evolution length from $Q_0^2$ to $Q^2$
(see eq.~\ref{eq:dll}).
The success of the GRV prediction came as a surprise
for many, as perturbative QCD should only be applicable for $Q$
not too close to
$\Lambda_{\rm QCD} \approx 0.2 \GeV$, because otherwise
$\as(Q^2) = 12 \pi / [(33-2n_f) \ln(Q^2/\Lambda_{\rm QCD}^2)]$ (LO)
diverges.

To conclude this section,
it is quite satisfying that the evolution of the
\ftwo data can be described
with parton densities following standard QCD evolution.
In the next section the results of the NLO DGLAP QCD analyses will be given.
However, the goal remains to calculate the measured
magnitude of the growth $(1/x)^\lambda$ from QCD,
rather than tuning it with the
starting point $Q_0^2$ of the DLL evolution.

 \section{QCD analysis of \mbox{\ftwo}
                                           \label{sn:qcda}} 
\begin{figure}[tbh]
   \centering
   \epsfig{file=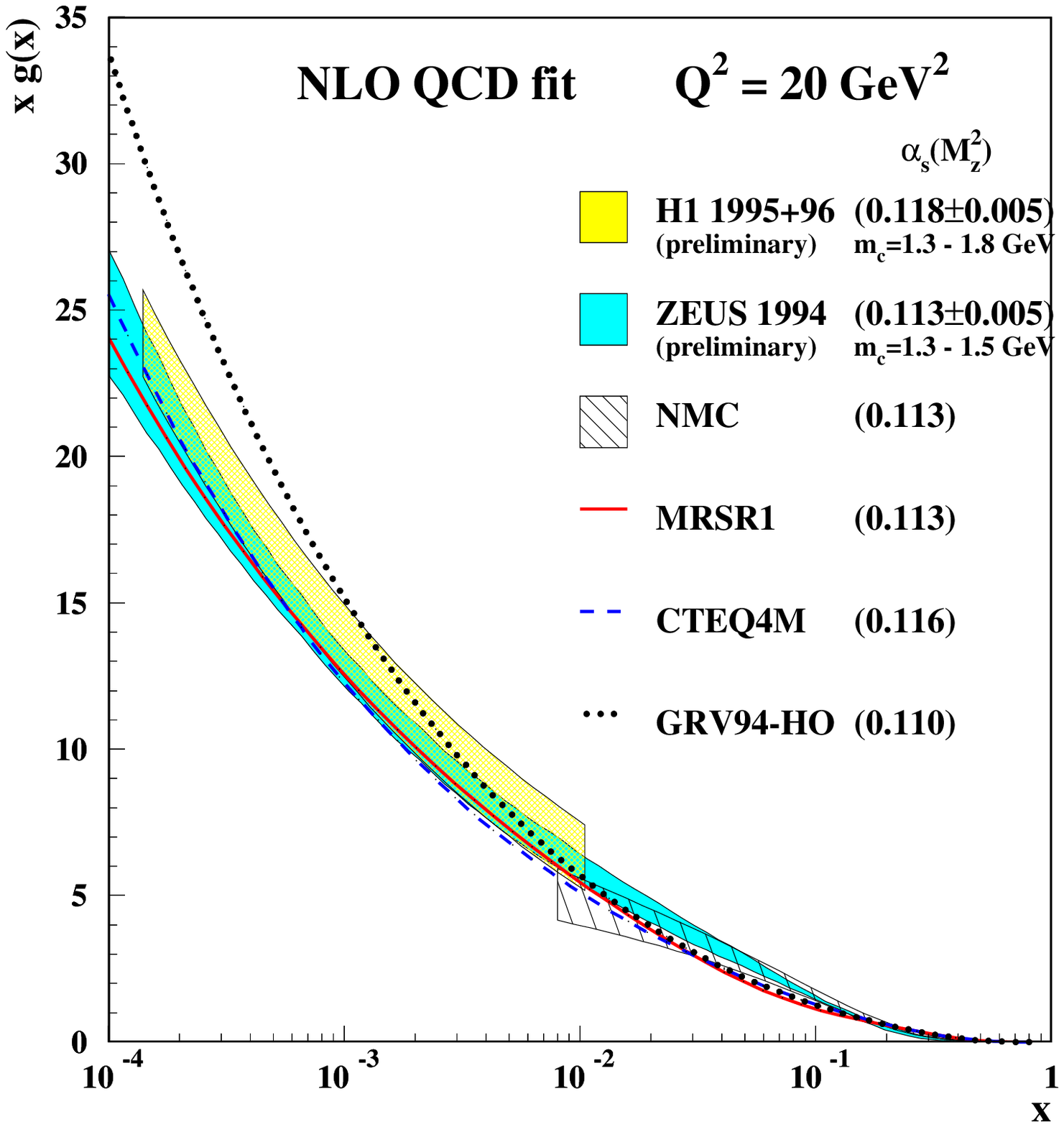,width=10cm,
   bbllx=62pt,bblly=195pt,bburx=497pt,bbury=670}
   \scaption{The gluon density
             $xg(x,Q^2)$ at $Q^2=20\GeVsq$ \cite{z:devenish},
             as extracted from NLO QCD analyses
             of the H1 \cite{h1:f2jlowx}
             and ZEUS \cite{z:botje2}
             \ftwo data
             together with the NMC result \cite{o:nmcglue}.
             Shown are the $\pm 1 \sigma$ error bands of the analyses.
             The values for the charm mass and for $\alpha_s(m_Z^2)$
             which were used in the analyses are given.
             The
             extracted gluon density is compared to the global analyses
             MRSR1 \cite{th:mrsr}, CTEQ4m \cite{th:cteq4} and
             GRV94-HO \cite{th:grv}.}
   \label{qcdana}
\end{figure}

From the DGLAP equations eq. \ref{eq:dglap} it is clear that
the scaling violations of \ftwo depend on both \as and
the gluon density. In fact, for $x<0.01$ and in lowest order
one can derive the approximate formula \cite{th:prytz}
\begin{equation}
  \frac{\dif F_2(x/2,\Qsq)}{\dif \ln \Qsq} \approx
       \frac{10}{27} \frac{\alpha_s (Q^2)}{\pi} xg(x,Q^2),
\end{equation}
because at small $x$ the proton is dominated by gluons, and the
scaling violations arise from quark pair creation from gluons.
The full NLO QCD analyses now employed at HERA are of course
more involved.
Fig. \ref{qcdana}a shows
the gluon density $x\cdot g(x,Q^2)$
extracted from NLO QCD fits to the \ftwo data \cite{z:devenish}.
Previous data from NMC cover $x>0.01$, and the HERA data
extend down to $x=0.0001$. In that region the gluon density
increases sharply towards small $x$.

Clearly, the density cannot increase forever; eventually
saturation effects will set in
\cite{lowx:levin,lowx:mueller}.
The effective transverse size of a gluon that is probed by a photon of
virtuality $Q$ is $\sim 1/Q$. Fluctuations below that scale happen,
but cannot be resolved.
When such gluons of transverse size $\sim 1/Q$
fill up the whole transverse
area offered by the proton, they will start to overlap and recombine.
This would be a novel
and very interesting situation indeed: high parton density,
but \Qsq large enough for $\alpha_s(Q^2)$ to be small!
Given the size of the proton $\approx 1 {\rm ~fm}$, the critical
condition for saturation effects to turn on can be estimated
\cite{lowx:mueller}
as
$x_{\rm crit} g(x_{\rm crit},Q^2) \approx \frac{1 {\rm fm}^2}{1/Q^2}
                  \approx 25 \frac{Q^2}{\GeVsq}$.
This value is by far
not reached by the measured gluon density (Fig.~\ref{qcdana}a).
It could be however that saturation does not set in uniformly over
the proton's transverse area, but starts locally in so-called
hot spots \cite{lowx:hotspots}.
In this case $x_{\rm crit}$ would be larger.
The inclusive structure function data however
do not require any saturation correction.
Even a conspiracy of two new effects,
a sharp rise of \ftwo from BFKL evolution dampened
by saturation effects, cannot be ruled out.
Probably saturation effects
will be first seen in hadronic final state
data \cite{lowx:hotspots}.

From an analysis of the scaling violations
$\dif \ftwo / \dif \ln \Qsq \sim \alpha_s$, or
equivalently, from a QCD fit to the \ftwo data,
the strong coupling constant can in principle be determined.
From an analysis of the 1993 data
$\alpha_s(m_Z^2) = 0.120 \pm 0.005 ({\rm exp.}) \pm 0.009({\rm theor.})$
was obtained \cite{th:ballforte}.
It is estimated \cite{z:botje1} that ultimately, with an integrated HERA
luminosity of 500 \pbinv, \as can be extracted by a NLO QCD analysis
from the HERA
structure function data with an experimental error of
$\Delta \as = 0.001-0.002$,
and a theoretical uncertainty of $\approx 0.006$.
The theoretical errors could be reduced by higher than NLO calculations.



\section{The Longitudinal Structure Function $F_L$ \label{sn:fl}}
One can express the longitudinal structure function \fl
in terms of the cross section for the absorption
of longitudinally polarized photons (see section \ref{sn:dis}),
\begin{equation}
F_L=\frac{Q^2 (1-x)}{4\pi^2\alpha} \cdot \sigl
\approx \frac{Q^2}{4\pi^2\alpha} \cdot \sigl.
\end{equation}
Longitudinal photons have helicity 0 and can exist only virtually.
For transverse photons with helicity $\pm 1$ the spin is parallel
to the direction of propagation, and the field vector perpendicular to it.
In the QPM,
helicity
conservation at the electromagnetic vertex yields
the Callan-Gross relation ($F_L=0$)
for scattering on quarks with spin $1/2$.
This does not hold when the quarks acquire transverse momenta
from QCD radiation.
Instead, QCD yields the Altarelli-Martinelli equation \cite{th:am}
\begin{equation}
  F_L(x,Q^2)=\frac{\as}{4\pi} x^2 \int_{x}^{1}\frac{dz}{z^3}
      \left[\frac{16}{3}F_2(z,Q^2)
       +8\sum_i e_{q_i}^2(1-\frac{x}{z})\cdot zg(z,Q^2)\right],
\end{equation}
exposing the dependence of \fl on the strong coupling and
the gluon density. At small $x$, the second term with the gluon
density is the dominant one.
In fact
$F_L(x,Q^2) = 0.3 \cdot [4\as/(3\pi)] \cdot xg(2.5x,Q^2)$ is not
a bad approximation for $x<10^{-3}$ \cite{th:fl}.

\begin{figure}[tbh]
   \centering
   \epsfig{file=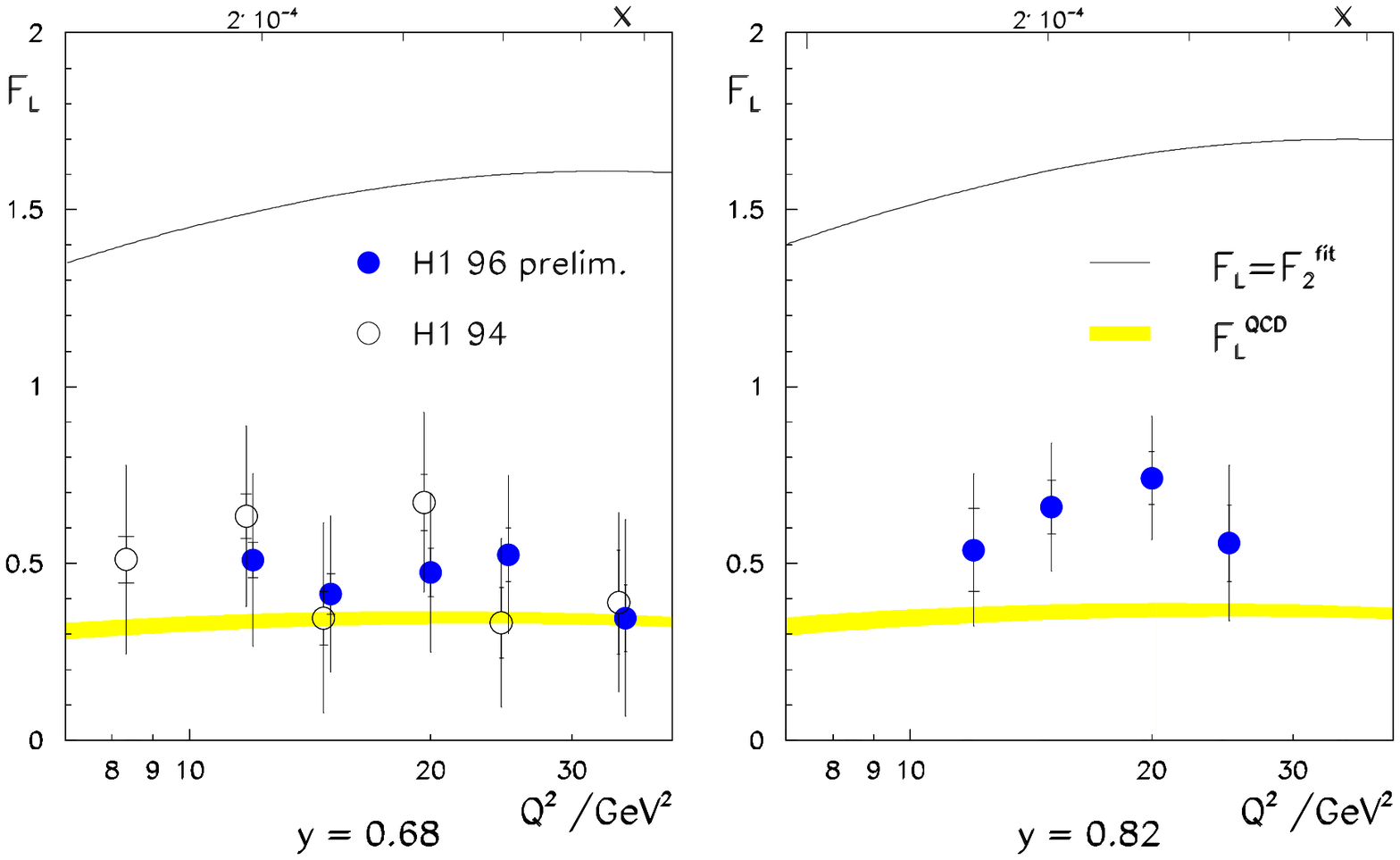,width=13cm,
   bbllx=24pt,bblly=221pt,bburx=570pt,bbury=567,clip=}
   \scaption{
              The longitudinal structure functions $F_L$
              extracted from a QCD analysis of high $y$ inclusive
              scattering data \cite{h1:fl,h1:f2jlowx} at two different
              $y$ values.
              The shaded bands give the range of expectations for
              \fl from a QCD analysis of structure function data with
              $y<0.35$. The full line represents the solution \fl=\ftwo.}
   \label{fl}
\end{figure}

The extraction of \ftwo from the cross section measurement
(eq. \ref{eq:dsig})
so far had to make an assumption
for $F_L$, because there existed no direct $F_L$
measurements in the HERA regime.
At large $y$, $y\approx 0.7$, this is a 10\% correction.
The argument can be turned around, and $F_L$ can be extracted
at large $y$ from a measurement of the cross section, assuming
that \ftwo follows a QCD evolution and can be extrapolated from
measurements at smaller $y$.
This procedure has been carried out by H1 \cite{h1:fl,h1:f2jlowx}
(Fig.~\ref{fl}).
The extracted \fl excludes the extreme possibilities
\fl=\ftwo and \fl=0
and implies
$R \equiv \sigma_L/\sigma_T=F_L/(F_2-F_L) \approx 0.5$,
since
$F_2\approx 1.5$. This
is self consistent with the gluon density
extracted from the H1 QCD fit.
It should be noted however that the \fl data points at high $y$
are somewhat above the expectation from the QCD fit.
More
precise measurements are necessary to clarify the situation.
A measurement of $F_L$ without theoretical assumptions on the evolution
of $F_2$ will be possible by measuring the DIS cross section at HERA at two
different centre of mass energies
\cite{hera:flfuture},
because these
data will allow to vary $y$ in eq. \ref{eq:dsig}
while keeping $x$ and \Qsq
fixed.

 \section{ $\sigtot$ and the Transition between DIS
            and Photoproduction \label{sn:transi}}   
\begin{figure}[tbh]
   \centering
   \epsfig{file=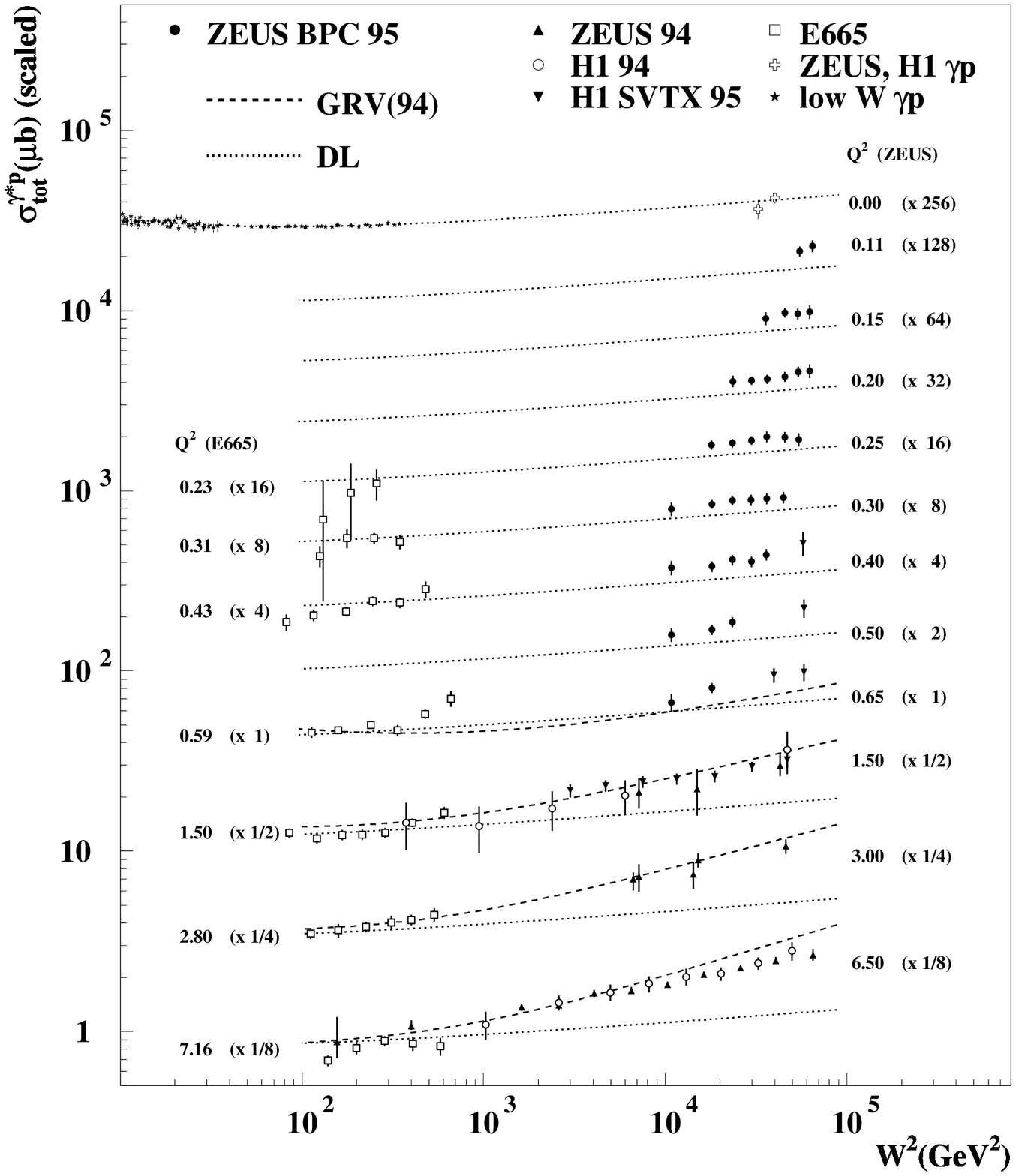,
   width=13cm}
   \scaption{The total $\gamma^\ast p$ cross section as a function
             of $W^2$ for different values of \Qsq \cite{z:f2bpc}.
             The data for virtual photon-proton scattering
             \cite{h1:f2of94,h1:f2svtx95,z:f2,z:f2bpc,o:e665} and
             photoproduction \cite{h1:gp,z:gp,o:gp}
             are compared to the predictions by the
             soft Pomeron model of DL \cite{th:dola2} and the perturbative
             QCD model by GRV \cite{th:grv}.}
   \label{sigt}
\end{figure}

The total $\gamma^\ast p$ cross section is shown as a function of $W^2$
in fig.~\ref{sigt}. The cross section increases with $W^2$, and the
slope increases with $Q^2$.

The data at $Q^2=0$ are well described by
the soft Pomeron parametrization by Donnachie-Landshoff (DL) \cite{th:dola1},
$\sigtot \propto W^{2\cdot0.08}$.
With increasing $Q^2$, the cross section grows faster than the DL
prediction.
There appears to be a smooth but rather fast transition between the
soft and the perturbative regions.
Already at $Q^2\approx 1~\GeVsqx$, a distinctively fast rise of the
total cross section $\sigtot \propto W^{2\cdot\lambda}$ with
$\lambda > 0.08$ is observed.
The H1 fit result $\ftwo \propto (1/x)^\lambda$ translates into
$\sigtot\propto W^{2\lambda}$ with $\lambda \approx 0.2-0.4$ for
$Q^2>1~\GeVsqx$.

The perturbative GRV mechanism \cite{th:grv} utilizing NLO DGLAP evolution
provides an adequate description of the
data down to $\Q^2 \approx 1 \GeVsq$.
It has not been possible to apply NLO DGLAP at lower \Qsq to describe
the data.
For example,
below $Q^2\approx0.4\GeVsq$ the GRV curves would turn over at small $x$,
reflecting the valence-like behaviour of the GRV input
partons at a low scale, by far failing to account for
the data \cite{z:f2bpc}.


\chapter{Models for Hadron Production in DIS \label{ch:models}}  

  \section{Reference Frames \label{sn:frames}}   
In this section the different reference frames and the
variables used to describe the hadronic final state are
introduced.
Rough expectations for the event properties
are derived mainly from phase space arguments.
They will be refined by perturbative QCD and
specific hadronization models.


\subsubsection{The laboratory frame}

The HERA experiments are performed in the
``laboratory frame'', in which the detector is at rest, and
the incoming electron and proton
beams are collinear.
The $z$ axis is defined by the proton beam direction.
After the collision, the transverse momentum
of the scattered electron is balanced by the one of the hadronic system.
The hadronic final state is however better studied in a frame where
the transverse boost is removed, such
that the virtual photon
and the incoming proton
are collinear.
Photon and proton 4-vectors are denoted by
$q=(E_{\gstar},\vec{q})$ and
$P=(E_p,\vec{P})$.
With $E_p \gg m_p$ the proton mass can be neglected.

\subsubsection{The hadronic CMS}

\begin{figure}[tbh]
   \centering
   \epsfig{file=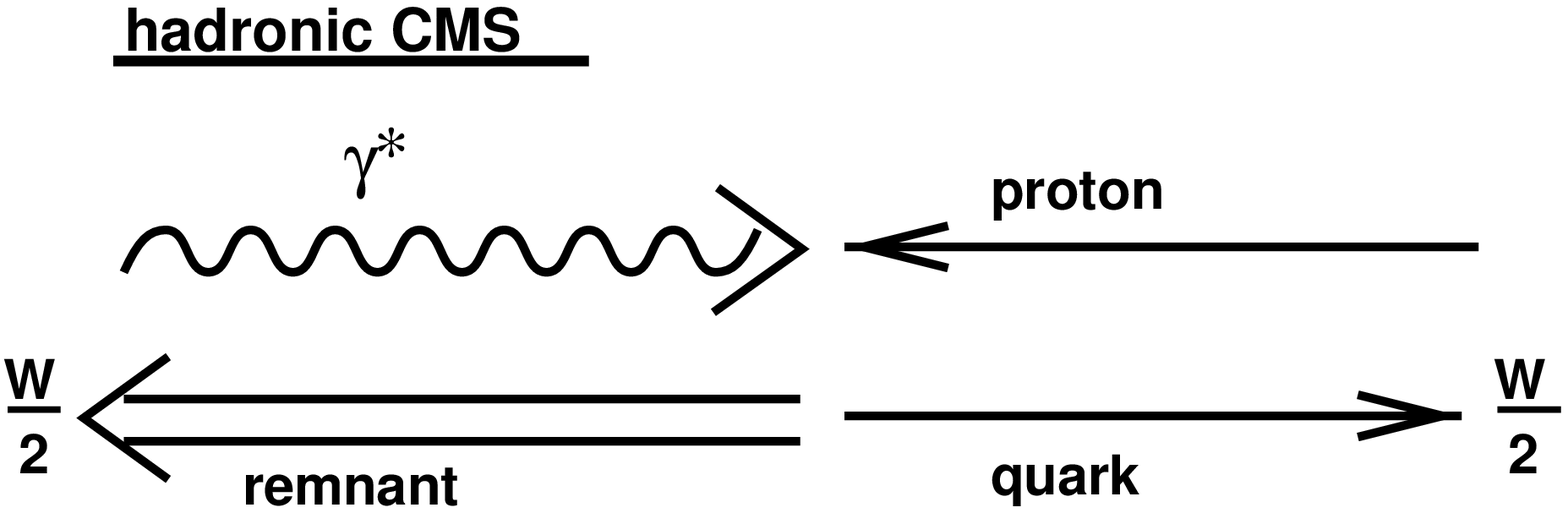,width=8cm}
   \scaption{The hadronic centre of mass system (CMS).}
   \label{cms}
\end{figure}

The hadronic centre of mass system (CMS) is defined by the condition
$\vec{P}+\vec{q}=0$.
The invariant mass of the hadronic system is given by $W^2=(P+q)^2$.
The positive $z$-axis is usually defined by the direction of
the virtual boson, $\vec{q}$. All hadronic final state particles with
4-momentum $p=(E,p_x,p_y,p_z)$ which have $p_z>0$ are said
to belong to the current hemisphere, and all particles with
$p_z<0$ are assigned to the target or proton remnant hemisphere.
Target and current systems are back to back and carry energy
$W/2$ each. They are not necessarily collinear with the incoming proton.
In the QPM however,
a quark in the proton with 4-momentum $xP$ absorbs the
virtual photon and is scattered back with 4-momentum $xP+q$.
The proton remnant retains the 4-momentum $(1-x)P$.
With these assumptions, the scattered quark and the proton remnant
are both collinear with the incoming virtual photon and the proton
(fig.~\ref{cms}).
This is no longer the case if one considers intrinsic transverse
momenta of the partons in the proton, or generates transverse momenta
perturbatively by radiation.

\subsubsection{The Breit frame}

\begin{figure}[tbh]
   \centering
   \epsfig{file=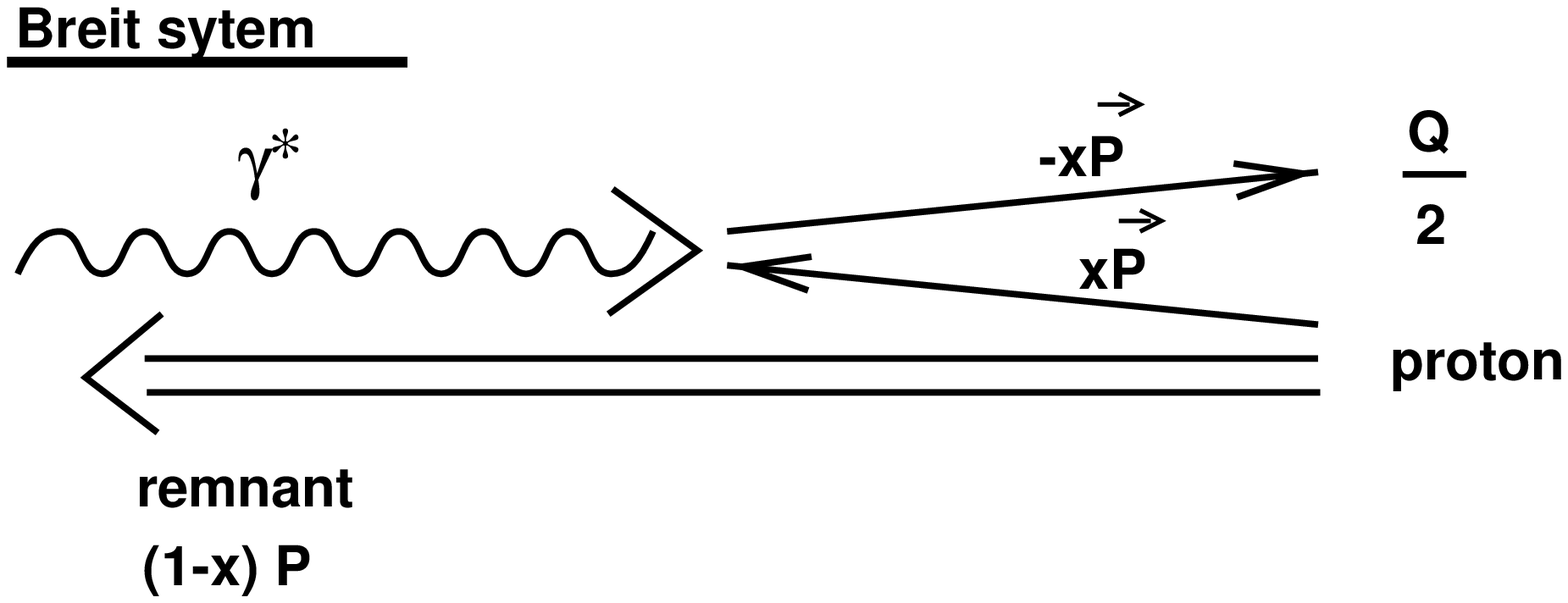,width=8cm,
   bbllx=15pt,bblly=280pt,bburx=565pt,bbury=516,clip=}
   \scaption{The Breit frame.}
   \label{breit}
\end{figure}

The Breit frame (BF) is defined by the conditions that
proton and virtual photon are collinear, and that the
virtual photon does not transfer energy, just momentum.
Since $q^2=-Q^2$,
$q=(0,0,0,Q)$
with the same orientation as in the CMS,
and $P=(E_p,0,0,-E_p)$.
In the QPM, the incoming quark absorbing the virtual photon
does not change its energy, but reverses its longitudinal
momentum with magnitude $Q/2$ (fig.~\ref{breit}).
The Breit system is also
called the brick wall system, because in this picture the
quark bounces back like a tennis ball off
a brick wall.
Neglecting the quark mass and transverse
momenta, its 4-momentum is $xP=(Q/2,0,0,-Q/2)$ before
and $(Q/2,0,0,Q/2)$ after the scattering.
The incoming 3-momentum vector $x\vec{P}$ is merely reverted.

Again, all particles with $p_z>0$ are assigned
to the Breit current hemisphere.
The CMS and BF are connected via a longitudinal boost.
The division into current and target hemisphere is frame dependent.
In both systems, the remnant diquark is a mere spectator in the
sense that its momentum remains unchanged.

  \section{Kinematics and Variables \label{sn:kinevar}}

Let us consider a particle with 4-momentum
$p=(E,\vec{p})=(E,p_x,p_y,p_z)$
and mass $m$, which travels under an angle $\theta$
with respect to the $z$ axis.
We define the transverse momentum
$\pt = \sqrt{p_x^2+p_y^2}$, the transverse mass
$\mt = \sqrt{m^2+p_T^2}$ and the transverse energy
$\et = E \cdot \pt/|\vec{p}|$.

The rapidity \cite{th:collins} of a particle is defined as
\begin{equation}
y := \frac{1}{2} \ln \frac{E+p_z}{E-p_z}
  = \ln \frac{E+p_z}{\mt} = \coth \frac{p_z}{E}.
\end{equation}
The rapidity transforms under
a boost in the $z$ direction with velocity $\beta$ as
\begin{equation}
y \rightarrow y'=y - \coth \beta.
\end{equation}
Hence the shape of the rapidity distribution is invariant against
longitudinal boosts. A useful relation is
\begin{equation}
  \dif y / \dif p_z = 1/E.
\end{equation}

Often the mass of a measured particle is not known. One therefore defines
the pseudorapidity
\begin{equation}
  \eta := - \ln \tan \frac{\theta}{2}.
\end{equation}
For $E \gg m$, $\eta \approx y$.

Energy and momentum of a particle are conveniently scaled
by their kinematically allowed maximum, leading to the
definitions
\begin{equation}
  \xf := p_z / \pzmax \hspace{1cm} \xp := |\vec{p}| / \pmax
  \hspace{1cm} \xe := E/\emax.
\label{eq:xf}
\end{equation}
For $E\gg m$ and $p_z \gg p_T$, these definitions converge:
$\xf\approx \xp \approx \xe$.
We shall use $z$ as a generic symbol
for the variables defined in eq.~\ref{eq:xf}.

The scaled longitudinal momentum \xf is known as Feynman-$x$.
In the CMS,  $\xf \approx 2\cdot p_z/W$, with  $-1 \leq \xf \leq 1$.
Similarly, in the Breit frame current hemisphere
$\xp = 2 |\vec{p}|/Q$
with $0 \leq \xp \leq 1$.
None of the variables defined in eq.~\ref{eq:xf}  is Lorentz invariant.
A Lorentz invariant that is currently not being used at HERA is
$z_h:=(Pp)/(Pq)$. In the proton rest frame it gives the
fraction of the energy transfer taken by the produced hadron,
$z_h=E/\nu$. For large $W$ and $\xf,z_h \gtrsim 0.1$ the differences
between $z$ and \xf are small.

\begin{figure}[tbh]
   \centering
 \epsfig{file=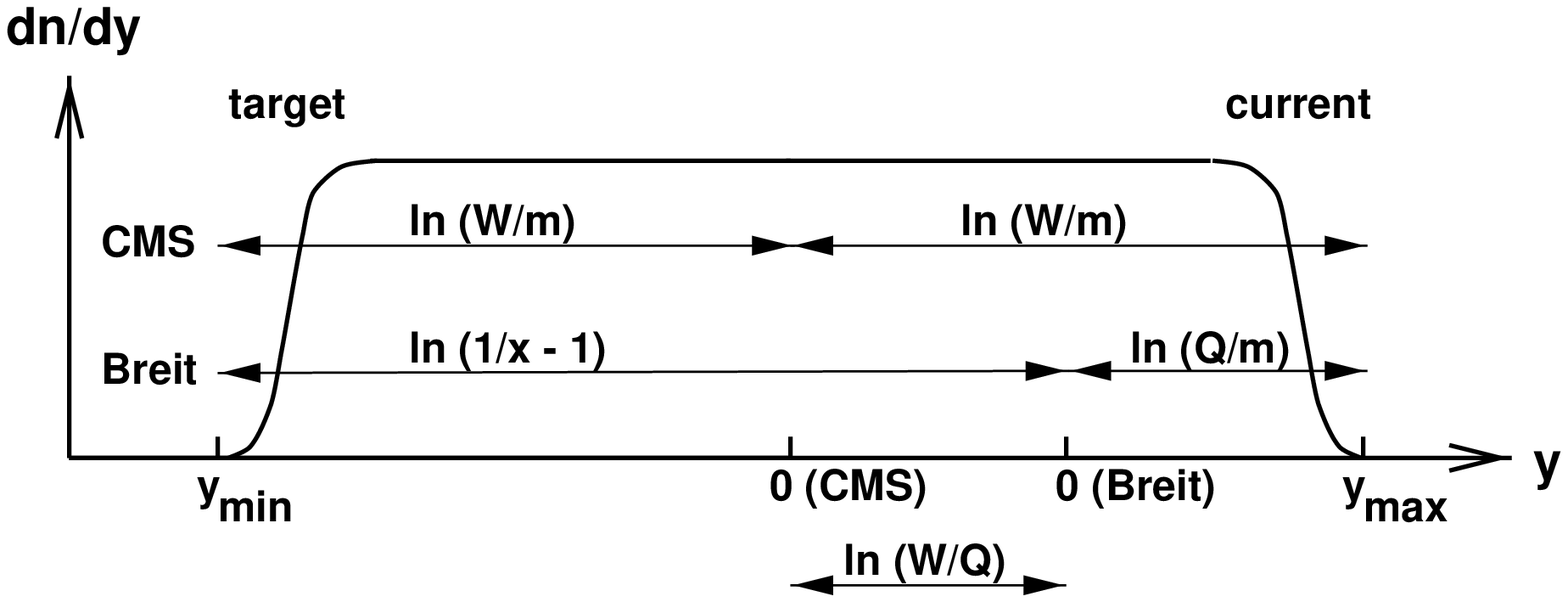,%
       width=10cm,bbllx=35pt,bblly=287pt,bburx=578pt,bbury=497,clip=}
   \scaption{The rapidity plateau. Indicated are the sizes of
    the target and current regions in the Breit and CM systems.
    The dimensions are chosen for typical HERA events with $W=120 \GeV$,
    $Q^2=30\GeVsq$ and for $m=m_\pi$.}
   \label{plateau}
\end{figure}

The rapidity can take any values with $\ymin \leq y \leq \ymax$
(see fig.~\ref{plateau}),
with
\begin{equation}
\ymax = \ln \frac{2\pzmax}{m} \hspace{1cm} \ymin = \ln \frac{2\pzmin}{m}.
\end{equation}
In the CMS, $-\ymin = \ymax \approx \ln W/m$. At HERA with
$W \leq 300~\GeVx$, pions can be produced  with $-7.7 \leq y \leq 7.7$.
Neglecting transverse momenta,
the CMS rapidity $y$ is related to \xf (for $x_F>0$) by
\begin{equation}
  \ymax - y \approx \ln (1/\xf).
\end{equation}
The difference of rapidities measured in the Breit and the CM systems is
\begin{equation}
 y^{\rm BF} - y^{\rm CMS} =
\ymax^{\rm BF} - \ymax^{\rm CMS} = \ln (Q/m) - \ln (W/m) = \ln (Q/W).
\end{equation}

  \section{Simple Mechanisms for Hadron Production
               \label{sn:fragf}} 

Much can be learnt about
the gross event features already
from phase space considerations and a few simple assumptions.
We start with the simple model of independent fragmentation,
applied to the scatterd quark and the remnant. In the following
sections, this naive picture will be refined with perturbative
QCD radiation and more sophisticated hadronization models.

The Lorentz invariant cross section for hadron production
may be written as
\begin{equation}
 E \frac{ \dif^3 \sigma}{\dif^3 p} =
   \frac{ \dif^3 \sigma}{\dif \phi \dif y p_T \dif p_T},
\end{equation}
since $\dif y / \dif p_z = 1/E$.
For a distribution that is isotropic in azimuth $\phi$,
the $\phi$ integration yields
\begin{equation}
 \int_0^{2\pi} E \frac{ \dif^3 \sigma}{\dif^3 p} \dif \phi =
   4 \pi \frac{ \dif^2 \sigma}{\dif y \dif p_T^2}.
\end{equation}

We describe
the inclusive densities
to find a hadron $h$ with energy fraction
$z:=E/E_q$ from the fragmentation of a quark $q$ with
fragmentation functions $D_q^h(z)$.
($D_q^h(z)$ is not a probability density, because its integral is not 1.)
They scale approximately, that is they depend only on the fractional
hadron energy $z$ and not on the quark energy $E_q$ \cite{th:fscal}.
The resulting hadron spectrum
from quark fragmentation in the reaction $ep\rightarrow e^\prime qX$ is then
\begin{equation}
\frac{1}{N}\frac{\dif n_h}{\dif z} =
\frac{1}{\sigtotal} \frac{\dif \sigma}{\dif z} =
\frac{ \sum_q e_q^2 f_q(x)D_q^h(z)}
     { \sum_q  e_q^2 f_q(x)}.
\end{equation}
Here \sigtotal is the total event cross section,
$N$ the number of produced events, and $n_h$ the number of
produced hadrons of type $h$. $f_q(x)$ is the quark density function
for flavour $q$ in the proton.

\begin{figure}[htb]
   \centering
   \epsfig{file=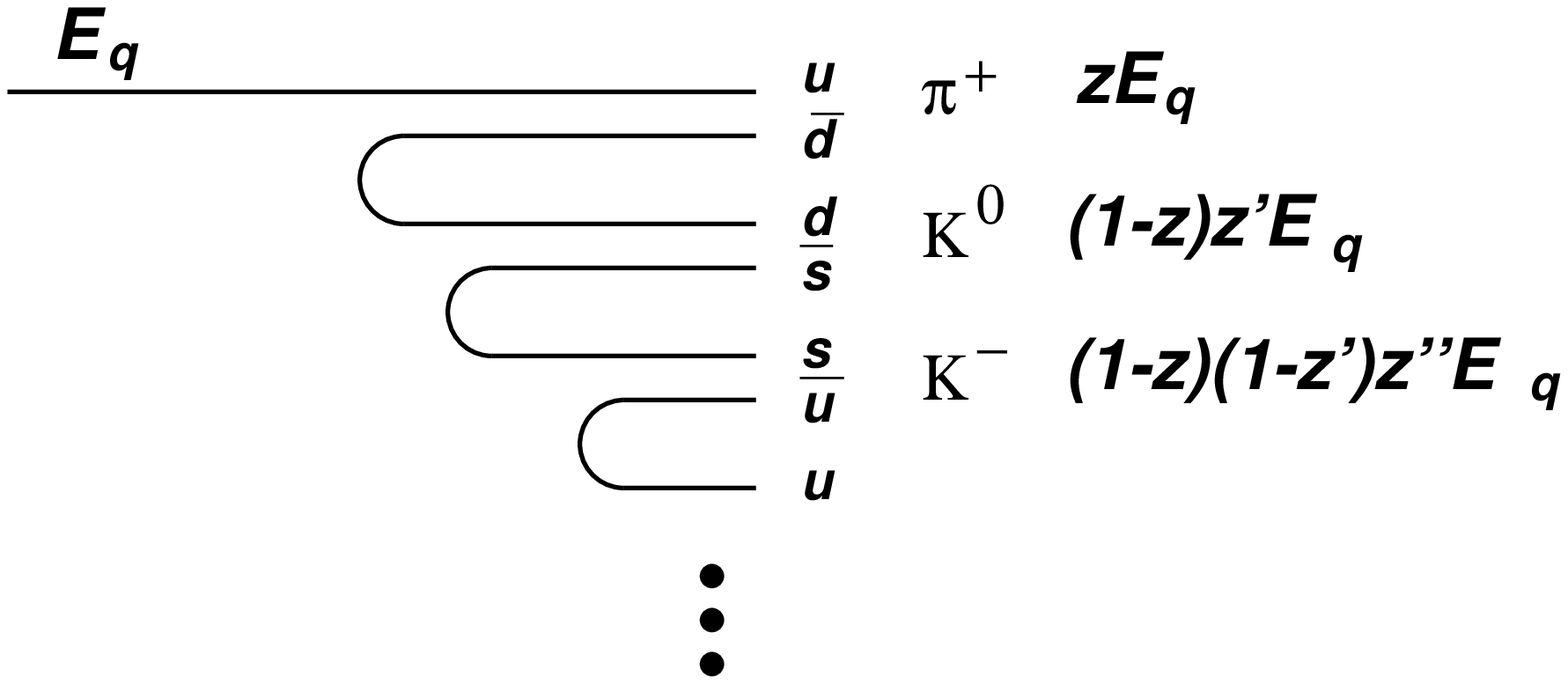,%
    width=8cm}
   \scaption{Independent fragmentation \`{a} la Feynman-Field \cite{th:ff}.
             A $u$ quark with initital energy $E_q$ branches
             into a $\pi^+$ meson with energy $E=zE_q$, and
             a $\ol{u}$ quark with energy $(1-z)E_q$, and so on.}
   \label{indep}
\end{figure}

In the simple model of independent quark fragmentation \cite{th:ff},
a quark with energy $E_q$ fragments into a hadron with energy $E$
according to a distribution function $f(z)$, where $z=E/E_q$,
see fig.~\ref{indep}.
The process is iterated with a quark carrying the remaining
energy $(1-z)E_q$ until there is no energy left to produce a hadron.
It is assumed that in the fragmentation process only limited
transverse momenta are produced, usually parameterized with
a falling exponential distribution in $p_T^2$ with
$\av{p_T^2}\approx 0.44$ \GeVsqx.

\begin{figure}[tbh]
   \centering
\begin{picture}(0,0) \put(0,0){{\bf a)}} \end{picture}
   \epsfig{file=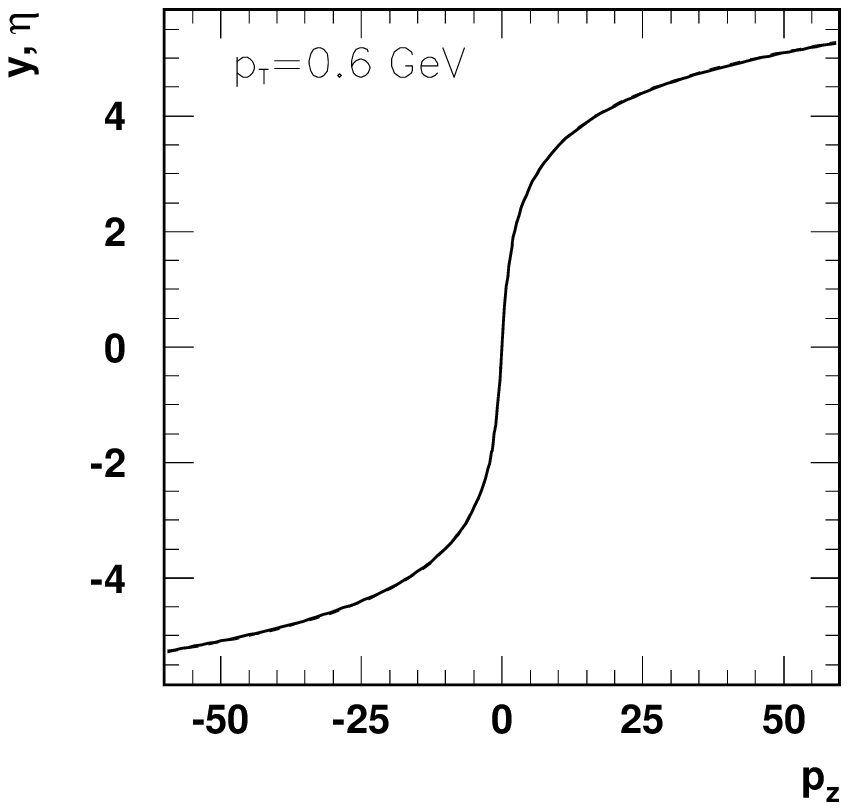,%
      bbllx=25pt,bblly=407pt,bburx=275pt,bbury=650,width=6cm,clip=}
   \hspace{1cm}
\begin{picture}(0,0) \put(0,0){{\bf b)}} \end{picture}
   \epsfig{file=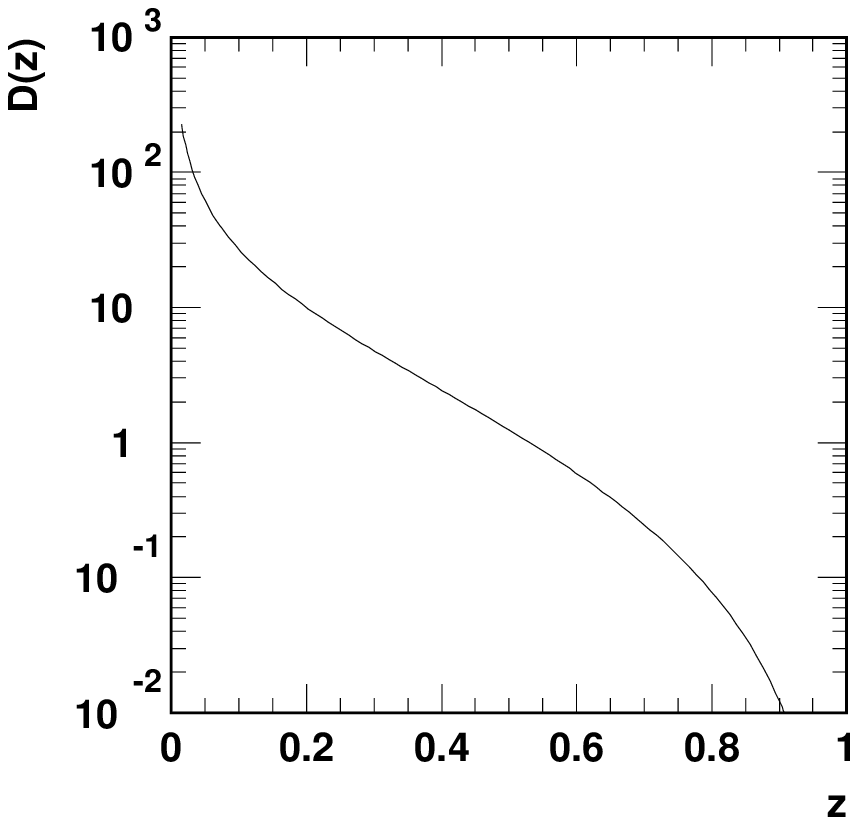,%
      bbllx=25pt,bblly=407pt,bburx=275pt,bbury=650,width=6cm,clip=}
   \scaption{{\bf a)} Relation between rapidity $y$
   and longitudinal momentum
   \pz for a particle with $m=m_\pi$ and $p_T=0.6 \GeV$ at $W=120\GeV$,
   typical for HERA.
   The curves for pseudorapidity $\eta$ and rapidity $y$ are indistinguishable.
   Differences between $y$ and $\eta$ become noticeable for
   smaller \pt; for $\pt=0.1\GeV$ it is at most 0.3 rapidity units.
   \ymax is given by $\pzmax \approx W/2$.
   {\bf b)} The fragmentation function $D(z)=\frac{3.5}{z}(1-z)^{2.5}$.
 }
   \label{rapidity}
\end{figure}

The model describes the gross features of fragmentation \cite{th:ff},
namely energy independence,
and for small $z$ a behaviour
\begin{equation}
  D_q^h(z) \propto 1/z.
\end{equation}
Since  $\dif y \approx  z \dif z$, this gives a uniform
number density of hadrons in rapidity,
the so-called central (in the CMS) rapidity plateau:
\begin{equation}
  \frac{1}{N}\frac{\dif n_h}{\dif y}
  \approx  \frac{1}{N} \frac{\dif n_h}{\dif z} \cdot z
   \propto D_q^h(z) \cdot z = {\rm const}.
\end{equation}
Fig.~\ref{rapidity}a gives the relation between the longitudinal
momentum $\pz=z \cdot \pzmax \approx z\cdot W/2$ for a typical HERA situation.
$|\xf|<0.1$ translates roughly to the rapidity region $|y|<3$
(fig.~\ref{rapidity}a).

A convenient parametrization for the fragmentation functions is
\begin{equation}
  D_q^h(z) = a \frac{1}{z} (1-z)^b.
\end{equation}
Energy-momentum conservation requires
\begin{equation}
  \sum_h \int_0^1 z  D_q^h(z) \dif z = 1,
\end{equation}
where the sum runs over all hadron species $h$.
If we do not distinguish between hadron species,
we can define $ D_q(z) := \sum_h D_q^h(z)$.
The normalization $a$ is then fixed by
\begin{equation}
  \int_0^1 z  D_q(z) \dif z = a/(b+1) = 1.
\end{equation}
A useful parametrization for light quarks $q=u,d,s$,
accurate to $20\%$, is \cite{th:webberscal}
\begin{equation}
  D_q(z) = 3.5 \frac{1}{z} (1-z)^{2.5},
\end{equation}
see fig.~\ref{rapidity}b.
The average number of hadrons per unit rapidity in the plateau
region ($z$ small) is then
\begin{equation}
 \frac{1}{N}\frac{\dif n}{\dif y}
  \approx  \frac{1}{N} \frac{\dif n}{\dif z} \cdot z
   = D_q(z) \cdot z \approx 3.5,
\end{equation}
of which roughly 2/3 will be charged.
The rapidity plateau is depicted in fig.~\ref{plateau}. There
will be a kinematic fall-off for $y\rightarrow \ymax,\ymin$.
The available rapidity range
$\ymax-\ymin$ increases logarithmically with the quark energy
(longitudinal phase space),
and so does the average total hadron multiplicity.

The average total hadron multiplicity in a quark jet is given by
\begin{equation}
  \av{n} = \int_{\zmin}^1  D_q(z) \dif z,
\end{equation}
where $\zmin = m / E_q$ is determined by a typical hadron mass $m$.
From
\begin{equation}
  \dif \av{n} / \dif \zmin  = -D_q(\zmin) \approx -a/\zmin
\end{equation}
we obtain a logarithmic scaling law for the multiplicity
\begin{equation}
   \av{n} =   a \ln \frac{E_q}{m} + c,
\end{equation}
where $c$ is the integration constant.

Though the independent fragmentation picture describes crudely the
data on e.g.
$\epem \rightarrow \qqbar \rightarrow {\rm hadrons}$,
it has its limitations.
It cannot be applied consistently
for small $z$, $z\lesssim 0.1$,
it is not Lorentz invariant
(the constants $a$ and $c$ are frame dependent),
and energy momentum conservation has
to be enforced by hand after both quarks have fragmented.
The lack of Lorentz invariance even leads to contradictions.
Let $W$ be the total CM energy. In the CMS, one calculates for
the total multiplicity
$\av{n} = 2 \cdot (c+a \ln \frac{E_q}{m})$ with $E_q=W/2$.
In a frame where one quark takes almost all of the energy, and
the other one only very little, one gets a different value
$\av{n} = c+a \ln \frac{E_q}{m}$ with $E_q=W$

The discussion so far covered only the lowest order processes,
where hadron production is determined by
the assumption of limited transverse momenta and
longitudinal phase space as a simple hadronization model.
Though simple, it provides a good guideline for hadron production.
QCD radiation modifys this simple
picture and will be discussed next (section \ref{sn:meps}).
Afterwards improved hadronization models are introduced that do not
suffer from the problems with the
independent fragmentation model (section \ref{sn:hadron}).

  \section{Perturbative QCD Radiation   \label{sn:meps}}

The perturbative
part of QCD radiation is treated either with
fixed order matrix elements, or with parton showers, or
a combination of both.
It depends on the application which treatment
is most appropriate.
In general, the matrix element will become more important with
increasing hardness of the interaction.
Perturbative QCD makes predictions for {\it partonic} final states,
but observed
are {\it hadronic} final states.
To make contact with the experimentally accessible
world, hadronization has to be taken into account.
Models for the non-perturbative effects of hadronization
are discussed in section \ref{sn:hadron}.
Excellent reviews on perturbative
QCD evolution and on hadronization are
\cite{rev:sjoestrand,rev:webber}. Further information specific to
DIS can be found in \cite{mc:generators}.
We start the discussion of perturbative QCD with matrix elements
and parton showers. An alternative technique which is not based
on Feynman diagrams is the dipole radiation approach, to be discussed
afterwards.

\subsubsection{Matrix elements}

\begin{figure}[tbh]
   \centering
   \epsfig{file=sumdiag_tot.ps,
   width=4cm}
   \hspace{2cm}
   \epsfig{file=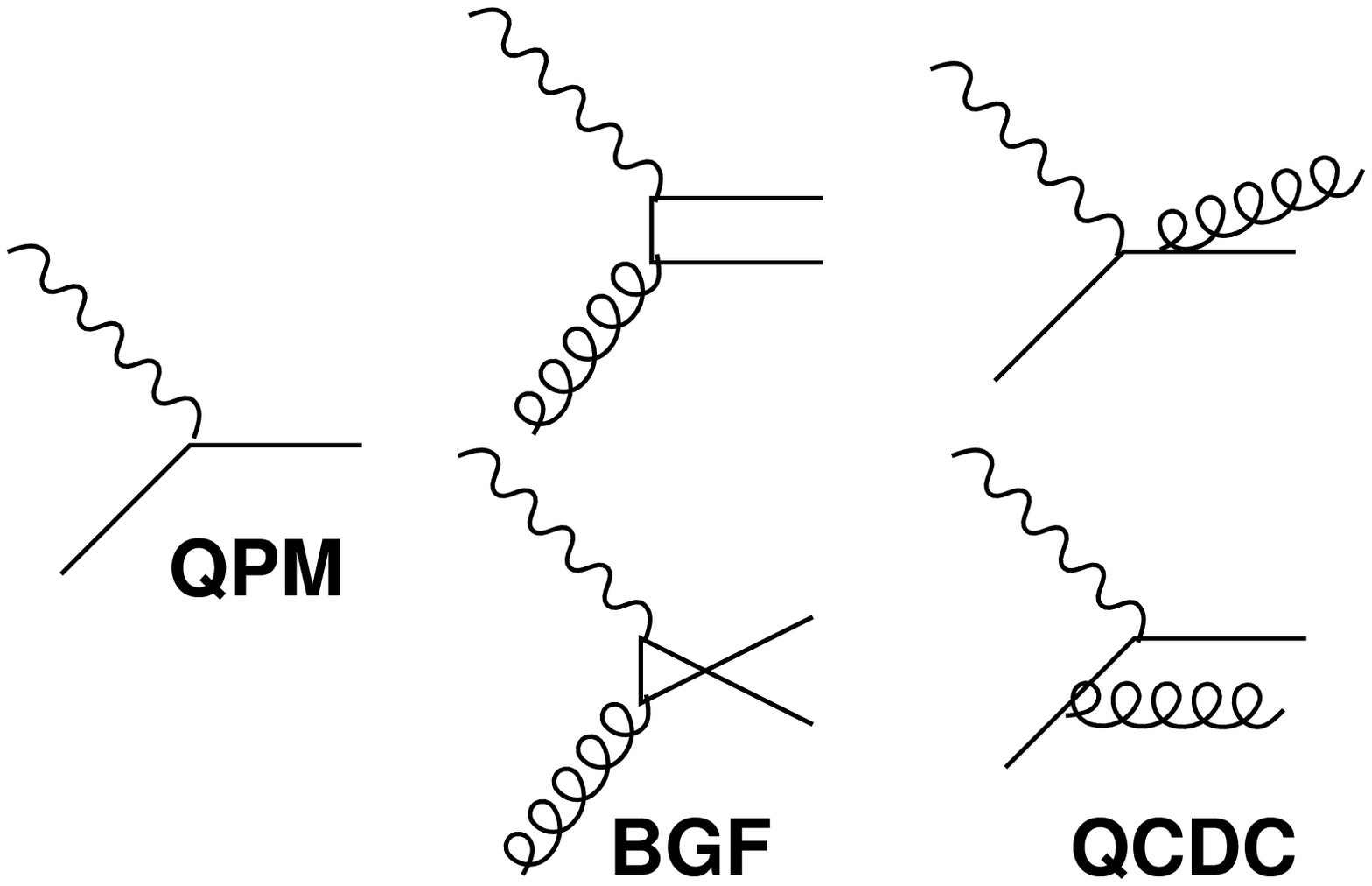,
   width=6cm}
   \scaption{Matrix elements in DIS. In leading order,
             scattering on a parton $i$ can proceed via
             the graph that corresponds to the
             simple quark parton model (QPM), or the QCD analogue
             of Compton scattering (QCDC), or the boson-gluon fusion
             graph (BGF) in case the incoming parton $i$ is a gluon.
             The graphs on the right hand are the LO realizations
             of the ``blob'' in the left graph.}
   \label{me}
\end{figure}

In fixed order perturbation theory,
the incoming parton flux is folded with
the matrix element \cite{th:mecalc} for
electron-parton scattering which leads to some final state parton
configuration.
In LO possible final state
configurations for $eq$ scattering are (apart from the scattered electron)
$q$ (QPM), $qg$ (QCDC),
and for electron gluon scattering (BGF) \qqbar~ (fig.~\ref{me}).
In NLO, one additional parton
can be emitted, and so on.

The calculations are ususally done numerically.
For state of the art
NLO calculations the programs
DISENT \cite{mc:disent}, MEPJET \cite{mc:mepjet} and
DISASTER++ \cite{mc:disaster} are available.
Originally such programs could only be used for jet analysis.
The present programs are more flexible and allow also calculations
of for example event shape variables.

\subsubsection{Leading Log parton showers}

A complete calculation to an order higher than NLO
appears prohibitive.
For many applications, where higher orders are important, the
parton shower ansatz is used.
For example, higher orders can be summed to all orders
in the leading log approximation.
In the DGLAP (leading $\log Q^2$) approximation, they result
in splitting functions which give
parton emission probabilities.

When a parton with 4-momentum $p$ radiates, it changes
its virtuality $T^2:=-p^2$. The evolution
parameter $t:=\ln (T^2/\Lambda^2)$ increases in the initial state
parton shower (space-like shower) by successive emissions,
and decreases in the final state parton shower (time-like shower).
The probability ${\cal P}$ that a branching $a\rightarrow bc$
will take place during a small change $\dif t$ is given by the evolution
equation \cite{th:dglap}
\begin{equation}
   \frac{\dd {\cal P}_{a \rightarrow bc}}{\dd t}
 = \int_0^1 \dd z \frac{\alpha_s(Q^2)}{2\pi} P_{a \rightarrow bc}(z).
\end{equation}
The functions $P_{a \rightarrow bc}(z)$ are just the Altarelli-Parisi
splitting functions in eq. \ref{eq:splitting},
$P_{a \rightarrow bc} = P_{ba}$.

Such calculations are usually done with Monte Carlo event generators,
where the parton shower evolution
is simulated step by step according to the emission probabilites,
until the whole event is generated.
Perturbative evolution is stopped
at some small scale $T_0^2$ of \order{1 \GeVsqx},
when it becomes unsafe to apply perturbation theory due to the growth
of \asx.

\begin{figure}[tbh]
   \centering
   \epsfig{file=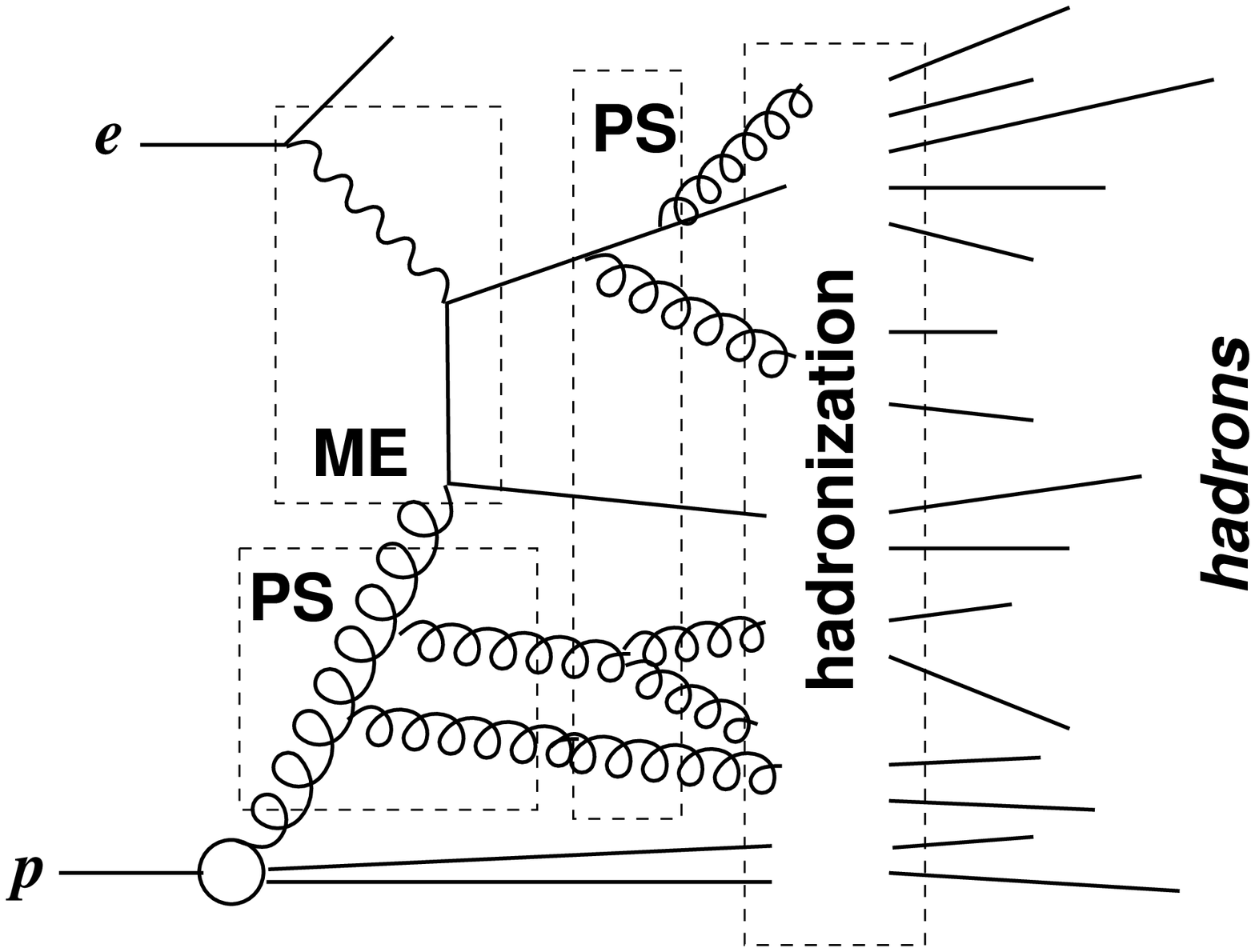,
   width=10cm,%
   bbllx=58pt,bblly=208pt,bburx=555pt,bbury=584pt}
   \scaption{Elements of an $ep$ event generator.
             As an example an event
             is shown with the BGF matrix element (ME).
             The initial and final state parton showers (PS) produce
             additional emissions. Finally, the partonic final state
             (the parton level) is hadronized to yield the observable
             hadrons (the hadron level).}
   \label{meps}
\end{figure}

When combined with the exact fixed order matrix element to take care
of very hard emissions that are not properly covered with leading logs,
the Monte Carlo event generators can be expected to provide a good
representation of what is
actually happening in DIS
events\footnote{
Remembering the double slit experiment, we should feel a bit
uneasy about a statement of what is actually happening.} (fig.~\ref{meps}).
Ambiguities result from the way
the matrix element and parton shower are combined (``matching''),
coherence effects are
treated, divergencies of the matrix element are cut-off,
and from other approximations in the implementation.
Unfortunately todays generators implement the matrix element and
parton showers only to LO.

Examples of DIS Monte Carlo generators that are based on leading log
DGLAP parton showers are
LEPTO \cite{mc:lepto}, HERWIG \cite{mc:herwig} and RAPGAP \cite{mc:rapgap}.
They are expected to be valid where the DGLAP approximation is valid.
As DGLAP based models, the initial state parton shower
is strongly ordered in \kt, increasing from the proton towards
the matrix element.

The leading log approximation (LLA) can also be applied directly for
specific observables, without utilizing an event generator.
Most relevant is the modified leading log approximation (MLLA)
which takes into account destructive interference between
soft gluons, in conjunction with the assumption that the resulting
parton spectra resemble the observable hadron spectra, up to
a constant factor (local parton hadron duality, LPHD).

\subsubsection{The colour dipole model}

\begin{figure}[tbh]
   \centering
\begin{picture}(0,0) \put(0,0){{\bf a)}} \end{picture}
   \epsfig{file=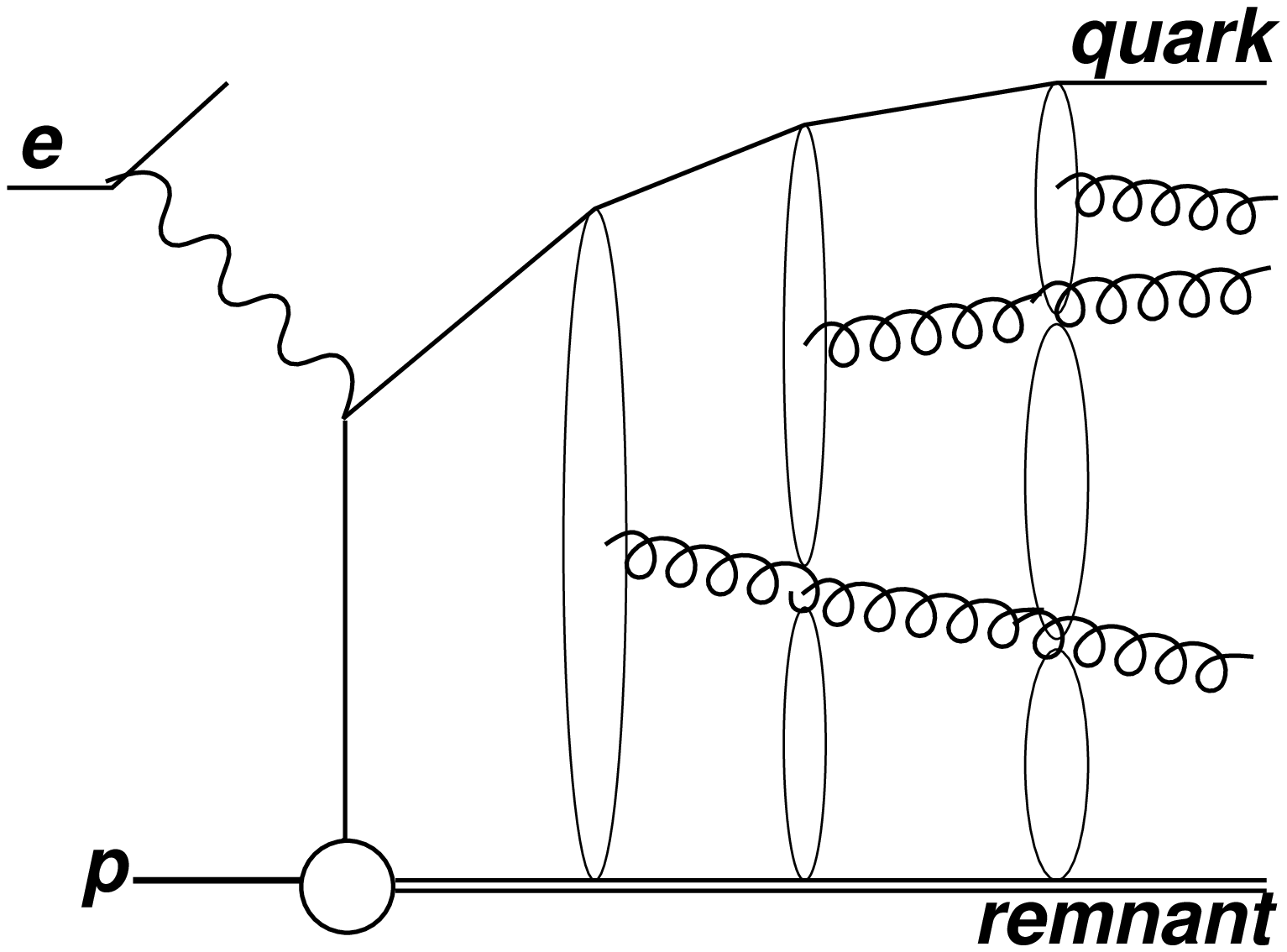,width=6cm}
   \hspace{1cm}
\begin{picture}(0,0) \put(0,0){{\bf b)}} \end{picture}
   \epsfig{file=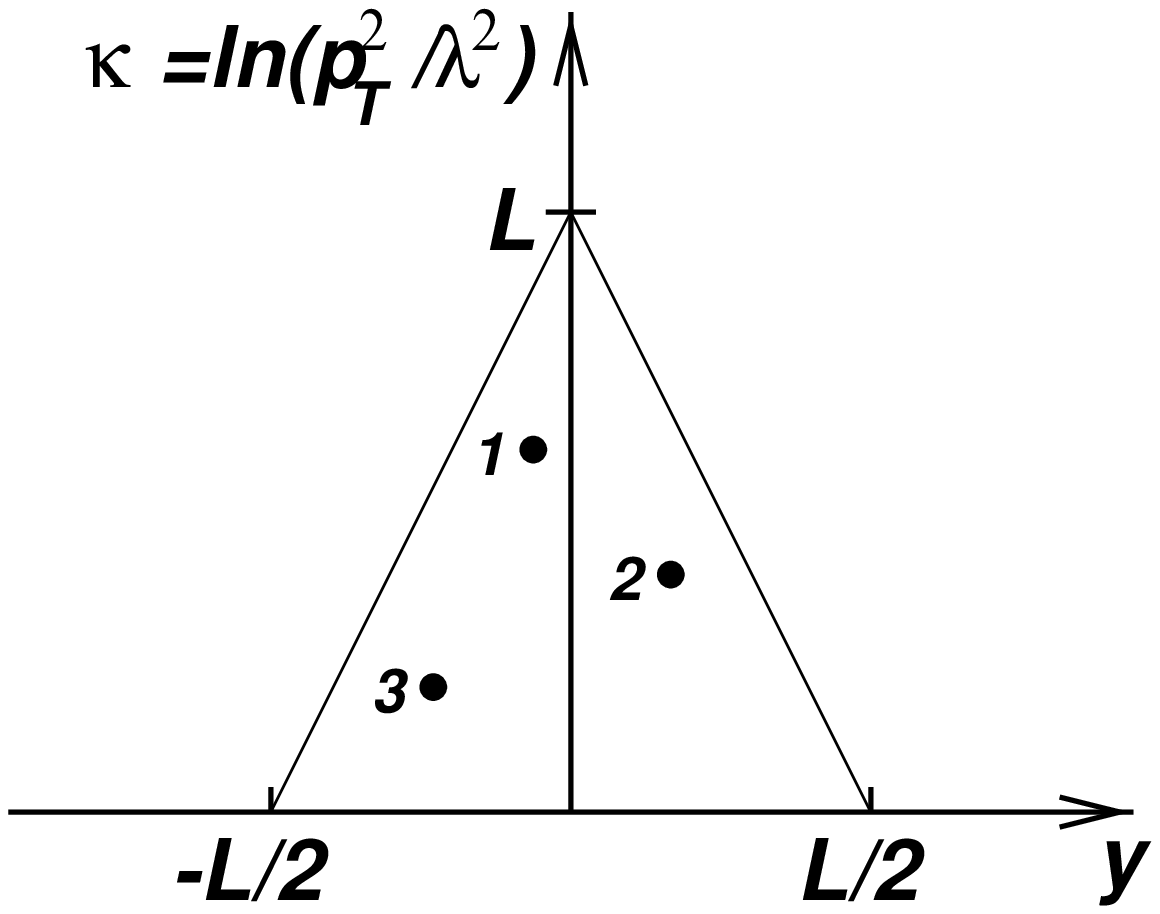,width=6cm}
   \scaption{
             {\bf a)}
             CDM in DIS: gluons are emitted
             from successively built colour dipoles.
             {\bf b)}
             The phase space for the emission of gluons with $p_T>\lambda$
             in the variables $\kappa=\ln(p_T^2/\lambda^2)$ and $y$.
             The phase space is bounded by
             the triangle with height $L=\ln(W^2/\lambda^2)$.
             Subsequent emissions have decreasing \pt, $p_{Ti+1}<p_{Ti}$,
             but no ordering in $y$. If the emissions are reordered
             in $y$, they lose the \pt ordering.
             }
   \label{triangle}
\end{figure}

Another type of parton shower model is the colour dipole model (CDM)
for QCD radiation \cite{mc:dipole}.
The colour charges of the scattered quark and the remnant are assumed
to form a colour dipole, from which gluons can be radiated
(fig.~\ref{triangle}a).
Subsequent gluon radiation emanates from
dipoles spanned between
the newly created colour charges and the others, and so on.
To good approximation it can be assumed that these dipoles radiate
independently.
The CDM uses
the LO cross section for the emission
of a gluon with transverse momentum \pt at rapidity $y$
in the soft gluon approximation \cite{books:halzen},
\begin{equation}
\dif \sigma = \frac{n_c \as}{2\pi} \frac{\dif p_T^2}{p_T^2} \dif y.
\end{equation}
The cross section is uniform in rapidity and $\ln p_T^2$.
Kinematically, the phase space is bounded by $|y|<\ln(W/p_T)$,
where $W$ is the total energy of the radiating system.
Gluons with \pt above a certain cut-off, $p_T>\lambda$, lie
inside the triangle in fig.~\ref{triangle}b.

In DIS one of the colour charges is not pointlike, the proton
remnant.
The suppression of radiation with short wavelength from
an extended source is taken into account with a special parameter
of \order{1 \fm} that describes the extendedness of the remnant
remnant\footnote{In the recent version also the colour charge of the
scattered quark is spread out, depending on the ``size'' ($\propto 1/Q$)
of the
final state quark that can be resolved with the photon of virtuality \Qsqx.}.
The CDM is implemented in the program
ARIADNE \cite{mc:ariadne}.
In ARIADNE the QCDC graphs are covered by dipole radiation
(and corrected to yield the exact LO matrix element contribution), but
for the BGF graph the matrix element is used. Further radiation
is then according to the CDM.

In contrast to the LL parton shower in the CDM
it is not possible to distinguish between initial and final state
radiation, or to reconstruct an ``evolution path'' for the
parton in the proton that is hit by the virtual photon.
Kinematically, the \kt phase space for radiation is bounded by
the \kt of the previous emission. In rapidity the emission
probability is uniform. Therefore the final gluon configuration
is not ordered in $k_T$, if one arranges them according to increasing rapidity
(see fig. \ref{triangle}b). Rapidity is directly related to $\ln x_i$
for fixed $W$ and $k_{Ti}$ (see section \ref{sn:lowxi}).
In that respect the CDM is similar to what can be expected from BFKL
evolution \cite{mc:bfklcdm}, though perhaps in a somewhat unconventional
fashion \cite{lowx:rathsman}.
Whereas the LL parton showers are
connected with the intuitive picture of an evolution path,
it is probably fair to say that the significance of the dipole model
is not yet understood very well.
We shall see that this model provides effortlessly overall the best
description of most final state data.
This is probably the main
reason why it is being discussed, whether or not it has a deeper
physical justification not yet appreciated by the community.

\subsubsection{The Linked Dipole Chain Model - LDC}

The CCFM approach for the hadronic final state has been reformulated
in a picture similar to the CDM, giving rise to the Linked Dipole
Chain model (LDC) \cite{th:ldc}.
This model is particularly interesting, as it should converge
to the DGLAP and BFKL predictions in their respective regions
of validity, and handle properly cancellations between
real and virtual emissions for the final state.
In the LDC also the case is treated where the largest \pt of
the process is not attached to the virtual photon vertex.
Some promising results from a Monte Carlo implementation  of the
LDC have been obtained \cite{mc:ldc}, but the program is not
yet publicly available.

\subsubsection{Coherence}

Quantum mechanical interference is an
important issue for the hadronic final state.
With matrix elements that sum and square amplitudes
interference effects are taken into
account naturally.
Coherence effects in parton showers that rely on emission
probabilities, not amplitudes, require special attention.
One may distinguish
a) interference between initial and final state radiation; and
b) interference between partons emitted either in the
   initial state or in the final state parton shower.

In the CDM there is no distinction between initial and final
state radiation, so in the dipole approximation problem a) does not arise.
Also soft gluon interference is automatically taken into account by
the dipole mechanism.
In other generators
such interference effects can be realized by phase space restrictions
for parton emission.
For example, the effect of
destructive interference between
subsequent emissions in a parton shower can be approximated by
imposing a posteriori ``angular ordering'': only emissions
are allowed with a smaller opening angle than the previous one,
$\theta_{i+1}<\theta_i$.

\begin{figure}[htb]
   \centering
\begin{picture}(0,0) \put(0,0){{\bf a)}} \end{picture}
   \epsfig{file=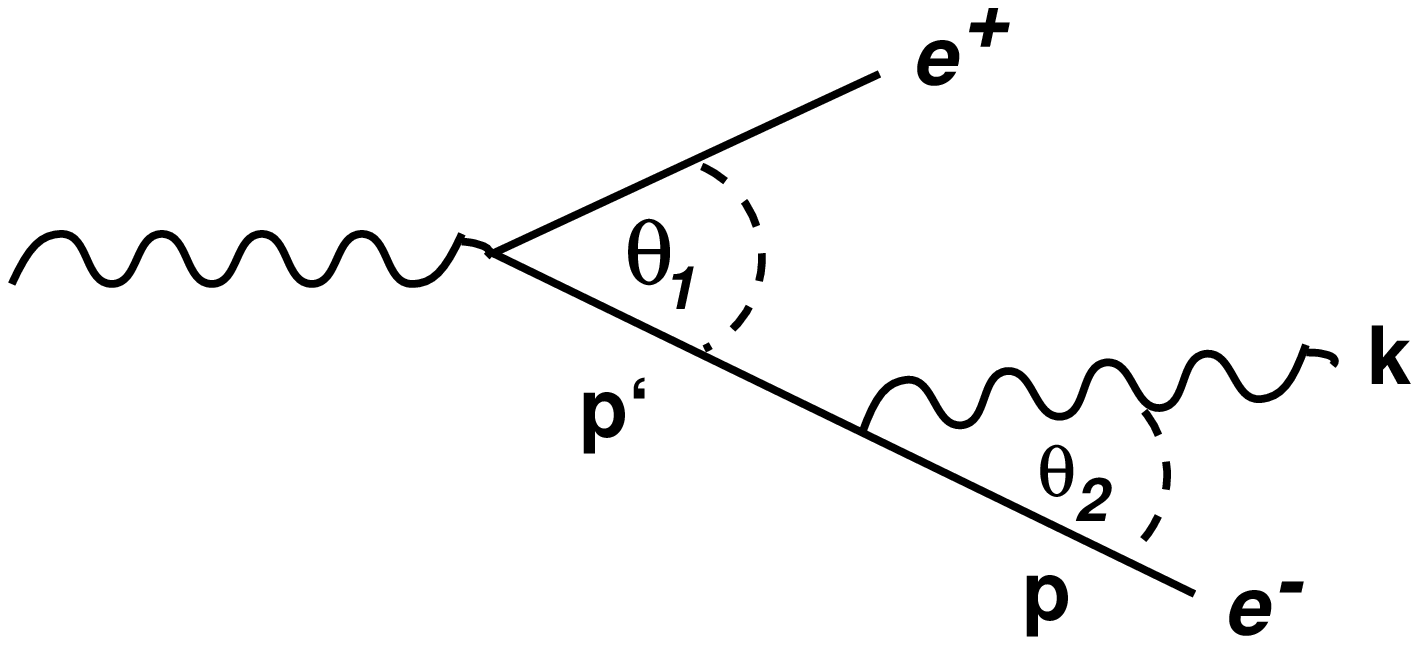,width=6cm}
   \hspace{1.0cm}
\begin{picture}(0,0) \put(0,0){{\bf b)}} \end{picture}
   \epsfig{file=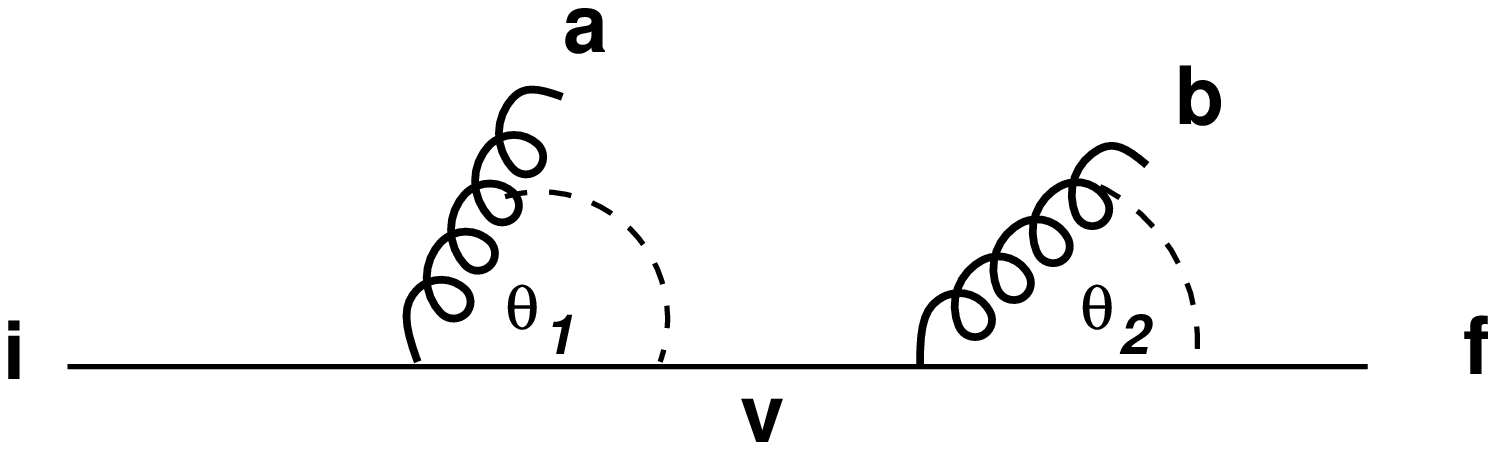,width=7cm}
   \scaption{{\bf a)} Angular ordering in an electromagnetic cascade.
             The 4-momenta of the outgoing and intermediate
             electron are $p=(E,\vec{p})$, $p'=(E',\vec{p}')$,
             and of the outgoing photon $k=(E_\gamma,\vec{k})$.
             {\bf b)} Angular ordering in a QCD cascade.}
   \label{coh}
\end{figure}

The physical reason for this condition is that large
wavelength quanta cannot be emitted from dipole sources with
small transverse dimensions.
This can be seen easily for \epem~ cascades \cite{books:dokshitzer}.
Consider the example fig.~\ref{coh}a, where
a virtual photon branches into an \epem~ pair, and the
$e^-$ radiates a photon. The formation time of the
final $ee\gamma$ state can be estimated
as the lifetime of the intermediate $e^-$ from its off-shellness $M$,
$\tau=\gamma \cdot \tau_0 = E'/M \cdot 1/M$.
We have
$M^2=(p+k)^2\approx 2EE_\gamma (1-\cos\theta_2)$, and
$1-\cos\theta_2 \approx \theta_2^2/2$ for small angles,
Since for soft photons, $E_\gamma \ll E'\approx E$,
$\tau \approx 1/(E_\gamma \theta_2^2)$. During this time
the \epem~ pair have separated a transverse distance
$\delta r \approx \tau \theta_1$. The transverse wavelength
of the emitted photon is $\lambda_T = 1/\kt$ with $\kt=E_\gamma\theta_2$.
If $\lambda_T$ is larger than $\delta r$, the photon cannot
resolve the individual charge of the $e^-$, it rather sees the
combined charge of the \epem~ pair, which is 0. For allowed
emissions it follows $\delta r > \lambda_T$, or $\theta_1>\theta_2$,
angular ordering.

In QCD the situation is a bit more complicated, because
the incoming parton is colour charged (fig.~\ref{coh}b).
Parton $b$ cannot
be radiated incoherently from parton $v$ when $\theta_2>\theta_1$.
In that case
it would see the combined colour charge of $a$ and
$v$, which is the colour charge of the incoming parton $i$.
Parton $b$ with $\theta_2>\theta_1$
can therefore be treated as effectively being radiated from
$i$ before $a$ branches off, thus restoring angular ordering.

  \section{Hadronization        \label{sn:hadron}} 
Experimentalists follow two different philosophies how to
prepare their hadronic final state data.
\begin{itemize}
\item
In one approach, the data are corrected just for detector effects in
order to remove any experimental bias and be able to compare the
data to other experiments.
The data are corrected ``to the hadron level'', they represent a measurement
of observable hadrons.
The data can also be compared to
QCD predictions that include hadronization effects,
but not directly to perturbative QCD calculations
(unless one is interested to study the difference between
observable hadrons and partonic calculations).
\item
In the other approach, a hadronization model is used to
correct the data for hadronization effects back ``to the parton level''.
The data can then be compared
directly to perturbative QCD calculations.
A problem of principle arises: partons are not observable, so
strictly speaking an observable ``at the parton level'' does not exist.
With great care one may proceed nevertheless.
One has to specify
the parton level in perturbative QCD
(like NLO, LLA, including cut-offs etc.),
and correct the data for all other effects beyond that:
higher order corrections plus hadronization effects.
The observable thus becomes scheme dependent.
There is another price to pay: the data have picked up a theoretical
bias due to the assumptions on higher order and hadronization effects.
In practice this bias is studied by comparing different Monte Carlo
generators. The encountered differences are treated as systematic error
of the measurement.

\end{itemize}

We summarize the main approaches to hadronization, some of which
have already been mentioned.
It has to be emphasized that the boundary between perturbative
QCD evolution and hadronization is not uniquely defined.
Ideally, predictions for final state observables would not
depend on how the boundary is defined, for example on the
lower cut-off for virtualities in the parton shower.

We start with hadronization models that are implemented in
Monte Carlo generators (independent fragmentation, string and
cluster model).
Other approaches that model only specific
aspects of hadron production are discussed afterwards.

\subsubsection{Independent fragmentation \cite{th:ff}}

From the fragmenting parton mesons are produced,
carrying a certain fraction $z$
of the original energy, see section \ref{sn:fragf} and fig.~\ref{indep}.
The branching is repeated with the remaining
energy until all the energy is used up.
In the physical picture for independent quark fragmentation
the quark forms a meson with an antiquark from a \qqbar~ pair,
with the remaining quark continuing fragmentation.
It is assumed that
the distribution of $z$ is energy independent, leading to scaling
fragmentation functions.
Furthermore,
all partons
of an event fragment independently. That leads to inconsistencies like
non-conservation of energy-momentum,
that have to be cured by hand.

Transverse momentum components are assumed
to be Gaussian
distributed\footnote{
The transverse momentum \pt
which a primary hadron receives from the fragmentation process
can be described by Gaussians in $p_x$ and $p_y$ with widths $\sigma$,
\begin{equation}
 \frac{\dd^2 n}{\dd p_x \dd p_y} = \frac{1}{\sigma \sqrt{2\pi}}
                                    \exp\left(\frac{-p_x^2}{2\sigma^2}\right)
                                    \frac{1}{\sigma \sqrt{2\pi}}
                                    \exp\left(\frac{-p_y^2}{2\sigma^2}\right).
\label{eq:ptgauss}
\end{equation}
This leads to an exponentially falling distribution in $p_T^2$
with $\av{p_T^2}=2\sigma^2$,
\begin{equation}
 \frac{\dd n}{\dd p_T^2} = \frac{1}{2\sigma^2}
                 \exp\left(\frac{-p_T^2}{2\sigma^2}\right).
\end{equation}
It follows,
that the distribution in $p_T$ is neither Gaussian nor exponential,
\begin{equation}
 \frac{\dd n}{\dd p_T} = \frac{p_T}{\sigma^2}
                   \exp\left(\frac{-p_T^2}{2\sigma^2}\right),
\end{equation}
with $\av{p_T}=\sqrt{\frac{\pi}{2}}\sigma
   = \frac{\sqrt{\pi}}{2}\sqrt{\av{p_T^2}}$.
The default parameter in
the Monte Carlo program JETSET \cite{mc:jetset} is $\sigma=0.36\GeV$.
}
\cite{mc:bo1},
which results in an
exponentially falling distribution in $p_T^2$.
From the parametrizations used in fragmentation models
\cite{mc:bo1,mc:jetset},
$\av{\ptfrag^2}\approx(0.3\GeV)^2$ is obtained for {\it primary} hadrons.

\subsubsection{String fragmentation \cite{mc:string}}

\begin{figure}[tbh]
   \centering
\begin{picture}(0,0) \put(0,0){{\bf a)}} \end{picture}
   \epsfig{file=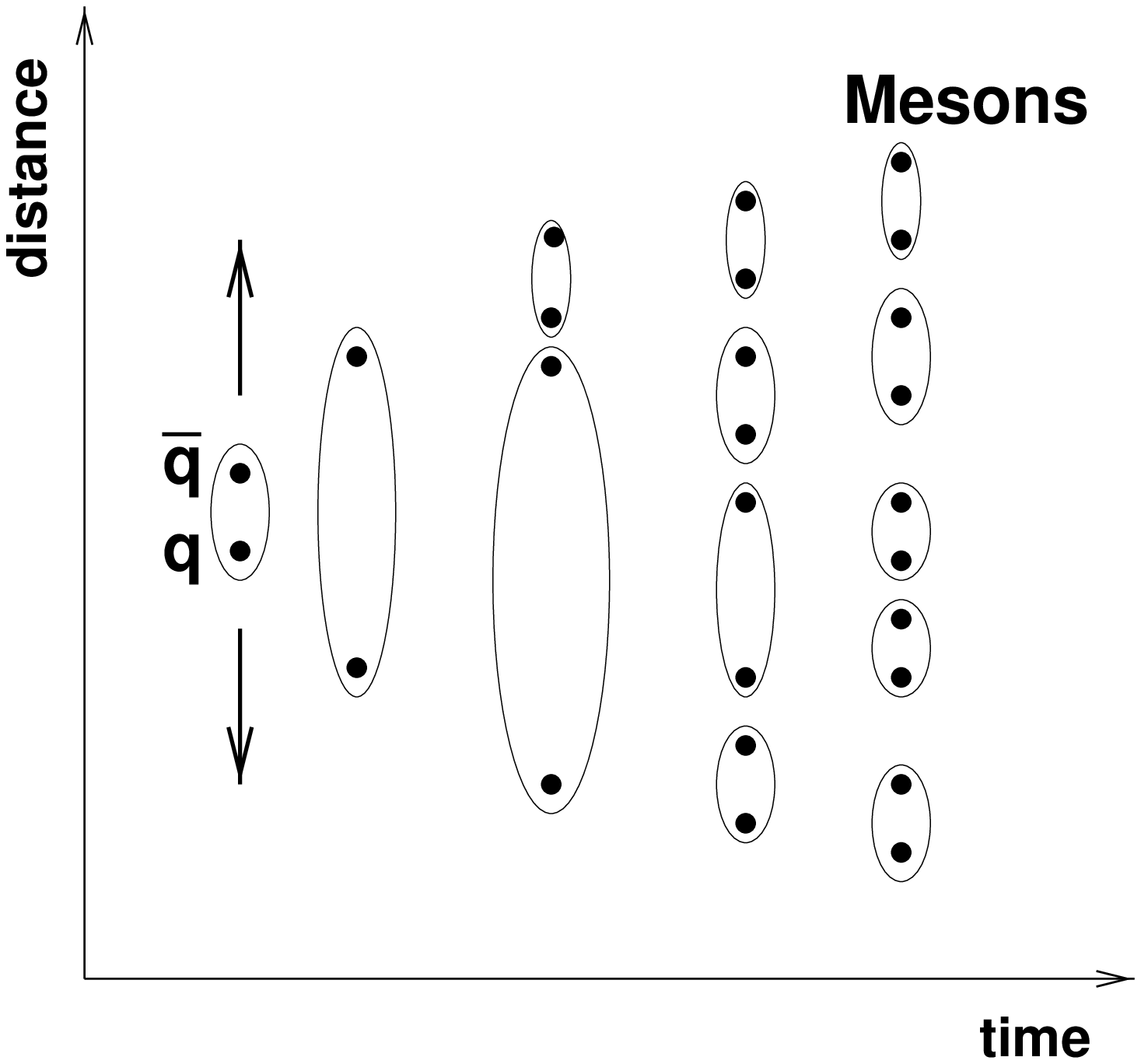,%
          width=5cm}
   \hspace{1.0cm}
\begin{picture}(0,0) \put(0,0){{\bf b)}} \end{picture}
   \hspace{0.5cm}
   \epsfig{file=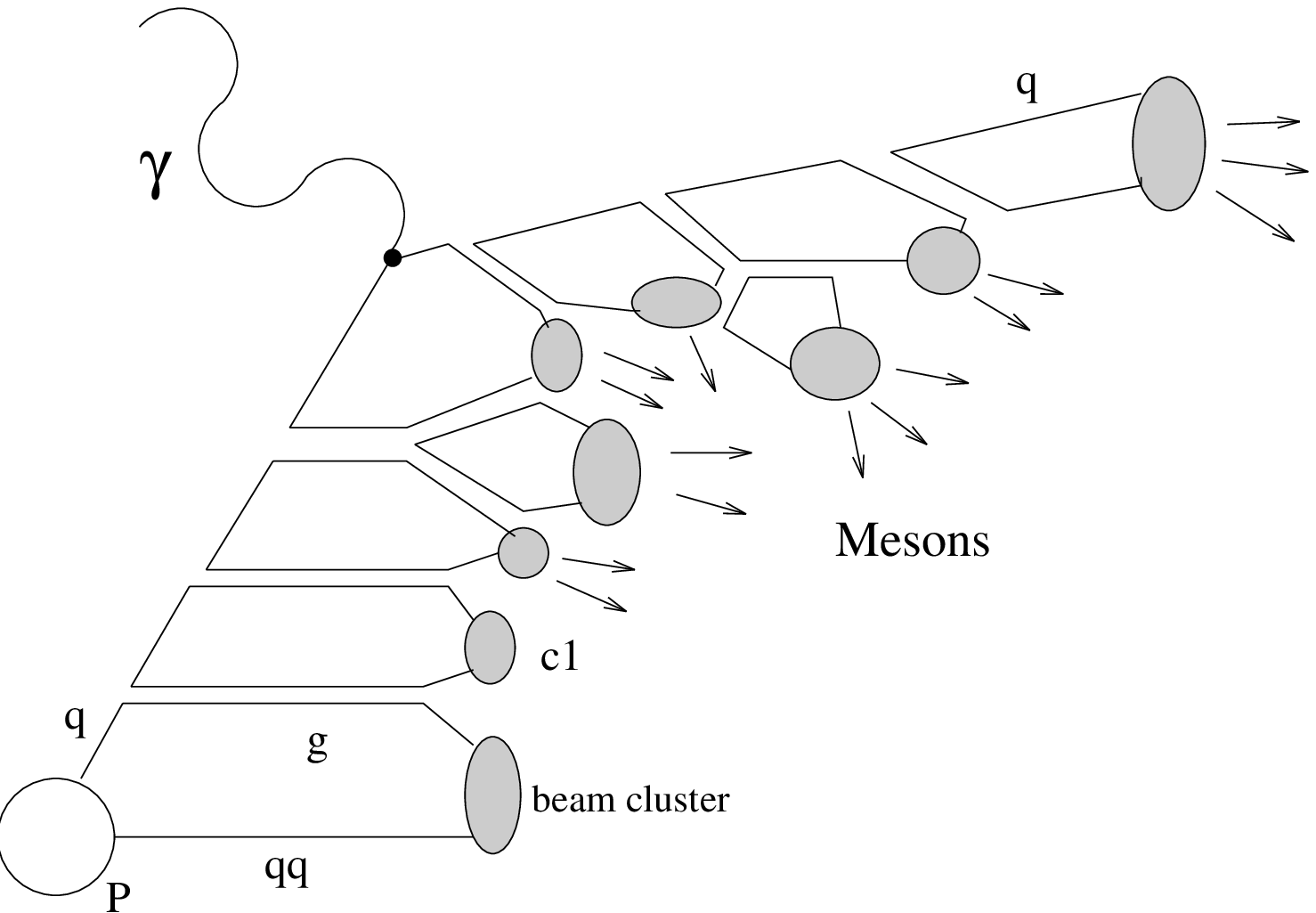,%
          width=7cm}
   \scaption{
             {\bf a)} String fragmentation. \qqbar~ pairs are created when
                the stretched string between the separating \qqbar~ pair
                breaks up. Mesons are formed when the string energy
                is too small to create further \qqbar~ pairs.
             {\bf b)}
             Cluster fragmentation. Gluons are represented by double lines,
             corresponding to the colour and anti-colour they carry.
             After the perturbative evolution gluons are split
             non-perturbatively into \qqbar~ pairs,
             from which colour neutral clusters are
             formed. They decay into mesons.
             }
   \label{fragmodel}
\end{figure}

In the Lund string model, a quark and an antiquark moving apart stretch a
colour field between them.
The field is thought to be string-like
with constant energy density per unit length of \order{1 \GeV / {\rm fm}},
and with transverse dimension of \order{1 \fm}.
When the stored energy becomes large enough, the string can break up
and create a \qqbar~ pair from the vacuum (fig.~\ref{fragmodel}a).
The newly created $q$ and
$\ol{q}$ terminate the loose string ends, such that the new string
pieces are themselves colour neutral.
The process is iterated with the new strings until all the available
energy is used up.
Transverse momentum components result from a
tunneling process; they are parametrized
as in eq. \ref{eq:ptgauss}.
The string model gives rise to scaling fragmentation functions,
a rapidity plateau with uniform particle density, and a logarithmic
increase of particle multiplicity $\av{n} = c+a \ln W$, where
$W$ is the string invariant mass.

String fragmentation can be formulated in a covariant
way, it does not suffer from the mentioned shortcomings of independent
fragmentation.
Furthermore, some quantum mechanical interference
effects in gluon radiation are taken into account with string fragmentation,
due to the way gluons are treated.
Carrying two colour charges, gluons are always the endpoints of two strings.
In a $q\ol{q}g$ configuration the gluon is realized as a kink
in the colour connection between the \qqbar~ pair.
Destructive gluon interference leads to a suppression
of radiation (and finally hadron production)
in the region opposite to the gluon.
Because in the string model there is no string
spanned directly between the $q$ and the $\ol{q}$
this ``string effect'' \cite{o:stringeffect}
appears naturally in the Lund model.

\subsubsection{Cluster fragmentation \cite{mc:cluster}}

According to the idea of ``preconfinement'' \cite{th:preconfinement},
colour connected partons tend to be close in phase space
towards the end of the perturbative phase.
This property is exploited in the cluster fragmentation model.
After the perturbative evolution all gluons
are forced to split into \qqbar~ pairs.
Colour connected
\qqbar~ pairs are combined into
low mass colour neutral clusters, which
decay into mesons according to phase space (fig.~\ref{fragmodel}).
Due to the small cluster mass, relatively few parameters
are needed to describe the cluster decay.
Nevertheless, large mass clusters may occur; they are split
longitudinally in a fashion similar to the string model before
they are allowed to decay.

\subsubsection{Local Parton Hadron Duality (LPHD)}

The hypothesis of local parton-hadron duality (LPHD) \cite{th:lphd}
is made
to allow predictions for hadron spectra from perturbatively calculated
parton spectra.
It states that
any hadron cross section depending on a quantity $\zeta$ is
related to the corresponding partonic cross section simply
by a constant factor,
$\dif\sigma_h/\dif \zeta_h = c \cdot \dif\sigma_p/\dif \zeta_p$.
The idea behind this seemingly surprising ansatz is that if
perturbative evolution is used for multi-parton emissions
down to a very low scale, there is not much energy left
for hadronization to change the shape of the distribution.
This hypothesis could be applied successfully to a variety of
phenomena, but one should keep its limitations in mind \cite{th:ochs}.
For
example, when a quark and antiquark separate and do not radiate
any gluons
(unlikely, but possible), LPHD would predict a gap in rapidity
in between, devoid of any hadrons.
A more physical assumption is that the
colour field stretched between quark and antiquark leads
to the production of hadrons filling the gap.
At HERA, the LPHD hypothesis has been applied to charged particle
spectra, see sections \ref{sn:multi} and \ref{sn:xp}.

\subsubsection{Fragmentation functions}

\begin{figure}[tbh]
   \centering
   \epsfig{file=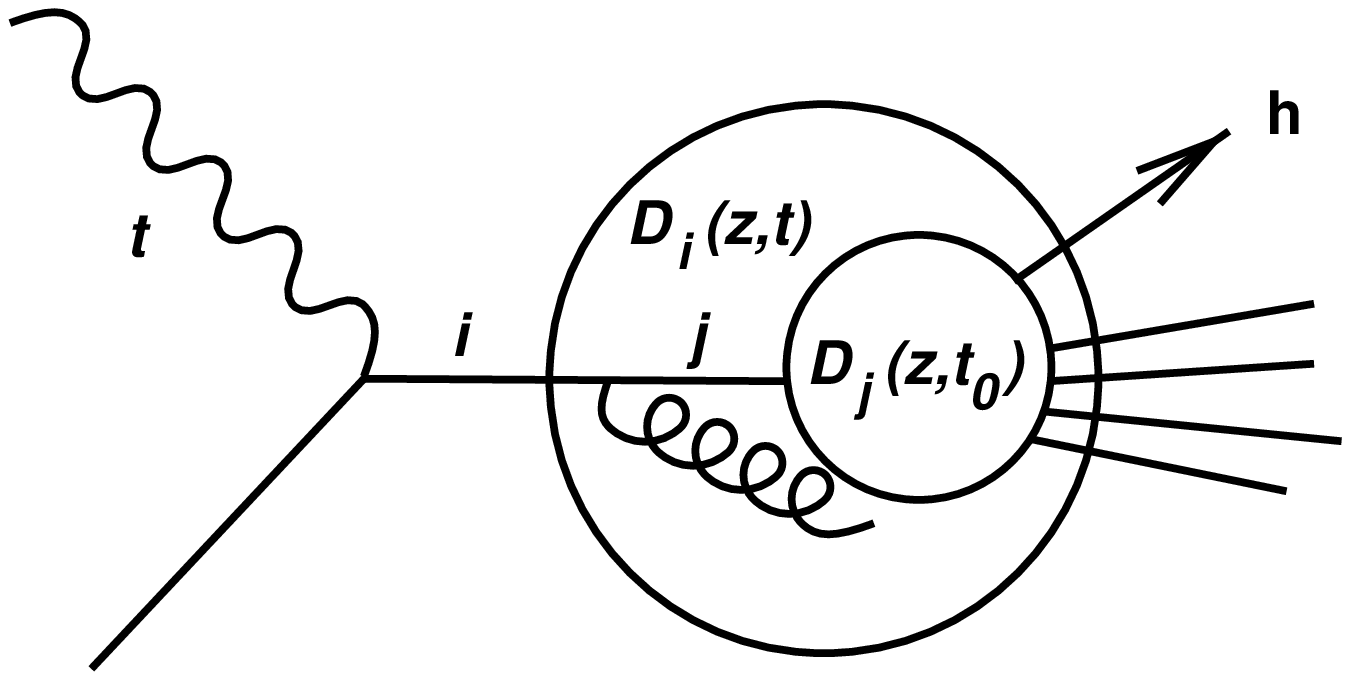,%
    width=8cm}
   \scaption{Scaling violations of fragmentation functions
             In this case the scale is given by the virtuality
             of the photon, $t=Q^2$.}
   \label{frag}
\end{figure}

The number density to find a hadron $h$ with energy fraction
$z:=E/E_q$ from the fragmentation of a quark $q$ is described by a
fragmentation function $D_q^h(z)$, and similarly for a gluon $g$.
They scale approximately, that is they depend only on the fractional
hadron energy and not on the quark energy \cite{th:fscal}.
QCD effects like gluon radiation off the quark
make the fragmentation functions
scale dependent: $D_q^h(z) \rightarrow D_q^h(z,t)$, where the scale $t$ is
provided by the hard process (fig.\ref{frag}).
While the fragmentation functions
themselves are not calculable in perturbative QCD,
their scale dependence can be calculated in the same way as the
scaling violations of the structure functions in DIS \cite{books:ellis}.

The fragmentation functions
obey an evolution equation
\begin{equation}
  \frac{ \partial D_{h/i}(z,\mu^2_D) }{\partial \ln \mu^2_D}
   = \sum_j \frac{ \alpha_s(\mu^2_D)}{2\pi}
    \int_z^1 \frac{\dif u}{u} P_{ji}(u,\alpha_s(\mu^2_D))
    D_{h/j}(\frac{z}{u},\mu^2_D).
\end{equation}
$t=\mu_D^2$ is the factorization scale between the parton
production process and the parton fragmentation process.
The splitting functions $P_{ji}$
for branchings $i\rightarrow j$ are in leading order
the same as for the parton densities in DIS, eq. \ref{eq:splitting},
but differ in higher order \cite{books:ellis}.
They have been calculated up to
next-to-leading order \cite{th:fragnlo}.

The fragmentation functions are universal, that is
process independent. For example, quark fragmentation functions
measured in $\epem \rightarrow {\rm hadrons}$
can be used to predict the hadron
distribution in the quark jet for $ep \rightarrow e^\prime q X$.

\subsubsection{High energy and power corrections}

Due to the limited \pt generated in hadronization,
it appears plausible that with increasing energy $Q$
the relative importance of hadronization effects diminishes for
suitably chosen observables.
It is expected that for infrared safe quantities
(see section \ref{sn:finst})
like
jet observables and some event shape variables
their perturbative prediction
approaches the value that is observable with hadrons.
The difference, due to higher orders and hadronization,
should decrease like an inverse power of the energy
$\propto (1/Q)^a$, a ``power correction'' \cite{th:powercorr}.
Power corrections have been investigated for thrust and other
event shape variables at HERA, see section \ref{sn:shapes}.

  \section{Rapidity Gaps  \label{sn:gaps}}       At HERA there exists a class of events
known as ``rapidity gap events''
\cite{z:gap1,z:gap2,h1:gap1,h1:gap2}.
They have given rise to tremendous activity, which would
deserve a review in its own right (recent summaries are
\cite{rev:diffr}).
Here as much is covered as is relevant for our purpose.

\begin{figure}[tbh]
   \centering
\begin{picture}(0,0) \put(0,0){{\bf a)}} \end{picture}
   \epsfig{file=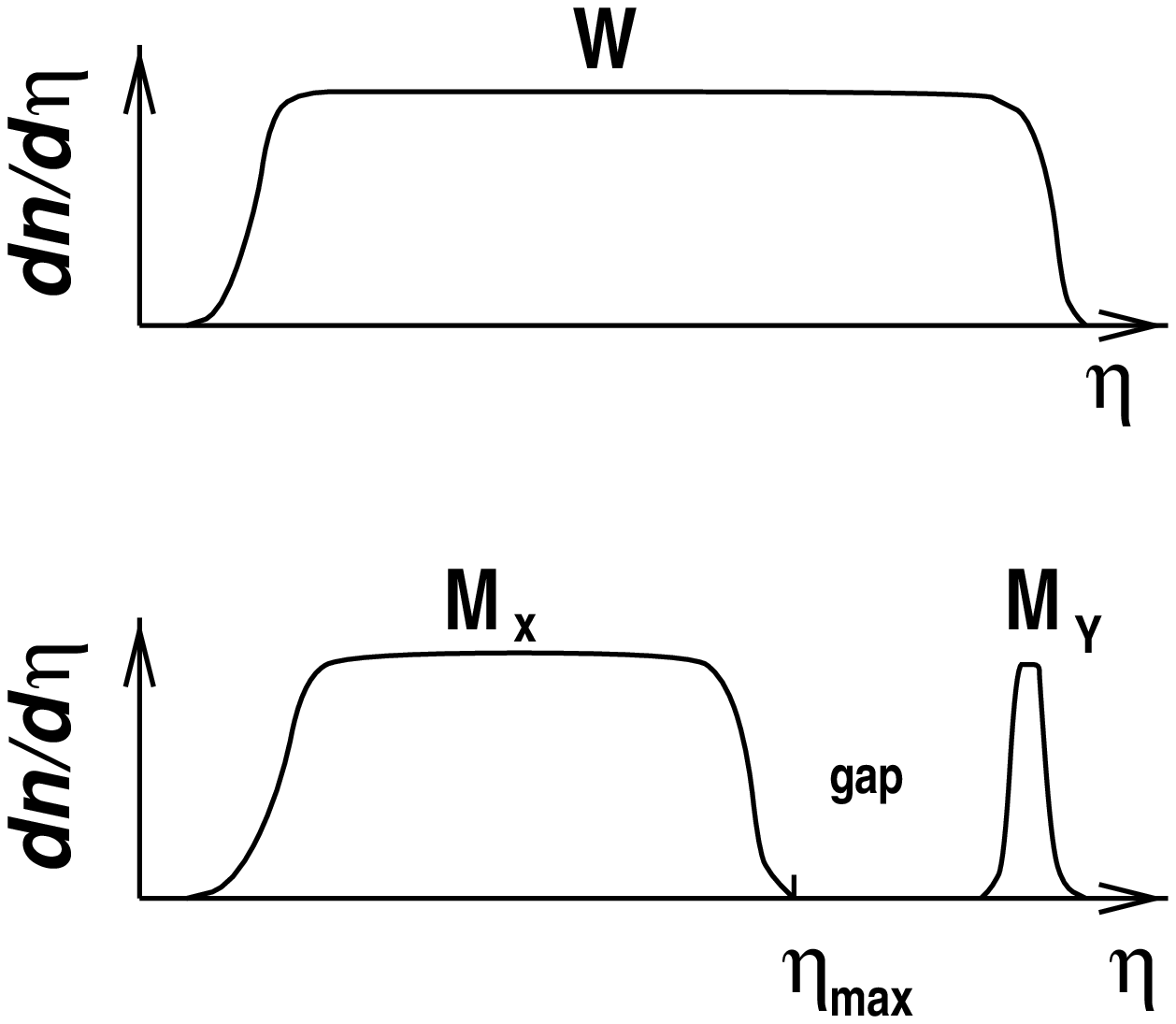,
   width=5cm}
   \hspace{1cm}
\begin{picture}(0,0) \put(0,0){{\bf b)}} \end{picture}
   \epsfig{file=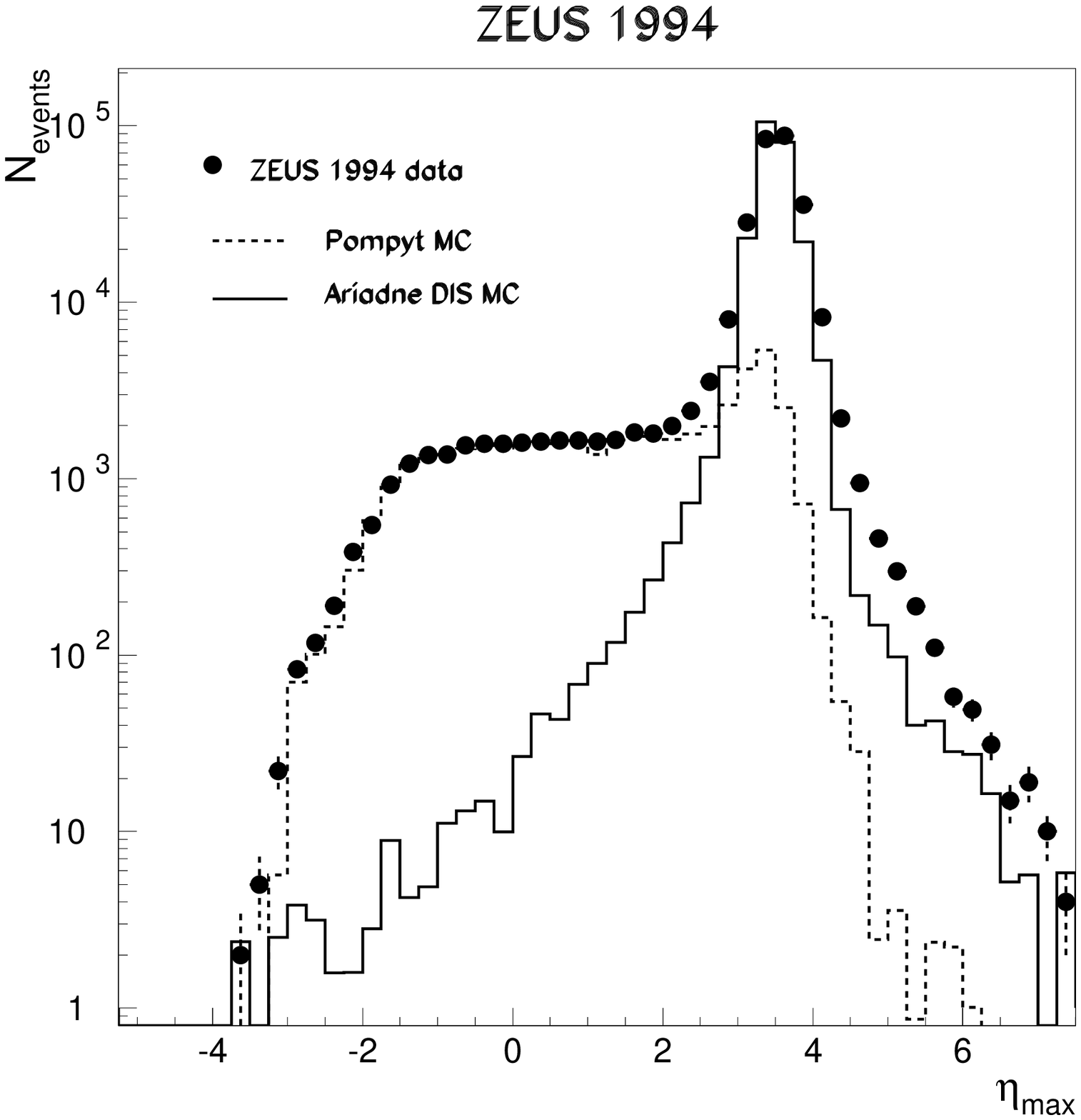,
   width=7.5cm, bbllx=21pt,bblly=143pt,bburx=536pt,bbury=687pt,clip=}
   \scaption{   {\bf a)}
                Rapidity distribution for normal events (top)
                and for rapidity gap events (bottom).
                $M_X$ and $M_Y$ are the invariant masses
                of the two systems $X$ and $Y$ separated by the largest
                gap in the event. The system $Y$ is the proton or its
                remnant.
                The proton direction is to the right.
                $\eta_{\rm max}$ is the
                forward  rapidity edge of the system $X$.
                A large gap implies a small mass $M_X$.
             {\bf b)}
                $\eta_{\rm max}$ distribution from ZEUS (uncorrected)
                in the laboratory system.
                The data are compared to a normal DIS model
                (ARIADNE \cite{mc:ariadne}) and to a model for
                rapidity gap events based on Pomeron exchange
                (POMPYT \cite{mc:pompyt}).
                For most events  $\eta_{\rm max}\approx 3.5$, given by
                the forward edge of the main calorimeter. Rapidity gap events
                are found in the tail at small $\eta_{\rm max}$.
             }
   \label{etamax}
\end{figure}

Rapidity gap events exhibit
a sizeable region of rapidity without any hadronic
activity -- the rapidity gap, see fig.~\ref{etamax}.
In normal events a colour field forms
between the scattered quark and the proton remnant, leading
to a homogeneous distribution of hadrons populating the
available rapidity range.
Large
gap sizes between neighbouring
hadrons from normal fragmentation
are expected to be unlikely (though possible).
Large gaps can be produced however in processes where
two systems are produced that are not colour connected.

\begin{figure}[tbh]
   \centering
\begin{picture}(0,0) \put(0,0){{\bf a)}} \end{picture}
   \hspace{1cm}
   \epsfig{file=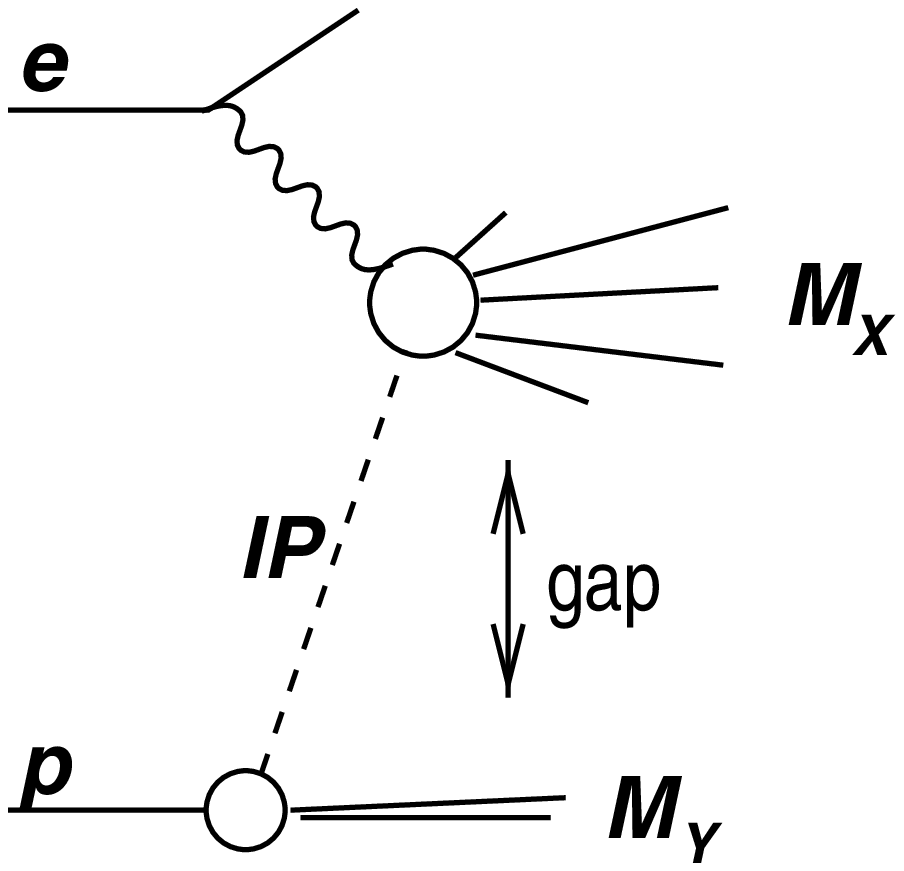,
   width=4cm}
   \hspace{2cm}
\begin{picture}(0,0) \put(0,0){{\bf b)}} \end{picture}
   \epsfig{file=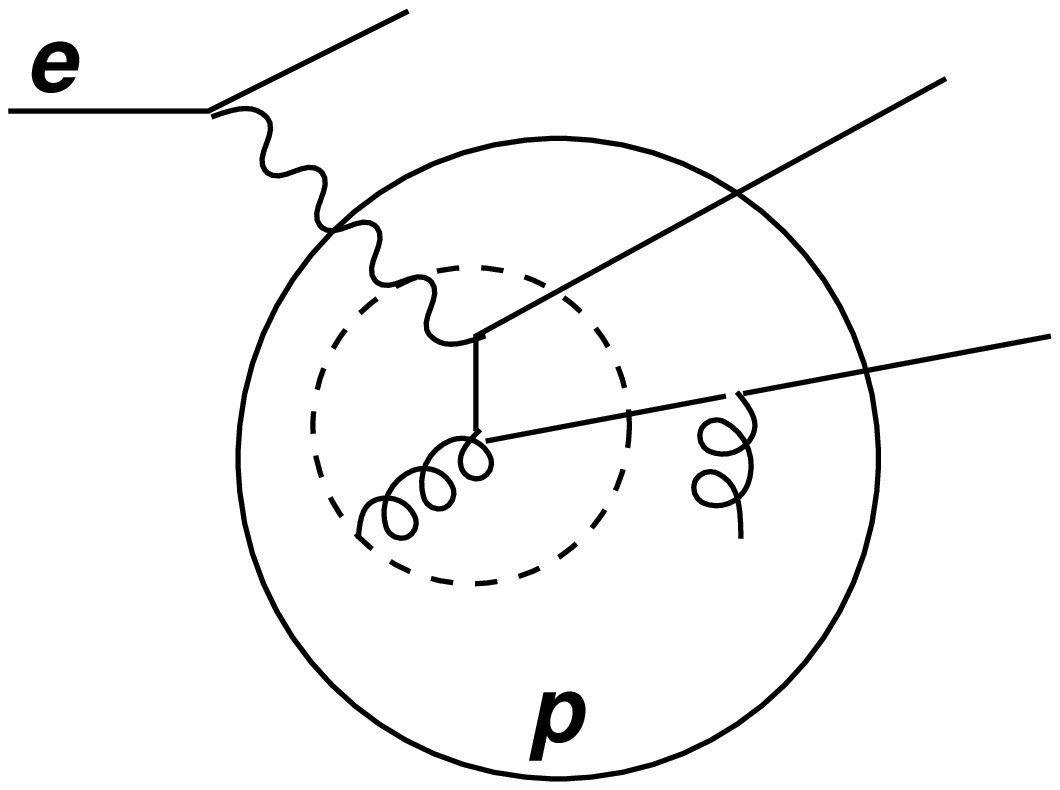,
   width=5cm}
   \scaption{
             a) DIS with Pomeron exchange. The proton momentum fraction
                carried by the Pomeron, \xpom,  can be reconstructed
                by $\xpom=(M_X^2+Q^2)x/Q^2$. The variable
                $\beta:=x/\xpom=Q^2/(M_X^2+Q^2)$ gives the
                Pomeron momentum fraction carried by the parton
                that is hit by the virtual photon.
             b) DIS with colour rotation. After the hard interaction
                (BGF, broken circle) a soft gluon is interchanged with
                the proton.
             }
   \label{gaps}
\end{figure}

Events with large rapidity gaps had been anticipated for
HERA from processes with Pomeron exchange \cite{th:predgap},
though the large rate came as a surprise: about 10\% of all DIS
events have a large rapidity gap\footnote{Often these events are
called ``diffractive'', or being due to a diffractive process.
The meaning of these terms is not yet uniquely defined amongst
the experiments and theorists.
We therefore prefer
the term ``large rapidity gap'', because the existence
of a gap of a certain size is well defined event by event.}.
In the Ingelman-Schlein model \cite{th:is} a Pomeron is exchanged
in the $t$ channel between the proton and the virtual photon,
see fig.~\ref{gaps}a.
The proton may be left intact, or fragment into a low mass system
$M_Y\lesssim 1.6 \GeV$.
In the Ingelman-Schlein model, the
virtual photon interacts with a parton in the Pomeron,
thus
probing the structure
of the exchanged Pomeron.
The resulting system with mass $M_X$ is not colour connected
with the proton remnant system, because the Pomeron carries no colour.
A rapidity gap between the two systems is the consequence.

Suppose the
above picture to be valid, the {\it partonic structure} of the
elusive Pomeron can be
probed! In fact, the evaluation of the rate of rapidity gap events
suggests that the exchanged object is predominantly gluonic,
and that most of the Pomeron momentum is carried by one
single gluon \cite{rev:diffr}.
The analyses suggest furthermore that
the Pomeron intercept in DIS is somewhat larger than expected from
the soft Pomeron, and possibly increasing with \Qsq \cite{rev:diffr}.

\begin{figure}[tbh]
   \centering
\begin{picture}(0,0) \put(0,0){{\bf a)}} \end{picture}
   \epsfig{file=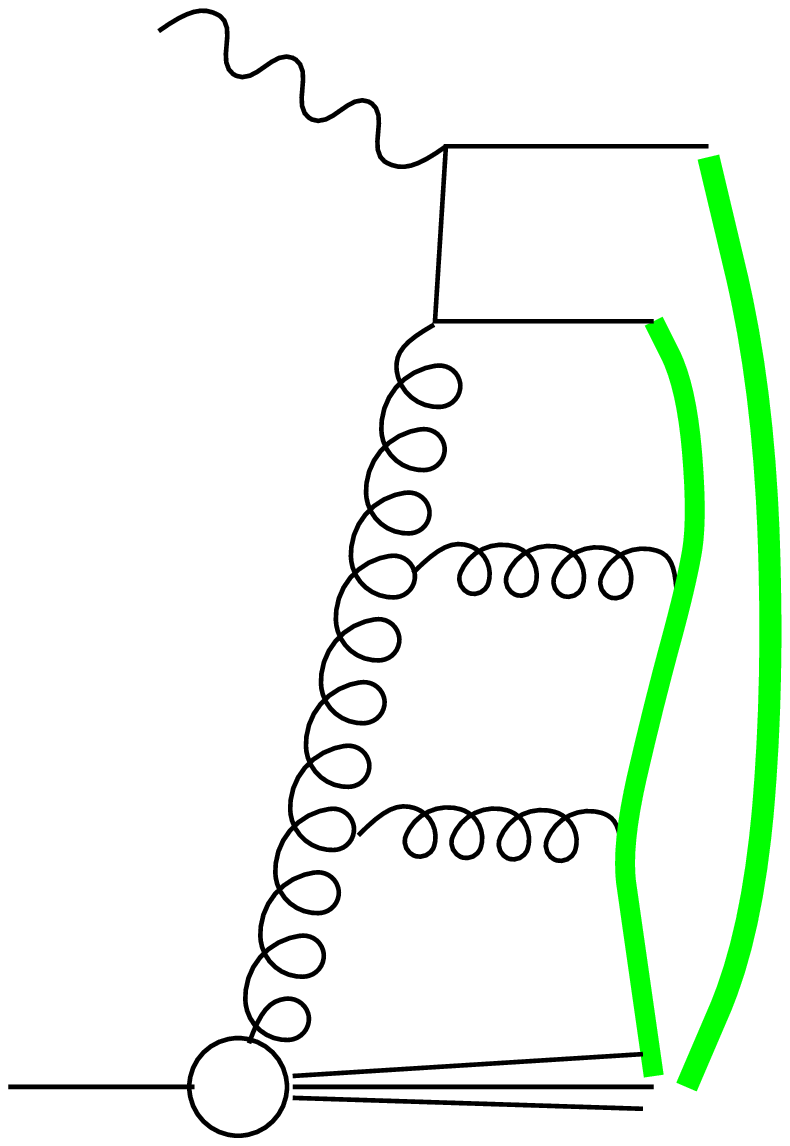,width=3cm,
           bbllx=185pt,bblly=222pt,bburx=436pt,bbury=577pt,clip=}
\begin{picture}(0,0) \put(0,0){{\bf b)}} \end{picture}
   \epsfig{file=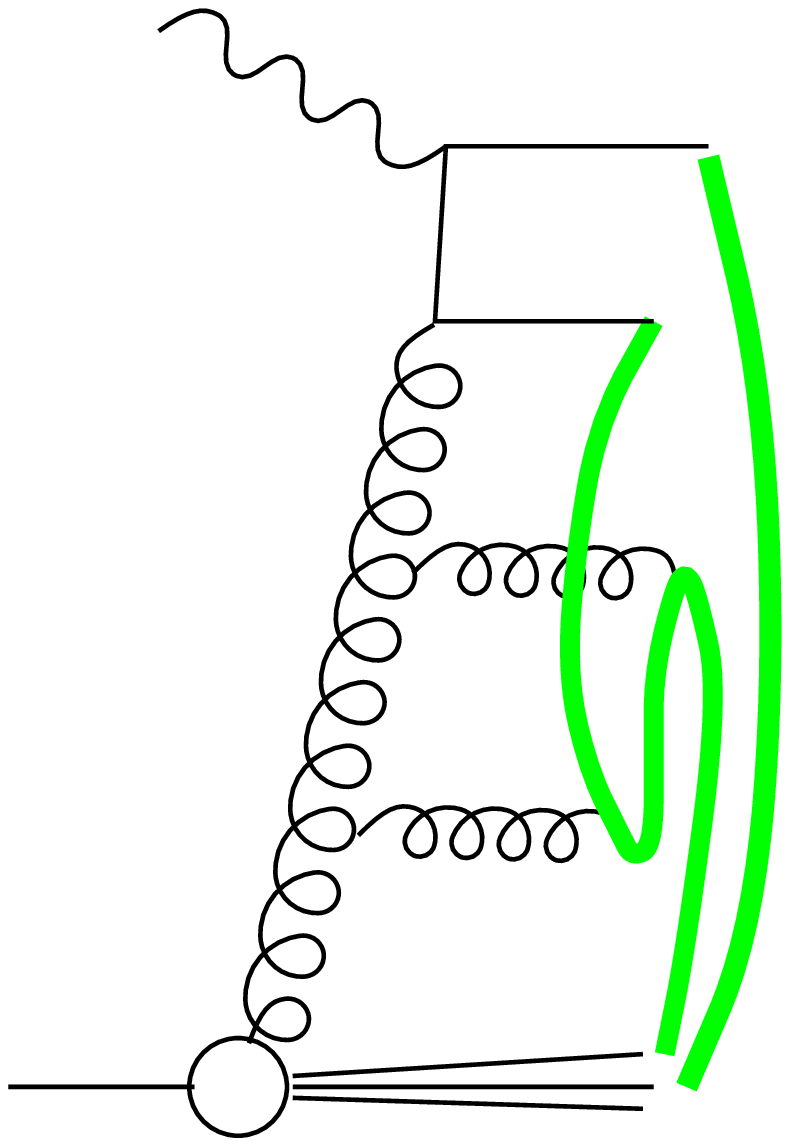,width=3cm,
           bbllx=185pt,bblly=222pt,bburx=436pt,bbury=577pt,clip=}
\begin{picture}(0,0) \put(0,0){{\bf c)}} \end{picture}
   \epsfig{file=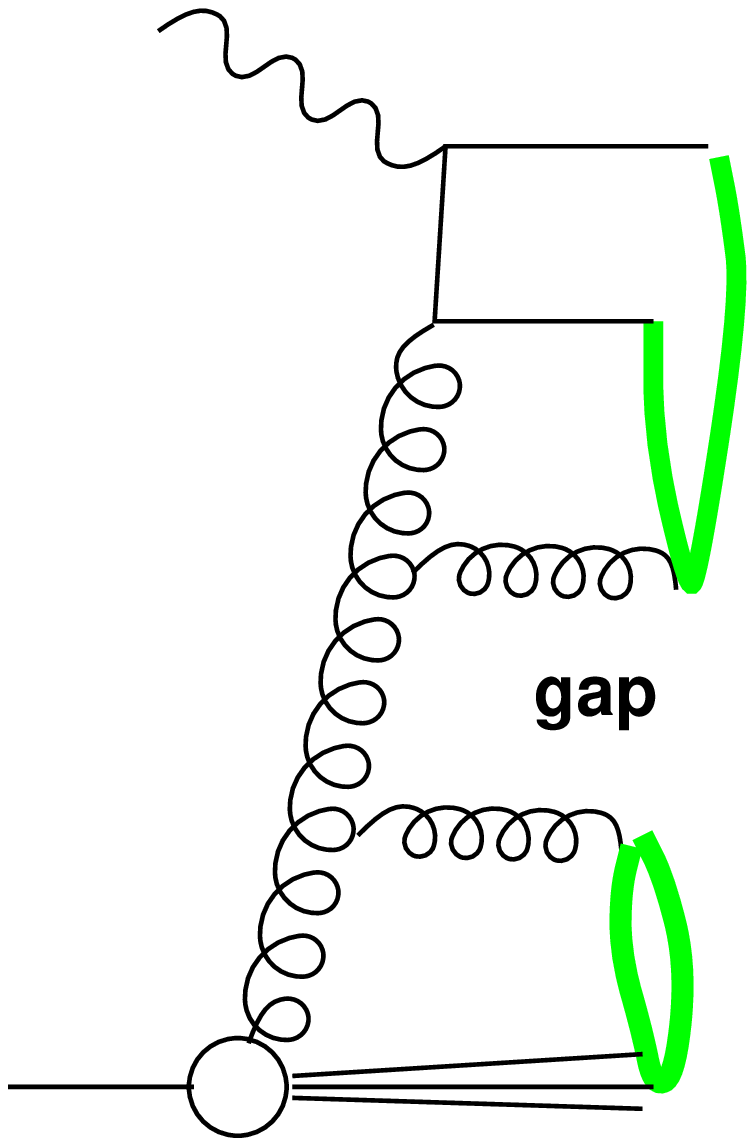,width=3cm,
           bbllx=185pt,bblly=222pt,bburx=436pt,bbury=577pt,clip=}
   \scaption{String configurations for the normal case (a),
             and after soft colour interactions, leading to either
             longer strings (b) or a rapidity gap (c)}
   \label{scistrings}
\end{figure}

Other models assume that rapidity gaps are due to changes in
the colour structure of the event {\it after} the hard interaction
has taken place \cite{mc:sci,th:buchmueller}. At low $x$, most
events are due to photon-gluon fusion, resulting in a colour
octet - antioctet configuration in the final state.
In the string model, there would be two fragmenting strings spanned
(fig. \ref{scistrings}a).
It is assumed that while the produced quark and antiquark travel to leave
the proton, soft gluons are exchanged
between them and the proton remnant, see fig.~\ref{gaps}b.
These gluons do not change the momenta significantly, but may rotate
the colour configuration of the quark - antiquark pair. The final
configuration may turn out to be a colour singlet,
in which case there
would be no colour field connecting it with the remnant
(fig.~\ref{scistrings}c).
The attractive feature of this model is an explanation of the absolute
rate of rapidity gap events. If soft gluon exchange results in
a random colour configuration, the ratio of colour singlet to
colour octet configurations would be $1:8$, and $1/9 \approx 10\%$!
In a related semiclassical approach, partonic fluctuations of the
virtual photon are scattered by the colour field of the proton
\cite{th:buchmueller2}.

Both models describe at least qualitatively the HERA data on
rapidity gap events. Detailed studies will be necessary to
discriminate them (for example
in the final state of the system $X$ \cite{h1:tgap}),
or perhaps discover them to
be not exclusive at all\footnote{Pomeron exchange where
most of the momentum is carried by one gluon in the Pomeron is not so much
different from boson-gluon fusion, where the hard dynamics is
determined by the incoming gluon, and soft gluons are only needed
for colour neutralization}.
For the discussion of the hadronic final state in DIS it will be
important to keep the existence of rapidity gap events
and possible production mechanisms for them
in mind.
Some analyses
\cite{h1:flow2,h1:flow3,h1:flow4,h1:mult,h1:k0,h1:pt}
explicitly exclude
rapidity gap events,
others don't, and some study the difference between the two samples.
The possible existence of
a mechanism to change the colour configuration will have a bearing on
the interpretation
of certain final state data at small $x$ (section ~\ref{sn:lowxi}).
Colour reconnections
are also being discussed in connection with $WW$ fragmentation at LEP
\cite{th:scilep,lep:lep2}.

  \section{Monte Carlo Generators  \label{sn:gen}}    

We shortly summarize the DIS Monte Carlo generators.
They incorporate the QCD evolution in
different approximations and utilize phenomenological models
for the non-perturbative hadronization phase.
The events are generated according to the electroweak cross section
$\frac{\dd^2 \sigma}{\dd x \dd Q^2}$ with
experimentally determined input parton densities.
QED radiative effects are treated with special programs
\cite{mc:heracles,mc:django}.

The Monte Carlo results depend on the model parameters, which
can be adjusted to the data.
Information on generator tuning can be found in
\cite{mc:heratune,mc:hztool,mc:carli,mc:generators}.
In general, the generators now
give an overall satisfactory description of the data.
None of the generators however is able to describe all aspects of
the data.

\subsubsection{LEPTO}

LEPTO \cite{mc:lepto} is based on the LO QCD matrix element with
leading log DGLAP parton showers for soft emissions. Therefore the
model is often called MEPS (Matrix Elements + Parton Showers).
Angular ordering is imposed to model colour coherence.
The rather sophisticated Lund
string model \cite{mc:string} as implemented in JETSET \cite{mc:jetset}
is used for hadronization.
A special feature are
``soft colour interactions'' (SCI) \cite{mc:sci}
in the hadronization phase. They lead to string rearrangements.
This feature is now default, because it describes at least
qualitatively events with a large ``rapidity gap'' that are seen in
the data.

\subsubsection{RAPGAP}

Originally, RAPGAP \cite{mc:rapgap} was created for rapidity gap events.
They are described by scattering on
a colour neutral object, the Pomeron, which is emitted from
the proton according to a flux factor, and whose internal structure
is described by parton density functions.
The present version includes also normal DIS event generation,
very similar to LEPTO.
In the latest version it is also possible to simulate a resolved
(not point-like) component of the virtual photon.

\subsubsection{HERWIG}

HERWIG \cite{mc:herwig}
is based on LL parton showers, with additional corrections to
describe properly the hard interaction. Final state radiation
is angular ordered, initial state radiation is ordered in
$E\cdot \theta$, where $E$ and $\theta$ are energy and angle of
the radiated parton.
The HERWIG philosophy is to
model the perturbative phase as accurately as possible, and
to hadronize with a relatively simple
cluster fragmentation model \cite{mc:cluster}.
Though there is no mechanism foreseen for rapidity
gap events, it turns out that they are generated
at a surprisingly large rate.
Also with HERWIG a resolved component of the virtual photon can be
simulated.

\subsubsection{ARIADNE}

In ARIADNE \cite{mc:ariadne} perturbative QCD radiation is modelled
with radiating colour dipoles according to the CDM.
The BGF graph is added by hand with
its LO matrix element.
Hadronization is performed with the
string model as implemented in JETSET.
ARIADNE allows to model rapidity gaps either by scattering on
a Pomeron, or by colour reconnections, but these options
are by default not activated\footnote{It was found that colour
reconnections in combination with the colour dipole model do not
give a good description of rapidity gap events at HERA \cite{th:lonnblad}.}.

%

\chapter{General Event Properties \label{ch:properties}}     

  \section{Energy Flow  \label{sn:flow}}         

The first analysis of the hadronic final state in DIS at HERA
was based on 88 events recorded in 1992 \cite{h1:flow1}.
The
measured energy flows and particle spectra were in rough
agreement with what was expected from some models including QCD
radiation and fragmentation, and exluded certain other
radiation scenarios.

The flow of transverse energy $E_T$
as a function of pseudorapidity $\eta$ provides a very simple,
global characterization of the hadronic final state.
It is measured with the calorimeters and includes all
produced particles, except the scattered electron.
The \et flow as a function of pseudorapidity
is defined as
\begin{equation}
  \frac{1}{N} \frac{\dif E_T}{\dif\eta} =
  \frac{1}{\sigma}
  \sum_h
  \int E_T \frac{\dif ^2 \sigma_h}{\dif \eta \dif E_T} \dif E_T ,
  \label{eq:flow}
\end{equation}
where the sum extends over all particle species $h$ with cross sections
$\sigma_h$. The definition is analogous for the \et flow as a function
of azimuth $\phi$.
The distribution is normalized to the total number of events $N$,
corresponding to the total event cross section $\sigma$.

\begin{figure}[tbh]
   \centering
  \epsfig{file=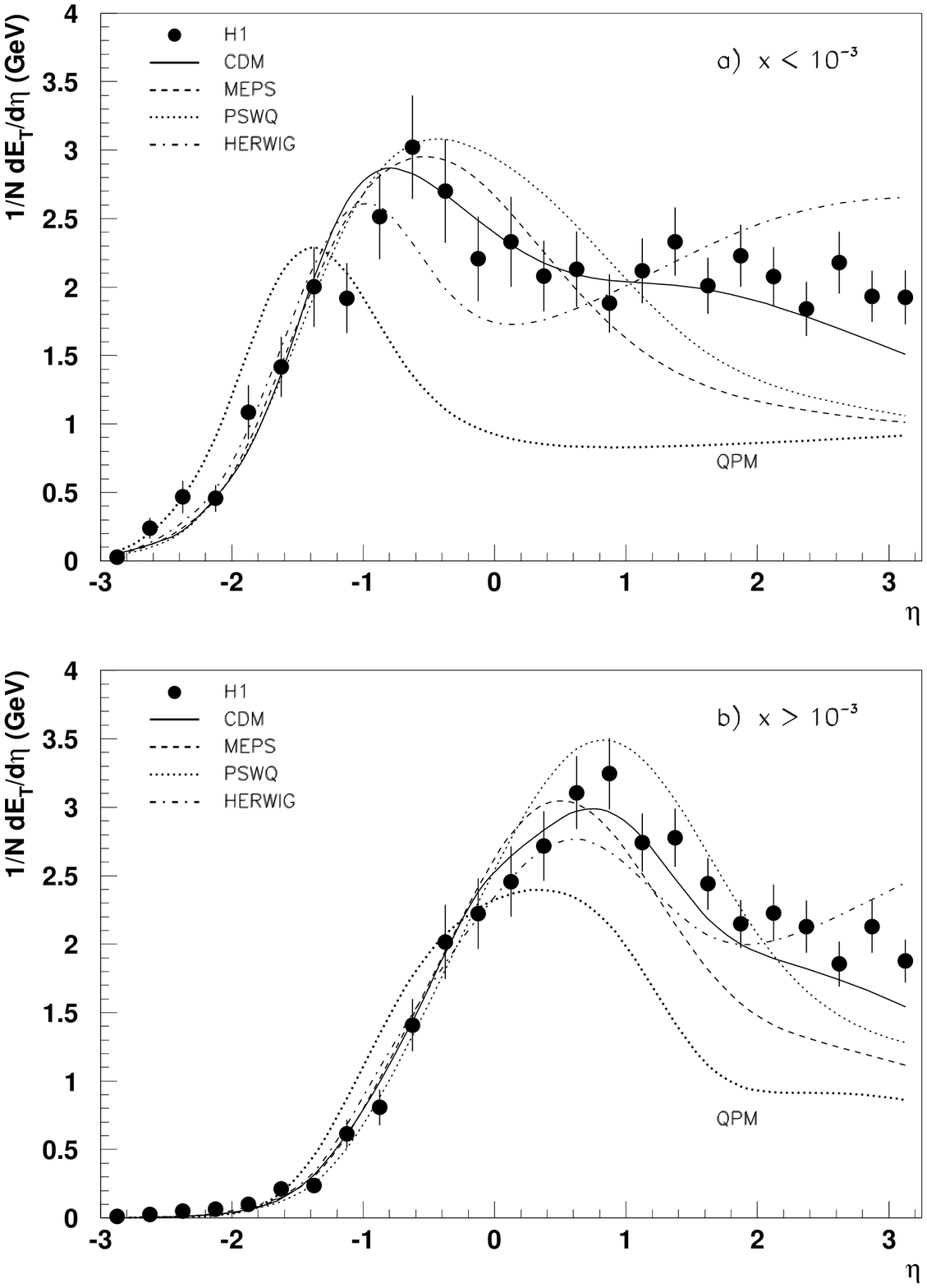,%
          width=7.5cm,bbllx=0pt,bblly=364pt,bburx=491pt,bbury=695,clip=}
  \epsfig{file=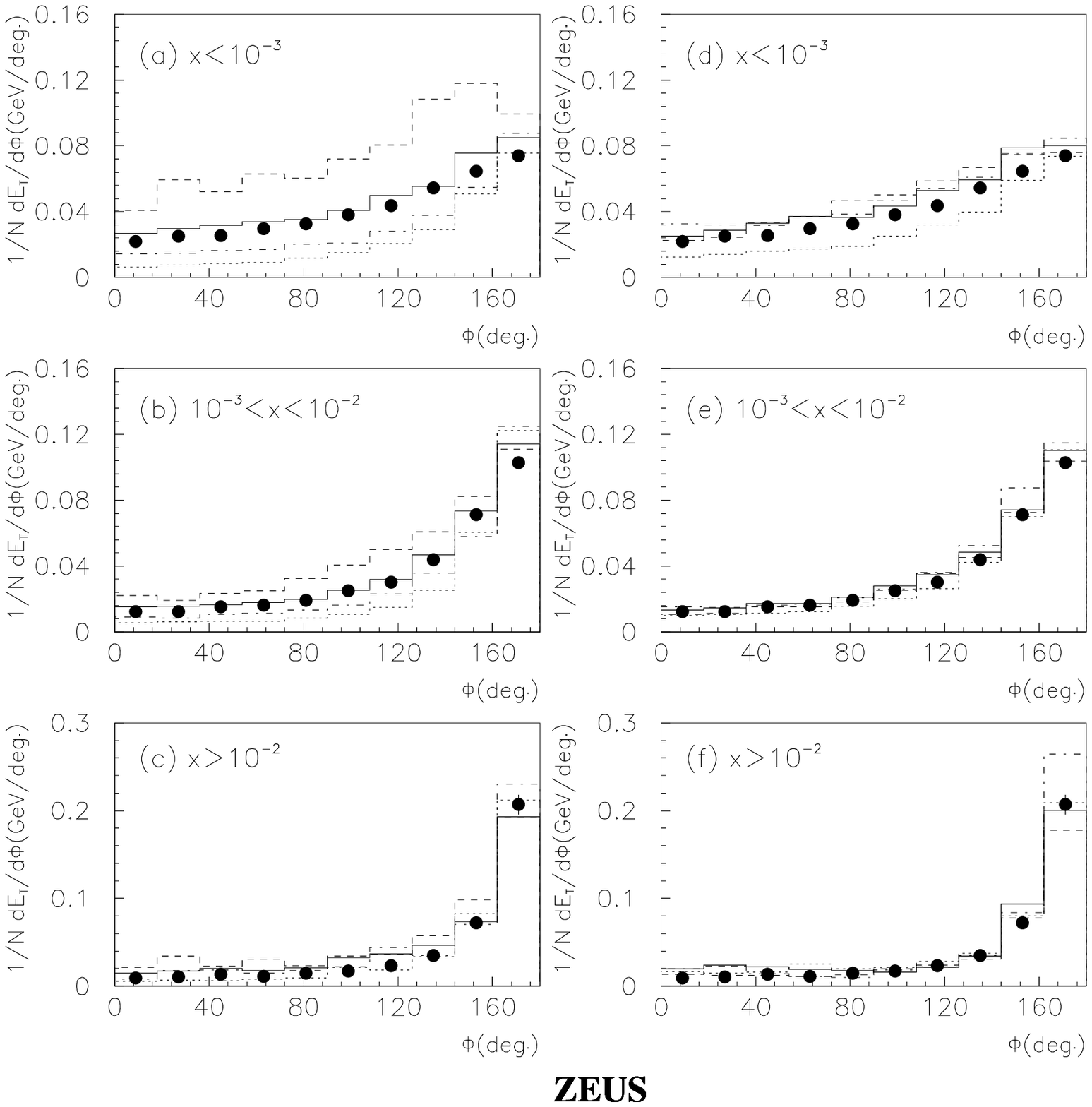,%
          width=7cm,bbllx=0pt,bblly=352pt,bburx=258pt,bbury=519,clip=}
   \scaption{Transverse energy flow in the laboratory frame for $x<10^{-3}$.
      The plots are normalized to the number of observed events $N$.
   Left) \et as a function of pseudorapidity $\eta$
       by H1 \cite{h1:flow2}.
       The proton beam direction is to the right.
       Shown are statistical and systematic errors added in quadrature,
       except for a 6\% scale error. The data are compared to
       the QPM without QCD radiation, and to models including
       QCD radiation: CDM (colour dipole model, ARIADNE 4.03),
       MEPS (matrix element
       plus parton showers, LEPTO 6.1),
       PSWQ (parton shower only, LEPTO 6.1, with
       maximal virtuality in the parton shower $W\cdot Q$), and
       HERWIG (HERWIG 5.7).
   Right) \et as a function of azimuthal angle $\phi$ with respect
      to the electron scattering angle in the transverse plane for
      clusters with $\theta>10\degr $ by ZEUS \cite{z:flow1}.
      Comparisons are made
      to different model predictions:
      matrix element only (ME, dash-dotted),
      matrix element plus parton shower (MEPS, full line),
      parton showers only (PS) with either $W^2$ (dashed)
      of $Q^2$ (dotted) as maximum for the virtuality in the PS.
      The H1 data are corrected for detector effects. The ZEUS
      data are uncorrected, and compared to the models including
      detector simulation.}
   \label{etlabo}
\end{figure}

It is instructive to examine the flow of hadronic transverse energy
as seen in the HERA detectors in the laboratory frame, fig.~\ref{etlabo}.
In this frame one measures according to eq. \ref{eq:flow}
the energy flow transverse to the beam line
as a function of pseudorapidity $\eta=-\ln\tan(\theta/2)$
and azimuth $\phi$. Here $\theta$ is the angle with
respect to the proton
beam axis (laboratory $+z$), and $\phi$ is
the angle between the energy deposition and the scattered
electron in the plane transverse to the beam line.
One sees the current jet from the scattered quark emerging opposite
to the scattered electron in $\phi$ (fig.~\ref{etlabo} right).
The rapidity region from the current jet
towards the remnant is filled with particles due to the colour field
stretched between the scattered quark and the remnant (the main calorimeter
acceptance ends at typically $\eta=3.5$), see fig.~\ref{etlabo} left.

Significantly more transverse energy is observed between the
current jet and the beam pipe than is expected from the QPM
without QCD radiation, but including fragmentation.
The QPM predicts
an \et of roughly 1~\GeV per unit pseudorapidity;
one arrives at a similar number
from the simple considerations in section \ref{sn:fragf}, assuming
3.5 hadrons per unit rapidity with an average \pt of 0.35 \GeV.
The extra \et can be attributed to hard and soft gluon radiation.
The models for these processes varied a lot in their predictions,
but have evolved since.

\begin{figure}[tbh]
   \centering
   \epsfig{file=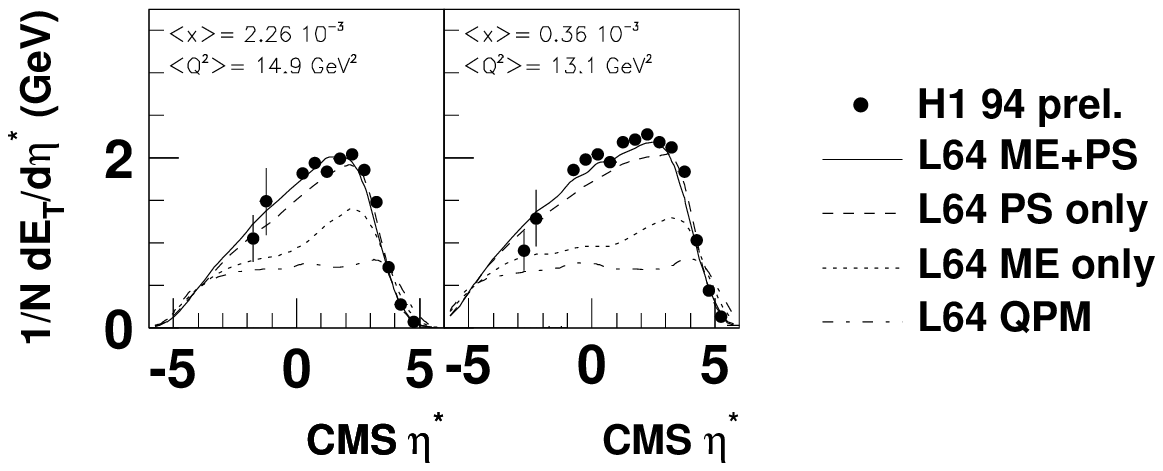,
           width=12cm,bbllx=44pt,bblly=527pt,bburx=370pt,bbury=672,clip=}
   \scaption{
             The \et flow from H1 \cite{h1:flow4} at
             $\av{x}\approx$ 0.002 (left) and 0.0004 (right)
             for $\av{Q^2}\approx 14 \GeVsqx$.
             The data are in the hadronic CMS (proton direction to the left).
             Shown are statistical errors only.
             The data are
             compared to LEPTO 6.4
             with different treatments of perturbative QCD effects:
             QPM: no radiation;
             ME: matrix element only;
             PS: parton showers only;
             ME+PS: matrix element plus parton shower (default).}
   \label{mepsflow}
\end{figure}

In fig.~\ref{mepsflow} the latest preliminary H1 data on energy flows
in the hadronic CMS \cite{h1:flow4} are compared to the model LEPTO 6.4
with different treatments for perturbative QCD radiation.
Without QCD radiation, a flat rapidity plateau with
$\et\approx0.7\GeV$ per unit rapidity is expected.
QCD radiation as expected from the hard matrix element generates
\et in the current hemipshere only. It is shown below that the \et
in the current hemisphere depends on \Qsq.
With parton showers only, a reasonable description of
the data is achieved. For this observable, the parton shower
covers already most of the relevant phase space, including
the matrix element.
The best description of the data is obtained by combining
radiation according to the LO matrix element with parton showers.

\begin{figure}[tbh]
   \centering
  \epsfig{file=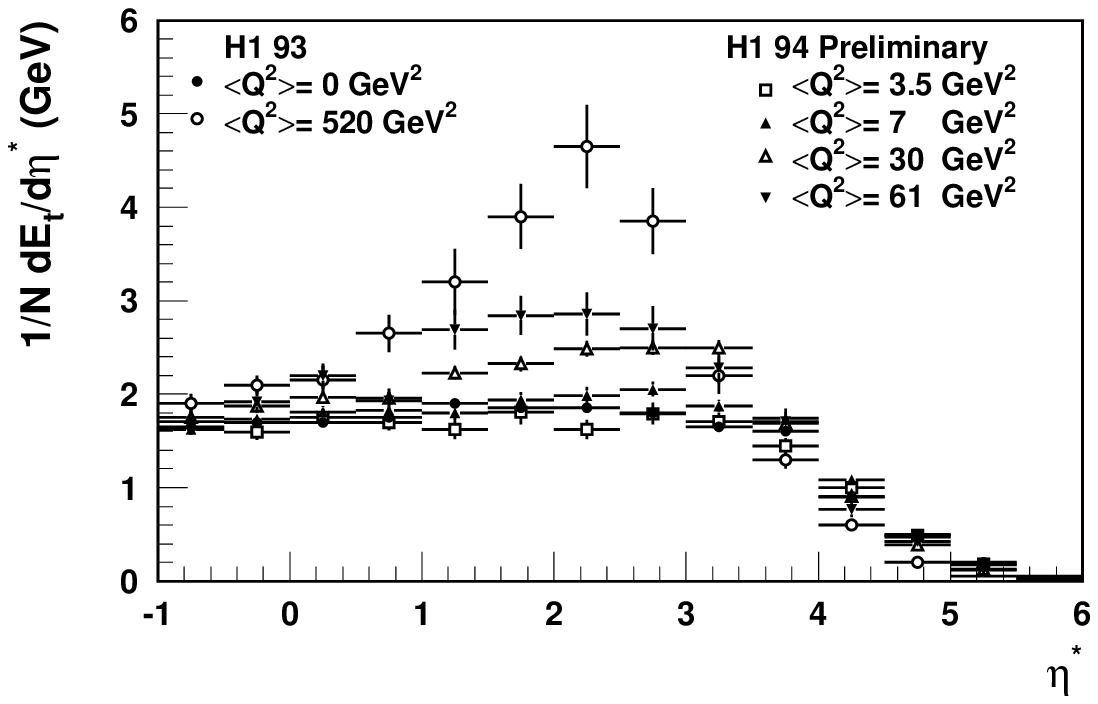,%
          width=10cm}
   \scaption{Transverse energy production in the CMS.
       \et as a function of pseudorapidity $\eta$
       by H1 \cite{h1:efldisgp,h1:flow4}.
       The DIS data for different \Qsq bins
       are compared to photoproduction data at $Q^2\approx 0$.
       The proton beam direction is to the left.
       Shown are statistical errors only, except for
       the two foremost data points measured with the plug calorimeter.}
   \label{etrosti}
\end{figure}

The energy flow has also been studied as a
function of \Qsq (fig.~\ref{etrosti}). At low \Qsq the rapidity
plateau is almost flat. With increasing \Qsqx, the \et
in the ``photon fragmentation region''
(loosely defined as far in the current hemisphere)
increases considerably. In that region, the virtuality of the
photon governing hard interactions plays an important r\^{o}le.
At central rapidity, the \Qsq dependence is comparatively small.
This supports qualitatively the picture that particle production
at central rapidity is independent of the type of the colliding
particles, be they $\pi, K, p$ or a virtual photon $\gamma^\ast$
\cite{th:bj_slac}.

In sect.~\ref{sn:eflows} the \et flow is discussed in more
detail in connection with low $x$ physics.
Further data on hadronic energy production are available from
H1, namely energy-energy correlations \cite{h1:flow2,h1:flow4},
and the distribution of transverse energy at central rapdity \cite{h1:flow4}.

 \section{Charged Particle Multiplicities \label{sn:multi}}


\subsubsection{Average multiplicity}

\begin{figure}[tbhp]
   \centering
   \epsfig{file=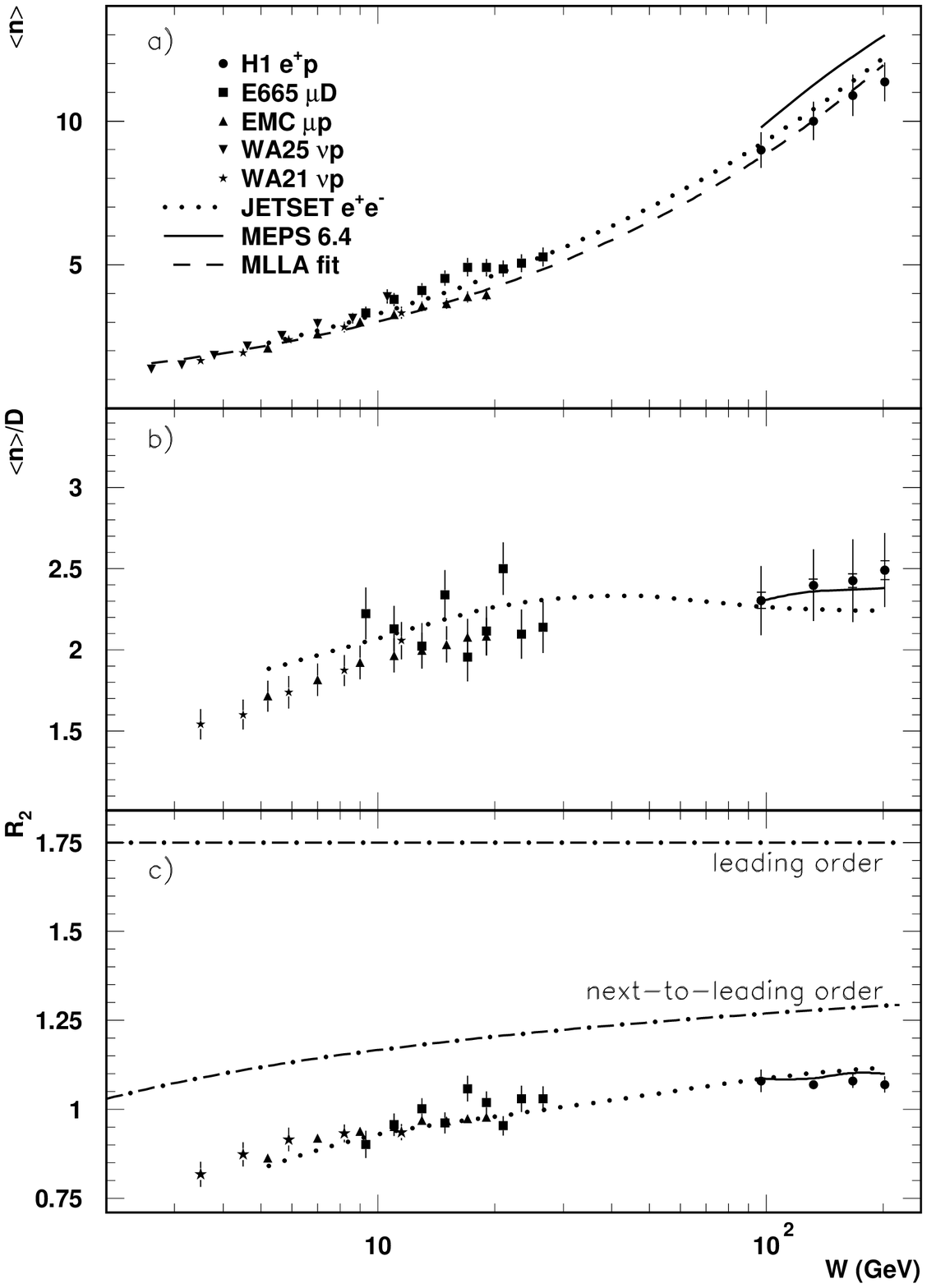,%
          width=12cm}
   \scaption{
             {\bf a)} The average charged multiplicity $\av{n}$,
             {\bf b)} the ratio $\av{n}/D$, and
             {\bf c)} the factorial moment $R_2$
             for charged particles in the CMS current
             hemisphere as a function
             of $W$.
             The DIS data are compared to the DIS QCD model MEPS,
             to the \epem model JETSET for $u,d,s$ quarks, and to
             a MLLA fit, and to the LO and NLO predictions for $R_2$.}
   \label{multi}
\end{figure}

The
multiplicity of charged particles in the CMS current hemisphere
($x_F>0$) has been measured by H1 \cite{h1:mult} with their central
tracking chamber for events\footnote{Rapidity gap events had been
excluded from this analysis.}
at $\av{\Qsq}\approx 23 \GeVsq$
in the kinematic range $80\GeV<W<220 \GeV$
(fig.~\ref{multi}a), thus
reaching fragmentation energies not yet accessible at LEP.
From phase space arguments alone, one would expect $\av{n} \propto \ln W$,
(see section \ref{sn:fragf}). Scale breaking due to QCD radiation should
lead to a faster rise of the multiplicity, however dampened
by colour coherence and the running of $\alpha_s$.
A faster than logarithmic rise is observed
when
comparing the H1 data to lower $W$ lepton-nucleon scattering experiments.
The QCD model LEPTO somewhat overestimates the multiplicity
(see fig.~\ref{multi}).
The DIS data are also compared to the JETSET prediction
(light quarks only)
for \epem data, which is known to give a good representation of the
\epem data.
The behaviour of the total multiplicity in the CMS current hemisphere
in DIS is similar to \epem~ annihilation data.
Differences are found however for the distribution in rapidity, see below.

The growth of multiplicity with energy
has been calculated via
the QCD evolution of fragmentation functions for running \as
in NLO
\cite{books:ellis,th:webberfrag}, yielding
\begin{equation}
  \av{n} = a \cdot\left[\alpha_s(W^2)\right]^b
           \cdot e^{c/\sqrt{\alpha_s(W^2)}}
           \cdot\left[1+d \sqrt{\alpha_s(W^2)}\right].
  \label{eq:webber}
\end{equation}
The two-loop expression eq. \ref{eq:asrun}
for \as is used with $W^2$ as scale.
The constants $b$ and $c$ are given by the theory as
\begin{equation}
  b=\frac{1}{4} + \frac{10n_f}{27\beta_0} \hspace{1cm}
  {\rm and} \hspace{1cm}
  c = \frac{\sqrt{96\pi}}{\beta_0},
\end{equation}
with $n_f = 3$ active flavours and $\beta_0=11-(2/3)n_f$.
$a$ and $d$ are free parameters.

Predictions for hadron spectra can also be derived
from the parton multiplicity in a jet, which is calculable in QCD
by summing up all the branchings in the cascade down to a
cut-off $Q_0$.
The hypothesis of local parton-hadron duality (LPHD)
(see sect. \ref{sn:hadron}) is made
for the transition from partons to hadrons: the hadron spectra
should have the same shape as the perturbative prediction, merely
differ by a normalization constant.
The prediction will depend on the cutoff $Q_0$, because
the charged particle multiplicity
is not an infrared safe observable, see section \ref{sn:finst}.
If one assumes that the shower stops at a scale given by a typical
hadron mass $Q_0 \approx m_\pi$, the $Q_0$ dependence is eliminated.
One can then compare the perturbative prediction
directly to the data.

Most important for the average multiplicity is the growth
at small fractional momentum
due to soft gluon radiation.
Quantum mechanical interference leads to a suppression
of soft gluon radiation (soft colour coherence).
The shower evolution in the leading-log approximation (LLA)
is modified (hence called MLLA)
to take into account destructive interference
of soft gluons \cite{books:dokshitzer}. The modification
consists of changing the evolution variable such that
angular ordering \cite{books:dokshitzer,th:angord}
between subsequent emissions $i$ and $i+1$
is fulfilled. That is, their opening angles satisfy
$\theta_i > \theta_i + 1$ (see section \ref{sn:meps}).

The MLLA+LPHD prediction for the hadron multiplicity for
running \as is \cite{books:dokshitzer,th:ochs}
\begin{equation}
  \av{n} = c_1 \frac{4}{9} N_{LA} + c_2,
  \label{eq:multimlla}
\end{equation}
with
\begin{equation}
   N_{LA}(Y) = \Gamma(B) \left(\frac{z}{2}\right)^{(1-B)} I_{1+B}(z),
\end{equation}
where $z=\sqrt{\frac{48}{\beta_0}Y}$, $Y=\ln(W/2Q_0)$,
$a=11+2n_f/27$ and $B=a/\beta_0$. $\Gamma$ is the Gamma function
and $I_\nu$ the modified Bessel function of order $\nu$.
$Q_0=0.27\GeV$ is used \cite{th:ochs}.
Eq. \ref{eq:multimlla} reduces to eq. \ref{eq:webber} for large $z$.

The data can be fit by either eq.~\ref{eq:multimlla} or
eq.~\ref{eq:webber}.
For fixed \as one would expect a faster growth than eq.~\ref{eq:multimlla},
namely a power law \cite{th:ochs}
\begin{equation}
   \av{n}=a(W/W_0)^{2b^\prime}-c,
   \label{eq:plaw}
\end{equation}
which is still consistent with the data. In fact, all these
fits would be indistinguishable in fig.~\ref{multi}.
The fit with eq.~\ref{eq:webber} yields
$a=0.034\pm0.005$ and $\Lambda=0.190\pm 0.060~\GeVx$.
In this fit the correction term $d\sqrt{\as}$ has been neglected,
because it had been found insignificant ($d=0.2\pm0.3$).
It has to be pointed out that $\Lambda$ extracted from eq. \ref{eq:webber}
need not be identical with \lambdams, though they are expected to be
similar
if $d$ is small \cite{th:webberfrag}.
The fit results are summarized in table \ref{tab:multi}.

\begin{footnotesize}
\begin{table}[h]
\begin{center}
\begin{tabular}{|l|c|c|c|}
\hline
ansatz        &  \multicolumn{2}{c|}{fit parameters} & fixed parameters  \\
\hline
power law, eq. \ref{eq:plaw}
                        & $a = 1.40\pm 0.04$     & $b^\prime=0.20 \pm 0.01$
                        & $c=0.5$, $W_0=1\GeV$  \\
MLLA+LPHD, eq. \ref{eq:multimlla}
                        & $c_1=1.21 \pm 0.05$    & $c_2=0.81\pm 0.08$
                        & $Q_0=0.270 \GeV$                  \\
NLO resummed, eq. \ref{eq:webber}
                        & $a=0.041\pm 0.006$     & $d=0.2\pm 0.3$
                        & $\Lambda = 0.263 \GeV$             \\
same, but $\Lambda$ free
                        & $a=0.034 \pm 0.005$    & $\Lambda=0.190\pm0.060 \GeV$
                        & $d=0$                \\
\hline
\end{tabular}
\end{center}
\scaption{The parameters of fits to the dependence of
the average charged multiplicity $\av{n}$ in the CMS current hemisphere as
a function of $W$ \cite{h1:mult}.
}
\label{tab:multi}
\end{table}
\end{footnotesize}

\subsubsection{Multiplicity Distributions}

\begin{figure}[tbh]
   \centering
   \epsfig{file=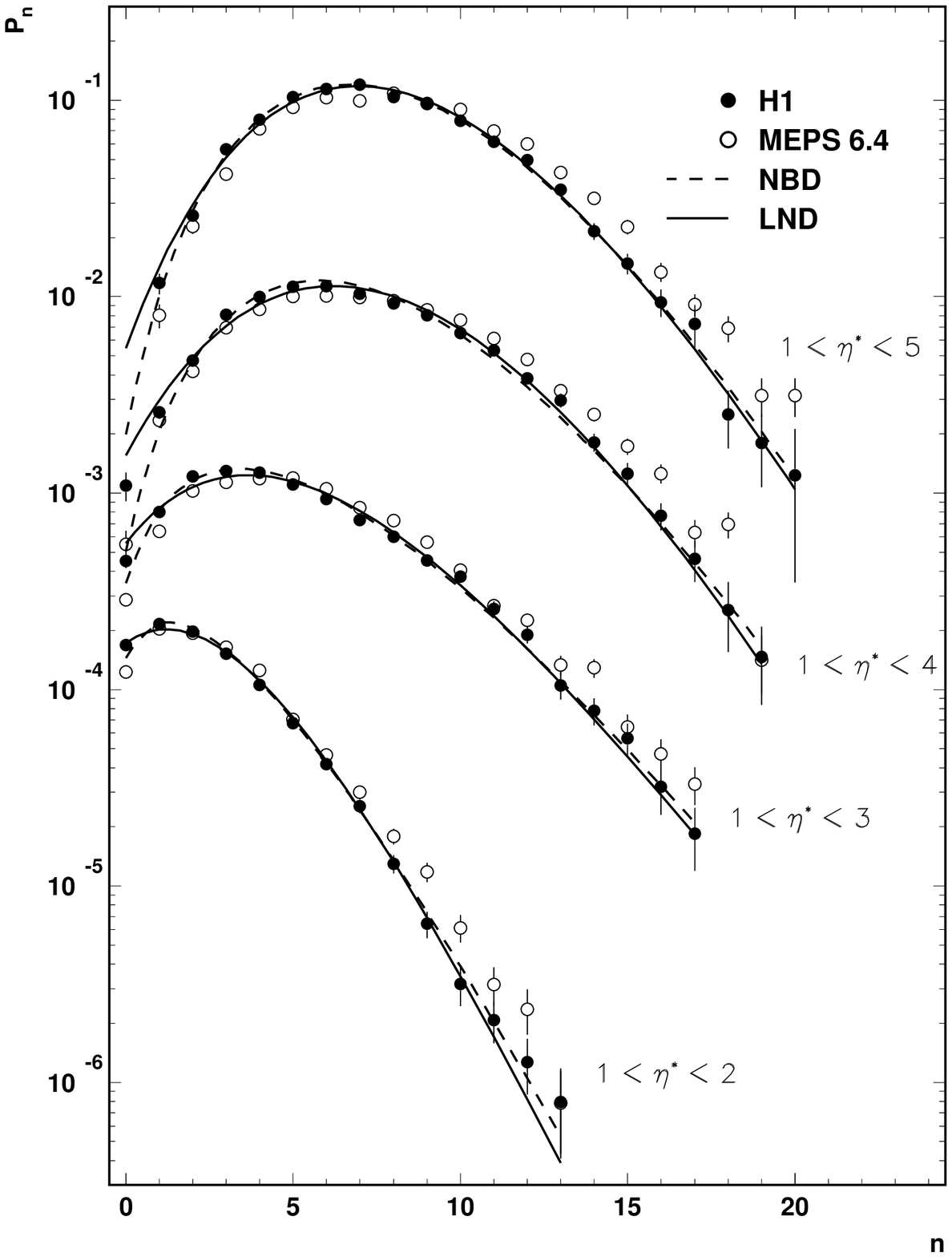,%
          width=10cm}
   \scaption{The charged particle multiplicity distribution $P_n$
             for $115\GeV<W<150\GeV$ in indicated CMS pseudorapidity
             domains. The data \cite{h1:mult} are compared to
             LEPTO (MEPS 6.4)
             and to the NBD and LND fit results. The data
             for $1<\eta^\star<5$ are shown at the true scale, the others
             are scaled down by factors of 10 with respect to the previous
             one. Shown are statistical errors only.
             }
   \label{multidisa}
\end{figure}

H1 has also measured the multiplicity distribution
$P_n$, giving the probability to observe in a given event $n$
charged particles,
as a function of $W$, \Qsq and for varying pseudorapidity ranges.
In general, the data can be parametrized with the
Negative Binomial Distribution (NBD),
\begin{equation}
P_n(k,\ol{n})=\frac{k(k+1) \cdot\cdot\cdot (k+n-1)}{n!}
              \left(\frac{\ol{n}}{\ol{n}+k}\right)^n
              \left(\frac{k}{\ol{n}+k}\right)^k.
\end{equation}
The two parameters $k,\ol{n}$ can be expressed in terms of
the average multiplicity and the dispersion $D$:
\begin{equation}
  D^2=\av{ (n-\av{n})^2 } \hspace{1cm} \av{n}=\ol{n}
 \hspace{1cm}
 \frac{D^2}{\av{n}^2} = \frac{1}{\ol{n}} + \frac{1}{k}.
\end{equation}
One arrives at the NBD in many phenomenological models \cite{th:nbd}
for hadron production.
A large body of data can be parametrized
by the NBD. In QCD the LLA also leads to the NBD for
the gluon multiplicity in a quark jet \cite{th:nbdlla}.
An alternative function is the Lognormal Distribution
(LND),
derived for multiplicative branching processes \cite{th:lnd}.
A variable is distributed according to the LND, if its
logarithm is Gaussian distributed.
The H1 data (fig~\ref{multidisa})
are equally well described by both parametrizations.
The MEPS model deviates from the data both
for very small and large multiplicities.
In particular the origin
of the excess of 0-prong events in the data over the
MEPS model is not clear.
For fixed $W$,
a \Qsq dependence of $\av{n}$ or $D$ in the range
$10\GeVsq<\Qsq<1000 \GeVsq$
was not found, within errors.

Hard gluon radiation would lead to a
superposition of different NBDs, hence deviations from a single NBD
for the hadron multiplicity distribution.
A multi-jet induced shoulder structure
as observed at LEP \cite{lep:shoulder} is however not visible in
the H1 data.

\begin{figure}[tbh]
   \centering
   \epsfig{file=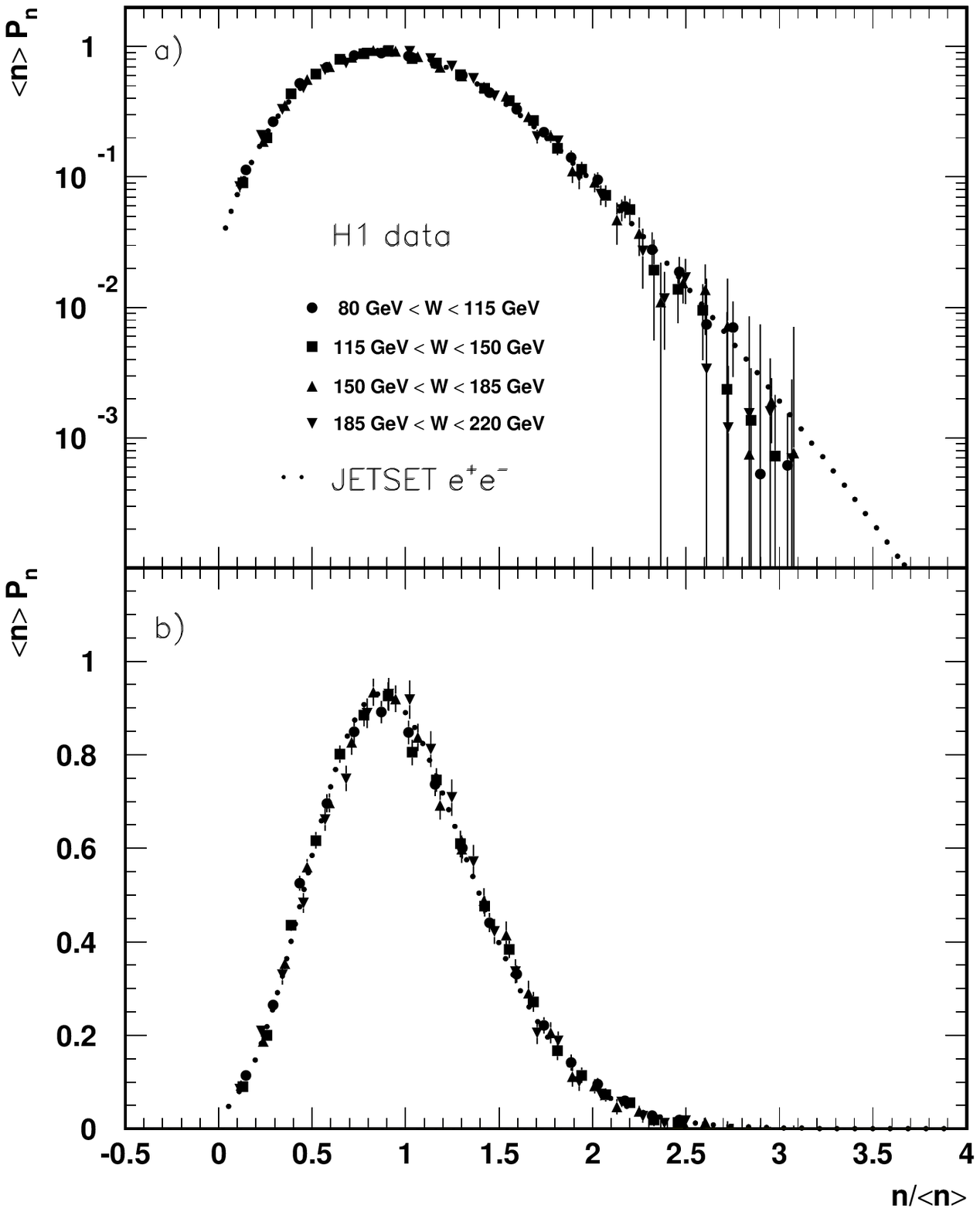,%
          width=11cm}
   \scaption{
                The multiplicity distribution  \cite{h1:mult} for
                $1<\eta<5$ in KNO form for different $W$ values,
                and
                compared to
                the JETSET prediction for \epem~ annihilation.
                Shown are statistical errors only.
             }
   \label{multidisb}
\end{figure}

In fig.~\ref{multidisb} the multiplicity distribution is plotted in the
scaling KNO (Koba, Nielsen and Olesen)
form \cite{th:kno}, $\av{n} P_n$ as a function of $n/\av{n}$.
In a broad class of production mechanisms, based upon scale invariant
stochastic branching processes, $\av{n} P_n$ is expected to depend
only on $n/\av{n}$, and not on energy \cite{th:polyakov}. The
H1 data for pseudorapidity $1<\eta<5$
for different $W$ bins exhibit KNO scaling, and agree
with the JETSET result for \epem~ annihilation.

In fig. \ref{multi}b and c the
energy dependence for the
ratio $\av{n}/D$ and the normalized
second order factorial moment \cite{rev:dewolf} $R_2:=\av{n(n-1)}/\av{n}^2$
are shown (further data on the moments of the multiplicity
distribution can be found in \cite{h1:mult}).
For strict KNO scaling, they should be energy independent,
which is the case for HERA energies at large $W$, but not for data at
smaller $W$. The data are rather well reproduced by the JETSET \epem~
model, and by the MEPS DIS model. A QCD calculation approaches the
$R_2$ measurements when NLO corrections are taken into account
\cite{th:malaza}, but deviates still significantly from the data.
The LO result in the double logarithmic approximation (DLA),
$R_2=7/4$, is reduced in NLO\cite{th:malaza} to
$R_2= (1-\chi\sqrt{\alpha_s}) \cdot 7/4$,
where $\chi\approx 0.55$.
Missing
higher order effects, which are taken into account with parton showers
in MEPS, may be responsible for the disagreement.

\subsubsection{Rapidity Distribution}

\begin{figure}[tbp]
   \centering
   \vspace{-1cm}
   \epsfig{file=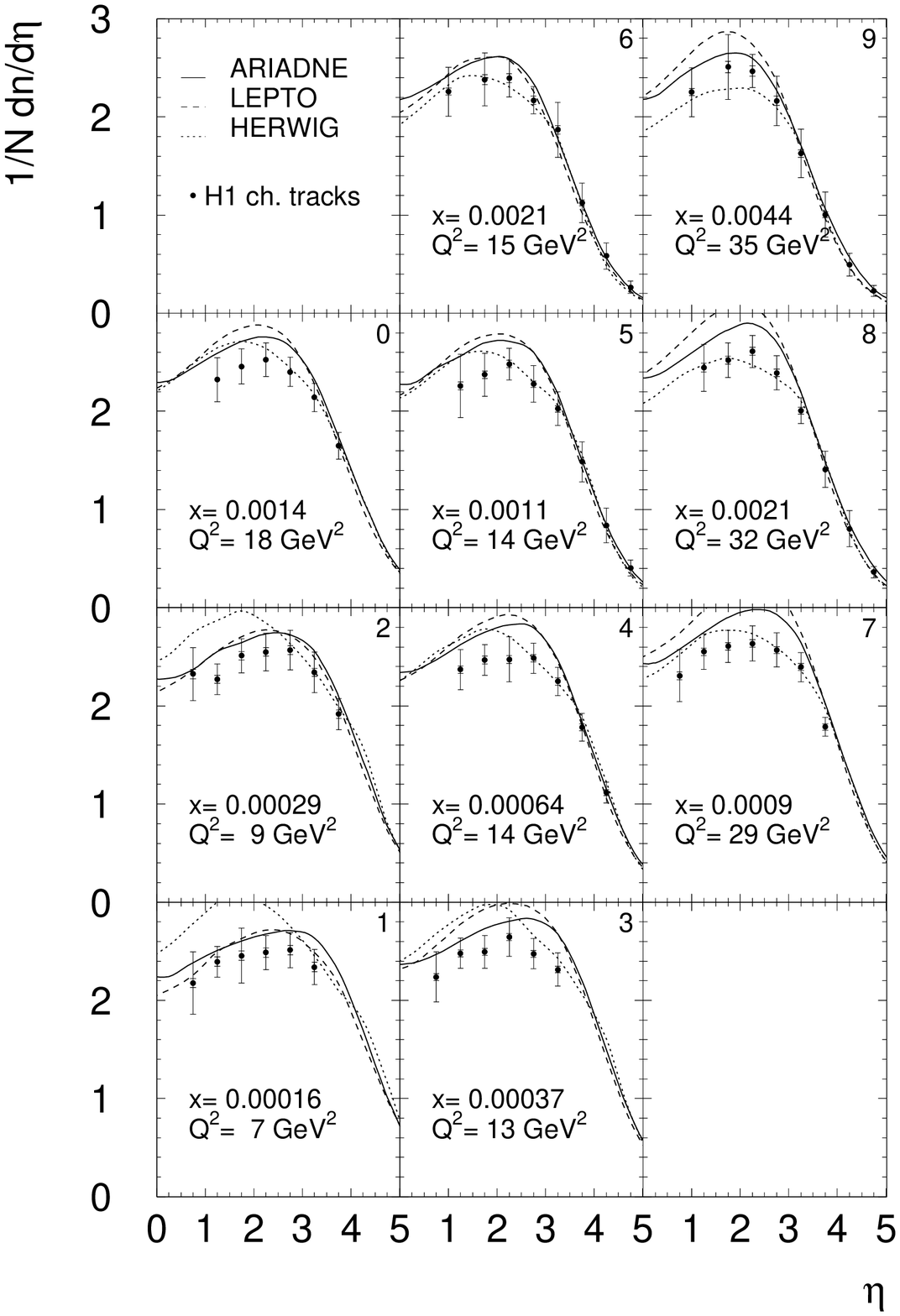,%
    width=14cm,bbllx=33pt,bblly=23pt,bburx=487pt,bbury=689pt,clip=}
   \scaption{
            The CMS pseudorapidity distribution
            of charged particles \cite{h1:pt}. The proton direction is to
            the left.
            Data are shown for nine different kinematic bins
            with the mean values of
            \xb and \Qsq as indicated,
            plus the combined sample (bin 0).
            The models
            ARIADNE 4.08, LEPTO 6.4 and
            HERWIG 5.8 are overlayed.}
   \label{eta_charged}
\end{figure}

H1 has measured the charged multiplicity as a function of pseudorapidity
\cite{h1:pt} for different kinematic regions at low \Qsqx,
see fig.~\ref{eta_charged}.
In the plateau region about 2.5 charged particles per $\eta$ unit are
observed.
The QCD models describe roughly the data, but are far from perfect.
In particular HERWIG overshoots the data at low $x$ towards the central
region.
We mention already here that quite substantial differences
between the models and the data emerge at small $x$ when only hard particles
($\pt>1~\GeVx$) are considered
(between 0.2 and 0.3 such particles are found per $\eta$ unit).
This effect will be discussed in detail in connection with
low $x$ physics in section \ref{sn:ptspectra}.

\begin{figure}[tbh]
   \centering
   \epsfig{file=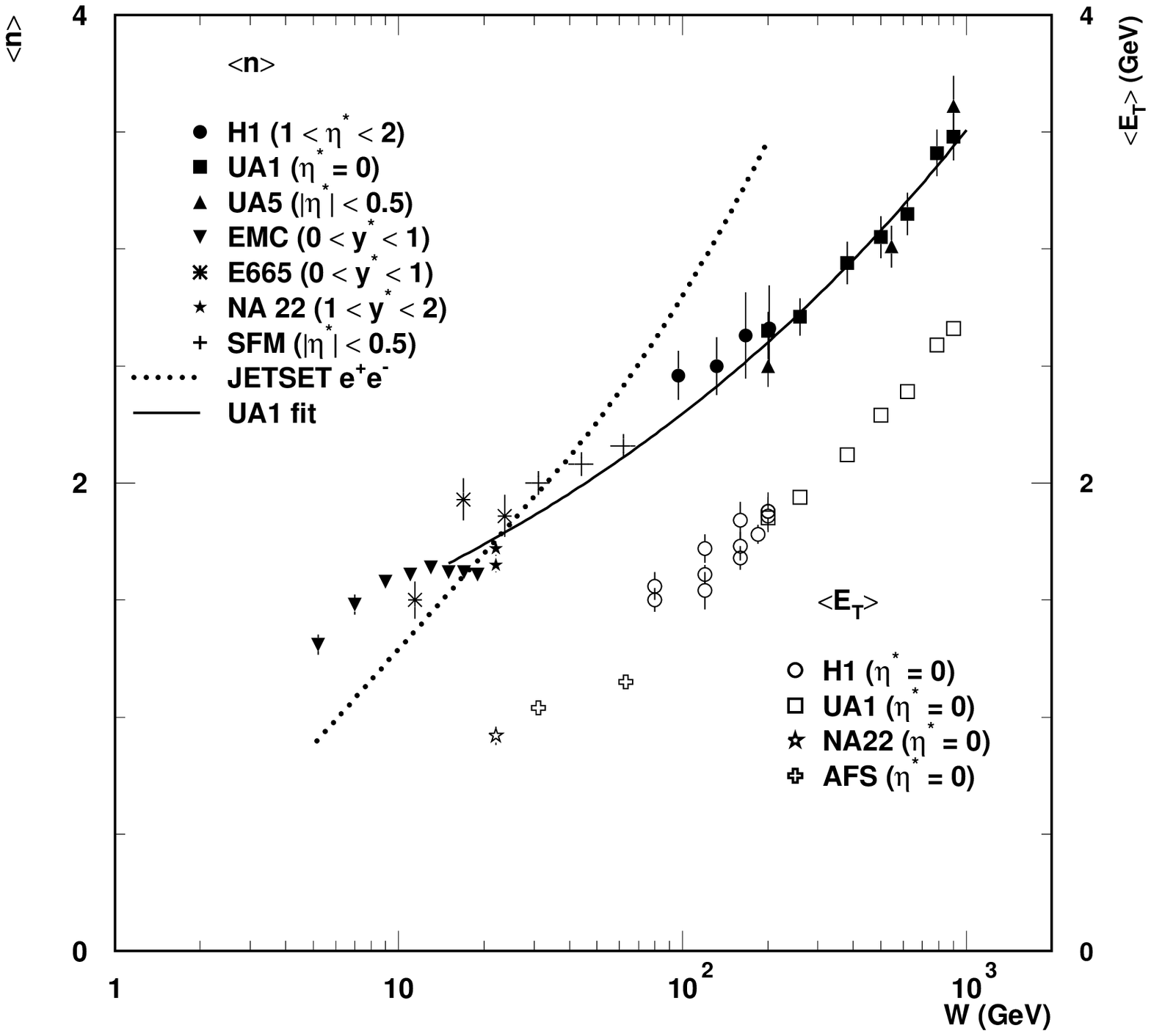,%
          width=12cm}
   \vspace{-0.5cm}
   \scaption{Solid symbols:
             $W$ dependence of charged particle density
             ($\av{n}=$ \# charged particles/unit pseudorapidity,
             left scale).
             Open symbols:
             $W$ dependence of the mean central
             transverse energy \av{\et} per unit pseudorapidity (right scale).
             The dotted line is the JETSET result
             for $\av{n}$ in \epem~ reactions, and the full curve a
             parametrisation $\av{n} = 0.35 + 0.74 (W^2)^{0.105}$
             \cite{coll:ua1}.
             The DIS multiplicity data \cite{h1:mult,o:e665n,o:emcn}
             and transverse energy data \cite{h1:efldisgp}
             are compared to
             hadron-hadron collisions
             \cite{coll:breakstone,coll:deroeck,coll:ua1,coll:na22}.
             The DIS \et data are discussed in section \ref{sn:eflows}.
             }
   \label{wdep}
\end{figure}

The charged particle density measured by H1 \cite{h1:mult}
close to the central region $1<\eta<2$,
$\av{n}\approx 2.5$ per $\eta$ unit,
is consistent with data from hadron-hadron interactions,
see fig.~\ref{wdep}. When compared with other DIS experiments,
a clear increase of the particle density with $W$ is seen.
The increase of the total multiplicity with $W$ is not entirely
due to the increasing longitudinal phase space (rapidity), but
also due to an increase of the number of particles per unit rapidity,
i.e. the height of the rapidity plateau increases with $W$.
The average \et at central rapidity (see section \ref{sn:eflows})
in DIS and in hadron-hadron collisions increases with a similar slope
(fig.~\ref{wdep})

The particle density at HERA is
significantly lower than for \epem~ data, represented by the JETSET
result.
Also the $W$ dependence in \epem~ is steeper than in
hadron-hadron interactions. The DIS data appear to behave as one would
expect from hadron-hadron data. That is also true for the transverse
energy (see section \ref{sn:flow})
measured at central rapidity \cite{h1:efldisgp}, see fig. \ref{wdep}.

Why are there for the same CM energy $W$
less charged particles in DIS than in \epem~
at central rapidity?
A possible explanation is the ``antenna effect'' \cite{th:antenna},
that leads to
suppressed radiation from the smeared out colour charge of the proton
remnant.
Another possible explanation is that scaling violations
leading to larger multiplicities are stronger in \epem interactions
than in $ep$ reactions due to the larger scale,
$Q^2_{\epem} = W^2_{\epem} = s_{\epem} \gg Q^2_{ep}$.
The \xf spectra in the CMS and their scaling violations are presented
in the next section.

  \section{Charged Particle Momentum Spectra
                                         \label{sn:charged}}

\begin{figure}[tbh]
   \centering
   \epsfig{file=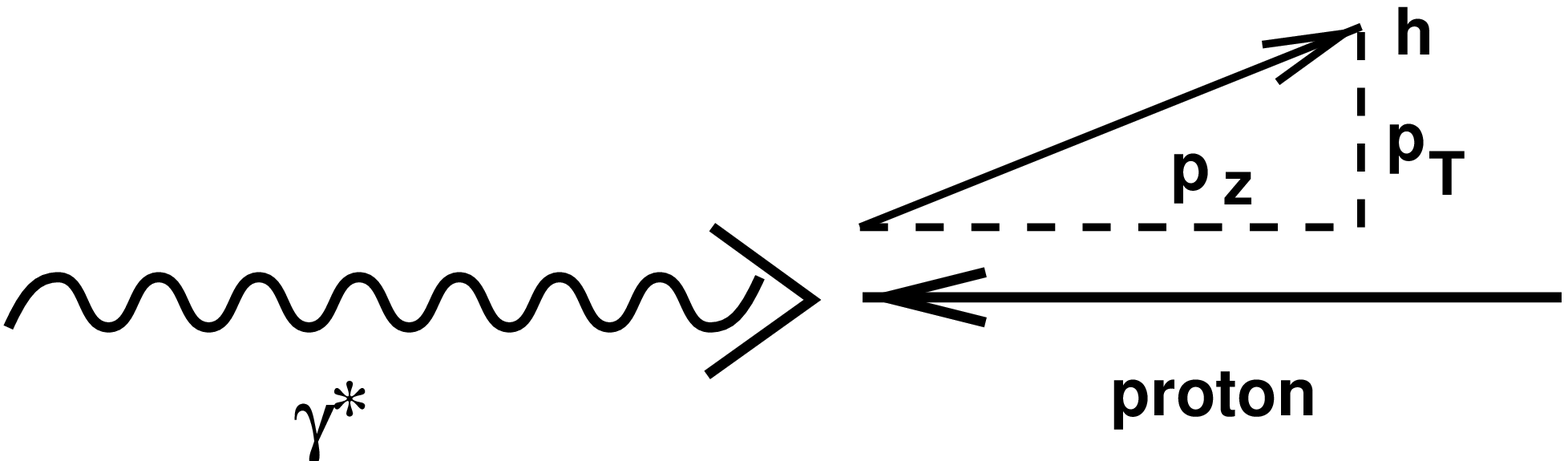,width=7cm}
   \scaption{
     The longitudinal (\pz) and transverse (\pt) momentum components
     of a hadron $h$ in the $\gamma^\star p$ CMS.}
   \label{ptpl}
\end{figure}

In this section inclusive charged particle spectra
measured by H1 and ZEUS  \cite{h1:flow2,z:xf}
with
their central
tracking chambers \cite{h1:flow2,z:xf} are presented.
This limits the acceptance to the CMS current
region, $\xf>0$.
Measured are longitudinal and transverse
momentum spectra with respect to the virtual photon direction
(fig.~\ref{ptpl}).
The kinematic region covered is $10<Q^2<100\GeVsq$ and
$55<W<200 \GeV$ for H1, and
$10<Q^2<160\GeVsq$ and $75<W<175\GeV$ for ZEUS.

\subsubsection{\xf spectra in the CMS}

\begin{figure}[tbh]
   \centering
   \epsfig{file=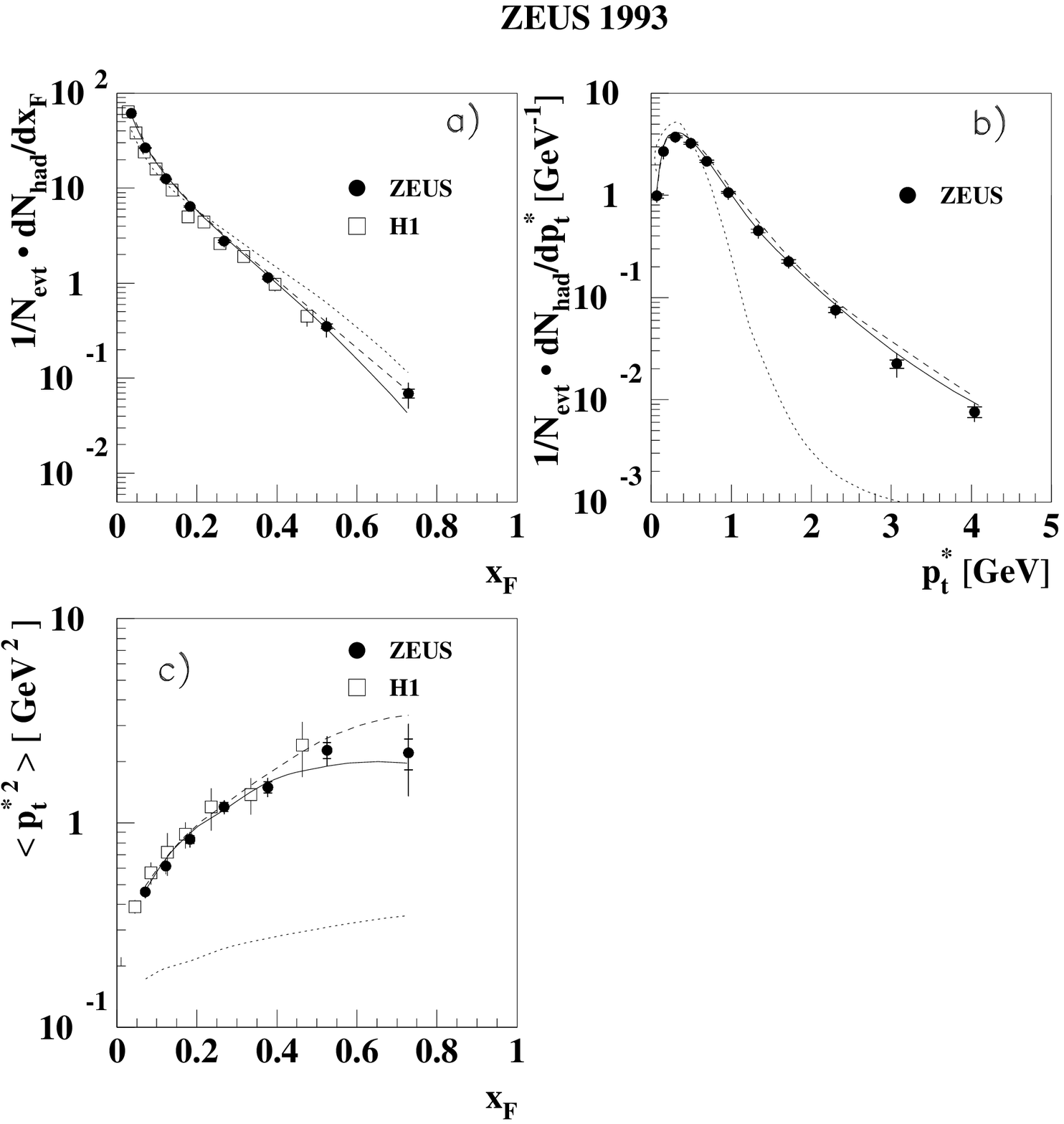,%
          width=14cm}
   \scaption{Charged particle distributions in the CMS current hemisphere
                from H1 \cite{h1:flow1}
                and ZEUS \cite{z:xf}.
             {\bf a)} The Feynman $x$ ($=x_F$) spectrum.
             {\bf b)} The \pt spectrum for $\xf> 0.05$.
             {\bf c)}  The $\av{p_T^2}$ vs. \xf (seagull plot for $x_F>0$).
             The data are compared with
             the QPM (dotted line)
             without QCD radiation, and the QCD models MEPS
             (LEPTO 6.1, full line) and
             CDM (ARIADNE 4.0, dashed line).}
   \label{ptzeus}
\end{figure}

The Feynman-$x$ ($\xf=2\pz/W$) spectra
are steeply falling with \xf, in fact
more steeply than what would be expected from the
naive QPM, that is quark fragmentation
without QCD radiation (fig.\ref{ptzeus}a).
The data can be described when
QCD radiation is taken into account.
Due to gluon radiation, more than one
parton share the available energy for fragmentation; therefore less
fast particles are produced, and the multiplicity of soft particles
increases (fig.~\ref{frag}).

\begin{figure}[tbh]
   \centering
   \epsfig{file=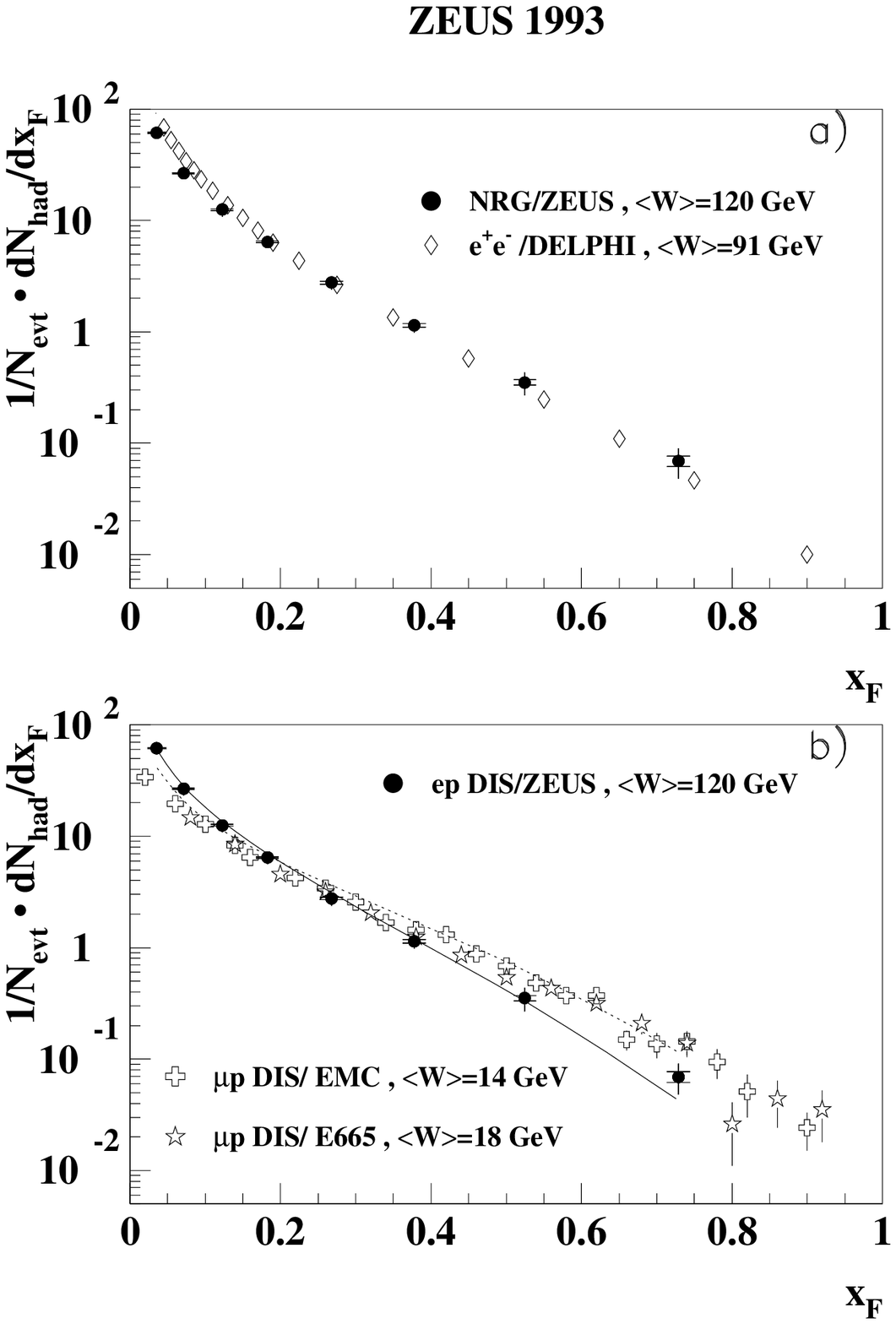,%
          width=9cm}
   \scaption{
             {\bf a)} The \xf spectrum from ZEUS \cite{z:xf}
                compared with DELPHI data
                \cite{lep:xf}. The DELPHI data are divided by 2.
             {\bf b)} The \xf spectra measured at HERA \cite{h1:flow2,z:xf},
                compared with
                the QPM (dotted line)
                without QCD radiation, the MEPS model
                (LEPTO 6.1, full line), and
                with fixed target $lN$
                data at lower $W$ from EMC  \cite{o:emcs1}
                and E665 \cite{o:e665s1}.
             }
   \label{xf1}
\end{figure}

The spectra can in principle depend on the available
phase space, determined by $W$, and on the virtuality $Q^2$.
A dependence on $x\approx Q^2/W^2$ is implied and important, because
the parton composition of the proton changes with $x$.
The following table \ref{tab:scales_xf} gives the scales involved.
\begin{footnotesize}
\begin{table}[h]
\begin{center}
\begin{tabular}{|l|c|c|c|c|c|}
  \hline
                    & $\av{W}$ (GeV) & $\av{\Qsq}$ (\GeVsq) &
                      $\sqrt{s}$ (GeV) &  $\av{x}$ \\
  \hline
  LEP \epem ($\sqrt{s}=m_Z$)  &  91    & 8300  & 91 & --         \\
  HERA $ep$                   & 120    &  28 & 300   & $10^{-3}$     \\
  fixed target $lN$ (EMC/E665)&  14/18 &  10 & 24/30 & $5\cdot 10^{-2}$ \\
  \hline
\end{tabular}
\end{center}
\scaption{
           Approximate kinematics and
           scales for the \epem, $ep$ and $lN$ data which are discussed
           here.
           }
\label{tab:scales_xf}
\end{table}
\end{footnotesize}
The DIS \xf spectra measured at HERA \cite{h1:flow2,z:xf}
are softer than the spectra measured at lower energy fixed target
experiments by EMC \cite{o:emcs1} and E665 \cite{o:e665s1},
see fig.~\ref{xf1}b.
From this comparison it is however
not clear whether the difference is due to the different values
of $x$, \W or \Qsqx.

In fig. \ref{xf1}a
the HERA data are compared with LEP data from DELPHI \cite{lep:xf}
(divided by 2 to account for the two hemispheres) at similar, though
somewhat smaller centre of mass energy $\sqrt{s_{e^+e^-}}=W=m_Z$.
For $\xf\gtrsim 0.1$ fragmentation univerality appears to hold,
since the HERA and LEP spectra agree.
At smaller $x_F$, the multiplicity is smaller in DIS than in \epem~
interactions.
(The different energy dependence of multiplicities at central rapidity
in DIS and \epem is discussed in section \ref{sn:multi}, see
fig. \ref{wdep}.)
The following circumstances may be held responsible for the difference:
1) mass effects
   due to the different flavour compositions;
2) the BGF
   contribution to DIS, which is absent in \epem~ interactions;
3) the scale $Q^2$, which is smaller at HERA than at LEP;
4) the antenna effect due to the extended colour charge of the proton
   remnant (though $x_F=0.1$ is already 2 units of rapidity away
   from the target region $\eta<0$).
Further studies are needed to identify the reason(s) for the difference.

A recent analysis by E665 \cite{o:e665_xfpt}
of the \W and \Qsq dependence of
the \xf spectra in the range $7.5 < W < 30 \GeV$ and
$0.15<\Qsq<20\GeVsq$ reveals that at low \xf they depend
mainly on $W$, and at large \xf mainly on \Qsq.
The increase with \W at small \xf is essentially due to the
increase of longitudinal phase space. The decrease at large \xf
has been attributed to the scale dependence of the fragmentation function.

\subsubsection{QCD predictions with fragmentation functions}

\begin{figure}[tbh]
   \centering
\begin{picture}(0,0) \put(0,0){{\bf a)}} \end{picture}
   \epsfig{file=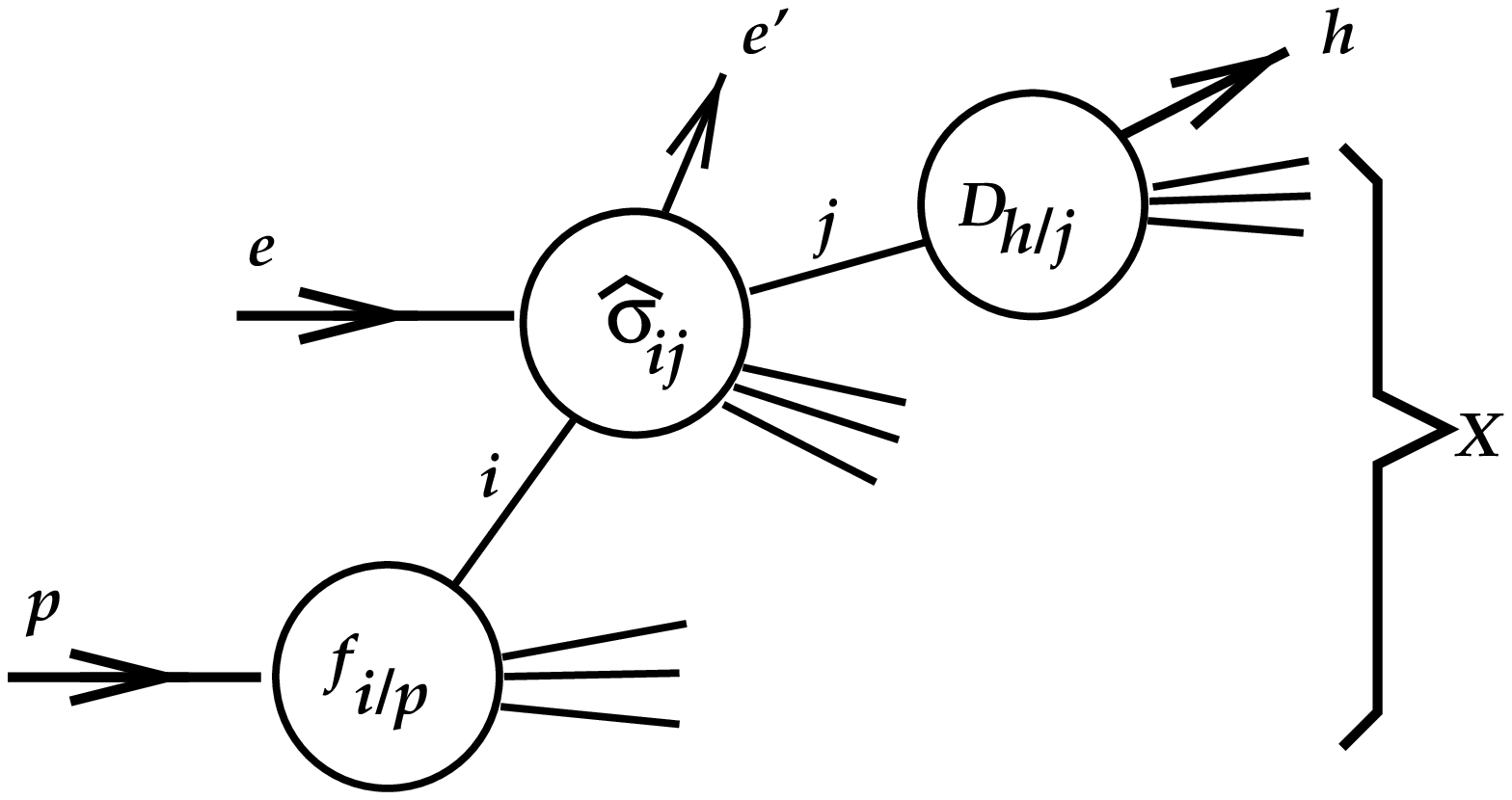,width=8cm}
   \hspace{2cm}
\begin{picture}(0,0) \put(0,0){{\bf b)}} \end{picture}
   \epsfig{file=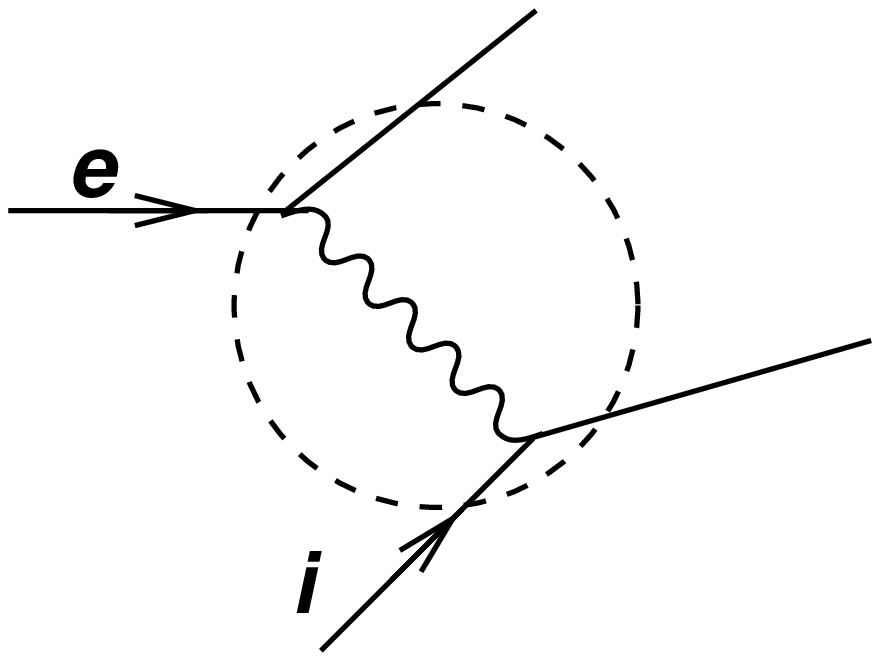,width=4cm}
   \scaption{
             {\bf a)} The process $ep\rightarrow e'hX$.
             In lowest order the blob
             labelled $\hat{s}_{ij}$
             represents electron-quark scattering via
             the exchange of a virtual photon {\bf (b)}.
            }
   \label{xf}
\end{figure}

The hadron production cross section (see fig.~\ref{xf})
$\dif \sigma /\dif z$
can be written as a convolution of the parton production cross section
with the fragmentation function $D_{h/j}(z,\mu_D^2)$,
which gives the number density
to observe a hadron $h$ with
momentum fraction $z$ from the fragmentation of parton $j$.
The parton production cross section itself is a convolution
of the parton density function $f_{i/p}(\xi,\mu_F^2)$ for
parton species $i$ in the proton with momentum fraction $\xi$,
with the hard scattering cross section $\hat{\sigma}_{ij}$
to produce
parton $j$ from electron scattering on that parton $i$, $e+i \rightarrow j+X$.
Finally, the cross sections need to be
summed over all parton species $i,j$:
\begin{equation}
\frac{\dif \sigma}{\dif z} = \sum_{i,j}
         f_{i/p}(\xi,\mu_F^2) \otimes
          \hat{\sigma}_{ij} \otimes  D_{h/j}(z,\mu_D^2).
\label{eq:scviol}
\end{equation}
The parton densities and the
fragmentation functions depend upon their respective
factorization scales, $\mu_F^2$ and $\mu_D^2$.
The hard scattering cross
section $\hat{\sigma}_{ij}=
\hat{\sigma}_{ij}(\mu_F^2,\mu_D^2,\mu_R^2,\xi)$
depends in addition upon the
renormalization scale $\mu_R^2$ and \as.
The scales $\mu_R^2,\mu_D^2,\mu_F^2$ are in principle arbitrary;
here they are taken to be \Qsq (not $W^2$~!),
the physical scale related with
the hard scattering in DIS \cite{th:scviol}.
The situation in DIS with $Q \ll W$ is different to \epem,
where $\Q=W=\sqrt{s}=2E_q$.

\begin{figure}[tbh]
   \centering
   \epsfig{file=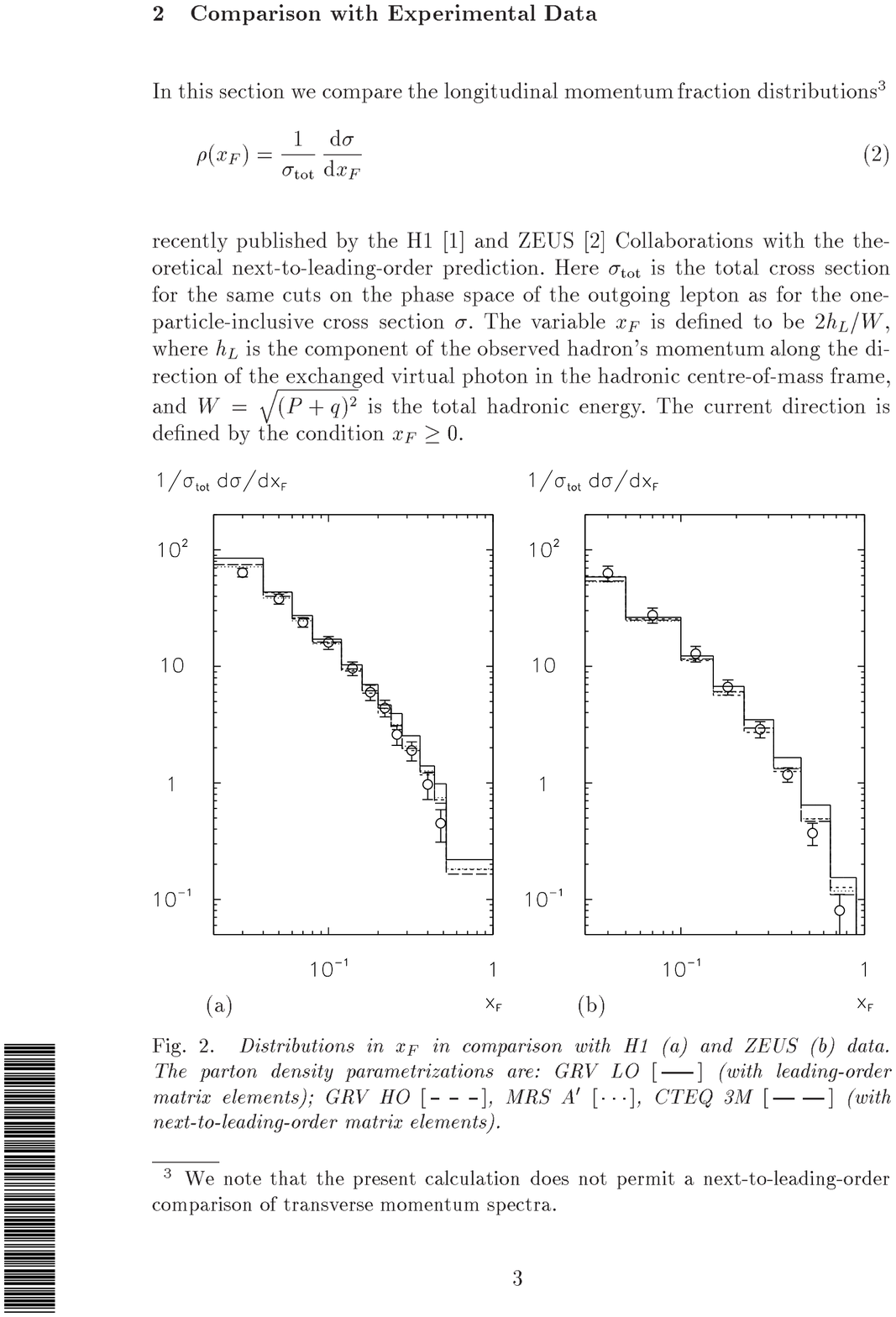,%
          width=14cm,bbllx=84pt,bblly=169pt,bburx=485pt,bbury=473,clip=}
   \scaption{The calculation by Graudenz \cite{th:scviol}
             for different parton density parametrization
             compared
             to the HERA \xf data from {\bf a)} H1 and {\bf b)} ZEUS.
             The spectrum is calculated in LO
             with the GRV LO parametrization (full line),
             and in NLO with the parametrizations
             GRV HO (dashed), MRSA' (dotted) and CTEQ3M (long dashed).
             }
   \label{graudenz}
\end{figure}

With experimentally determined parton densities and
fragmentation function parametrizations
from other, mainly \epem, reactions \cite{th:ffct},
the NLO QCD calculation for the HERA \xf spectrum
is quite
satisfactory (fig.~\ref{graudenz}) \cite{th:scviol}.
The LO calculation predicts a spectrum that is too hard compared with the data.
When moving out of the
current fragmentation region, i.e. towards small and negative
$x_F$, these calculations are expected to break down due to
fragmentation effects that are not related to the scattered parton.

Since the scaling violation of the parton density and fragmentation
functions as well as the hard cross section depend on $\alpha_s$,
a measurement of the scale dependence of $\dif \sigma /\dif z$
allows in principle a measurement of $\alpha_s$.
Due to the \Qsq range covered at HERA, it would be possible
to measure \as from the scaling violations of the fragmentation
functions in a single experiment.
For an integrated luminosity
of 250 \pbinv a competitive statistical error for $\alpha_s(m_Z^2)$ of
$\pm 0.0007$ is expected.
The theoretical uncertainties are
presently estimated to be much larger \cite{th:scviol}:
$\pm 0.005$ from the parton densities
and $\pm 0.011$ from the scale uncertainty.
A better knowledge of parton densities would reduce the corresponding
error. Possibly the theoretical situation is more favourable in
the Breit system, where the scale of the hard interaction
also sets the energy of the fragmenting quark, $Q/2=E_q$, similar
as in \epem. First indications of scaling violations in the Breit frame
are presented in section \ref{sn:xp}.

\subsubsection{Transverse momenta}

The transverse momentum spectrum of charged particles
from ZEUS
with $\xf>0.05$ is shown in fig.~\ref{ptzeus}b.
The data exhibit a high \pt tail over the expectation from
the QPM without QCD radiation, which is well described by
the QCD models.
\pt can be generated in the QCD cascade, where partons are emitted
before and after the hard scattering.
Somewhat arbitrarily one may distinguish radiation from the
hard matrix element (BGF and QCDC) and soft gluon radiation
(parton shower).

In general, contributions to the \pt of primary hadrons are expected
also from other sources than QCD radiation (\ptrad)
\cite{mc:bo2,rev:schmitz}:
from an intrinsic \pt of the initial parton in the proton (\ptintr),
and from fragmentation (\ptfrag).
These contributions can be summed up\footnote{
An intrinsic \pt of a parton in the proton
leads to a tilt
of the fragmentation axis with respect to the $\gamma^\ast$ axis.
A daughter hadron with momentum
fraction \xf will therefore inherit a transverse momentum
$\xf \cdot \ptintr$. Models assume a
Gaussian distribution as in eq.~\ref{eq:ptgauss} with
$\av{\ptintr^2}=(0.39\GeV)^2$ \cite{mc:lepto} in accord with
the uncertainty principle applied to
a parton confined to the proton volume.
The experimental value is
$\av{\ptintr^2}=[0.29^{+0.05}_{-0.07}({\rm stat.})
 ^{+0.14}_{-0.18}({\rm syst.}) \GeV]^2$ \cite{o:emcs1}.

\pt from fragmentation is also assumed to be Gaussian distributed
(eq.~\ref{eq:ptgauss}) \cite{mc:bo1}.
A good choice for the description of the HERA and LEP data with
the Lund string model is
$\av{\ptfrag^2}\approx(0.3\GeV)^2$ \cite{mc:lepto,mc:leptune,mc:heratune}.
$\av{\ptfrag^2}=(0.41\pm0.02 \GeV)^2$ has been measured
with fixed target data \cite{o:emcs1}.
},
\begin{equation}
  \av{p_T^2} \approx x_F^2 \av{\ptintr^2}
               + \av{\ptfrag^2} + \av{\ptrad^2}.
\end{equation}
The \pt of the finally observed hadrons will be diluted due
to decays of the primary hadrons.
At HERA, the relative importance
of \ptrad is bigger than at the
lower energy DIS experiments. This is best studied in the so-called
seagull plot.

\begin{figure}[tbh]
   \centering
   \epsfig{file=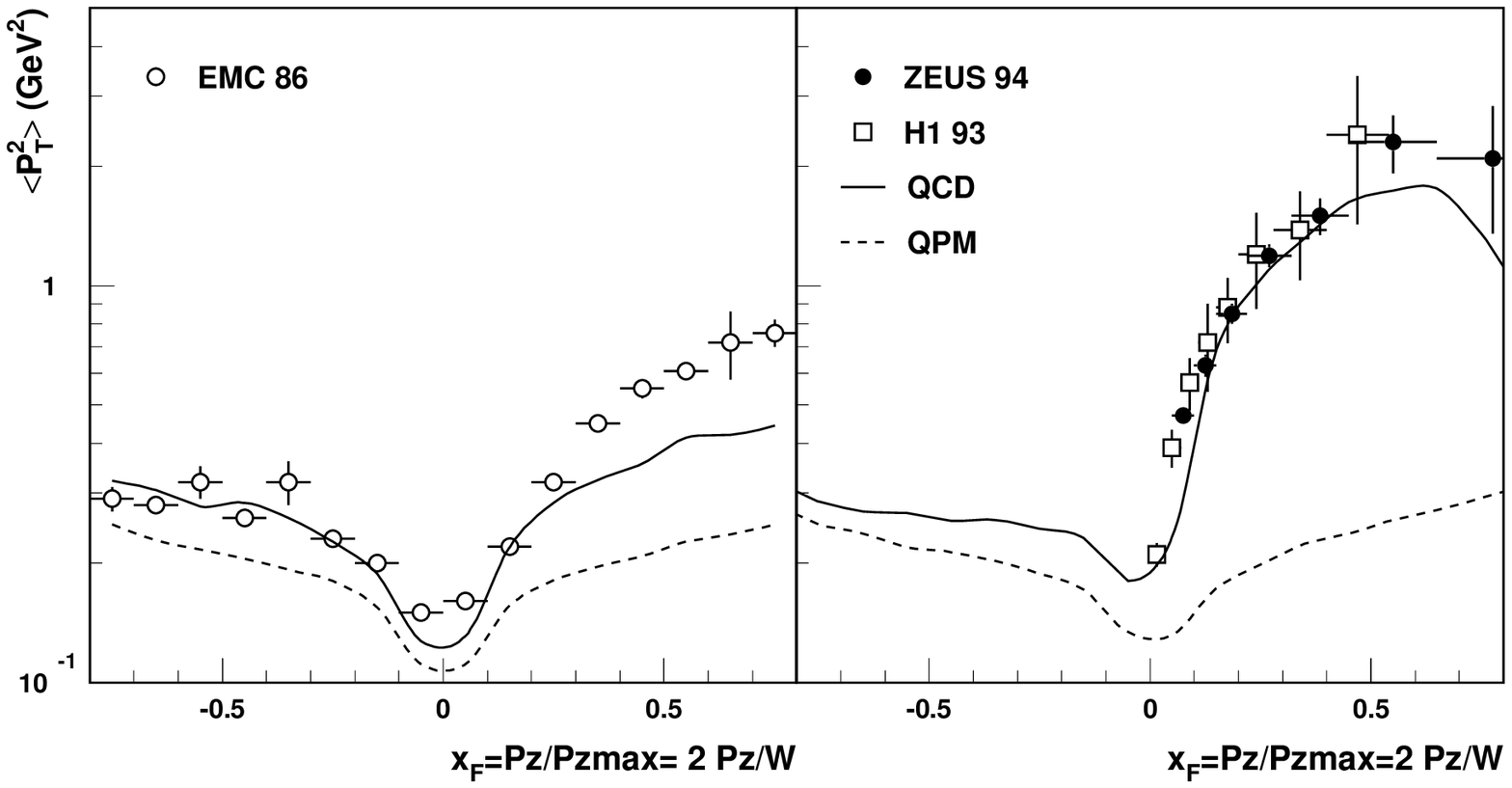,%
              width=12cm}
   \scaption{The seagull plot, $\av{p_T^2}$ vs. \xf for charged
             hadrons in the CMS \cite{h1:carli}.
             $\xf>0$ defines the current hemisphere,
             $\xf<0$ the target hemisphere.
             The data from H1 \cite{h1:flow2}, ZEUS \cite{z:xf} and
             EMC \cite{o:emcs1}
             are compared to a QCD model (CDM, ARIADNE 4.08) and the QPM.}
   \label{sg}
\end{figure}

In the seagull plot the mean $p_T^2$ of charged hadrons is
plotted as a function of $x_F$.
The data from EMC \cite{o:emcs1},
H1 \cite{h1:flow2} and ZEUS \cite{z:xf} are shown in
figs.~\ref{sg} and \ref{ptzeus}c.
The shape of this distribution
is reminiscent of a seagull in the sky\footnote{
The dip at \xf=0 is mainly a phase space effect. Consider
massless hadrons distributed according to the longitudinal
phase space model,
\begin{equation}
   \frac{\dd^2\sigma}{\dd y \dd p_T^2} = a \cdot \exp(-p_T^2/b).
\end{equation}
Because $\dd y/\dd p_z = 1/E$, we have
\begin{equation}
   \frac{\dd^2\sigma}{\dd x_F \dd p_T^2} =
   \frac{W/2}{E} \frac{\dd^2\sigma}{\dd y \dd p_T^2}.
\end{equation}
For small longitudinal momentum, $E=\sqrt{p_z^2+p_T^2}\approx p_T$.
So for $x_F \rightarrow 0$, the cross section is weighted with
a factor that grows with $1/p_T$ for small $p_T$, giving a large
weight to small $p_T$ hadrons in the average $p_T^2$.
}.
Due to the limited
acceptance to $x_F\gtrsim 0$, at HERA only one wing of the
seagull plot is measured.

The HERA data are
well described by QCD models.
Without QCD radiation (the QPM)
the prediction falls far below the data.
While the QPM predictions are quite similar
for EMC and HERA energies, QCD radiation effects
are much larger at the higher energy experiments.
The $\av{p_T^2}$ increases with $W$ \cite{z:xf}.

The fixed target data also cover the target fragmentation
region, $\xf<0$. In the target region QCD radiation is
suppressed as compared to the current region.
This suppression
can be explained with the smeared out colour charge in the proton
remnant, if one considers radiation from a colour dipole.
Perturbative QCD evolution as implemented in LEPTO yields a similar
asymmetry.
The description of the EMC data by the CDM (ARIADNE 4.08)
is certainly not satisfactory. It would be interesting to
try to tune the QCD models to both data samples simultaneously.

\begin{figure}[tbh]
   \centering
   \epsfig{file=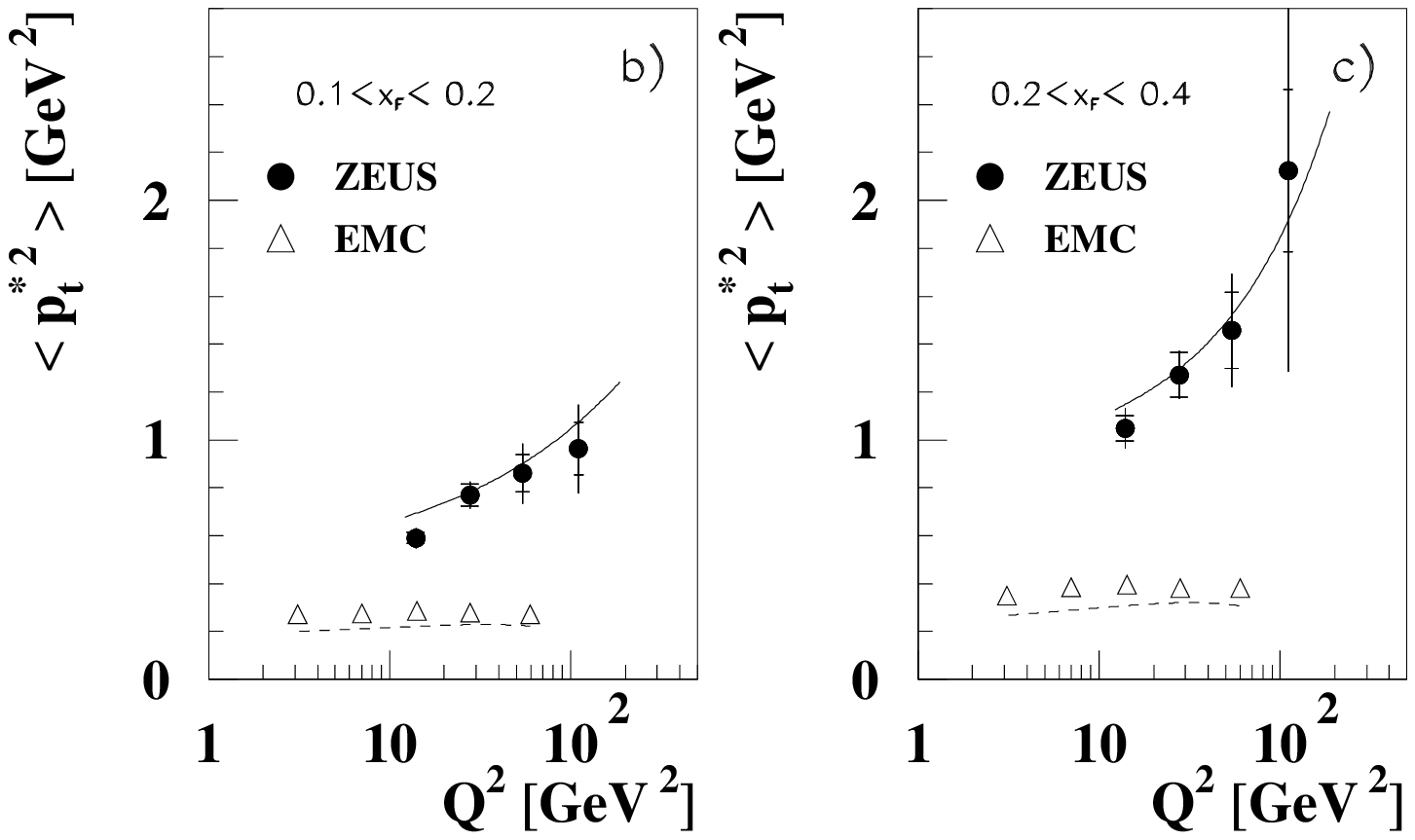,%
          width=12cm}
   \scaption{The \Qsq dependence of the charged particle $\av{p_T^2}$
             for different values of \xf. The
             ZEUS \cite{z:xf} data at $\av{W}=120\GeV$ are compared
             to EMC data \cite{o:emcs1} at $\av{W}=14 \GeV$. Also
             shown are the MEPS predictions for the two values of $W$.}
   \label{ptq2}
\end{figure}

At HERA energies an increase of $\av{p_T^2}$ with
\Qsq is observed at fixed $x_F$, whereas at
the lower energy fixed target data from EMC \cite{o:emcs1}
almost no \Qsq dependence was found, see fig.~\ref{ptq2}.
With very precise data in the kinematic region $7.5<W<30\GeV$,
$0.15<Q^2<20\GeVsq$ and $1.5 \mmmm < x < 0.6$
the $\mu p$ experiment E665 has
recently seen a $W$ as well as a \Qsq dependence for the
mean $p_T^2$ \cite{o:e665_xfpt}.

\begin{figure}[tbh]
   \centering
   \epsfig{file=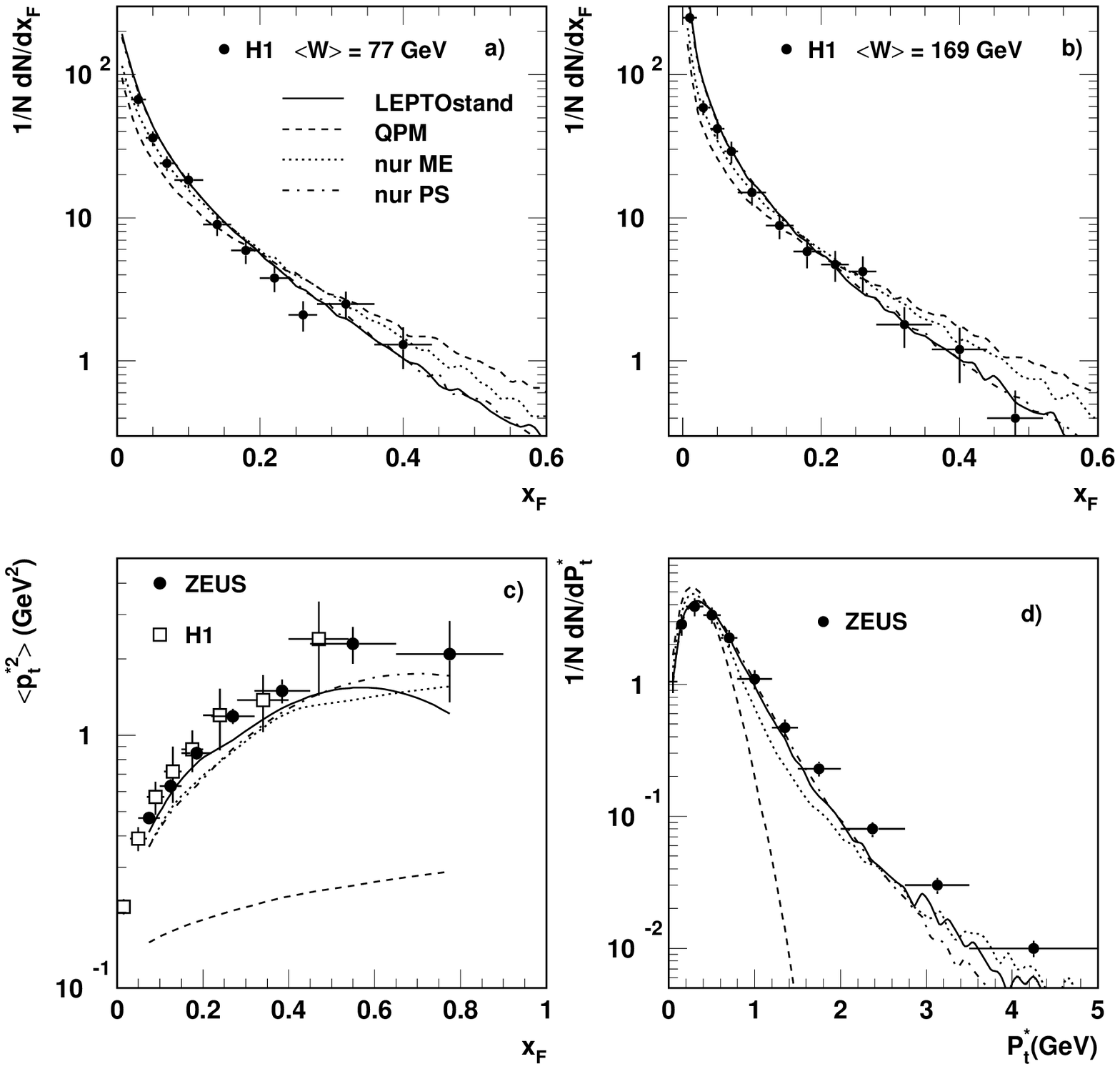,%
              width=14cm}
   \scaption{QCD radiation effects
             according to the LEPTO 6.4 model \cite{h1:mohr}.
             The data are compared to the LEPTO model without
             QCD radiation (QPM), QCD radiation from the LO
             matrix element alone (dotted, ME),
             QCD radiation from parton showers alone (PS, dash-dotted),
             and the full model (full line). }
   \label{sgmohr}
\end{figure}


We can use a QCD model to investigate the different contributions
to the softening of the \xf spectrum and to the increasing $p_T$.
In the MEPS model (LEPTO 6.4) radiation from the LO matrix element
at the photon vertex (BGF and QCDC) can be identified as the
main cause; additional soft gluon radiation is relatively
unimportant (fig.\ref{sgmohr}). On the other hand, the
QCD parton shower as implemented in LEPTO appears to cover
already a large part of the LO matrix element contribution.
Note that the parton shower changes when the
matrix element cut-off changes (or is switched off)
in order to cover the maximal phase space without double counting
(this adjustment is sometimes called ``matching'').

  \section{Strangeness Production \label{sn:strange}} 


Strange particles may be produced in DIS when the scattered quark
is a strange quark (either from the sea, or produced in BGF events),
of from $s\ol{s}$ pairs created during parton showering and hadronization,
from heavy quark decays, or
from non-standard processes like instantons.
Strange particles produced directly
from the scattered quark are expected to carry a
relatively large fraction
of the scattered quark's energy,
whereas most of the strange particles
are expected from hadronization with small fractional energy.
In the Lund hadronization model, $q\ol{q}$ pairs are
created in the colour field of the string from the vacuum.
Because $m_s > m_u \approx m_d$,
$s\ol{s}$ pair creation is suppressed with respect to the other
light quarks with a ratio $\ls : 1 : 1$.
Typically, $\ls = 0.2-0.3$, with 0.3 the Lund default value
representing the experimental mean from \cite{rev:strange}.

H1 \cite{h1:k0} and ZEUS \cite{z:k0}
identify and reconstruct $K^0$ and $\Lambda$
particles
via the
invariant mass of their
decay products in the decays
$K^0_s \rightarrow \pi^+\pi^-$,
$\Lambda \rightarrow p \pi^-$, $\ol{\Lambda} \rightarrow \ol{p} \pi^+$.
In the following, the notations $K^0$ and $\Lambda$ include
both particles and antiparticles.
The decays are
measured in the central drift chambers. This limits the
acceptance for strange particles
in the laboratory system to $-1.3<\eta<1.3$ and
$\pt>0.5~\GeVx$. The results for $K^0$ and $\Lambda$ yields are given
in table~\ref{tab:strange}.
Neglecting the slightly different kinematic selections
and averaging the ZEUS and H1 results,
for $-1.3<\eta<1.3$ and $\pt>0.5 \GeV$ the \knot yield is
$0.111\pm 0.005$ per event and unit of pseudorapidity,
and for $\Lambda$ baryons it is $0.042\pm0.0005$.

\begin{footnotesize}
\begin{table}[htb]
\begin{center}
\begin{tabular}{|l|c|c|}
   \hline
                &      H1                  &   ZEUS \\
  \hline
event selection &  no large rapidity gap      &        \\
                &  $10\GeVsq<p_T^2<70\GeVsq$ & $10\GeVsq<p_T^2<640\GeVsq$ \\
                &  $0.0001 < x < 0.01$     & $0.0003 < x < 0.01$      \\
                &  $0.05   < y < 0.6$      & $0.04 < y$               \\
  \hline
strange  &  $0.5\GeV < \pt <2.12\GeV$ & $0.5\GeV < \pt <3.5\GeV$  \\
particle selection   &  $-1.3<\eta<1.3$         &     $-1.3<\eta<1.3$  \\
  \hline
\# \knot/event  & $0.287\pm0.008\pm0.012$ & $0.289 \pm 0.015 \pm 0.014$ \\
\#\knot / \# charged $^*$ & $0.058\pm 0.002 \pm 0.003$ &
                       $0.077 \pm 0.006 \pm 0.008$         \\
\# $\Lambda$ \ event & $0.053 \pm 0.007 \pm 0.007$ &
                       $0.038 \pm 0.006 \pm 0.002$ \\
  \hline
\end{tabular}
\end{center}
\scaption{Strange particle production at HERA.
          A lower \pt cut of 0.15 GeV for H1 and
          0.2 GeV for ZEUS is applied for the charged particles in the
          ratio.
          $^*$ For the ratio \#\knot / \# charged both experiments exclude
          events with a
          large rapidity gap.
          }
\label{tab:strange}
\end{table}
\end{footnotesize}

\begin{figure}[tbh]
   \centering
   \epsfig{file=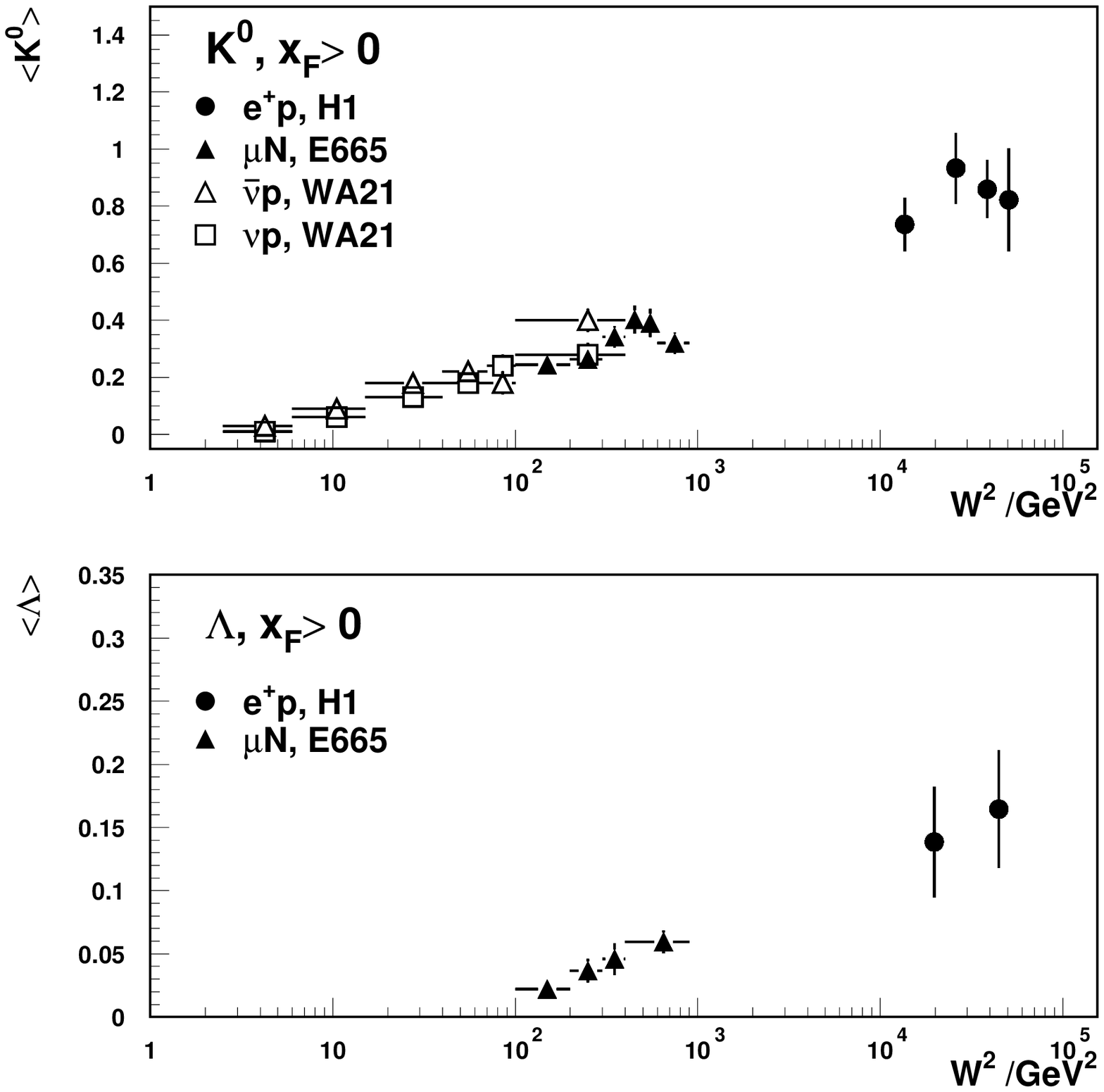,%
          width=10cm}
   \scaption{The average \knot (top)
             and $\Lambda$ (bottom)  multiplicity in the
             CMS current hemisphere as a function of $W^2$.
             The H1 data \cite{h1:k0} are compared to
             fixed target data from E665 \cite{o:e665k0} and
             WA21 \cite{o:wa21k0}.}
   \label{k0b}
\end{figure}

The strange particle yield has been studied as a function
of $x$, $W$, \Qsq, $\eta$, \pt and $x_F$. In general,
the same features are observed as for charged particles.
These are an almost flat distribution
in pseudorapidity, steeply falling \pt spectra, and
no significant dependences on the event kinematics other
than on $W$. The ratio of $K^0$ to charged particle yield
was found to be the same in events with and without a large rapidity gap,
within errors.
The average $K^0$ multiplicity in the CMS current hemisphere
increases logarithmically with $W$, see fig.~\ref{k0b}.
This behaviour is similar to what has been seen for charged particles,
though a faster than logarithmic
rise is not seen with $K^0$ mesons (compare~\ref{sn:multi}).

\begin{figure}[p]
   \centering
   \vspace{-1.5cm}
   \epsfig{file=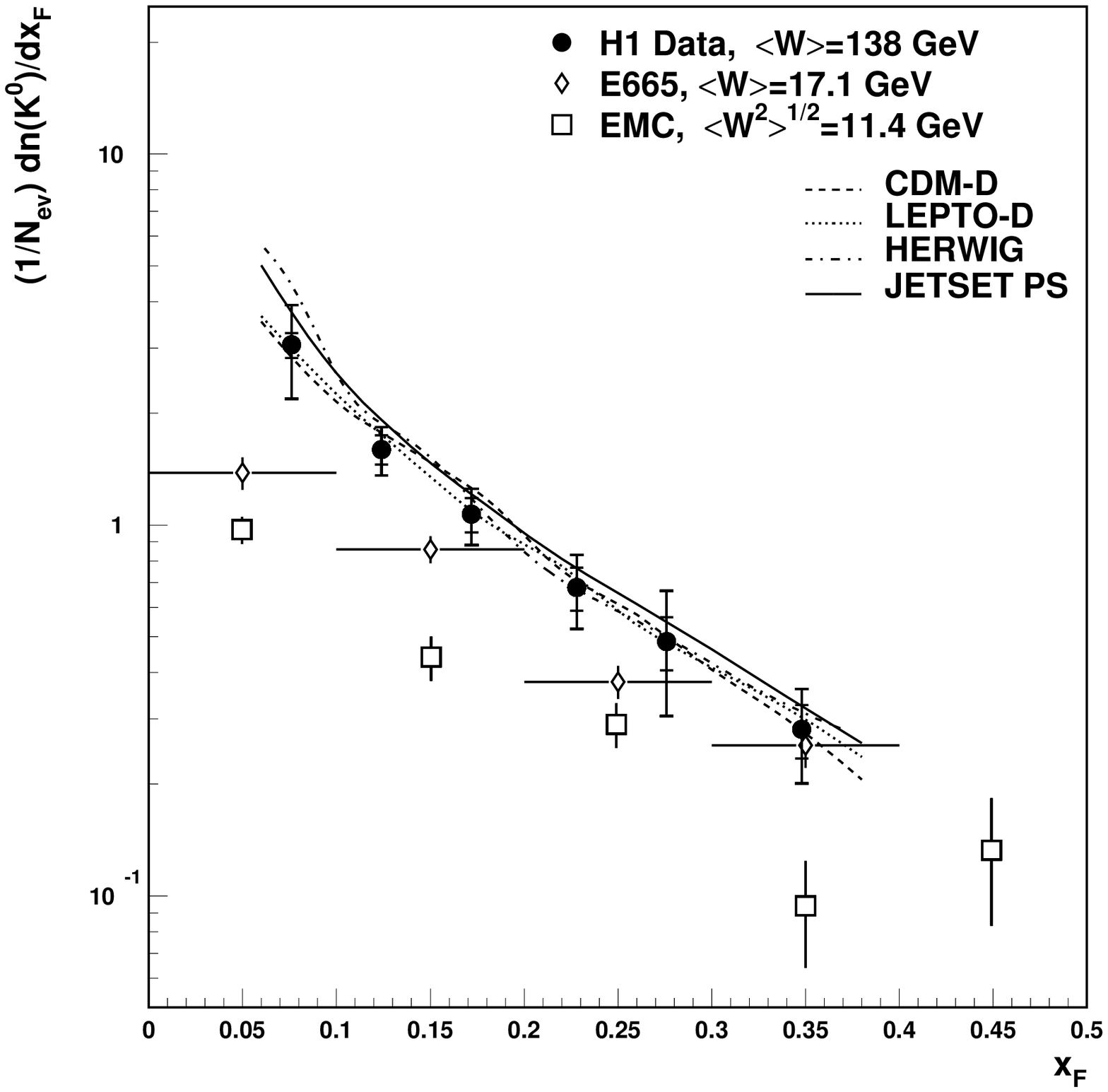,%
          width=10cm}
   \vspace{-0.5cm}
   \epsfig{file=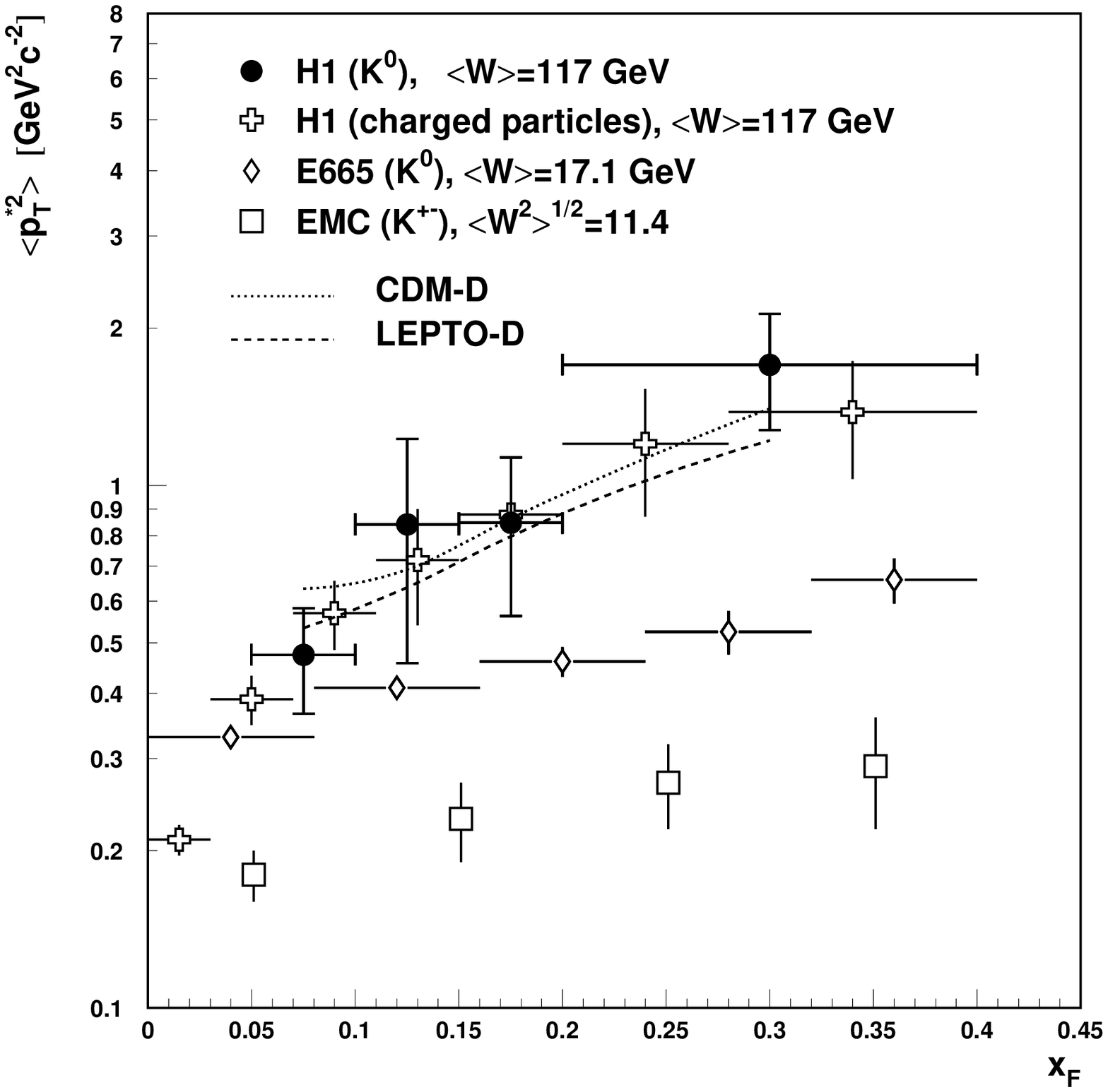,%
          width=10cm}
   \scaption{Top) \xf spectrum of \knot mesons.
             The H1 data \cite{h1:k0} are compared to
             fixed target data from EMC and E665 \cite{o:emck0,o:e665k0},
             and to the DIS event generators CDM, LEPTO and HERWIG.
             For CDM and LEPTO the DELPHI fragmentation parameter set
              \cite{lep:delphifrag} with \ls=0.23 were used.
             The curve JETSET is a simulation of \epem annihilation
             ($u,d,s$ quarks only) at $\sqrt{s}=138$ GeV.
             Bottom) The average $p_T^2$ as a function of \xf
             for \knot \cite{h1:k0} and charged particles \cite{h1:flow2}
             from H1, compared to fixed target data  \cite{o:emck0,o:e665k0}
             and QCD models.}
   \label{k0a}
\end{figure}

The data can be reasonably well described by
standard QCD models invoking Lund string fragmentation,
provided $\ls \approx 0.2$ is chosen.
As an example, the \xf spectrum and the seagull plot for
\knot mesons are shown in fig.~\ref{k0a}.
The larger $\av{p_T^2}$ observed when going from fixed target to
HERA energies can be understood with QCD radiation in the same
manner as for charged particles. About the same $\av{p_T^2}$
are observed for $K^0$s and for charged particles.
It is not yet clear why the supposedly
scaling \xf spectra of the fixed target data lie below the HERA data
for $\xf>0.1$, but this is also expected from Monte Carlo \cite{h1:milstead}.
Probably mass effects play a r\^{ole} at smaller energies, in contrast
to the HERA data.
The H1 data agree with the expectation for \epem~
annihilation to light quarks at the same CM energy, represented
by the JETSET generator.

  \section{Charm Production  \label{sn:charm}}
\subsubsection{The charm yield}

Charm production
in DIS has been investigated by detecting
$D^{\star +} \rightarrow D^0 \pi^+ \rightarrow K^-\pi^+\pi^+$
\cite{h1:charm,z:charm} and $D^0 \rightarrow K^-\pi^+$ decays
(and their charge conjugates). The decay products are detected in the
central drift chambers.
$D^0$ mesons are reconstructed by the $K\pi$ invariant masses.
$D^\star$ mesons are identified by the
kinematically tightly constrained mass difference
$\Delta m = m(D^0\pi^+_{\rm slow}) - m(D^0)$
between a $D^0\rightarrow K\pi$ candidate and the slow pion, and the
$D^0$ candidate mass.

\begin{figure}[tbh]
   \centering
\begin{picture}(0,0) \put(0,0){{\bf a)}} \end{picture}
\hspace{0.5cm}
   \epsfig{file=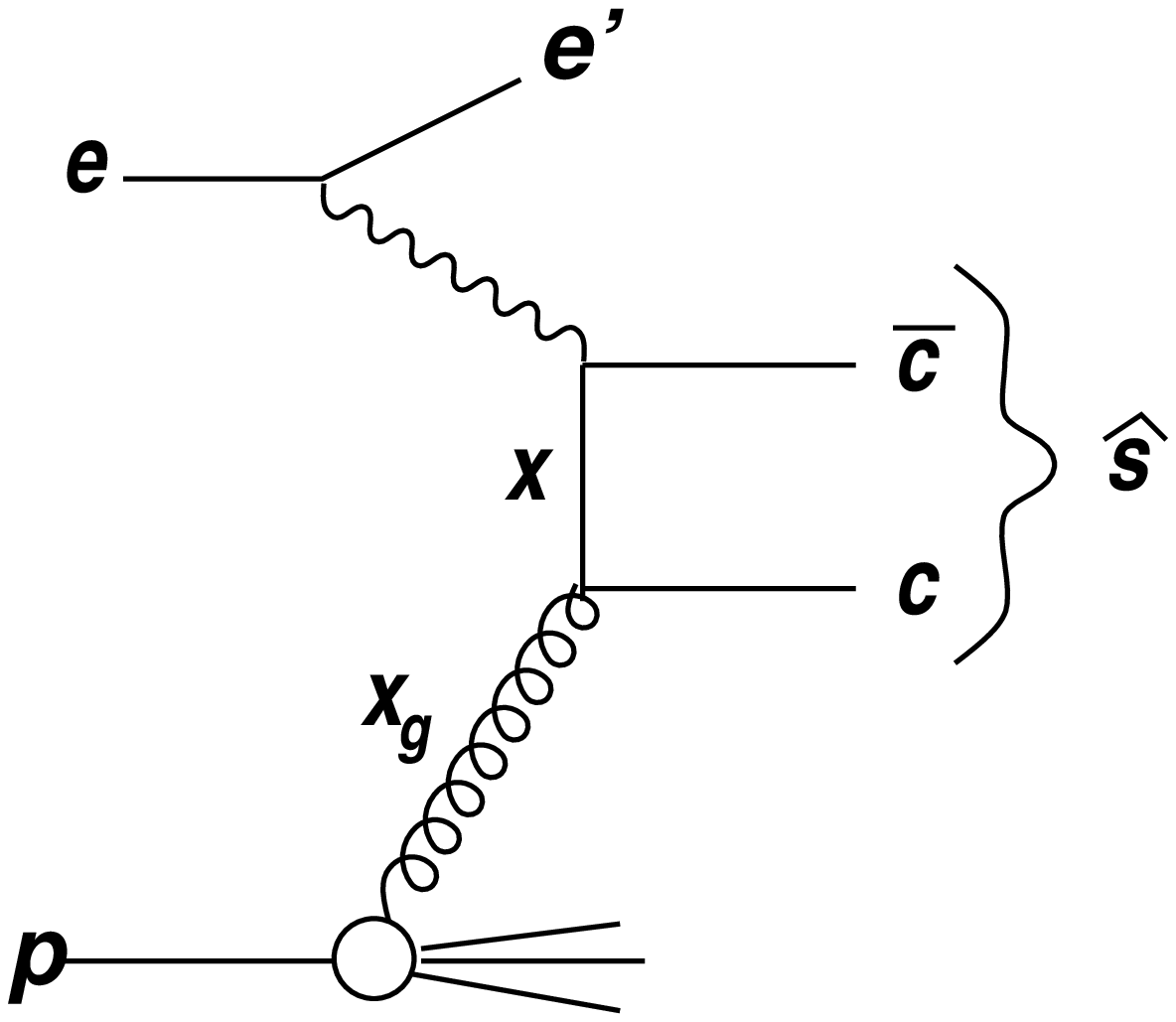,%
          width=5cm,bbllx=123pt,bblly=235pt,bburx=480pt,bbury=560,clip=}
\begin{picture}(0,0) \put(0,0){{\bf b)}} \end{picture}
   \epsfig{file=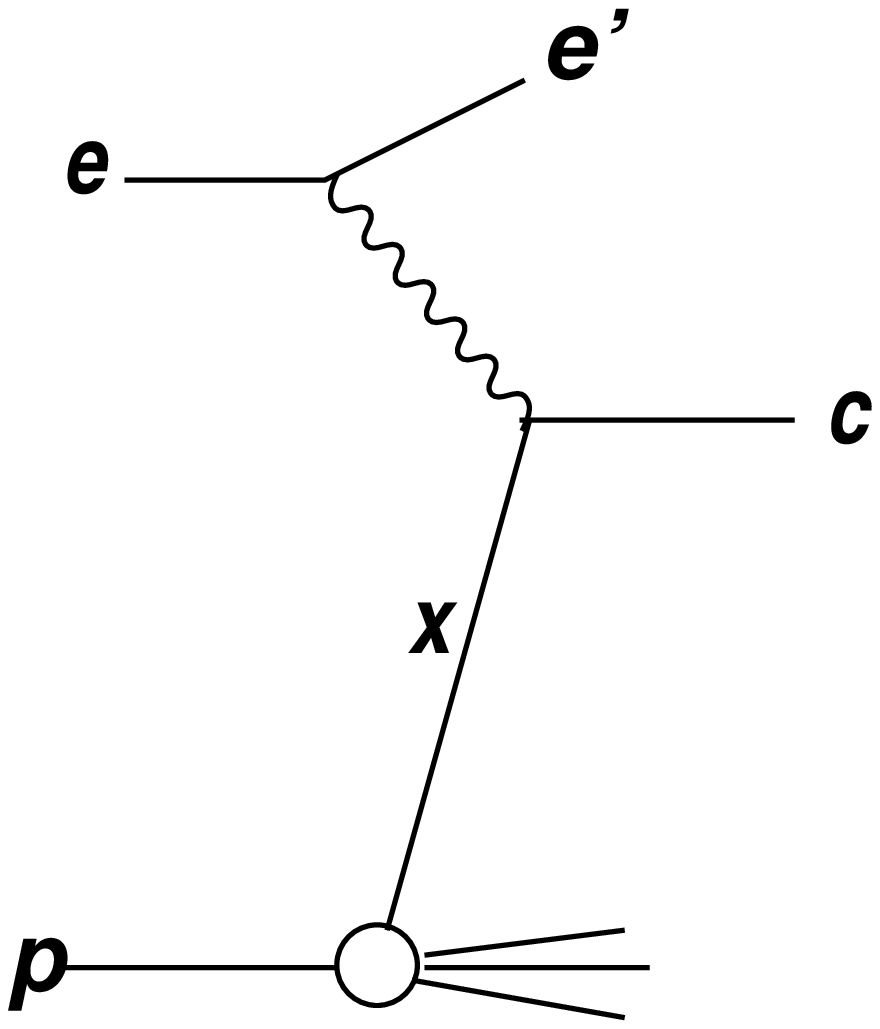,%
          width=5cm,bbllx=123pt,bblly=235pt,bburx=480pt,bbury=560,clip=}
   \scaption{Charm production in LO:
             {\bf a)} boson gluon fusion (BGF) and
             {\bf b)} from sea or valence quarks (QPM).
               In the BGF process the fractional gluon momentum $x_g$ is
               related to Bjorken $x$ by
               $x_g = x(1+\hat{s}/Q^2)$.}
   \label{charmgraph}
\end{figure}

Apart from possible
anomalous sources of charm (new particles, instantons, ...),
the main interest
in charm production stems from its sensitivity to the gluon content
of the proton at small fractional momentum $x_g$.
The two lowest order standard charm production processes
in DIS are boson gluon fusion (BGF) and scattering off a charm
sea quark in the proton
(QPM)\footnote{The boundary between QPM and BGF is subject to definition.},
see fig.~\ref{charmgraph}.
At high $x$ also an intrinsic (valence) charm component
to the proton at the few permil level has been discussed \cite{th:cintr}.
At small $x$ the BGF contribution is expected to dominate
due to the increasing gluon
content.
When the invariant mass
$\sqrt{\hat{s}}$ of the \qqbar~ system is
large enough
($\hat{s}\gg (2m_c)^2$)
so that mass effects can be neglected,
the quark flavours
are expected to be produced according to $u:d:s:c=4/9:1/9:1/9:4/9$ due
to their electric charges, provided one is below the $b$ threshold.
\bbbar~ production is suppressed by the large bottom mass.
It has been estimated \cite{h1:daum} that in DIS
$b:c$ production is $\approx 0.02$, integrated
over a large range in $x,Q^2$, but can be as large as 0.1 at small
$x$ and large $Q^2$, that is large $W$. There are no results on
open $b$ production from HERA yet.

The charm cross section\footnote{The
elastic (diffractive) cross section for $ep\rightarrow ep J/\psi$
is comparably small, $100\pm 20 \pm 20 $~pb for $Q^2>8\GeVsq$ and
$30\GeV<W<150\GeV$ \cite{h1:disjpsi}.}
is inferred from the $D$ production rate
by taking into account the charm fragmentation function and
the appropriate $D$ branching ratios, and by extrapolating
from the \pt and $\eta$ regions accessible to the measurement.
The H1 and ZEUS measurements
cover
$5 \GeVsq < \Qsq < 100 \GeVsq$ and are
summarized in tab. \ref{tab:charm}.
They are in reasonable agreement with each other, and with NLO
calculations \cite{th:cnlo}
based on current parton density parametrizations.
About 4\% of the DIS events contain a $D^{\star\pm}$ \cite{z:charm},
and about 25\% of the DIS events are due to charm production
\cite{h1:charm,z:charm}. The asymptotic value of 40\% expected from
the quark charges is not yet reached.
H1 finds for the ratio of $D^\star : D^0$
production $0.38\pm0.07\pm0.06$, in agreement with other experiments
(see compilation in \cite{h1:charm}). Note that the $D^0$s include
non-primary $D^0$s from $D^\star$ decays.

\begin{footnotesize}
\begin{table}[tbh]
\begin{center}
\begin{tabular}{|c|c|c|c|}
\hline
                             & H1            & ZEUS        & NLO \\
\hline
\hline
  \sigc  (nb)        & $0.01<y<0.7$ & $0<y<0.7$  &      \\
\hline
$5 \GeVsq<Q^2<10\GeVsq$   &
                                                      &
          $13.5\pm 5.2 \pm 1.8^{+1.6}_{-1.2}$            &
          9-13                                            \\
$10\GeVsq<Q^2<100\GeVsq$ &
          $17.4 \pm 1.6 \pm 1.7 \pm 1.4$               &
          $12.5 \pm 3.1 \pm 1.8 ^{+1.5}_{-1.1}$        &
          8-14                                            \\
\hline
\hline
                  & $8\mmmm < x < 8 \mmm$     &
          $2 \mmmm < x < 5 \mmm$ &
                                                    \\
\hline
  $\av{\ftwoc/F_2}$         &
          $0.237 \pm 0.021 ^{+0.043}_{-0.039}$ &
          $\approx 25\%$                      &
                                                    \\
\hline
\end{tabular}
\end{center}
\scaption{Charm production at HERA \cite{h1:charm,z:charm}.
The extrapolation of the H1 measurement
from the range $0.01<y<0.7$ to $0<y<0.7$ would modify the
measured cross section by approximately $+6\%$.
The NLO calculations \cite{th:cnlo},
depending on the assumed gluon distribution function, and on the
value of $m_c$ used, vary by the range given.}
\label{tab:charm}
\end{table}
\end{footnotesize}

\subsubsection{Momentum spectra of $D$ mesons}

\begin{figure}[tbh]
   \centering
\begin{picture}(0,0) \put(0,0){{\bf a)}} \end{picture}
   \epsfig{file=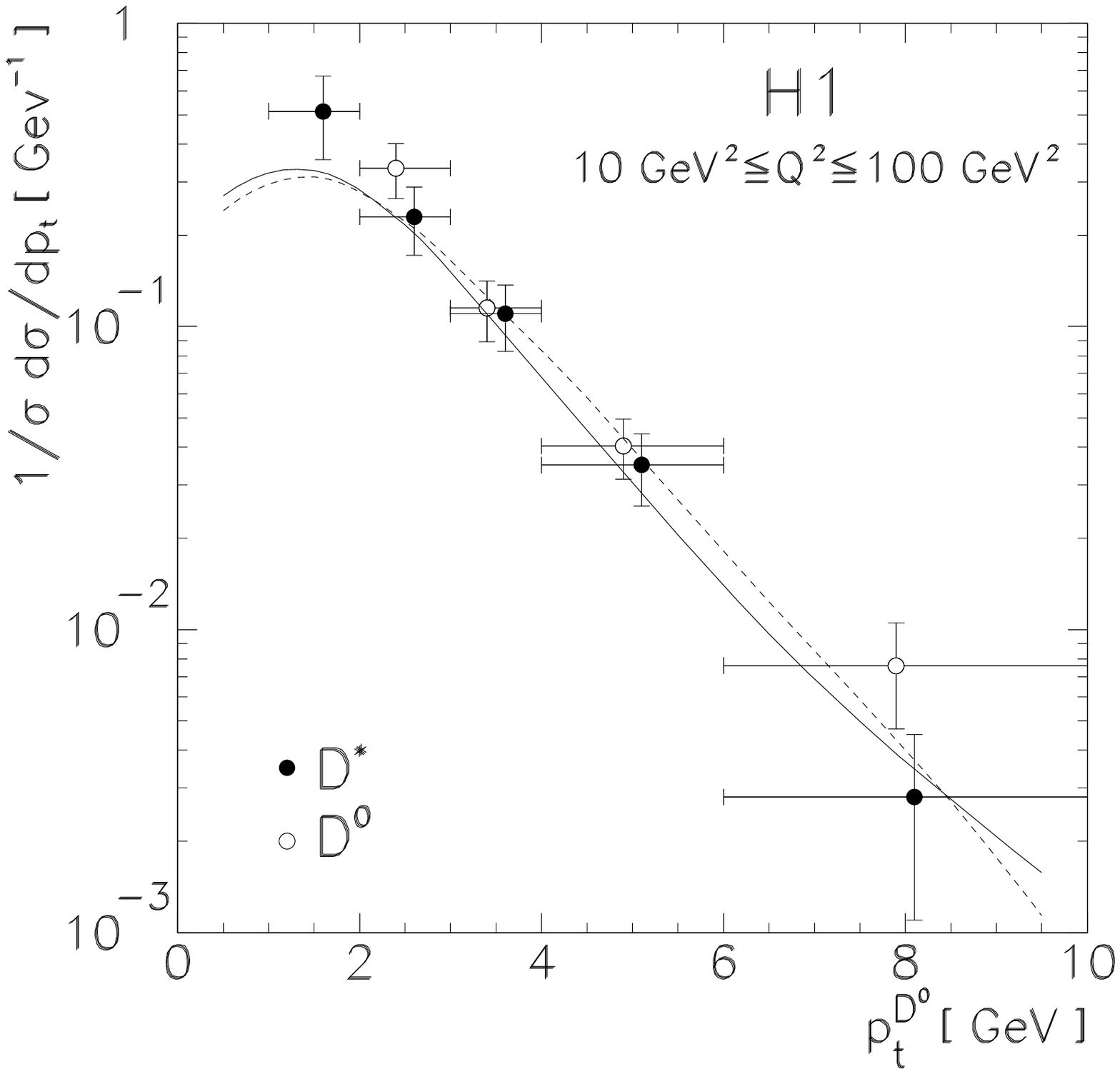,%
          width=7.1cm}
\begin{picture}(0,0) \put(0,0){{\bf b)}} \end{picture}
   \epsfig{file=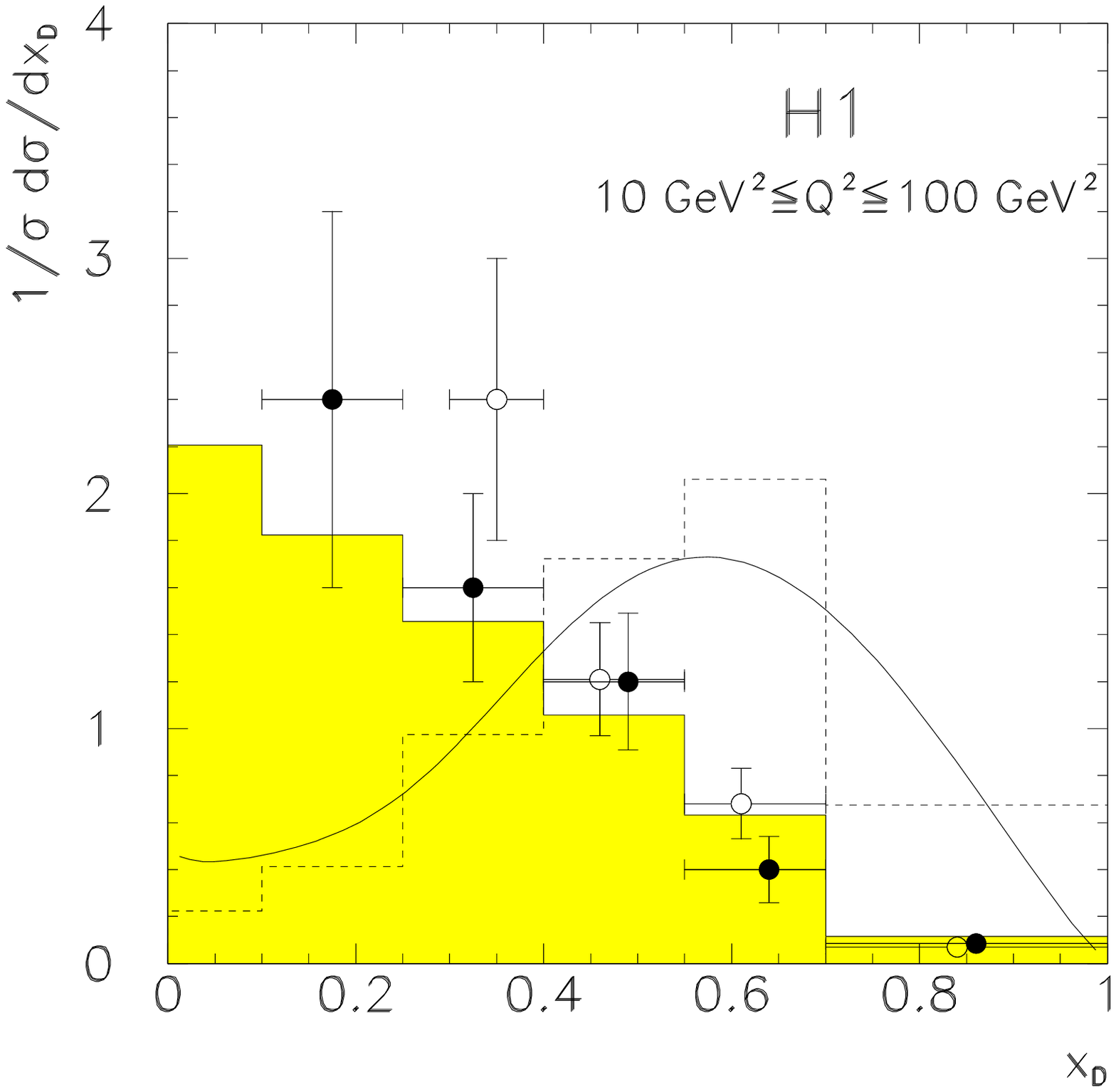,%
          width=7.1cm}
   \scaption{
             {\bf a)} \pt spectrum
                of $D^0$ and $D^{\star}$ mesons \cite{h1:charm},
                compared to the AROMA 2.1 \cite{mc:aroma}
                expectation for $m_c=1.3\GeV$ (full line)
                and $m_c=1.7 \GeV$ (dashed line). Only statistical errors
                are shown.
             {\bf b)}
                $x_D$ distribution for $D^0$ (open points)
                and $D^{\star +}$ (full points) mesons
                in the laboratory pseudorapidity range $-1.5<\eta_D<1.5$.
                The data are compared to the expectation for BGF processes
                (AROMA, shaded), for QPM-like contributions from a
                charm sea (dashed histogram), and an extrapolation
                from charged current $\nu N$ scattering
                off strange sea quarks \cite{o:strangesea} (full line).}
   \label{charm}
\end{figure}

In order to investigate further the charm production mechanism,
momentum spectra of the $D$ mesons
in the hadronic CMS
have been measured
\cite{h1:charm,z:charm}.
The \pt spectrum is well described
by the Monte Carlo generator AROMA \cite{mc:aroma} based upon the
BGF process (see fig.~\ref{charm}a), and also by NLO calculations
\cite{z:charm}.
The spectrum of the scaled momentum variable $x_D:=2|\vec{p}|/W$
falls with increasing $x_D$ (see fig.~\ref{charm}b), in agreement
with the BGF based charm generator,
and also with the NLO calculations
\cite{z:charm}.
In contrast, QPM-like events from charm sea quark interactions
would lead to an $x_D$ spectrum that is peaked at $x_D\approx 0.6$,
reflecting the hard charm fragmentation function.
H1 excludes
at 95\% C.L. a 5\% contribution from charm sea quarks to the total
charm production.

\subsubsection{The charm structure function \ftwoc}

\begin{figure}[bh]
   \centering
   \epsfig{file=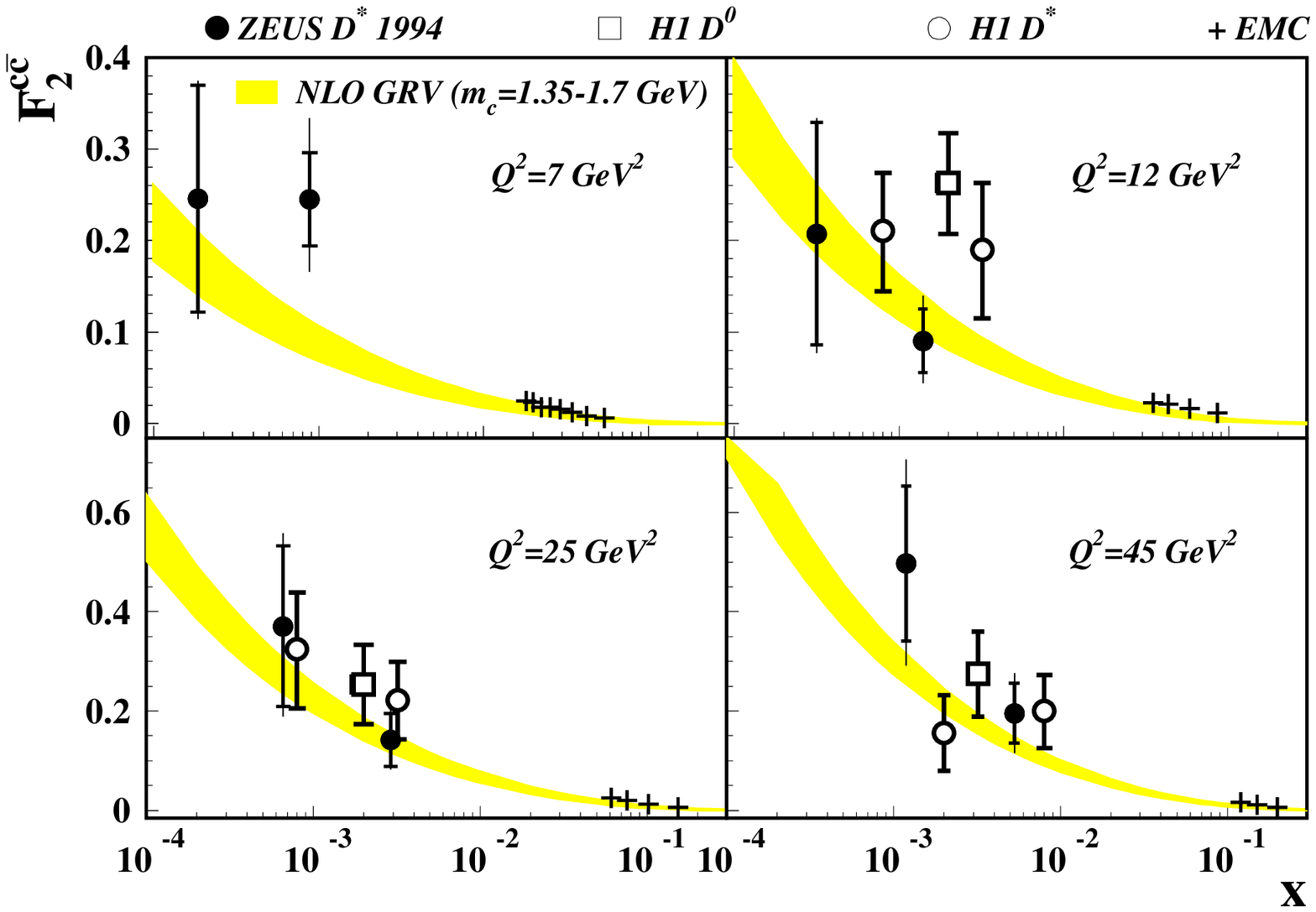,
          width=12cm}
   \scaption{The charm contibution \ftwoc~ to the proton
             structure function \ftwo from H1 \cite{h1:charm} and
             ZEUS \cite{z:charm}. Results from EMC \cite{o:emcf2c}
             are shown as crosses. The shaded band gives the NLO
             QCD prediction from the GRV NLO gluon density
             for charm masses between 1.35 \GeV (lower bound) and
             1.7 \GeV (upper bound).}
   \label{f2c}
\end{figure}

The cross section measurements in bins of $x$ and \Qsq can be expressed
in the form of
a charm structure function \ftwoc~ (fig.~\ref{f2c})
analogous to $F_2$. When compared with EMC data at
$x=0.02-0.3$, \ftwoc~ rises towards small $x$.
The data are consistent with NLO calculations based
exclusively on the BGF process with
gluon densities that describe the current HERA $F_2$ data.
In the calculation both the renormalization and factorization scales
are chosen to be $\mu=\sqrt{Q^2+4m_c^2}$.
The dominant uncertainty in the calculation arises from the
charm quark mass. The predicted \ftwoc~ changes by 15\% when $m_c$
is changed by 0.2~GeV.

The measurements establish the boson gluon fusion process as
the dominant one for charm production in DIS at $Q^2<100$ \GeVsqx,
and thus the sensitivity to the gluon
distribution function in the proton.
The data are consistent
with parton density parametrizations which were derived from
the inclusive $F_2$ measurements, providing an independent
cross check of the interpretation of the $F_2$ data in terms
of parton densities.




\subsubsection{The gluon density from charm production}

In a new preliminary analysis \cite{h1:charm95} to determine
the NLO gluon density from charm production
H1 measures the
visible \dstar cross section $\sigma_{\rm vis}^{D^\star}$
(corrected for detector effects)
in the range
$2<\Qsq<100$ \GeVsqx, $0.01<y<0.7$ for \dstar mesons with laboratory
$\pt>1.5\GeV$ and $|\eta|<1.5$. The measured cross section is
$\sigma_{\rm vis}^{D^\star} =
( 5.63 \pm 0.66 ^{+0.84}_{-0.68})$ nb.
The measured differential \dstar cross sections
as functions of the kinematic variables
\xb, \Qsqx, $p_T$, $\eta$ (not shown)
are well described by
the AROMA \cite{mc:aroma} charm generator and
by a NLO calculation (program HVQDIS 1.1 \cite{mc:hvqdis} with
the Peterson charm fragmentation
function \cite{mc:peterson}).

\xgluon~ is the proton momentum fraction carried by the gluon entering the
BGF process.
\xgluon~ can in leading order be reconstructed
from
\begin{equation}
  \xgp = x (1+\frac{\shat}{Q^2}) \mbox{\rm ~~with~~}
  \shat = \frac{p_{Tc}^{\ast 2} + m_c^2}{z(1-z)} \mbox{\rm ~~and~~}
   z:= \frac{P \cdot c}{P \cdot q} =
   \frac{(E-p_z)_c^{\rm lab}}{2yE_e}.
  \label{eq:ckin}
\end{equation}
Here $c$ is the charm quark's 4-vector and
$p_{Tc}^{\ast}$ it's CMS transverse momentum. $P$ is the proton 4-momentum,
and $E_e$ is the electron beam energy in the laboratory frame.
The Lorentz invariant $z$ can be
calculated from the charm quark's energy $E$ and longitudinal momentum
$p_z$ in the lab. frame.
For BGF events it can be correlated with the quark scattering
angle in the photon-gluon CMS, see section \ref{sn:jas} and eq. \ref{eq:zj}.

Measured are charm mesons though, not charm quarks.
Therefore the observable \xgobs~ is defined by
replacing $p_{Tc}^{\ast}$ with $1.2 \cdot p_{TD^\star}^{\ast}$
and $(E-p_z)_c^{\rm lab}$ with $(E-p_z)_{D^\star}^{\rm lab}$
in eq.~\ref{eq:ckin}.
\xgobs~ is well correlated with \xgluon~ \cite{h1:charm95}.

\begin{figure}[tbh]
   \centering
\begin{picture}(0,0) \put(0,0){{\bf a)}} \end{picture}
   \epsfig{file=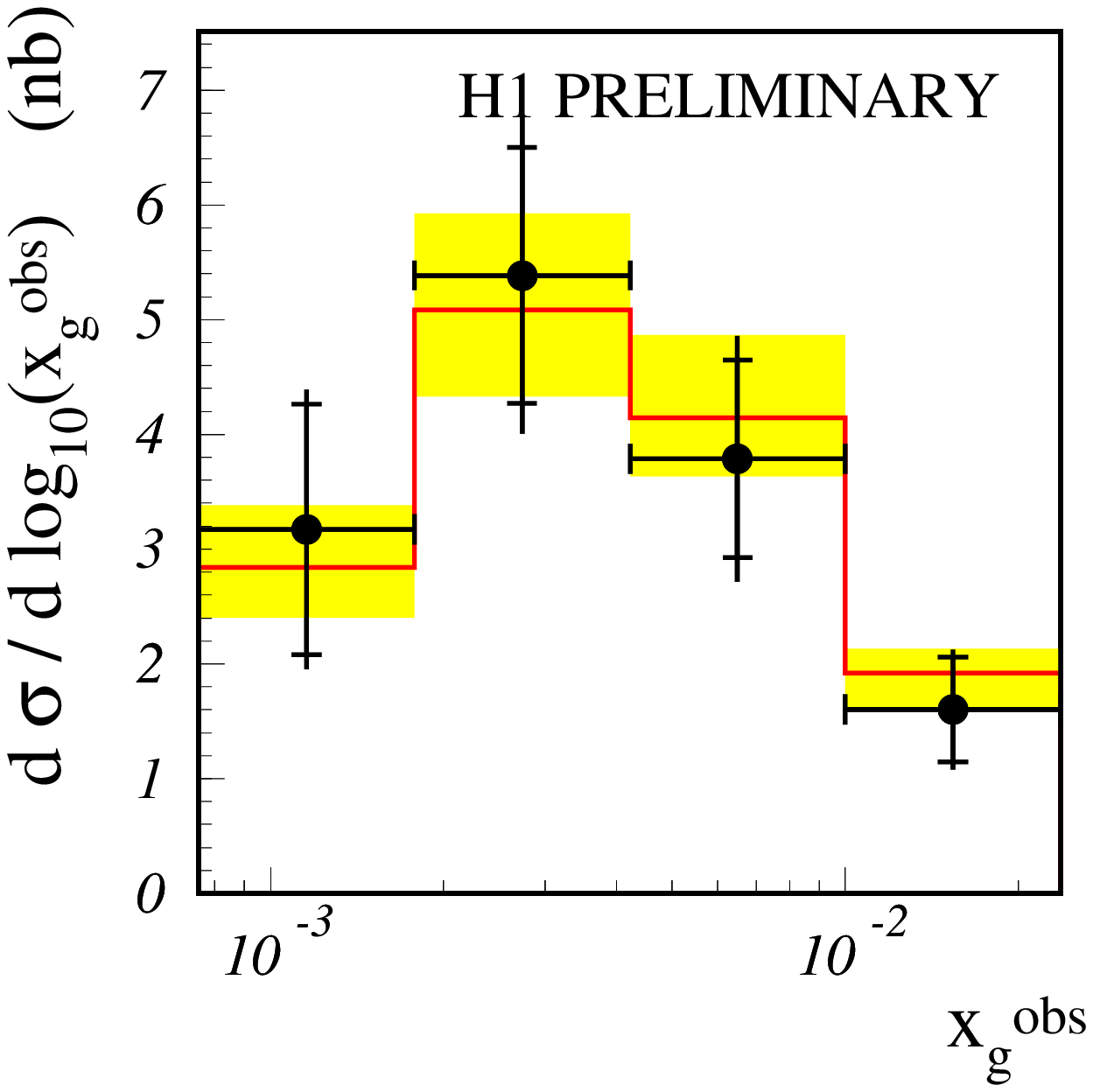,
          width=6cm}
   \hspace{1cm}
\begin{picture}(0,0) \put(0,0){{\bf b)}} \end{picture}
   \epsfig{file=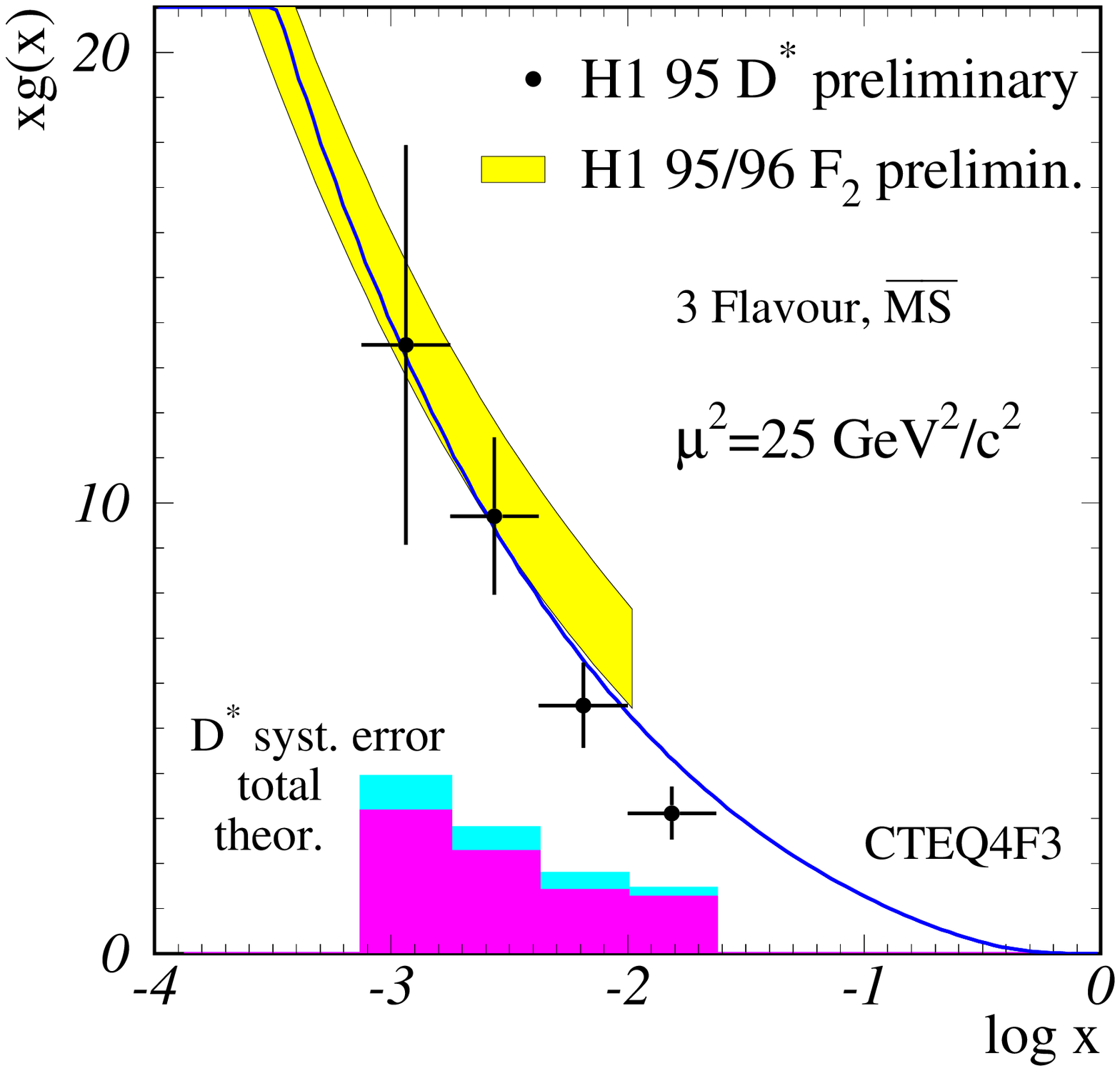,
          width=7.5cm}
   \scaption{
            {\bf a)} The visible \dstar cross section
                as a function of \xgobs~ (see text), corrected for
                detector effects.
                The preliminary H1 data \cite{h1:charm95} (data points)
                are compared to
                a NLO QCD calculation \cite{mc:hvqdis} using
                the GRV 94 HO \cite{th:grv} parton density function
                (histogram; the shaded bands represents the variation
                of the cross section when $m_c$ is varied between
                1.3 and 1.7 GeV).
             {\bf b)}
             The NLO gluon density $x\cdot g(x)$ at a scale
             $\mu^2= 25~\GeVsqx$,
             unfolded from the \dstar cross section \cite{h1:charm95}.
             The statistical errors are shown as error bars, the
             systematic errors as shaded histograms.
             The line shows the CTEQ4F3 parametrization \cite{th:cteq4},
             the shaded band the gluon density from a NLO QCD analysis
             of the H1 \ftwo data \cite{h1:f2jlowx}.}
   \label{xgcharm}
\end{figure}

The measured cross section as a function of \xgobs~ is shown in
fig.~\ref{xgcharm}a. It is well described by the NLO calculation.
The NLO calculation takes into account the gluon initiated processes
$\gamma^\ast g \rightarrow \ccbar, \ccbar g$ and the quark initiated
processes $\gamma^\ast q \rightarrow \ccbar q$,
$\gamma^\ast \bar{q} \rightarrow \ccbar \bar{q}$, where the \ccbar~
pair is produced from a radiated gluon.
Using the NLO calculation for the \xgobs~ cross section, the
gluon density can be unfolded from the data.
In the calculation $m_c=1.5 \GeV$ is used, and the factorization
scale is set to $\mu_F = \sqrt{4m_c^2 + Q^2}$. The resulting NLO
gluon density $xg(x,\mu^2)$ (fig.~\ref{xgcharm}b) at an average scale
$\mu^2=24\GeVsq$
reaches down to
$\xgp \approx 10^{-3}$, and agrees well with a recent indirect
extraction from the \ftwo data \cite{h1:f2jlowx}.
Smaller values of \xgp~ could be probed with
a larger angular acceptance for $D^\star$ decays in the backward region,
to be provided with silicon trackers.
The ultimate limit is given by  $\xgp > \shat/(sy) \approx 10^{-4}$.

By reconstructing event-by-event \xgluon, this measurement probes
the gluon density locally in \xgluon.
In contrast, with \ftwoc~ the charm contribution
is measured as a function of Bjorken $x$, integrating over
all $\xgluon>x$. The comparison of the directly measured gluon
density with the one extracted from the inclusive \ftwo, and the agreement
of \ftwoc~ with the prediction based upon such gluon distributions
constitutes an important test of QCD with universal parton (here: gluon)
densities. A similar test was obtained with dijet events in LO,
see fig.~\ref{logluon}.

  \section{Bose-Einstein Correlations \label{sn:be}}       

Due to Bose-Einstein statistics, identical bosons prefer to occupy
the same quantum state.
In particle physics, the Bose-Einstein effect
leads to an enhanced probability for identical bosons -- like sign pions --
to have similar momenta \cite{o:goldhaber}.
The shape and strength of their correlation function provide
information on the production process.

As a measure of distance in momentum space for two particles
with 4-momenta $p_1$ and $p_2$ with invariant mass $M=\sqrt{(p_1+p_2)^2}$
one defines\footnote{This variable is traditionally denoted as $Q^2$.
Here we use $T^2$ instead, because in DIS $Q^2$ is already used for
the virtuality of the exchanged photon.}
\begin{equation}
   T^2 := - (p_1-p_2)^2 = M^2-4m_\pi^2,
\end{equation}
assuming the particles to be pions.
One measures the number of like sign charged particle
pairs as a function of $T$.
This correlation function $\rho(T)$ is compared
to a function $\rho_{\rm ref}(T)$ obtained similarly, but
from a reference sample where Bose-Einstein correlations are absent:
\begin{equation}
  R(T):= \rho(T)/\rho_{\rm ref}(T).
\end{equation}
As reference sample may serve a) unlike sign charged particle pairs, or
b) pairs from different events (event mixed method).

If the particle emitting sources are
distributed in space-time
according to $\rho_s(\xi)$, a ratio
\begin{equation}
  R(T) = R_0(1+\lambda| \tilde{\rho}_s(T)|^2),
\end{equation}
is expected for the Bose-Einstein enhancement.
$\tilde{\rho}_s(T)$ is the Fourier transform of
$\rho_s(\xi)$ and $R_0$ a normalization constant. The correlation
strength $\lambda$ is 0 for completely coherent emitters, and 1 for
completely incoherent emitters.

The measurement at small $T$, where the Bose-Einstein effect is expected,
is notoriously difficult due to the finite double-track resolution of the
tracking devices and due to \epem~ pairs from photon conversions.
The construction of a reference sample from the data
introduces a residual bias. This effect
is estimated by Monte Carlo, and cancelled by building the
``double ratio''
\begin{equation}
  RR(T) = R(T)_{\rm data} / R(T)_{\rm MC}.
\end{equation}

\begin{figure}[htb]
   \centering
   \epsfig{file=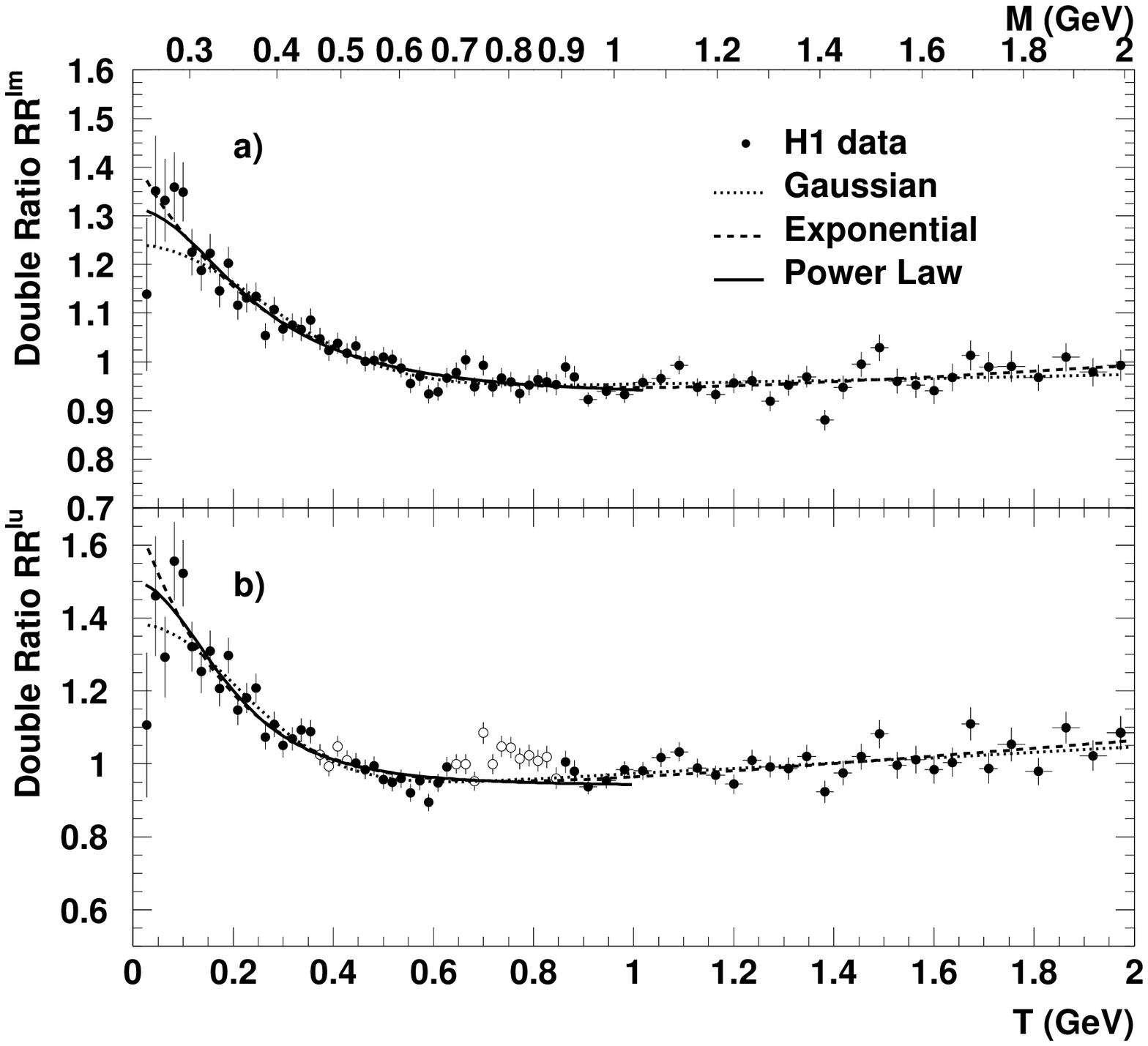,%
          width=10cm,bbllx=8pt,bblly=9pt,bburx=547pt,bbury=496,clip=}
   \scaption{Bose-Einstein correlations measured by H1 \cite{h1:be}.
             The double ratio $RR(T)$ is plotted as a function
             of $T$ and $M$.
             For the reference sample,
             tracks from different events are paired (top), or
             unlike sign pairs are used (bottom). The fits of
             the Gaussian, exponential and power law parametrizations
             (with extra linear terms to accomodate the sloping background) are
             shown. The open points in the bottom plot are excluded
             from the fits, because they are in a region which is affected by
             correlations due to $K^0_s\rightarrow \pi^+\pi^-$ and
             $\rho^0 \rightarrow \pi^+\pi^-$ decays.}
   \label{be}
\end{figure}

The H1 data \cite{h1:be} are shown in fig.~\ref{be}.
A clear enhancement at small $T$, attributable to the Bose-Einstein
effect, is visible.
The data cover $6<Q^2<100~\GeVsqx$, $\nnnn < x < \nn$ and
$65<W<240~\GeVx$.
No dependence of the Bose-Einstein effect
on the kinematic variables
$x$,
$Q^2$ or
$W$ was found, nor a difference between
the diffractive and non-diffractive data samples.

Different parametrizations of the effect have been
fitted to the data.
The traditional Goldhaber parametrization \cite{th:goldhaber}
\begin{equation}
   R(T) = R_0 \left[1+\lambda \exp(-r^2 T^2)\right]
\label{eq:goldhaber}
\end{equation}
is derived from the the assumption of a static, spherically symmetric
Gaussian source density $\rho_s(\xi) = \rho_s(0) \exp(-\xi^2/(2r^2))$.
The parameter $r$ gives the size of the production volume.
In high energy reactions,
the sources are moving relativistically \cite{th:bj_geometry},
and the assumption of a static source becomes invalid \cite{th:be_rel}.
In the relativistic string fragmentation picture \cite{th:be_string},
an approximately exponential shape is predicted,
\begin{equation}
    R(T) = R_0 \left[ 1+\lambda \exp(-r T)\right],
\end{equation}
where the parameter $r$ is related to the string tension \cite{th:be_lund}.
Alternatively, assuming a self-similar, scale-invariant
pattern for perturbative
QCD cascades, a power law is derived \cite{th:be_power,rev:dewolf}
\begin{equation}
    R(M) = A+B \cdot (1/M^2)^\beta.
\end{equation}
Compared to the data (fig.~\ref{be}),
all parametrizations are acceptable, though
the exponential and the power law ans\"atze give somewhat better
results at low $T$. A power law behaviour had been observed
previously \cite{o:be_power}.

In order to compare with other experiments, the following results
are obtained with the Gaussian (Goldhaber)
parametrization, eq. \ref{eq:goldhaber} (table \ref{tab:be}).
\begin{footnotesize}
\begin{table}[h]
\begin{center}
\begin{tabular}{|l|c|c|}
\hline
reference sample  &  $r$ (fm) & $\lambda$                 \\
\hline
event mixed       & $0.54 \pm 0.03 ^{+0.03}_{-0.02} $
                  & $0.32 \pm 0.02 ^{+0.06}_{-0.06} $              \\
unlike sign pairs & $0.68 \pm 0.04 ^{+0.02}_{-0.05} $
                  & $0.52 \pm 0.03 ^{+0.19}_{-0.21} $      \\
\hline
\end{tabular}
\end{center}
\scaption{The fit parameters from a Gaussian parametrization
of the Bose-Einstein effect seen in the H1 data \cite{h1:be}.
}
\label{tab:be}
\end{table}
\end{footnotesize}With the event mixed reference sample, a lower $r$ value is obtained
than with unlike sign pairs. Such a systematic effect has been observed
also in other analyses.
Other measurements
in \epem~ and lepton-nucleon interactions
scatter between $r=0.39 \fm$ and $r=0.97 \fm$,
and $\lambda=0.27$ and $\lambda=1.08$ (see the compilation in \cite{h1:be}),
but no pattern emerges.
At least the extracted values of $r$, including the HERA ones,
can be said to be in the ball park of the length scale at which
hadronization takes place, 1 fm.
H1 finds marginal evidence for a dependence of $r$ on the
charged particle multiplicity density $\ol{n}$, given as number of
particles per unit pseudorapidity, $\ol{n}=\dif n/\dif\eta$. A strong
multiplicity dependence of $r$, roughly $r=0.4 + 0.075 \cdot \ol{n}$
(with $\approx 20\%$ accuracy) for $2<\ol{n}<20$
was seen in \ppbar~ collisions \cite{o:be_ua1,o:be_e735}. This is
confirmed by the H1 data with a limited lever arm, $2<\ol{n}<5$.

\chapter{The Quark Fragmentation Region \label{ch:qfrag}}       

  \section{Charged Particle Spectra \label{sn:xp}}  

\subsubsection{Comparisons with \epem~ reactions}

\hyphenation{ex-pe-ri-ments}

The physics issues that have been addressed
with
charged particle spectra in the Breit frame
\cite{h1:breit,h1:breit2,z:breit,z:breit2,z:breit3}
are
the study of fragmentation properties,
QCD coherence effects, and scaling violations with
the goal of determining $\alpha_s$.
The advantage of $ep$ experiments over
\epem~ experiments at fixed beam energies is that
the scale variable $Q$ can be varied continuously.
Complications arise however from initial state parton radiation
and from boson-gluon fusion events,
which are absent in \epem.

With the Breit current hemisphere a relatively small subsystem
of the whole event is considered,
(for most events $Q \ll W$),
which should be not much affected
by the dynamics of the proton remnant
(see sections \ref{sn:frames} and \ref{sn:kinevar} with fig.~\ref{plateau}).
The Breit frame current hemisphere of an $ep$ event,
the quark fragmentation region,
is similar to one hemisphere of an \epem~ event.
Considering just
the lowest order process, in \epem~ interactions the
outgoing quarks have opposite momenta of equal magnitude.
The same is true for the incoming quark and the outgoing
quark in the Breit frame of DIS.
Furthermore, as in \epem,
the scale of the hard interaction, $Q$, also
determines the phase space for QCD effects,
given by the quark energy $E_q=Q/2$.
In contrast, in the CMS
$E_q=W/2 \gg Q/2$.
In the CMS it is therefore more
difficult to disentangle a kinematic effect from
a scale dependent QCD effect.


The scaled momentum variables of a hadron with momentum
$\vec{p}$ and energy $E$ are defined as (compare sect.~\ref{sn:kinevar})
\begin{equation}
x_p := \frac{|\vec{p}|}{\pmax}  \hspace{1cm}
\xi :=  \ln \frac{1}{x_p}
\end{equation}
with $0<\xp<1$.
In practice, one often approximates $\pmax\approx E_q=Q/2$
(massless kinematics).
With the variable $\xi$
the soft part of the momentum spectrum at small $x_p$, where
the particle density is the largest,
is expanded.

\begin{figure}[tbh]
   \centering
   \epsfig{file=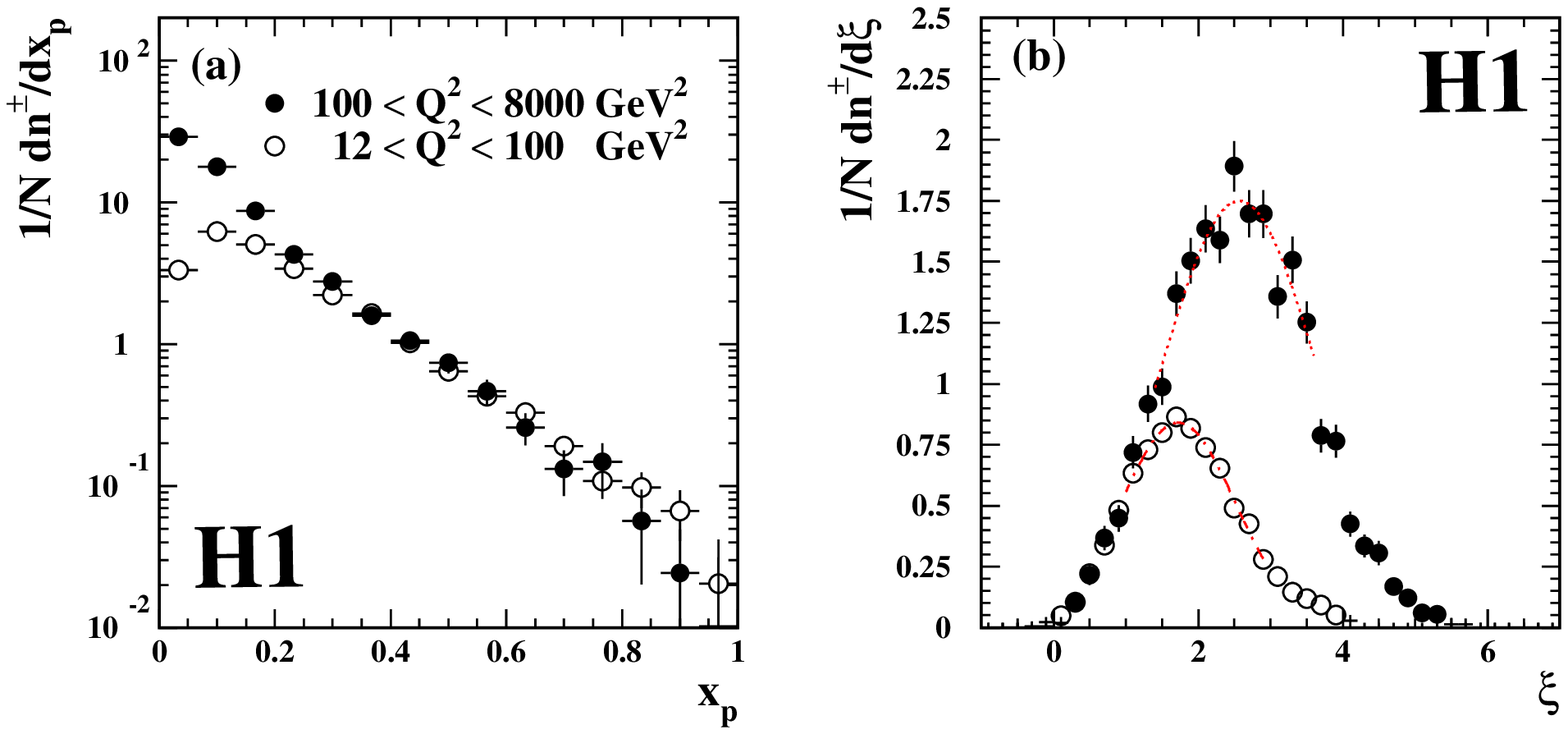,%
   width=12cm,bbllx=17pt,bblly=267pt,bburx=577pt,bbury=546,clip=}
   \scaption{
             The distributions $D(x_p):=1/N \dd n/\dd  x_p$ {\bf (a)} and
             $F(\xi):=1/N \dd n/\dd\xi$ {\bf (b)} for
             charged particles from the Breit current hemisphere
             for large and for small $Q^2$ from H1 \cite{h1:breit2}.
             The curves are Gaussian fits to the data.}
   \label{xpxi}
\end{figure}

Fig.~\ref{xpxi} shows the measured \cite{h1:breit}
$x_p$ and $\xi$ distributions of
charged particles in the Breit current hemisphere
for large and for small $Q^2$.
$D(\xp)$ is steeply falling, and
$F(\xi)$ is approximately
Gaussian\footnote{
Qualitatively the Gaussian shape can be understood assuming
that hadrons result from a chain of
branching processes $i=1,...k$ \cite{h1:dewolf}
(gluon coherence giving rise to a hump backed, approximately Gaussian, shape
is discussed later).
The branching stops, when there is not enough energy left to
produce a hadron of mass $m$. On average,
\begin{equation}
   \frac{m}{E_q} \approx  \av{z_i}^{\av{k}}.
\end{equation}
The $z_i$ give the energy fraction retained in each branching,
and are assumed to be randomly distributed according to some
splitting function $P_i(z_i)$.
We expect that the hadron multiplicity $n$ is proportional to the
number of branchings $k$, hence a
logarithmic growth of the average multiplicity with
energy follows,
\begin{equation}
\av{n} = c \cdot \ln (E_q/m).
\end{equation}

The energy left after $r$ branchings is
\begin{equation}
  E_q^r = E_q \cdot z_1 \cdot \cdot \cdot z_r = E_q\prod_{i=1}^{r} z_i.
\end{equation}
A hadron from this stage has energy $E^r = x_r \cdot E_q$ with
$x_r = z_r^\prime E_q^r/E_q$, where $z_r^\prime = 1-z_{r+1}$.
We define the random variable $\xi_r := \ln(1/x_r)$, which is
\begin{equation}
  \xi_r := \ln \frac{1}{x_r} =
   \ln \left(\frac{1}{z_r^\prime} \cdot \prod_{i=1}^{r} \frac{1}{z_i}\right)
    = \ln\frac{1}{z_r^\prime} + \sum_{i=1}^{r} \ln \frac{1}{z_i}.
\end{equation}
$\xi_r$ can be written as a sum of random variables, and is thus
Gaussian distributed for  $r\rightarrow \infty$ according to
the central limit theorem, regardless of $P_i(z_i)$.

The final hadron spectrum of all hadrons from all $k$ branching stages
would then be the result of the sum of
Gaussians $G_r(\xi)$ with $r>k$, where $k$ itself is a random variable.
For not too large $k$, this would give an
approximate Gaussian, resulting from fall-offs at the
kinematic boundaries at small and large $\xi$.
For large $k$ (large energy),
a plateau $F(\xi)=$const can develop,
which translates into a rapidity plateau $\dif n / \dif y =$const.
}.
When $Q$ increases,
soft particles (large $\xi$) are produced much
more copiously. The hard part of the spectrum (large $x_p$, small $\xi$)
is approximately invariant of $Q$, ``scaling''.
This can be understood naively.
If hadron production stems from
a branching process with energy independent branchings, scaling at high
\xp is expected.
At low $x_p$, the reduction of the phase space
forces a turnover of the spectrum,
which happens at lower \xp for larger initial energy.

\begin{figure}[tbh]
   \centering
   \epsfig{file=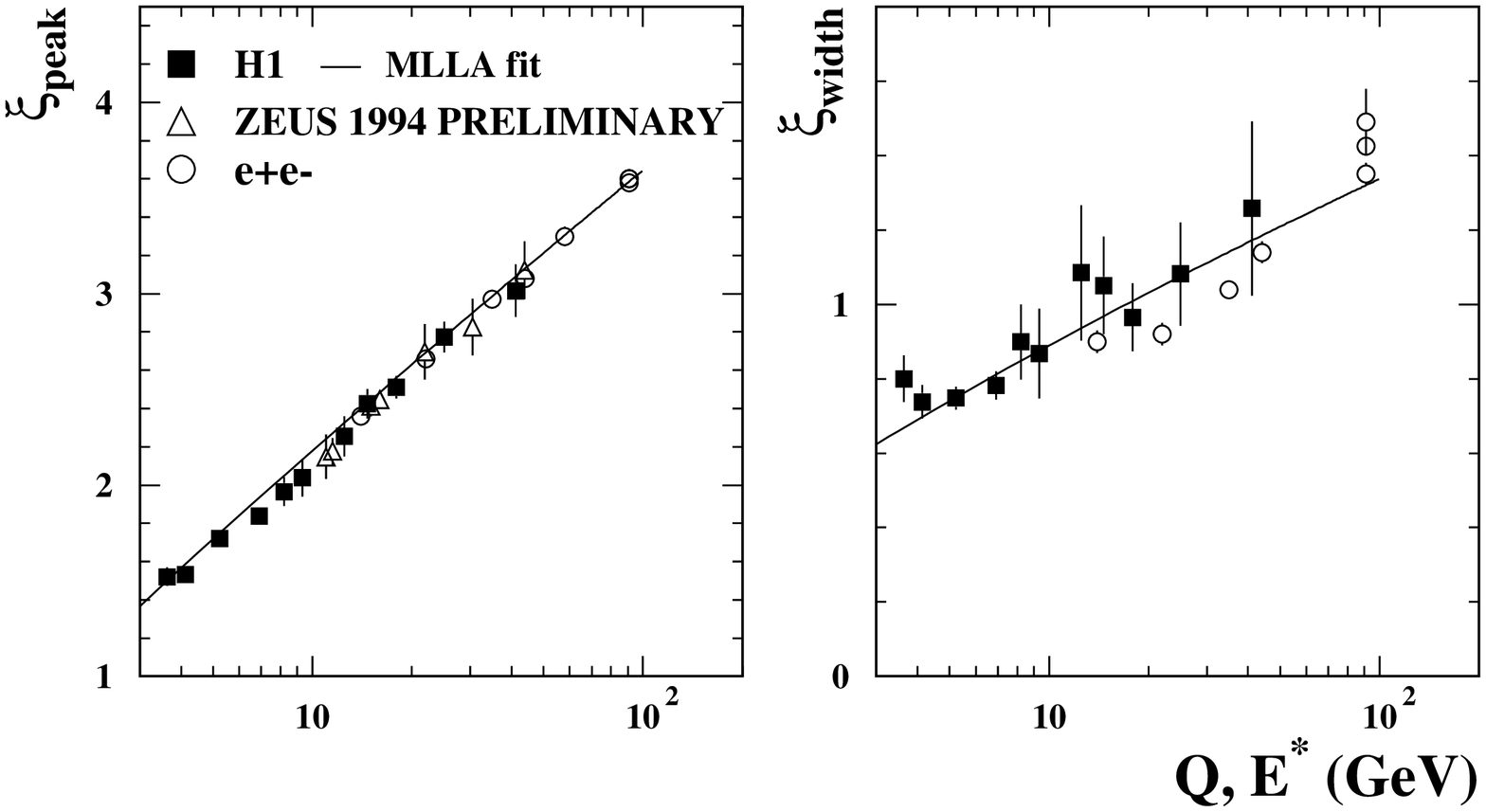,%
           width=12cm}
   \scaption{
             The evolution of of the peak position $\xi_{\rm peak}$ (left)
             and width $\xi_{\rm width}$ (right) of
             $F(\xi)$ with
             $Q$. Data from $ep$ \cite{h1:breit2,z:breit_kant}
             and \epem~ reactions
             are compared.
             Also shown is a MLLA fit
             to the H1 data \cite{h1:breit2}.}
   \label{xievol}
\end{figure}

\begin{figure}[tbh]
   \centering
   \epsfig{file=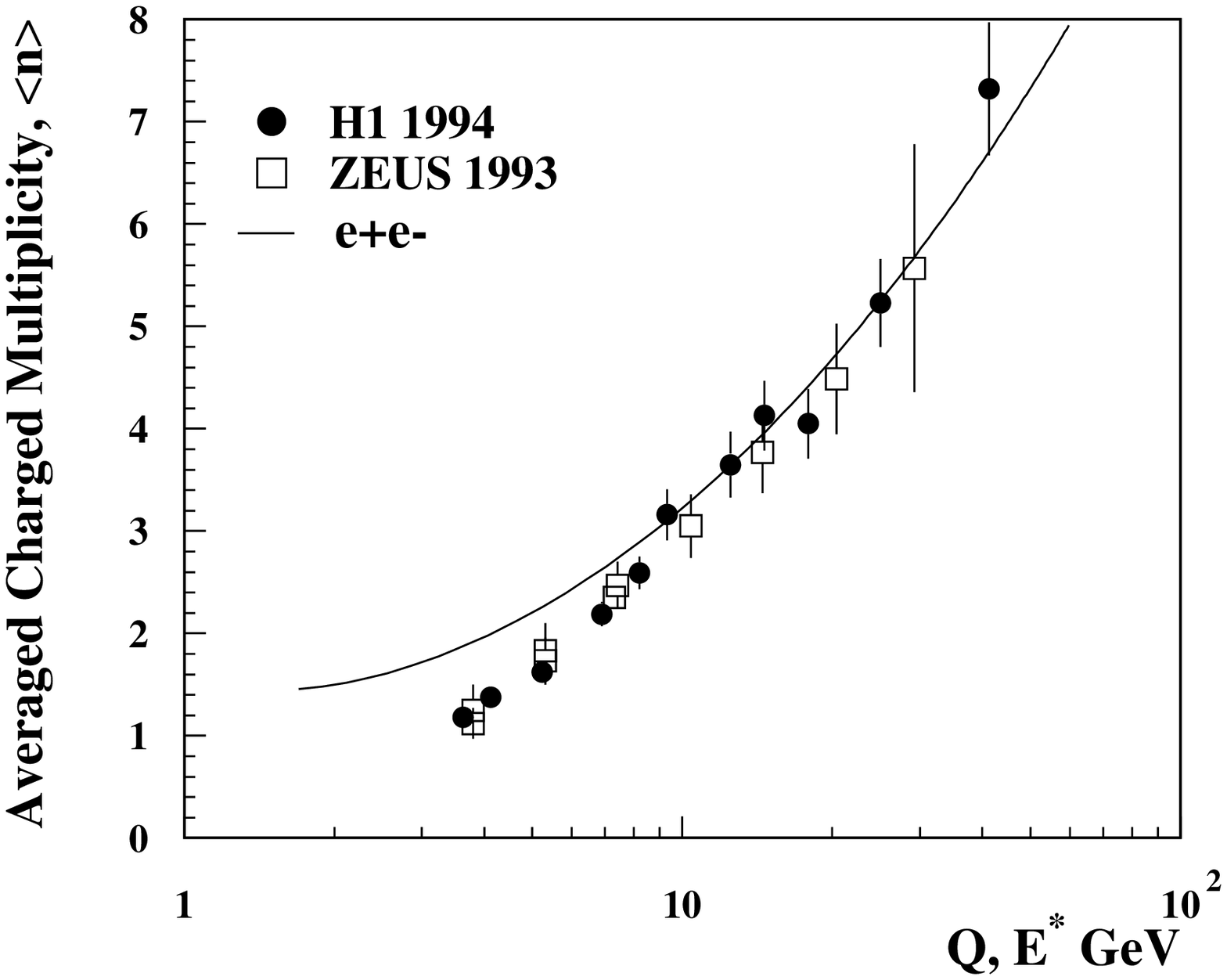,%
   width=8cm}
   \scaption{
             Average charged particle multiplicity in the Breit current
             hemisphere as a function of $Q$. The H1 \cite{h1:breit2}
             and ZEUS \cite{z:breit} data
             are compared to a curve representing half of the charged
             multiplicity of \epem annihilation events at CM energy
             $E^\ast$.}
   \label{nbreit}
\end{figure}

The evolution of $F(\xi)$ can be effectively summarized by the
$Q$ dependence of the peak position $\xi_{\rm peak}$
and width $\xi_{\rm width}$ of the Gaussian, shown in fig.~\ref{xievol}.
The average charged multiplicity \av{n}
(the integral over the fragmentation
function) is shown in fig~\ref{nbreit}.
An approximately logarithmic dependence of these variables on $Q$
is observed:
\begin{eqnarray}
  \av{n}         &  \approx & -1.7  + 4.9~  \cdot \log_{10} (Q/\GeV) \\
  \xi_{\rm peak} &  \approx & ~0.6  + 1.5~  \cdot \log_{10} (Q/\GeV) \\
  \xi_{\rm width}&  \approx & ~0.4  + 0.45  \cdot \log_{10} (Q/\GeV).
\end{eqnarray}

$\xi_{\rm peak}$  and $\xi_{\rm width}$ measured in \epem ~annihilation
at $\sqrt{s_{e^+e^-}}=E^{\ast}=Q$
are within errors consistent with the $ep$ data.
For $Q>10\GeV$, also the total multiplicities $\av{n}$
agree, but at smaller $Q$,
\av{n} is significantly smaller in $ep$ than
in \epem.

The deviation for $Q\lesssim 10 \GeV$
can be attributed largely to LO QCD processes in $ep$,
in particular to the boson-gluon fusion process which occur at
small $x$ (hence small $Q$) where the gluon density is large.
For these events, the scattered quark may not be collinear with
the photon, or may not even point into the Breit
current hemisphere \cite{th:streng}.
Its fragments are then emitted partly
into the Breit target hemisphere\footnote{Consider the subsystem
of the hard scattering subprocess with invariant mass \shat~
(QPM events with massless quarks have \shat=0).
One can easily verify that for $\shat>Q^2$
the $z$ momentum component of the hard subsystem
is negative, pointing into the target hemisphere.}.
For $Q\gtrsim 10 \GeV$ charged particle production in the $ep$ Breit
current hemisphere appears to be similar to one hemisphere
in \epem annihilation, supporting the notion of universal quark
fragmentation
\footnote{
In \epem~ data the tails of the distribution deviate from a Gaussian.
In the $ep$ data, no significant deviation from
a Gaussian has been found yet.}.

Predictions for the hadron spectra are available from
a) perturbative QCD in the modified leading log approxiation (MLLA),
   where the transition to hadrons rests on the hypothesis of
   local parton hadron duality (LPHD);
b) perturbative QCD in NLO, based on the QCD factorization theorem with
   universal, but scale dependent fragmentation functions;
c) event generators with their sophisticated schemes to model
   perturbative QCD evolution and hadronization.
They will be compared to the data in the following.

\subsubsection{MLLA + LPHD predictions}

The MLLA+LHPD derivation of the spectrum $F(\xi)$ is discussed in
\cite{books:dokshitzer,rev:khoze,rev:dokshitzer}.
The spectrum of partons in a parton cascade initiated by a
quark of energy $E_q$ (and related virtuality)
is calculated perturbatively in the
modified leading log approximation (MLLA), including colour coherence.
The cascade is calculated
down to a low cut-off $Q_0\approx \lambdaqcd$ for the parton virtuality.
It is then assumed
(the hypothesis of local parton hadron duality, LPHD)
that the final state hadrons follow the same spectrum, up to a
constant factor. The predicted spectrum $F(\xi)$ has a ``hump backed''
form of approximately Gaussian shape around the peak.
On the other hand, when coherence effects are neglected, and
$E_q \rightarrow \infty$
a plateau $F(\xi)$=const. should form in the soft region, until
the kinematic limit is reached at a
scale $\xi = \ln(1/x_p) \approx \ln (E_q/m)$.
Coherence suppresses soft parton radiation, thus
depleting the plateau at large $\xi$ --
a ``hump backed'' plateau is
formed\footnote{
A flat plateau $\dif n/\dif \xi$=const. implies also a flat rapidity plateau,
$\dif n/\dif y$=const., since $\dif y/\dif p_z = 1/E$. A hump backed
distribution  $\dif n/\dif \xi$ therefore results in a depletion
for $y\rightarrow 0$.
A dip in the rapidity distribution has been measured also in \epem~
annihilation \cite{o:tpc,lep:dip}.
The observation of a dip for 2-jet like events in \epem~ annihilation
provided evidence
for soft and collinear gluon radiation \cite{o:tpc}.}.

Analytical formulae in their
various approximations for the spectrum are in general quite lengthy
\cite{books:dokshitzer,rev:khoze,rev:dokshitzer}.
However, for large energies $E_q$ and
concentrating on the main part of the spectrum $F(\xi)$, without the tails,
the peak and width of the Gaussian are predicted to evolve with
$E_q$ as \cite{lep:xi}
\begin{eqnarray}
\xi_{\rm peak} &=& 0.5 \cdot \ln (E_q/\leff)
                 + c_2 \sqrt{\ln(E_q/\leff)} + \kappa  \\
\xi_{\rm width} &=& \sqrt{[\ln(E_q/\leff)]^{3/2}/(2c_1)}.
\end{eqnarray}
The constants are
$c_1=\sqrt{36 n_c/b}$,
$c_2=B\sqrt{b/(16n_c)}$ with
$b=\frac{11}{3}n_c-\frac{2}{3}n_f$ and
$B=\frac{1}{b}(\frac{11}{3}n_c+\frac{2}{3}n_f/n_c^2)$; they depend
on the number of colours, $n_c$, and the number of flavours, $n_f$
(here $n_f=3$).
$\kappa$ accounts for higher order corrections
and is expected to be of order 1. $\leff$ is an effective scale for
the evolution.

It is not a priori clear that the above MLLA+LPHD calculations for
showering quarks can be applied to $ep$ data, where BGF events
play a r\^{o}le at small $Q$.
A good combined fit to the H1 data is obtained though, yielding
$\leff = 0.21 \pm 0.02 \GeV$ and $\kappa=-0.43 \pm 0.06$, in agreement
with an analysis of combined \epem~ data, for which
$\leff = 0.21 \pm 0.02 \GeV$ and $\kappa=-0.32 \pm 0.06$ was obtained
\cite{lep:xi}.

It should be noted that the existence of a hump backed plateau,
fig.~\ref{xpxi},
does not {\it prove} colour coherence effects. At finite energy, a hump
backed, approximate Gaussian shape for the hadron spectrum
can also be obtained in Monte
Carlo simulations without coherence and with an independent fragmentation
model \cite{rev:khoze}.
One may argue that for an incoherent shower the turn over of the
plateau happens
at a scale $x_p \approx m/E_q$. This point would define $\xi_{\rm peak}$,
hence $\dif \xi_{\rm peak}/ \dif \ln E_q = 1$, in contradiction to
the slower growth of the
data and of the MLLA prediction.
However, in an incoherent shower model
with string fragmentation, also a slower growth close to the data
can be obtained \cite{z:breit}.
Furthermore, in \cite{rev:khoze} it is demonstrated
how string fragmentation can turn a flat plateau for partons into
a hump backed one for hadrons. String fragmentation appears to
compensate for negligence in the perturbative evolution.
On the other hand, the growth of the average charged multiplicity
with $Q$ is less well described by string fragmentation with an incoherent
parton shower
\cite{z:breit}.

\subsubsection{Invariant energy spectra and the soft limit}

\begin{figure}[tbh]
   \centering
   \epsfig{file=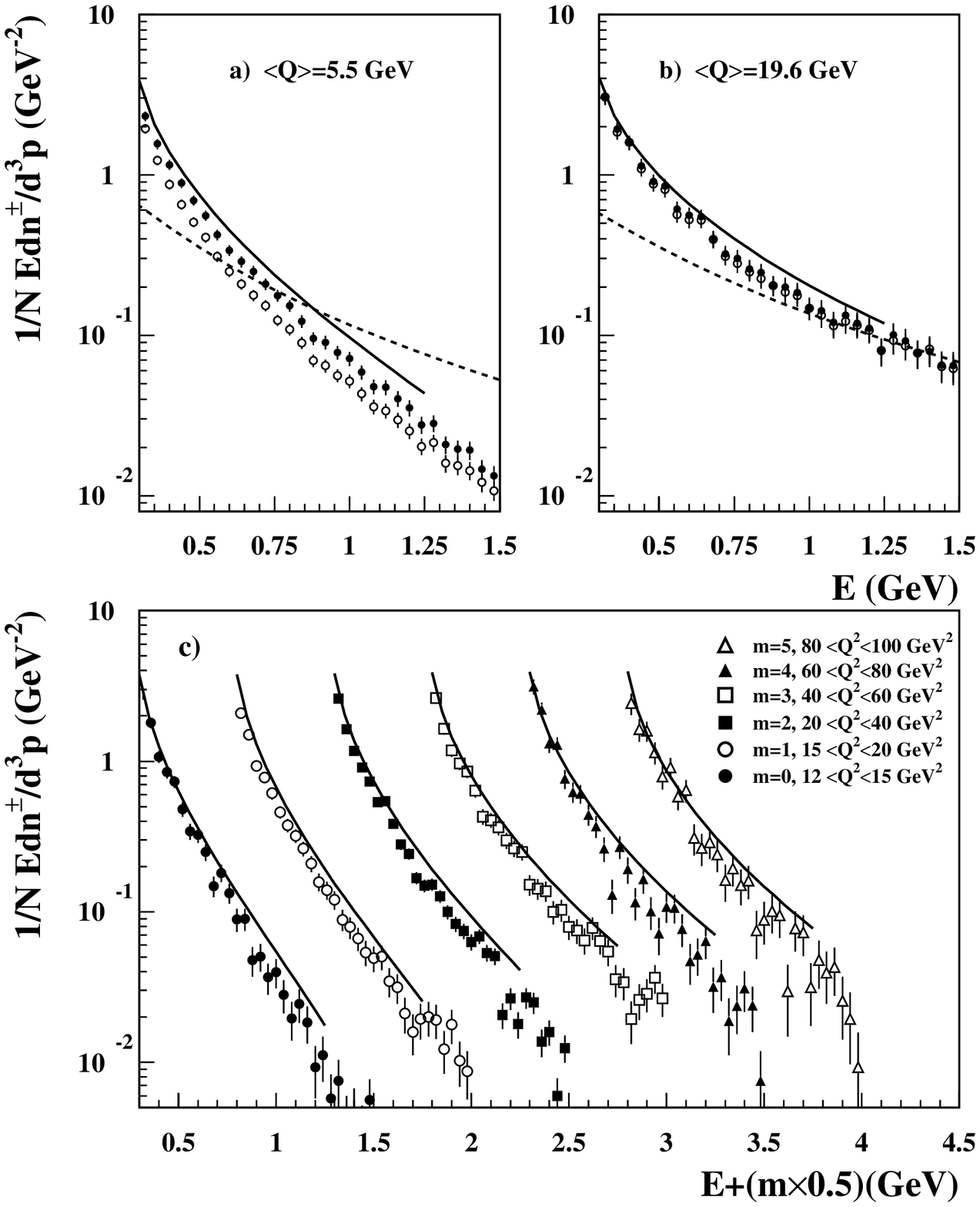,%
   width=10cm,bbllx=26pt,bblly=59pt,bburx=540pt,bbury=703,clip=}
   \scaption{Invariant charged particle spectra in the Breit
             current hemisphere \cite{h1:breit2}.
             In {\bf a)} and {\bf b)} the data at small and large $Q$ are
             compared to the MLLA+LPHD prediction \cite{th:khoze} with
             running \as (solid line), and with constant \as (dashed line).
             The open and full points are before and after the
             anti-BGF selection.
             In {\bf c)} the specta are shown for different \Qsq ranges.
             For better visibility, they are offset by an incremental
             spacing of $0.5 \GeV$. The full lines are the
             MLLA/LPHD expectations.}
   \label{invbreit}
\end{figure}

Based upon MLLA+LPHD, the spectra from fragmenting quarks
have been calculated also
in the Lorentz invariant form
\begin{equation}
  \frac{1}{N} E \frac{\dif n}{\dif^3 p}
\end{equation}
to study the small momentum limit \cite{th:khoze}.
The calculation agrees well with \epem~ data.
One salient feature is that for $p\rightarrow 0$
the calculated spectra do not
depend on the energy of the fragmenting parton.
The invariant spectra from H1 \cite{h1:breit2} are shown
in fig.~\ref{invbreit}.
The particle momentum $\vec{p}$ is measured in
the drift chamber.
Its  energy is set to $E=\sqrt{Q_0^2+|\vec{p}|^2}$,
following the presciption in \cite{th:khoze}, where the
``particle mass'' is given by the mass cut-off of the parton shower,
taken to be $Q_0=0.27\GeV (\approx \Lambda_{\rm QCD})$.

Comparison of the MLLA+LPHD prediction for fragmenting quarks,
which has been shown to agree with \epem~ data \cite{th:khoze},
with the $ep$ data is hampered by the presence of BGF events.
At low $Q$, the prediction is far off the H1 data \cite{h1:breit2},
see fig.~\ref{invbreit}a.
That effect can be reduced by a special anti-BGF
event selection \cite{h1:breit2}, which is based on the
calorimetric energy seen in the Breit current hemisphere.
That selection reduces the
discrepancy at low $Q$. At large $Q$, where the discrepancy
is smaller, there is little change after that cut, see
fig.~\ref{invbreit}b.
This study shows that in the $ep$ Breit frame
spectra there are fewer fast particles
than in \epem~ events at the same energy.
The difference is presumably due to BGF events that are absent in \epem.
For larger $Q$, where the BGF contribution decreases, also
the difference decreases.

The prediction of an energy (here: $E_q=Q/2$)
independent soft
spectrum is confirmed -- the measured spectra converge for
$|\vec{p}| \rightarrow 0$, independent of $Q$ (fig.~\ref{invbreit}c).
It can also be noted that
a calculation with fixed $\alpha_s$, instead of a running $\alpha_s$,
is clearly untenable (figs.~\ref{invbreit}a,b).
A meaningful extraction of \as is
however not possible yet, because the calculations
are available only in LO.

In the same H1 analysis \cite{h1:breit2},
the KNO mulitplicity distribution was studied in
the Breit current hemisphere. KNO scaling was found to be broken;
at low $Q$ the distributions broaden, and deviate the most from
the approximately scaling \epem~ data. Again, BGF events and the
limited rapidity range can be held responsible for this observation.
It would be interesting to test also KNO-G scaling \cite{th:knog},
which is a reformulation of the KNO form, valid for discrete distributions
(like multiplicities) at finite energies.

\subsubsection{Scaling behaviour}

The scaling behaviour
of the \xp spectra in the Breit current
hemisphere are best studied by displaying
the \Qsq dependence of the normalized hadron cross section
$\frac{1}{\sigtotal} \dif \sigma/\dif \xp$ for fixed \xp, where
\sigtotal is the event cross section.
With the large range of \Qsq accessible at HERA,
the scaling behaviour can be studied in a single experiment.
Ultimately, \as could be extracted
from such measurements \cite{th:scviol}.

\begin{figure}[tbh]
   \centering
\begin{picture}(0,0) \put(0,0){{\bf a)}} \end{picture}
   \epsfig{file=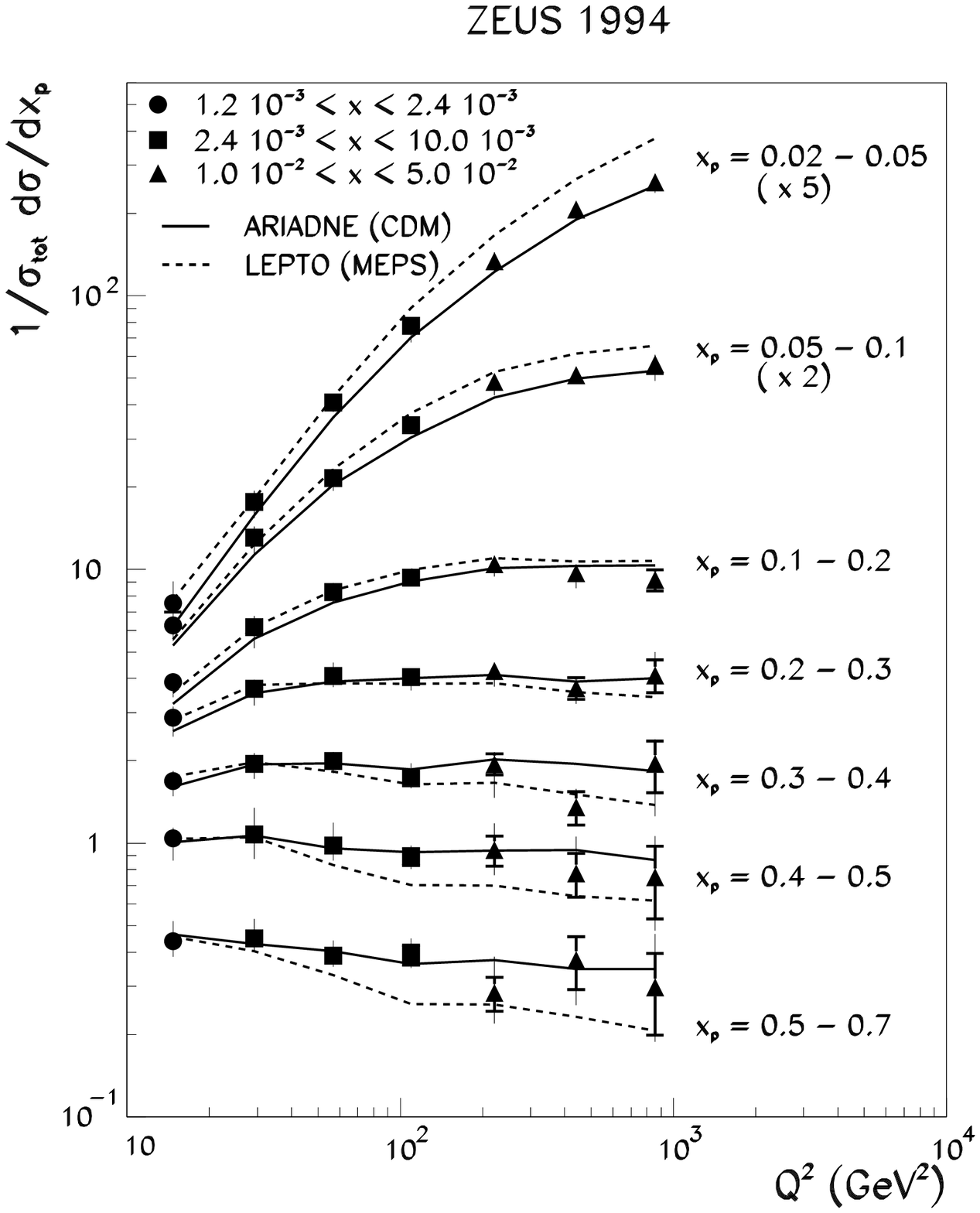,
           width=7cm}
\begin{picture}(0,0) \put(0,0){{\bf b)}} \end{picture}
   \epsfig{file=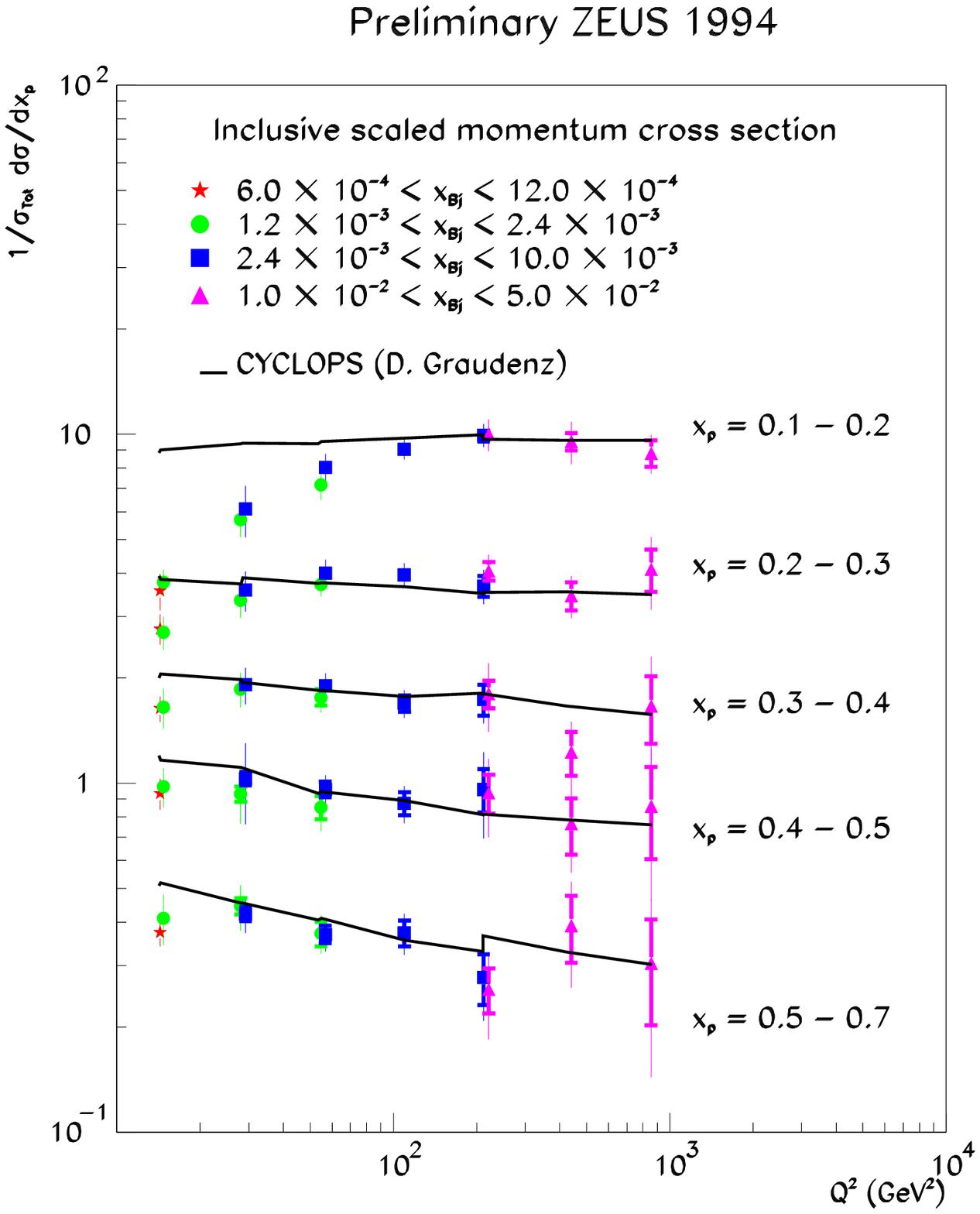,
           width=7cm}
   \scaption{Charged particle cross section as a function of \Qsq
             for different \xp bins. The data \cite{z:breit2}
             are compared to {\bf a)} ARIADNE 4.08 \cite{mc:ariadne}
             and LEPTO 6.5 \cite{mc:lepto};
             and {\bf b)}
             to the NLO calculation with
             CYCLOPS \cite{mc:cyclops}.
            }
   \label{xpmodels}
\end{figure}

The ZEUS data \cite{z:breit2}, covering $10<\Qsq<1280\GeVsq$
are shown in fig.~\ref{xpmodels}.
The spectra were found to be almost
independent of \xb for fixed \Qsq \cite{h1:breit,z:breit2}.
Approximate scale independence is observed for $\xp\approx 0.3$:
the cross section
$\frac{1}{\sigtotal} \dif \sigma/\dif \xp$ does not depend on \Qsq.
For smaller $x_p$, the cross section increases
strongly with \Qsq (compare also fig.~\ref{xpxi}a).
These effects are well described
by the ARIADNE 4.08 \cite{mc:ariadne} event generator;
in LEPTO 6.5 \cite{mc:lepto}
the scale dependencies are too strong, see fig.~\ref{xpmodels}.
Possibly the strong coupling parameter for the
final state parton shower in LEPTO can be adjusted to describe the data.

\begin{figure}[tbh]
   \centering
   \epsfig{file=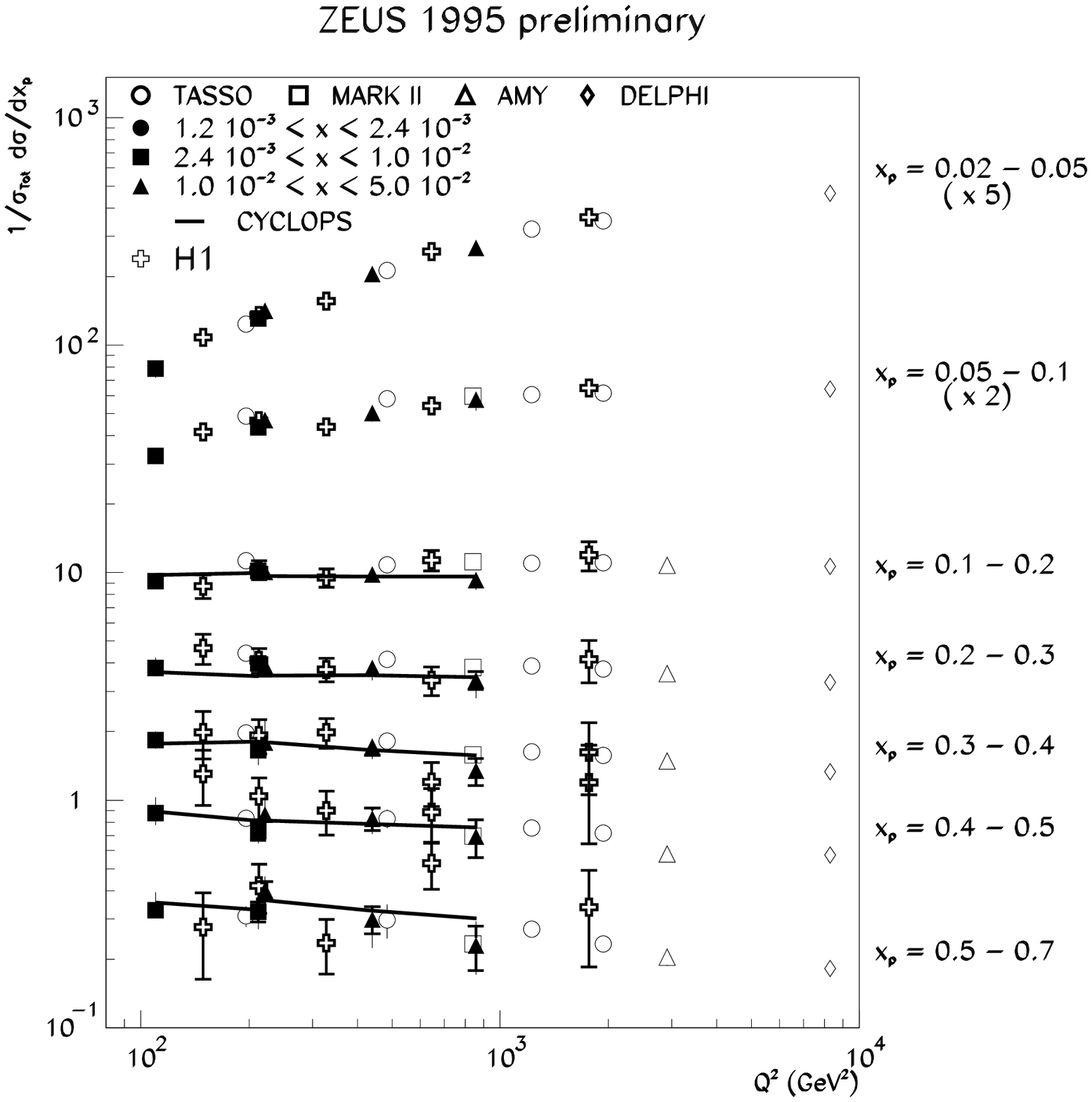,
           width=12cm}
   \scaption{Charged particle cross section as a function of \Qsq
             for different \xp bins.
             H1 data \cite{h1:breit2} and
             preliminary ZEUS data \cite{z:breit3}
             are compared to \epem~ data
             \cite{lep:tasso,lep:mark2,lep:amy,lep:delphi},
             and to a NLO
             calculation with the program CYCLOPS
             \cite{th:scviol,mc:cyclops}. }
   \label{xpviol}
\end{figure}

In fig.~\ref{xpviol} new preliminary ZEUS data (1995) \cite{z:breit3}
are shown together with
lower statistics H1 data (1994) \cite{h1:breit2} for $\Qsq>100~\GeVsqx$.
They agree very well. The $ep$ data agree also very well with
data from \epem~annihilation
(corrected for $K^0$ and $\Lambda$ decay products, and divided by 2),
supporting the notion of universal
quark fragmentation in this kinematic region.
The measured cross sections at small \xp exhibit clear scaling
violations with a positive slope. It is not clear however that
this effect is due to the expected scale dependence of the
fragmentation function in perturbative QCD.
These expectations
will be discussed now.

The spectra have been calculated \cite{th:graudenz2}
according to eq. \ref{eq:scviol}
in NLO perturbative QCD (program CYCLOPS \cite{mc:cyclops}) by folding
parton density functions (here \mrsap \cite{th:mrsa})
with the NLO matrix element for parton production and
a NLO fragmentation function \cite{th:binnewies}
for parton fragmentation into hadrons.
All of these ingredients are scale dependent.
$\Lambda_{\rm QCD} = 0.23 \GeV$ was used.

The calculation agrees well with the data in the kinematic region
shown in fig.~\ref{xpviol}, $\xp>0.1$ and $\Qsq>100~\GeVsqx$.
In this regime, the
expected scaling violations are small, and can hardly be
established with the present precision of the data.
With higher precision data, and with a larger lever arm
both at higher and smaller \Qsq, the predicted scaling violations
will be measurable. It will be very interesting to study to what extent
this effect is due to the scale dependence of the fragmentation
function, the matrix element, or the parton densities.

The fragmentation function picture has its limitations \cite{th:graudenz2}.
The fragments must not be too far separated in rapidity from
the parent parton, assumed to fragment independently.
It is estimated that the hadron rapidity
in the Breit system should be above some minimal value $y^\prime$,
with $y^\prime \approx 1$.
The condition $y>y^\prime$,
translates
into a lower bound on \xp,
\begin{equation}
   \xp > \frac{2 m_T}{Q}\frac{1}{\sqrt{1-\tanh^2 \ymin}}
  \label{eq:graud}
\end{equation}
with the transverse mass $m_T$, typically of \order{0.5\GeV}.
Furthermore, hadron masses are not taken into account with
fragmentation functions. The calculation thus must break down when
$\xp \cdot Q/2$ approaches ${m_\pi}$.
Indeed, the NLO calculation would by far overestimate the data
in a region where $\Qsq<100\GeVsq$ and $\xp \approx 0.1$ \cite{th:graudenz2},
see fig.~\ref{xpmodels}. We note however deviations also
at $Q^2\approx 15 \GeVsq$ and $\xp>0.5$, where the NLO
calculation should be reliable according to eq. \ref{eq:graud}.

  \clearpage
  \section{Event Shapes \label{sn:shapes}}     
\subsubsection{Definition of event shape variables}

Information on the shape of events,
- pencil-like, spherical, cigar-like, planar etc. -
is conveniently given by
simple functions of the hadron momenta $p_i$,
$F={\cal F}(p_i)$.
Measurements of such event shape variables
provide information about both perturbative and
non-perturbative aspects of QCD.
Due to the large kinematic range covered, HERA is particularly
well suited to study their scale dependence, for example on energy,
and by that means disentangle the two contributions.
Furthermore, provided the non-perturbative part can be estimated,
the strong coupling constant \as can be extracted.

First measurements at HERA of the event shape variables
thrust $T$,
jet mass parameter $\rho_C$,
and jet broadening parameter $B_C$ are reported by
H1 \cite{h1:shapes}.
The measurements are restricted to the Breit frame current
hemisphere in order to avoid complications from the
not so well understood region towards the proton remnant.
The energy of the scattered quark in the Breit frame (in the QPM),
$Q/2$, provides the scale for the process.
$Q$ values
between 7 and 100 GeV are covered.

The variables are defined by
\begin{equation}
   T := \max \frac{\sum_i |\vec{p}_i \cdot \vec{n}_T|}
                  {\sum_i |\vec{p}_i|}
   \hspace{1cm}
   B_C := \frac{\sum_i |\vec{p}_{Ti}|}
              {2\sum_i |\vec{p}_i|}
   \hspace{1cm}
   \rho_C := \frac{M^2}{Q^2} = \frac{(\sum_i \vec{p}_{i})^2}
                                    {Q^2}.
\label{eq:shapes}
\end{equation}
For the thrust calculation,
the unit vector $\vec{n}_T$, defining the direction of a
thrust axis, is
varied\footnote{H1 has also measured the thrust for
fixed thrust axis, defined by the virtual photon.}
 in order to maximize $T$,
the normalized longitudinal momentum sum of all particles
$i$.
$T=1$ for a collinear event configuration,
and $T=1/2$
for a spherical configuration.
Thrust measurements are usually expressed in $1-T$, so that
low values correspond to pencil-like events, as for the
other event shape variables.

In the H1
measurement calorimeter clusters are used in the sums eq. \ref{eq:shapes},
which do not have a one-to-one correspondence to incident particles.
However, these event shape variables are infrared safe quantities
(see section \ref{sn:finst}),
they do not change when an object in the sum is split:
${\cal F}(p_1,...p_i,...p_n) =
{\cal F}(p_1,...z p_i,(1-z)p_i,...p_n)$.

\subsubsection{Thrust measurements}

\begin{figure}[p]
   \centering
   \epsfig{file=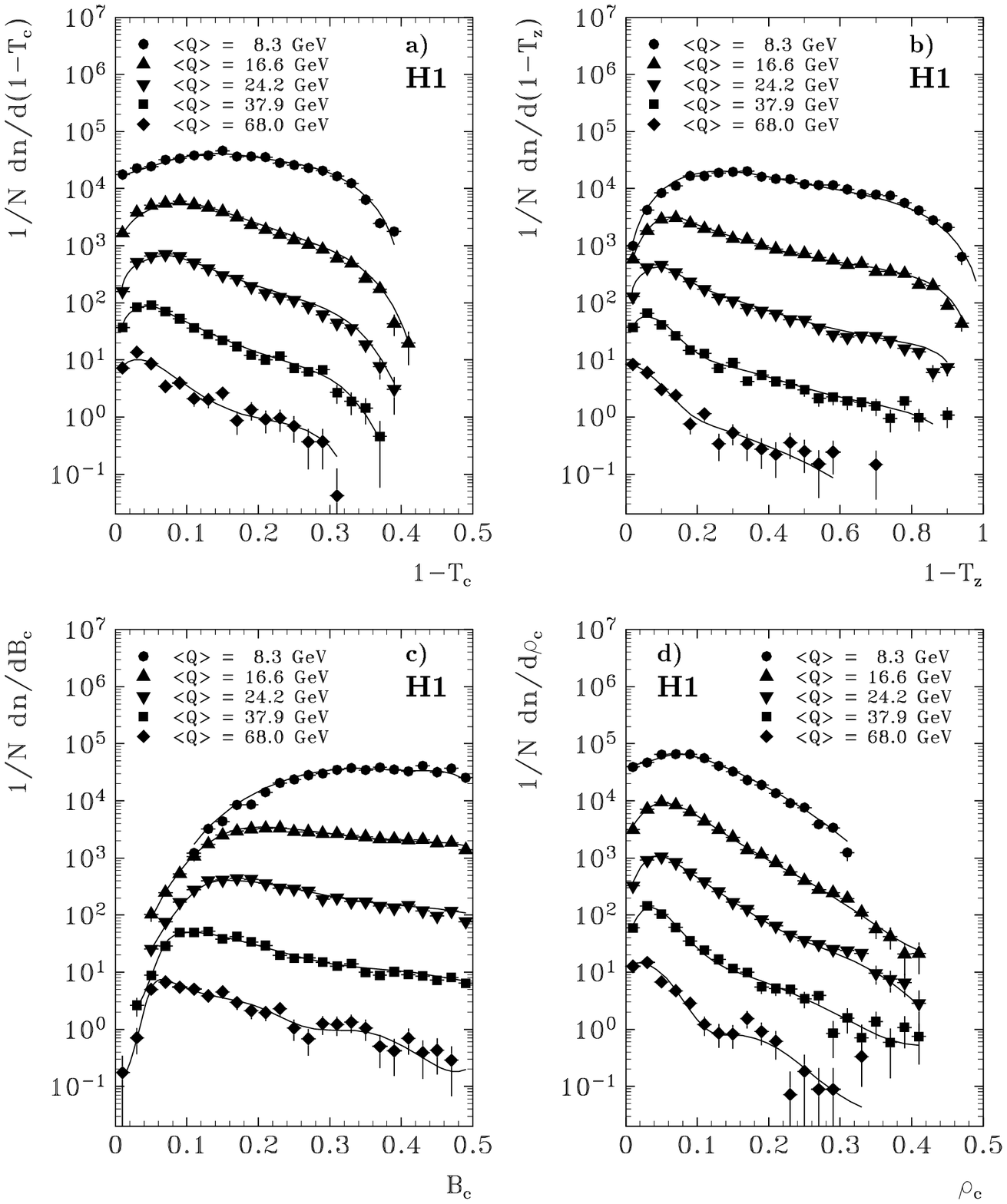,%
          width=7cm,bbllx=44pt,bblly=435pt,bburx=291pt,bbury=740,clip=}
   \epsfig{file=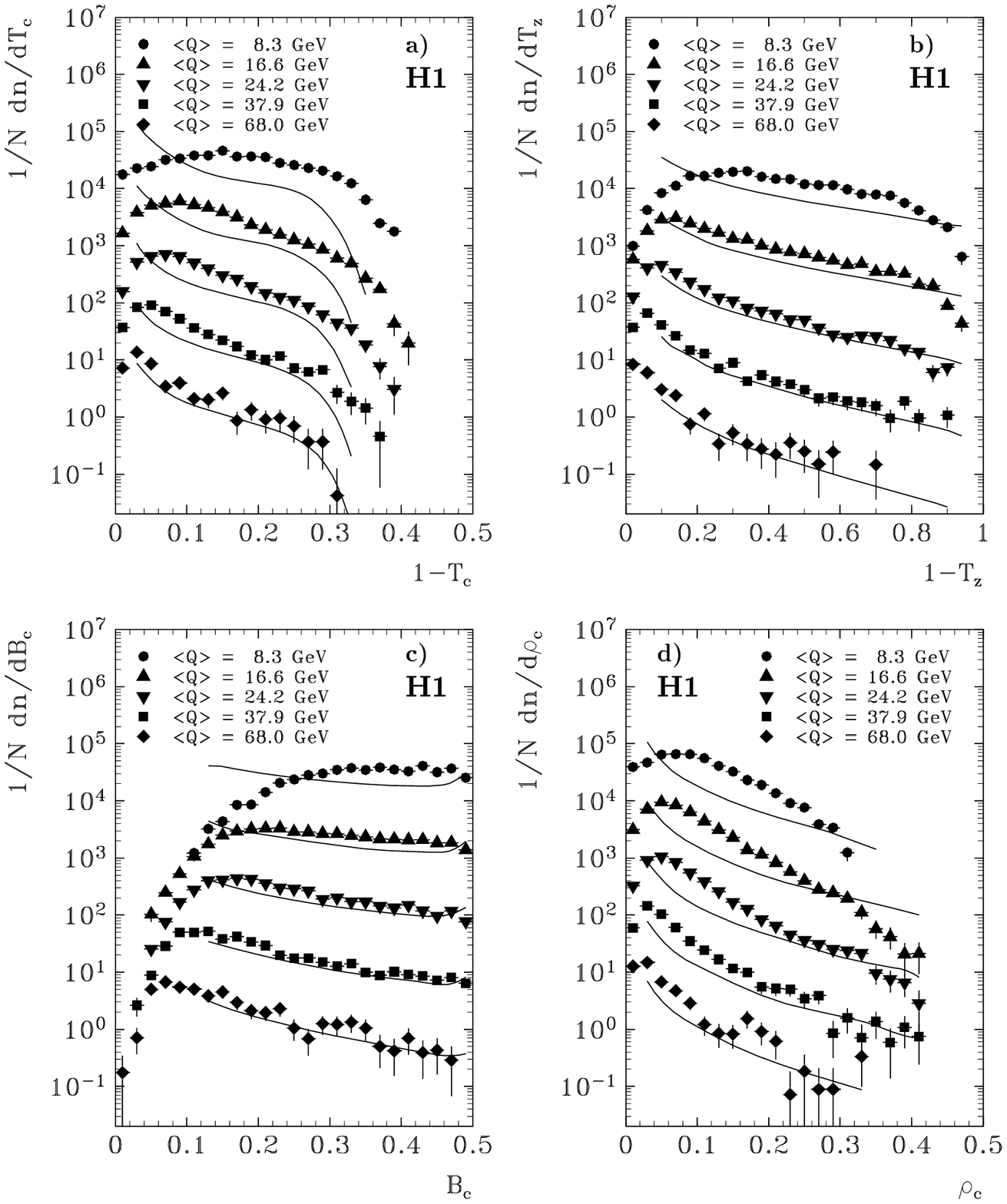,%
          width=7cm,bbllx=44pt,bblly=435pt,bburx=291pt,bbury=740,clip=}
   \epsfig{file=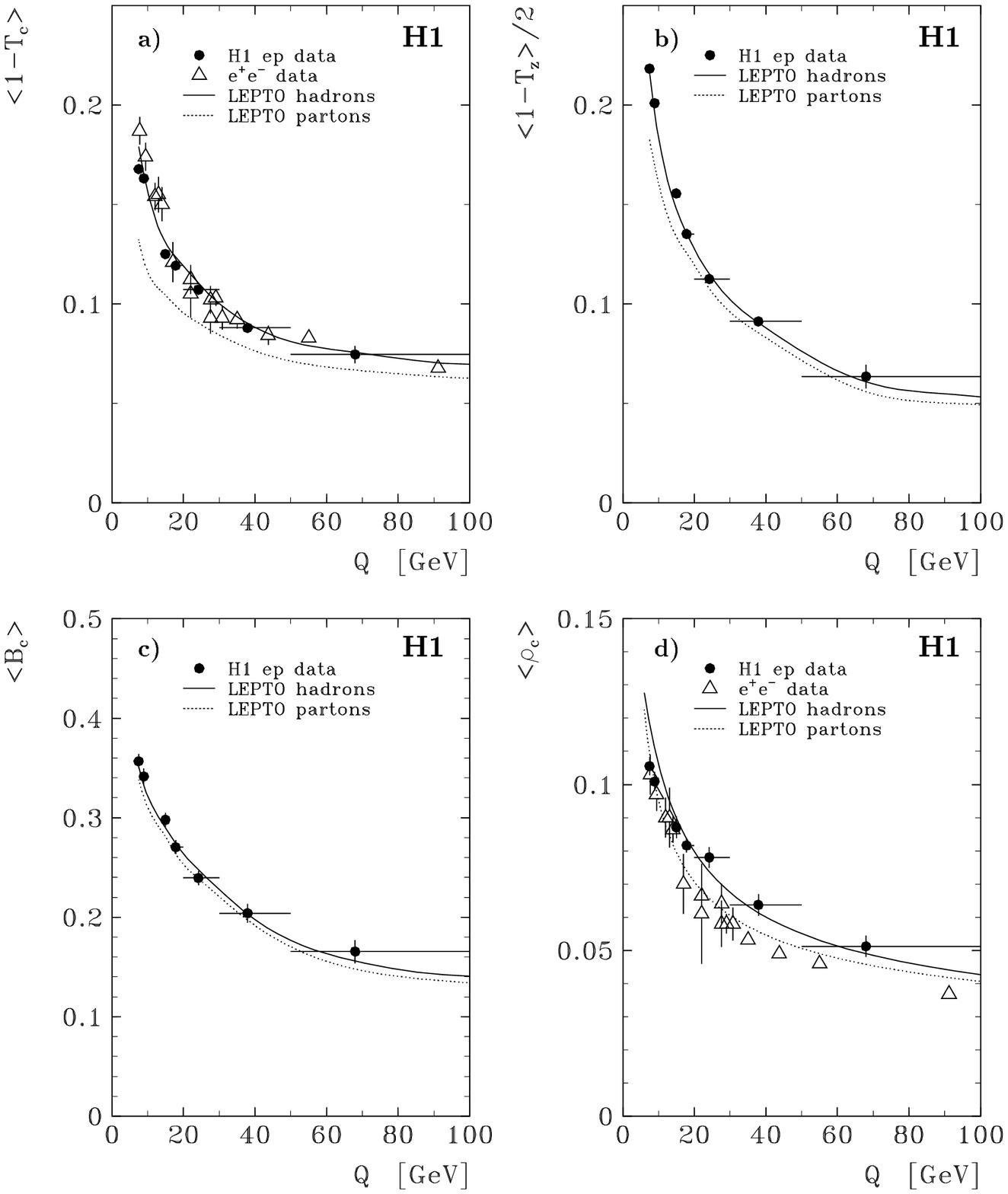,%
          width=7cm,bbllx=44pt,bblly=435pt,bburx=291pt,bbury=740,clip=}
   \epsfig{file=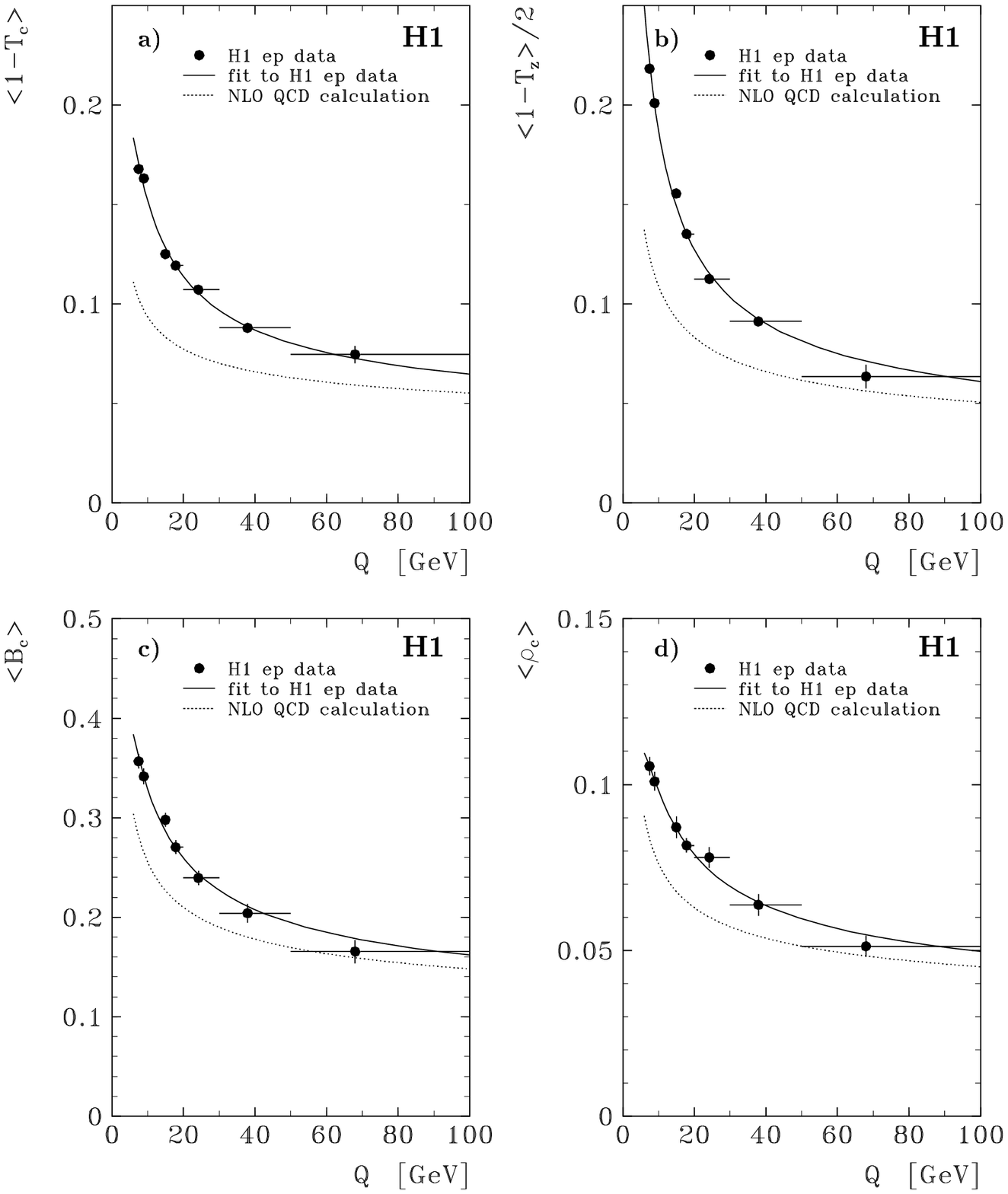,%
          width=7cm,bbllx=44pt,bblly=435pt,bburx=291pt,bbury=740,clip=}
   \scaption{Top: thrust distributions \cite{h1:shapes}.
             The differential thrust distribution
             $(1/N) \dd N/\dd (1-T)$,
             compared to LEPTO 6.1 (top left) and
             a NLO calculation (top right).
             The spectra for
             $\av{Q}=8.3-68 \GeV$ are multiplied by factors
             of $10^n, n=0,...4$.
             Bottom: the mean $(1-T)$ as a function of $Q$. The H1 data
             are compared to
             (bottom left) \epem~ data for the full event
             (two hemispheres) at $\sqrt{s}=Q$, and the LEPTO
             result for hadrons and partons; and
             (bottom right) to
             the NLO calculation, and the fit result including
             the power correction.}
   \label{shape1}
\end{figure}

Here we shall concentrate on the thrust measurements, see fig.~\ref{shape1};
the conclusions
from the other shape variable measurements are similar.
The measured
mean value $\av{1-T}$
decreases with increasing $Q$, the
events become more collimated.
The thrust distribution as well as the mean thrust and their energy dependences
are well described
by the LEPTO model (version 6.1) \cite{mc:lepto},
which incorporates a perturbative phase with the LO QCD matrix elements and
leading log parton showers, and the Lund string model for hadronization.

Qualitatively, data from \epem~ annihilation
\cite{lep:pluto,lep:amy,lep:tasso,lep:delphi,lep:mark2}
follow the same trend as the $ep$ data, but are systematically higher at small
energies, $Q<15$ GeV.
One would actually not expect exactly the same
distributions,
partially because in \epem~ the full event with two hemispheres is
used\footnote{Note that the
thrust for a system of two identical back to back jets
equals the thrust of a single jet.}.
Furthermore,
the flavour compositions differ, and in $ep$ there are $O(\alpha_s)$
processes contributing, namely initial state QCD radiation and
boson-gluon fusion, which are absent in \epem.
In fact, the other shape
variables ($\rho_C,B_C$)
show bigger differences than thrust between \epem~ and $ep$ data.

\subsubsection{Power corrections}

It has been argued \cite{th:powercorr} that for any infrared safe
variable the mean value (and higher moments) can be written as
\begin{equation}
   \av{F} =  \av{F}^{\rm pert.} +  \av{F}^{\rm pow.}.
   \label{eq:ansatz}
\end{equation}
The perturbative part $\av{F}^{\rm pert.}$
is calculable in fixed order perturbation theory, depending only
on \as and the choice of renormalization
scale\footnote{For the purpose of extracting a meaningful, process
independent $\alpha_s$, the calculation has to be done in fixed order
perturbation theory to at least 2nd order (NLO).}.
The only energy dependence comes from the running $\alpha_s$,
since there is no other intrinsic scale in
perturbative QCD than \lambdaqcd.
QCD radiation patterns would be identical at all energies,
if \as were constant.
If the shape variable is invariant against
momentum scaling with a constant $c$ (like $T$),
${\cal F}(p_i) = {\cal F}(c\cdot p_i)$,
the distribution of the event shape variable
depends on $Q$ only through the running \as:
\begin{equation}
 \frac{\dif}{\dif Q}           \frac{\dif N}{\dif F} =
 \frac{\partial}{\partial Q}   \frac{\dif N}{\dif F} +
 \left( \frac{\partial}{\partial \as} \frac{\dif N}{\dif F} \right)\cdot
 \frac{\partial \as}{\partial Q},
\end{equation}
because the first term on the right hand side vanishes.
In $ep$ reactions a $Q$ dependence may arise however via the
$Q$ dependent parton densities.

Higher orders and hadronization effects are collected in the
``power correction term'' $\av{F}^{\rm pow.}$,
which according to \cite{th:powercorr} scales approximately $\propto 1/Q$
(in general $\propto (1/Q)^p$ with the power $p$).
A plausibility argument for such a term is given below.
In contrast to the perturbative term, this term does depend explicitly on
the energy, because  hadronization introduces a scale, given by hadron masses.
The higher order perturbative effects are calculated
with running \as down to a
scale $\mu_I$, with $\lambdaqcd \ll \mu_I$
to avoid the divergence of \as for $Q\rightarrow \lambdaqcd$,
and  $\mu_I \ll Q$ to maximize the
perturbative evolution.
A conventional choice is $\mu_I=2\GeV$.
Below the scale $\mu_I$, the calculation uses a constant effective coupling
$\ol{\alpha}_0$, a free parameter, which has to be
determined by experiment, and which depends on the choice of $\mu_I$.

The NLO expressions with $Q$ as renormalization scale are \cite{th:powercorr}
\begin{equation}
 \av{F}^{\rm pert.} = c_1 \alpha_s(Q^2) + c_2 \alpha_s^2(Q^2)
 \label{eq:pert}
\end{equation}
\begin{equation}
  \av{F}^{\rm pow.} = a_F \frac{16}{3\pi} \frac{\mu_I}{Q}
                    \left[\ol{\alpha}_0(\mu_I^2)-\alpha_s(Q^2) -
                         \frac{\beta_0}{2\pi}
                     \left(\ln \frac{Q}{\mu_I} +
                          \frac{K}{\beta_0}+1\right) \alpha_s^2(Q^2) \right],
\label{eq:pow}
\end{equation}
where $\beta_0=11-(2/3)n_f$, $K=67/6-\pi^2-(5/9) n_f$ and
$n_f=5$ flavours\footnote{It is not clear what should be used
for
$n_f$ in this situation.
The lowest values of $Q/2$ are above the charm threshold, the larger ones are
above the bottom threshold.
On the other hand, for the parton shower with many soft emissions
only light flavours are expected to be active.
In fact, in the multiplicity analyses
in sections \ref{sn:multi} and \ref{sn:xp} $n_f=3$ was used.}.
The constant $a_F$ is predicted by the theory and depends on the
event shape variable $F$ under consideration (see tab.~\ref{tab:shapes}).
In the H1 analysis, the constants $c_1$ and $c_2$ are calculated in the
$\ol{MS}$ scheme with the program DISENT \cite{mc:disent}.
For e.g. $\av{1-T}$ they are $c_1=0.384\pm0.033$ and $c_2=0.57\pm0.21$.

With the ansatz eq. \ref{eq:ansatz} a good fit to the
H1 $\av{1-T}$ data (and \av{\rho_C}, \av{B_C} as well) is obtained
(see fig.~\ref{shape1}) with \as and $\ol{\alpha}$ as the only
free parameters.
The perturbative contribution decreases with
energy $Q$ due to the decreasing \asx.
The power correction is substantial, but decreases from
$\approx 50\%$ at $Q=8\GeV$ to  $\approx 20\%$ at $Q=70~\GeVx$.
The prediction from LEPTO for partons is higher
and closer to the data than the NLO calculation (see fig.~\ref{shape1}),
because the LEPTO parton shower includes QCD radiation
in principle to all orders in the leading log approximation.
In the ansatz \ref{eq:ansatz}, the higher orders are
part of the power correction term.
Whereas the differential thrust distribution $\dif N/\dif T$
is well described by the LEPTO model, the NLO calculation
(covering states with up to 3 final state partons) is
insufficient,
producing too narrow events at small $Q$, but approaches
the data for larger $Q$. This is not surprising, as
the large NLO correction to the LO in eq.~\ref{eq:pert}
(NLO/LO = $\as c_2/c_1$) suggests important higher order corrections,
which should decrease with increasing energy due to the running \as.

From the thrust fit
one obtains for the two free parameters
$\ol{\alpha}_0(\mu_I^2=4\GeVsq)=0.497\pm0.005^{+0.070}_{-0.036}$,
and
$\alpha_s(m_Z^2)=0.123\pm0.002^{+0.007}_{-0.005}$
for the strong coupling extrapolated to the $Z$
mass\footnote{
Note that eq.~\ref{eq:pow} is not any longer a pure power law $\propto 1/Q$,
but is modified by corrections up to $O(\alpha_s^2)$.
A pure power law fit of $\av{1-T}^{\rm pow}=2\lambda/Q$ would give
a worse $\chi^2$ than
the ``QCD improved'' expression eq.~\ref{eq:pow} \cite{h1:rabbertz}.}.
Similar values are obtained from the other event shapes,
see tab.~\ref{tab:shapes}\footnote{According to \cite{th:powercorr},
the power correction
term for the jet broadening parameter $B_C$ should be $\propto (1/Q)\ln Q$.
From a comparison of
the H1 data with the DISENT NLO calculation
the extra $\ln Q$ term cannot be supported \cite{h1:shapes}.}.
Assuming the power corrections to be universal
(same $\ol{\alpha}_0$ for all shape variables),
a combined fit to the thrust and jet mass data
yields
$\alpha_s(m_Z^2)=0.118\pm 0.001^{+0.007}_{-0.006}$ \cite{h1:shapes};
one has to admit
however that the small combined statistical error
disregards the fact that
the data samples that enter the different distributions are the same,
not at all statistically independent.
To be on the safe side, one should quote 0.003 as statistical error instead.
The obtained value agrees with the world average \cite{rev:pdg} with
an error of similar size as in other determinations
(see tab. \ref{tab:shapes} and \cite{rev:bethke}).

It is quite remarkable that the ansatz \ref{eq:pow} gives
consistent values for \asx,
and within 20\% the same constant $\ol{\alpha}_0$
for all investigated shape variables.
Furthermore, similar values are being obtained in shape analyses
of \epem data, see table ~\ref{tab:shapes}. These encouraging
results support the concept of universal power corrections, which
deserves further investigation of its origin in QCD,
and of its limitations.
Besides new insights into hadronization, a better understanding
would also improve the \as measurements.
Recently also ``power corrections''
for the event shape distributions, not
just the mean values, have become available \cite{th:powerdis}.
It would be very interesting to compare
these calculations
to the data.

\begin{footnotesize}
\begin{table}[tbh]
\begin{center}
\begin{tabular}{|c|c|c|c|}
   \hline
Observable & $a_F$ &  $\ol{\alpha}_0(\mu_I^2=4\GeVsq)$ & $\alpha_s(m_Z^2)$ \\
   \hline
H1 DIS         &       & &                                       \\
$\av{1-T}$     &  1    &  $0.497 \pm 0.005 ^{+0.070}_{-0.036}$
                       &  $0.123 \pm 0.002 ^{+0.007}_{-0.005}$  \\
$\av{B_C}$ & 2   &  $0.408 \pm 0.006 ^{+0.036}_{-0.022}$
                       &  $0.119 \pm 0.003 ^{+0.007}_{-0.004}$  \\
$\av{\rho_C}=\av{M^2/Q^2}$ & 1/2   &  $0.519 \pm 0.009 ^{+0.025}_{-0.020}$
                       &  $0.130 \pm 0.003 ^{+0.007}_{-0.005}$  \\
\hline
\epem data     &       &  &          \\
$\av{1-T}$     &  1    &  $0.519 \pm 0.009 ^{+0.093}_{-0.039}$
                       &  $0.123 \pm 0.001 ^{+0.007}_{-0.004}$  \\
$\av{M_H^2/Q^2}$ &1  &  $0.431 \pm 0.020 ^{+0.071}_{-0.030}$
                       &  $0.115 \pm 0.002 ^{+0.005}_{-0.003}$  \\
\hline
\end{tabular}
\end{center}
\scaption{Results of the event shape analyses \cite{h1:shapes}.
          Following \cite{th:powercorr}, different values for
          $a_F$ are used for the jet mass parameter in $ep$ and \epem.
          }
\label{tab:shapes}
\end{table}
\end{footnotesize}

\subsubsection{Naive model for power corrections}

The power behaviour $\propto 1/Q$ can be made plausible \cite{rev:webber}
with the simple tube or longitudinal phase space model \cite{books:feynman}
for hadronization.
In this model a colour connected pair
of partons with total energy $Q$
produces two back to back jets of hadrons with energy $E_j=Q/2$ each,
where the hadrons
are distributed uniformly in rapidity $y$ and according
to a probability density function $\rho(p_T)$ in transverse momentum.
The resulting hadron distribution function is then
$\Phi(y,p_T)=n_0\rho(p_T)$, where $n_0$ gives
the number of hadrons per unit rapidity.
The energy and longitudinal momentum of one jet are given by
\begin{equation}
  E_j = \int_{0}^{\ymax} \int_{0}^{\infty} n_0 \rho(p_T) E \dd p_T \dd y
      = \lambda \sinh \ymax
\end{equation}
\begin{equation}
  p_{jz} = \int_{0}^{\ymax} \int_{0}^{\infty} n_0 \rho(p_T) p_z \dd p_T \dd y
      = \lambda (\cosh \ymax -1 ) \approx E_j - \lambda,
\end{equation}
where $\lambda$ is related to the average hadron $p_T$ by
\begin{equation}
  \lambda := \int_{0}^{\infty} n_0 \rho(p_T) p_T \dd p_T = n_0 \av{p_T}.
\end{equation}
Used were $E=m_T\cosh y$, $p_z=m_T\sinh y$, $m_T\approx p_T$
for light hadrons, and the
quick convergence of $\cosh y$ towards $\sinh y$.
For a pencil like jet $p_{jz}=E_j$ would have been expected; hadronization
yields a negative correction of relative size $\lambda / E_j$,
a simple power of the energy.

In this model we can calculate hadronization corrections to
the otherwise vanishing $1-T_j$ and jet mass $M_j$ of the jet.
According to the definitions eq. \ref{eq:shapes},
\begin{equation}
  1-T_j = 1-\frac{p_{jz}}{E_j} \approx 1-\frac{E_j- \lambda}{E_j} =
         \frac{2\lambda}{Q}
\end{equation}
\begin{equation}
  \frac{M_j^2}{Q^2} = \frac{E_j^2-p_{jT}^2-p_{jz}^2}{Q^2} =
   \frac{E_j^2-p_{jz}^2}{Q^2} \approx \frac{2E_j \lambda}{Q^2} =
   \frac{\lambda}{Q}.
\end{equation}
The corrections are proportional to $1/Q$, and twice as large
for thrust as for the scaled jet mass squared, in accord with the
constants $a_F$ for $ep$ in tab. \ref{tab:shapes}.
The different values for $a_F$ in \epem derived in \cite{th:powercorr} appear
somewhat counterintuitive.
One has to take into account however that the
structure of DIS and \epem~ events is more complicated
than just two separating colour charges.

We can estimate the order of magnitude of the parameter
$\lambda=n_0 \av{p_T}$ from the data.
From the seagull plot at low energies, where QCD effects are small,
and for $x_F\rightarrow 0$ where intrinsic $k_T$ effects vanish, we estimate
$\av{p_T}\approx 0.3 \GeV$, see fig.~\ref{sg}.
From measured hadron rapidity distributions
one gets for the hadron density per unit rapidity
$n_0 \approx 2.0$ \cite{rev:schmitz}, compare also fig.~\ref{wdep}.
That yields $\lambda \approx 0.6~\GeVx$. In fact, a good description
of \epem~ event shape variables is obtained with power corrections
parameterized with $\lambda=0.5\GeV$ \cite{rev:webber}.
The fitted $\ol{\alpha}_0$ from the thrust analysis yields for the
$1/Q$ term in eq. \ref{eq:pow} a coefficient of similar order,
$2\lambda\approx 1.7$.

\chapter{Jets \label{ch:jets}}

  \section{Introduction \label{sn:jintr}}            
Amongst the spray of particles
emerging from a high energy reaction
the human eye can recognize
collimated subsystems
of hadrons, so-called jets (fig.~\ref{deth}).
Jets are fascinating objects; they allow to view
high energy quarks and gluons which are not observable as free particles
due to confinement.
As such they can be considered as the parton
images imprinted on the hadronic final state.
At short distances, partons can be treated as ``asymptotically free'';
their interactions can be calculated perturbatively.
When the partons emerge from the confinement volume,
they fragment into observable hadrons.
At high energies, the hadrons will be collimated around
the original parton direction.
These hadron jets still
carry information on the underlying partonic interactions.

Jets in DIS at HERA result from the scattered quark, and from
additional QCD radiation either in the initial or the final state.
Clean jet structures can develop due to
the large available phase space ($W$ up to 300 GeV).
The main experimental challenge in jet studies is twofold:
1) to measure jet quantities like
energy-momentum, charge etc., and 2) to relate them to
the corresponding parton quantities.
A good correlation between partonic and hadronic jet quantities
is vital to allow perturbative QCD to be tested.

Jets at HERA have been studied with different intentions.
The early measurements established clear
jet structures in DIS \cite{z:jetobs,h1:jetmulti}.
Fragmentation properties of quark and gluon jets have been
studied by measuring jet shapes \cite{z:jetshapes,h1:carlij}.
Measurements of jet rates allow to study the partonic
QCD mechanisms
and to
measure the parameters by which they are determined, like the
strong coupling strength \as \cite{h1:as,z:as},
or the density of gluons in the proton \cite{h1:jetgluon}.
It turns out that
parts of the phase space for jet production,
in particular
at small $x$ and
\Qsqx,
is not yet well understood
\cite{h1:jetlowq2,h1:dijet,z:jetprod,z:mikunas}.
The dedicated search for BFKL ``footprints'' with
forward jets is covered in a chapter on low $x$ physics
(section \ref{sn:fjets}).

  \section{Jet Algorithms \label{sn:jalgo}}          For quantitative measurements jets need to be defined:
a jet is the output of a jet algorithm.
A jet algorithm prescribes how to
combine objects (partons, hadrons, energy deposits, ...)
close in phase space to jets.
It hinges
on a resolution parameter below which jets cannot be resolved.
The jet algorithm needs to be implemented not only experimentally,
but also in the theoretical calculation for meaningful predictions.
There
exist several jet algorithms for different applications.

Clustering algorithms,
like the JADE \cite{mc:jade} or the \kt algorithm \cite{mc:kt},
combine {\it all}
final state particles into few jets, based upon a
measure of distance.
In the JADE algorithm, all particles $i,j$ with invariant
mass
\begin{equation}
m_{ij}^2=2E_iE_j(1-\cos\theta_{ij})<\ycut\cdot W^2
\end{equation}
are merged to form new
objects.
(Experimentally, at HERA a pseudoparticle carrying
energy and momentum of the unobserved proton remnant is introduced.)
Here \ycut is the
resolution parameter, and $W$ serves as reference scale.
The final
set of jets is obtained, when no further merging is possible.

In the \kt algorithm two objects are merged,
if
\begin{equation}
  2\cdot\min(E_i^2,E_j^2)(1-\cos\theta_{ij}) < k_{\rm cut}^2.
\end{equation}
Particles can be merged also with the remnant.
At HERA either
$\kcut=Q$ is chosen \cite{h1:as2}, or \kcut is set to a fixed
value, for example $\kcut=3\GeV$ \cite{h1:jetlowq2}.
For small angles, the merging requirement becomes
$\kt<\kcut$, where \kt is the transverse momentum of
the less energetic object with respect to the other one
(hence the name \kt algorithm).

For the cone algorithm \cite{mc:cone}
a distance measure in $\eta$ (pseudorapidity) -
$\phi$ (azimuth) is introduced:
$\Delta R = \sqrt{(\Delta \eta)^2+ (\Delta \phi)^2}$.
All particles inside a cone with radius $R$ (at HERA usually $R=1$)
are combined to form a jet, if the resulting jet \et with respect to
the proton axis
is above a certain cut-off (usually a few GeV).
The cone axis is chosen such that
the \et inside the cone is maximized.

The result of a jet algorithm depends on the way masses are
treated when combining objects (recombination scheme).
Measured jets consisting of more than one particle
always have a finite invariant mass, while most
jets in NLO calculations are massless
(they consist of just one massless parton).
For this reason it is often found that the final result (e.g. \as)
depends on the chosen recombination scheme.
A lot of effort is being spent on minimizing such effects.
For the cone algorithm it also matters in which order objects are combined.

  \section{Jet Shapes \label{sn:jshapes}}          

ZEUS has measured the internal structure of jets
in DIS and has compared them to jets in other reactions \cite{z:jetshapes}.
One expects gluon jets to be broader than quark jets due to the larger
colour charge of the gluon.
In the laboratory frame,
jets with $\et>14\GeV$ and $\eta \in [-1,2]$
were selected with the cone algorithm ($R=1$)
for events with $\Qsq>100\GeVsq$.
In most cases the current quark jet recoiling from the
scattered electron will be selected.

One measures
the average fraction $\psi(r)$ of jet transverse energy
$E_T(r)$ inside a cone of radius $r$ around the jet axis
as a function $r$, defined by
\begin{equation}
\psi(r):= \frac{1}{N_{\rm jets}} \sum_{\rm jets}
          \frac{E_T(r)}{E_T(R)}.
\end{equation}
$N_{\rm jets}$ is the number of jets in the event sample.
$\psi(R)=1$ by definition.
The jet shapes are corrected for detector effects to the hadron level.

It is found that
the jet shapes narrow as \et increases, as expected kinematically.
(A jet will look narrower with increasing boost.)
They show no significant $\eta$ dependence.
(In a preliminary analysis of 2+1 jet events in the Breit frame,
H1 found that these jets become broader towards the remnant \cite{h1:carlij}.)
The gross shape features
are reproduced by the models investigated,
PYTHIA 5.7 \cite{mc:pythia}, LEPTO \cite{mc:lepto} and
ARIADNE \cite{mc:ariadne}, but none of them
provides an accurate description
of the data over the entire phase space.

\begin{figure}[tbh]
   \centering
   \epsfig{file=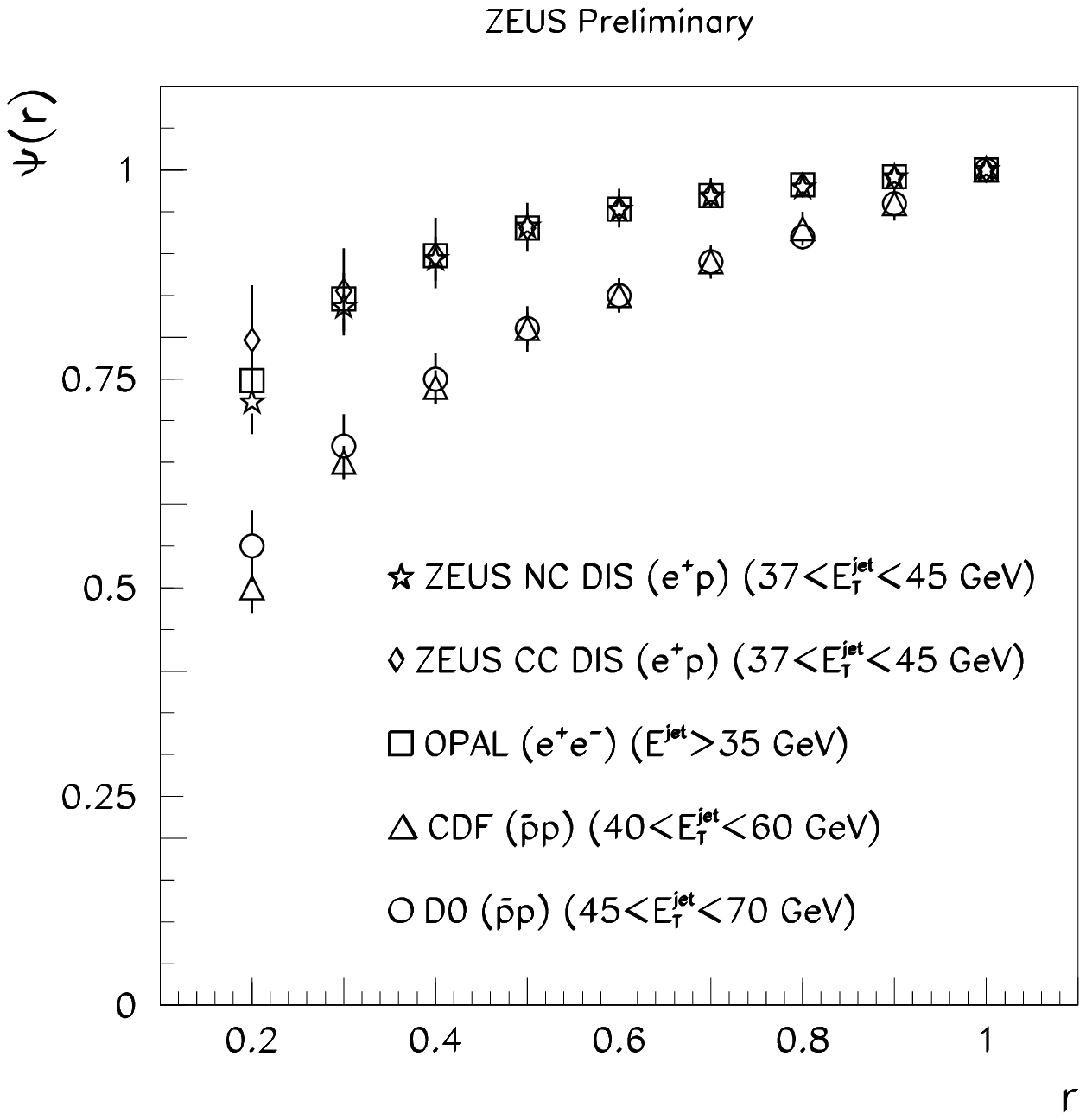,
   width=8cm,%
   bbllx=112pt,bblly=294pt,bburx=488pt,bbury=676pt}
   \scaption{The average jet shapes $\psi(r)$ for DIS charged
             (CC, $ep \rightarrow \nu X$ via $W$ exchange)
             and neutral current
             (NC, $ep\rightarrow eX$ via $\gamma,Z$ exchange)
              events \cite{z:jetshapes},
             compared to \epem~ jets \cite{lep:jetshapes} and
             \ppbar~ jets \cite{coll:jetshapes}.}
   \label{jetshapes}
\end{figure}

On average, DIS jets are seen to be narrower than jets from photoproduction.
This can possibly be
explained by the larger fraction of gluon jets in photoproduction,
mostly from resolved processes, where a quark from the photon scatters
with a gluon from the proton, $qg\rightarrow qg$.
(However, multiple interactions
between the proton and photon remnants
could also lead to a a jet broadening in photoproduction \cite{z:gpshapes}).
A similar observation is made when comparing DIS jets to
jets from \epem~ annihilation \cite{lep:jetshapes},
and to jets from \ppbar~ interactions \cite{coll:jetshapes}
with similar $E_T$, see fig.~\ref{jetshapes}.
DIS jets and \epem~jets are predominantly quark jets and have the same
universal shape within errors. They are narrower than the jets
measured in \ppbar~ interactions which contain
a significant fraction of gluon
jets.

  \section{The Strong Coupling $\alpha_s$ \label{sn:jas}} 
\subsubsection{Analysis method}

\begin{figure}
\epsfig{figure=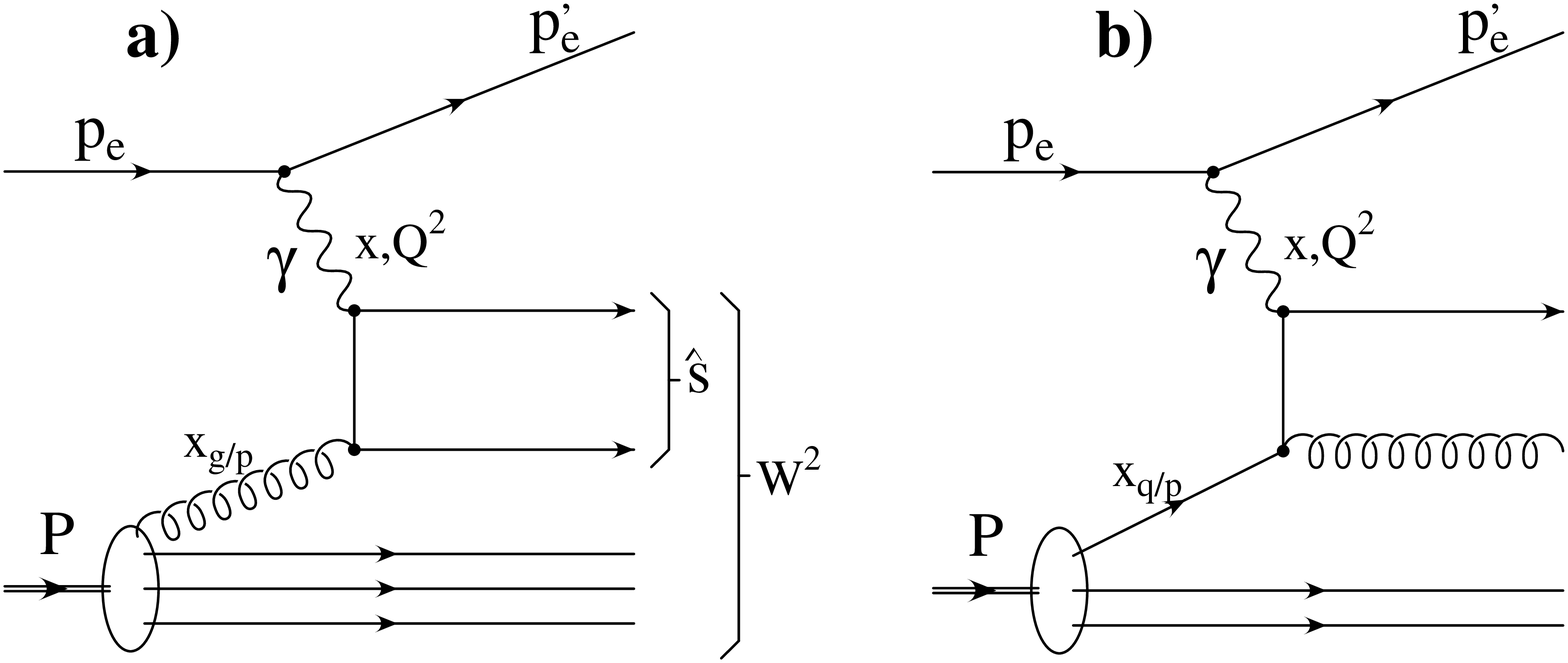,width=10cm}
\begin{tabular}{l}
 \vspace{-5.cm} \\
 DIS variables: \\
   $\Qsq= - q^2 $ \\
   $ x = \Qsq / (2 P \cdot q) $ \\
   $ W^2 = \Qsq (1 - x )/x $ \\
 $\hat{s} = {(j_1 + j_2)}^2$ \\
 $\xi = \xb (Q^2 + \hat{s})/Q^2$ \\
 $z_j = (P \cdot j)/(P \cdot q)$ \\
\end{tabular}

\scaption{Feynman diagrams for the production of $2+1$~jet events
to first order \as in \ep-collisions.
$q$ and $P$ denote the four-momenta of the photon and proton.
$j_1$ and
$j_2$ are the four-momenta of the jets associated
to the hard subprocess with invariant mass squared \shat.
$x_{g/p}$ and $x_{g/p}$ denote the proton fractional momenta
carried by the gluon or quark entering the hard subprocess.
\label{feynjets}}
\vspace{-0.5cm}
\end{figure}

Considering the LO graphs for dijet production (fig.~\ref{feynjets}),
the QCDC Compton (QCDC) and boson-gluon fusion (BGF) graphs,
it is clear
that the rate of events with 2+1 jets (the +1 refers to the
remnant jet which escapes largely unobserved down the beam pipe)
depends on the strong coupling constant \asx, and on the
parton densities in the proton.
Inspired by the great potential to measure
the behaviour of \as over the large
span of \Qsq accessible at HERA with high statistical precision,
jet studies have focussed on the
determination of \as right from the beginning of HERA data analysis.
For this purpose one assumes parton densities that are constrained
by other data.

The kinematics of the hard subprocess in 2+1 jet evenst is depicted in
fig.~\ref{feynjets}.
One defines the invariant
\begin{equation}
   \xi:=x\left(1+ \frac{\shat}{Q^2} \right) \approx \frac{\shat}{W^2},
\end{equation}
where $\shat:=(j_1+j_2)^2$ is the invariant mass squared of the 2-jet system.
In LO $\xi$ can be identified with
the momentum fraction of the proton
carried by the parton that enters the hard scattering,
\xqp~ or \xgp~ for a quark or gluon.

At present the strategy is to measure \as with
dijet data at relatively large $Q^2$,
where the QCDC process dominates with well known quark densities
at large  $\xi$.
Dijet data at small $Q^2$, where the BGF process is the dominant one,
are used to determine the less constrained gluon density at small $\xi$,
assuming a value for $\alpha_s$, see section \ref{sn:jgluon}.
Best use of the information in the data
will in the future be made from a combined determination of \as and
the gluon density.

In the measurements the ratio of the number of 2+1 jet events
$N_{2+1}$ to
the number of all
events
$N_{\rm tot}$,
\begin{equation}
   R_{2+1} := \frac{N_{2+1}}{N_{\rm tot}}
\end{equation}
is determined as a function of \Qsqx.
(H1 defines $N_{\rm tot}:=N_{1+1}+N_{2+1}$.)
In the \as analyses the JADE algorithm is applied in the laboratory frame with
resolution parameter \ycut=0.02.
Another approach uses the dijet rate differential in
the JADE resolution parameter \ycut,
but no \as value has been quoted yet \cite{h1:weber}.
$R_{2+1}$ is being corrected for a) detector effects and b) hadronization
effects to the parton level\footnote{The parton level needs to be
carefully defined, see section \ref{sn:hadron}.},
and is then compared to the QCD prediction in NLO:
\begin{equation}
  R_{2+1} = A \cdot \alpha_s(Q^2) +
            B \cdot (\alpha_s(Q^2))^2 + \order{\alpha_s^3}.
  \label{eq:r2p1}
\end{equation}
$A$ and $B$ are calculable coefficients that depend on the
kinematic conditions, the QCD renormalization and factorization scales,
and on the parton densities.
\as is determined from a comparison of the measured jet rate, corrected
to the parton level, with the NLO prediction of eq.~\ref{eq:r2p1}.

The analyses cuts are designed such that
a) the observed jet
   rates are well described by the Monte Carlo used for corrections;
b) the correction factors are reasonably small;
c) the NLO prediction for partonic jets resembles the partonic jets
   found in the Monte Carlo generator.
Jets from initial state parton showers complicate matters;
to a large extent they are not covered by the NLO calculation.

\subsubsection{Data analysis}

\begin{figure}[tbh]
   \centering
\begin{picture}(0,0) \put(0,0){{\bf a)}} \end{picture}
   \epsfig{file=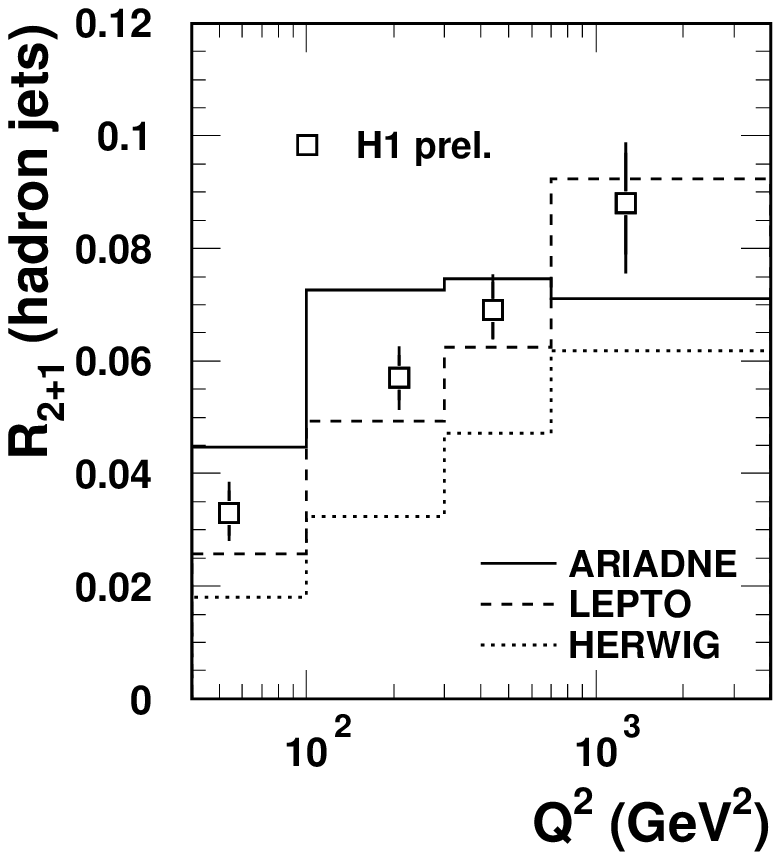,
   width=6cm,%
   bbllx=74pt,bblly=395pt,bburx=308pt,bbury=654pt}
   \hspace{1cm}
\begin{picture}(0,0) \put(0,0){{\bf b)}} \end{picture}
   \epsfig{file=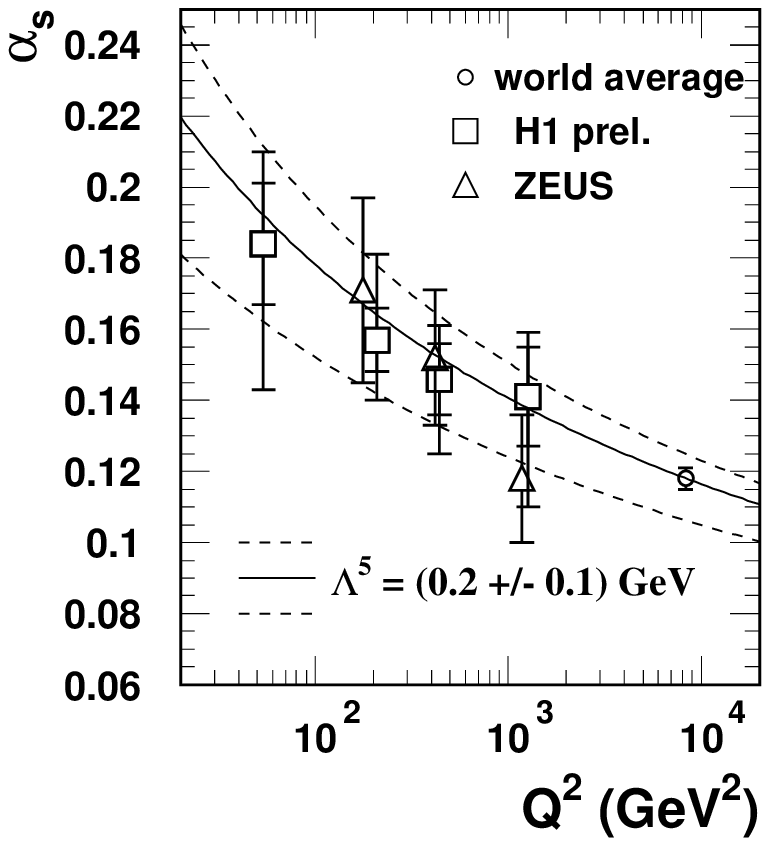,
   width=6cm,%
   bbllx=74pt,bblly=395pt,bburx=308pt,bbury=654pt}
   \scaption{
             {\bf a)} The jet rate $R_{2+1}$ as a function of \Qsq
             for hadron jets (corrected
             for detector effects), reconstructed with the JADE
             algorithm with $\ycut=0.02$.
             The preliminary
             H1 data \cite{h1:as2} are compared to the
             models ARIADNE 4.08 \cite{mc:ariadne}, LEPTO 6.5 \cite{mc:lepto}
             and HERWIG 5.9 \cite{mc:herwig}.
             {\bf b)} The strong coupling constant \as as a function of
             scale \Qsq. Shown are the HERA data \cite{h1:as2,z:as}
             together with the world average for
             $\alpha_s(m_Z^2)$ \cite{rev:pdg}.
             The lines give the expectation for the running \as
             with the QCD scales (for $n_f=5$ flavours)
             $\lmsfive=0.2 \pm 0.1$ GeV.}
   \label{alphas_hera}
\end{figure}

The corrections for detector and hadronization effects rely
mostly on the MEPS model (program LEPTO) \cite{mc:lepto},
where
the hard scattering at the photon vertex is calculated
with the LO matrix element; higher orders are taken into account with
leading log DGLAP parton showers.
Hadronization is performed with
the Lund string model \cite{mc:string}.
Other programs (ARIADNE \cite{mc:ariadne}, HERWIG \cite{mc:herwig})
are used as cross checks.
Fig.~\ref{alphas_hera}a shows $R_{2+1}$
\cite{h1:as2} as a function of \Qsq in
comparison to the models.
The data are corrected for
detector effects to the hadron level. LEPTO describes this rate
reasonably well. It is not understood why the
other models do
not describe this data at relatively large $Q^2$, where little
theoretical ambiguity is expected.

The QCD calculations in
fixed order perturbation theory are performed with the programs
PROJET \cite{mc:projet}, DISJET \cite{mc:disjet}, MEPJET \cite{mc:mepjet},
and DISENT \cite{mc:disent}.
PROJET and DISJET are no longer in use, because parts of the phase
space at small \Qsq and large $W$
were not fully included\footnote{The published ZEUS analysis \cite{z:as}
presented here was based on PROJET. It has been
verified though that due to the cuts applied the effect on the \as extraction
is negligible (less than 2 per mil) \cite{z:trefzger}.}.

In the following two paragraphs, typical features encountered
in the HERA \as analyses are reported.
One finds
$3-20 \%$ 2+1 jet events in the total event sample
\cite{h1:jetmulti,z:jetprod}, depending strongly on \ycut and
also on the event and jet selection.
The dependence of $R_{2+1}$ on \ycut for $\Qsq \gtrsim 100\GeVsq$
is well described by Monte Carlo (LEPTO 6.1) \cite{h1:jetmulti} and
by NLO calculations (PROJET) \cite{z:as}.
With increasing \ycut, fewer structures are resolved as jets.
According to the Monte Carlo (LEPTO 6.1), around 75\% of
matrix element 2+1 jet events
are due to BGF for $\Qsq<100\GeVsq$ (50\% for $\Qsq>100\GeVsq$), the
rest is due to QCDC \cite{h1:jetmulti}. In LEPTO, around 50\% of the
2+1 jet events originate not from the LO matrix element, but from the
parton shower,
that is mostly initial state radiation \cite{h1:as,h1:grindhammer}.
The parton shower contribution is reduced at high \Qsq \cite{h1:flamm}.

The NLO correction to the LO prediction is roughly 20\%,
depending on the analysis details, in particular on \ycut \cite{z:as}.
Migrations between jet classes due to hadronization effects can be
large: $\approx 60\%$ of 2+1 parton jet events
and $\approx 90\%$ of partonic 1+1 events are correctly classified
after hadronization \cite{h1:jetmulti}.
The resulting corrections for hadronization are sizeable,
around 30\% \cite{h1:as2}.
Similar numbers are obtained for
migrations due to detector effects.
In general, smearing effects become less severe for larger \Qsq.
Therefore the \as analysis apply lower \Qsq cuts, 120~\GeVsq for
ZEUS \cite{z:as} and 40 \GeVsq for H1 \cite{h1:as2}.

\begin{figure}[tbh]
   \centering
\begin{picture}(0,0) \put(0,0){{\bf a)}} \end{picture}
   \epsfig{file=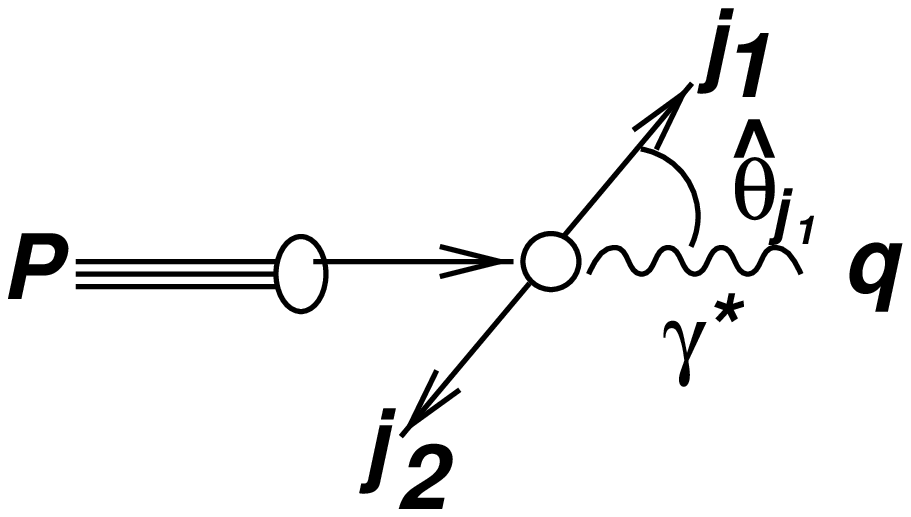,
   width=6cm,
   bbllx=160pt,bblly=308pt,bburx=475pt,bbury=489pt,clip=}
\begin{picture}(0,0) \put(0,0){{\bf b)}} \end{picture}
   \epsfig{file=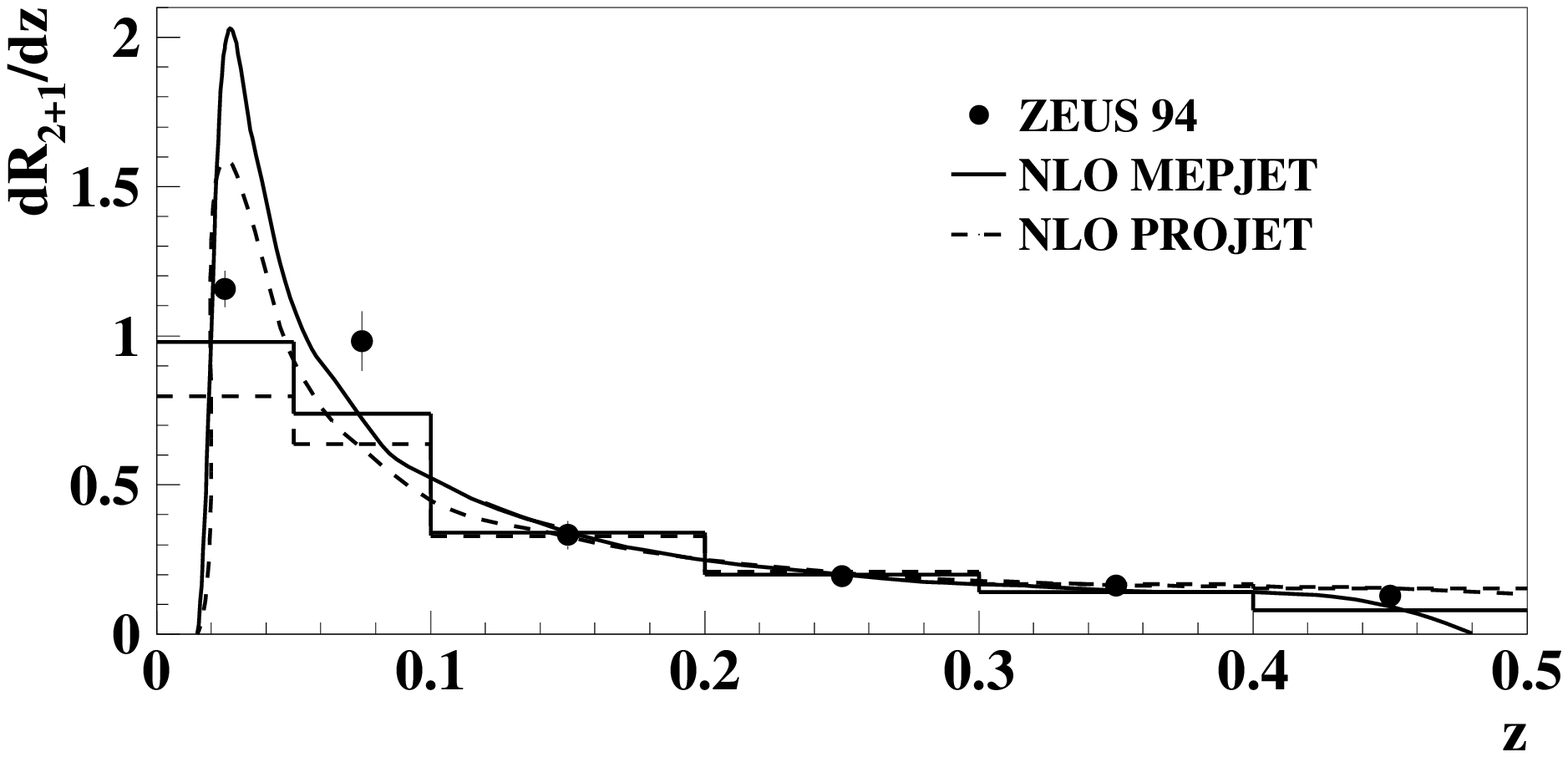,
   width=8.5cm,
   bbllx=1pt,bblly=72pt,bburx=550pt,bbury=347pt,clip=}
   \scaption{
    {\bf a)} Two jet production
    in the centre of mass system
    of the incoming parton and the virtual photon.
    The scattering angle $\hat{\theta}_j$ of jet $j$ is defined
    with respect to the incoming parton direction.
    {\bf b)} The $z$ distribution of one of the two
    non-remnant jets from the ZEUS analysis \cite{z:trefzger}.
    The ZEUS data \cite{z:as} are compared to the NLO calculations
    from PROJET \cite{mc:projet} (dashed line) and from MEPJET
    \cite{mc:mepjet} (full line).
    The histograms show the NLO results
    with the same binning as the data.}
   \label{zp}
\end{figure}

Two Lorentz invariants can be introduced to describe the jet kinematics,
for example $\shat=(j_1+j_2)^2$, the invariant mass squared of the two jets,
and
\begin{equation}
   z_j :=
          \frac{P\cdot j}{P\cdot q} \approx \frac{1}{2}
          (1-\cos \hat{\theta}_j).
  \label{eq:zj}
\end{equation}
The index $j$ denotes the jet from a particular parton.
$z_j \in [0,1]$ and
$z_{j_1} = 1 - z_{j_2}$.
The approximate relation with the scattering angle $\hat{\theta}_j$
in the parton-photon scattering frame
(the rest frame of the two jet system, fig.~\ref{zp}a)
holds for small jet masses and small \Qsq.
In the experiment one defines
$z:=\min (z_{j_1},z_{j_2})$ with $z\in[0,0.5]$.
$z$ can be reconstructed from the measured jet angles $\theta_j$ and
energies $E_j$ in the laboratory frame,
\begin{equation}
   z \approx \min_{j=j_1,j_2}
   \frac{E_j(1-\cos\theta_j)}{\sum_{i=j_1,j_2}E_j(1-\cos\theta_i)}
\end{equation}

The LO matrix element for the QCDC process diverges
$\propto \frac{1}{(1-x/\xi)(1-z_j)}$,
and for BGF
$\propto \frac{1}{z_j(1-z_j)}$
with the familiar soft ($\xi\rightarrow x$) and
collinear ($z_j \rightarrow 0,1$) divergencies.
It turns out that close to the collinear divergency ($z<0.1$),
more 2+1 jets are observed
than expected from either the NLO calculations (see fig. \ref{zp})
or the LEPTO simulation
\cite{z:jetprod,z:as}.
In this part of the phase space the NLO corrections are largest,
and according to the Monte Carlo, parton shower effects are most
severe (apart from the LO BGF and QCDC events, QPM events with
initial state parton showers can contribute up to 50\% to the
total 2+1 jet rate).
The region $z<0.1$, corresponding to small angles
$\thjet$ in the lab system
with respect to the proton remnant direction, is therefore cut out for the
determination of \as \cite{z:as,h1:as2}.
H1 applies an additional
cut $\thjet>10\degr$, and requires
$\thjet<145\degr$ in order to exclude
the backward detector region which was not instrumented with hadronic
calorimetry.

\subsubsection{Results}

From a comparison of the jet rate $R_{2+1}$, corrected to the parton level,
with the NLO expression of \ref{eq:r2p1}, \as is determined as
a function of \Qsq.
ZEUS has measured \as in three \Qsq bins from 120 to 3600 \GeVsqx,
and H1 (preliminary)
in four bins from 40 to 4000 \GeVsqx, see fig.\ref{alphas_hera}.
The data are consistent with the
decrease of \as with increasing
scale according to the renormalization group equation.
A fit to the data with the NLO expression for the running $\alpha_s(Q^2)$,
eq. \ref{eq:asrun},
with $\alpha_s(m_Z^2)$ as free parameter yields
the results in table \ref{tab:as}.
These \as measurement from jets at HERA agree
with the world average \cite{rev:pdg}.

\begin{table}[h]
\begin{center}
\begin{tabular}{|l|llll|}
\hline
ZEUS                 &    $\alpha_s(m_Z^2) =$                  &
                          $0.117 \pm 0.005 ~(\stat)$             &
                          $^{+0.004}_{-0.005} ~(\sys_{\rm exp})$    &
                          $\pm 0.007 ~(\sys_{\rm th}) $            \\
H1 prel.              & $\alpha_s(m_Z^2) =$                  &
                          $0.115 \pm 0.003 ~(\stat)$             &
                          $^{+0.008}_{-0.011} ~(\sys)$          &
                          $ + 0.006 ~(\rec) $           \\
world avg.            &     $\alpha_s(m_Z^2) =$        &
                          $0.118 \pm 0.003$                    &
                                                               &  \\
\epem           &     $\alpha_s(m_Z^2) =$        &
                          $0.122 \pm 0.006$                    &
                                                               &  \\
\hline
\end{tabular}
\end{center}
\scaption{\as measurements by H1 (prel.) \cite{h1:as2}
and ZEUS \cite{z:as} from 2+1 jet events. ZEUS quotes
experimental and theoretical systematic errors separately.
H1 finds that the use of alternative recombination schemes
would increase the measured \as value by the amount given (rec).
For comparison, the
world average from the PDG \cite{rev:pdg} is quoted.
The combined result from LEP and SLC hadronic final state analyses at the
$Z$ mass is also given \cite{rev:bethke}.
See \cite{rev:bethke}
for a recent compilation of other \as measurements.}
\label{tab:as}
\end{table}

For the ZEUS result, experimental and theoretical systematic errors are
given separately. H1 found that alternative recombination schemes than
the ones used would increase \as by the amount quoted.
The dominating systematic errors are the corrections for hadronization
effects and the dependence on the choice of renormalization and
factorization scales. The largest experimental systematic error is due
to the uncertainty of the absolute calorimeter calibration, which is
known to about $\pm 5\%$.

It is estimated \cite{h1:hadig} that with a large amount of
luminosity to be collected at HERA in the future, 250 \pbinv,
the errors on $\alpha_s(m_Z^2)$ can be reduced to
$\pm 0.0013$ (stat.) and $\pm 0.005$ (syst.).
Other algorithms than JADE may turn out to be advantageous in
the future \cite{h1:as2}. In any case,
a good understanding of perturbative QCD beyond LO and
hadronization will be mandatory to achieve a good precision on \asx.
For example, the puzzle why two otherwise respectable QCD models
fail to describe the jet rate at
the hadron level (fig.~\ref{alphas_hera}a) needs to be resolved.

%

  \section{The Gluon Density \label{sn:jgluon}}     
The gluon density at small fractional momentum \xgp~ ($\xgp< 0.04$)
had been extracted previously
only indirectly from inclusive structure function measurements.
A direct measurement of the gluon density is in principle possible
via the measurement of 2+1 jet events from the boson-gluon fusion (BGF)
process $\gamma^\ast g \rightarrow q\ol{q}$.
The 2+1 jet cross section can in leading order QCD be described
as the sum of quark initiated and gluon initiated processes,
QCDC and BGF. Schematically,
\begin{equation}
  \sigma_{2+1} = \alpha_s(Q^2)\left[ A_{\rm QCDC} \cdot q(\xqp,Q^2) +
                  A_{\rm BGF} \cdot g(\xgp,Q^2) \right].
\end{equation}
Here \xqp~ and \xgp~ are the fractional proton momenta carried
by the incoming quark and gluon.
They are given in LO by
$\xi=x(1+\frac{\shat}{\Qsq}) \approx \frac{\shat}{W^2}$,
see fig. \ref{feynjets}.
$A_{\rm BGF}$ and $A_{\rm QCDC}$ are coefficients that
can be calculated perturbatively.
The experimental challenge is to reconstruct $\xi$ by measuring \shat.
That requires to
separate in the hadronic final state the hard subprocess from
initial state parton showers and the proton remnant.

For the extraction of the gluon density $g(\xgp,Q^2)$,
$\alpha_s(Q^2)$
is assumed to be known and taken from the PDG world average.
The contributions from the competing QCD Compton process (QCDC)
$\gamma^\ast q \rightarrow qg$, which is calculable using known
quark density functions $q(\xqp,Q^2)$, and from wronlgy classified
QPM events
have to be subtracted statistically from the
observed 2+1 jet rate.

Of particular interest is the gluon density at small $\xgp$, where
no direct measurements exist.
In order to access small values of $\xi$ ($=\xgp$ in LO),
small invariant jet-jet mass $\sqrt{\shat}$ need to
be resolved. The JADE jet algorithm resolves 2-jet events
with $m_{ij}^2/W^2 > \ycut$. That defines the
accessible phase space, and implies $\xi\approx \shat/W^2>\ycut$.

To reach lower values of $\xi$ than given by the canonical
$\ycut=0.02$, the H1 analysis \cite{h1:jetgluon}
uses the cone jet algorithm ($R = 1$)
applied in the hadronic CMS.
For the jets $\pt>3.5\GeV$ is required. Since for massless
objects
\begin{equation}
       p_T^2 = \shat z (1-z),
\end{equation}
values of
$\sqrt{\shat}>2 \pt$ can be obtained.
With the large $W^2>4400~\GeVsqx$,
much lower $\xi$ values can be probed than with the JADE algorithm.

From DIS events at small \Qsq ($12.5 < Q^2 < 80$ \GeVsq),
2+1 jet events are selected with $10\degr < \thjet < 150\degr$ in the
laboratory frame. Their distance in (lab) pseudorapidity had
to be smaller than $|\Delta \eta| = 2$. In the hadronic CMS, that
cut translates roughly to $z>0.1$. These angular cuts remove events
affected by parton showers.
$\sqrt{\shat}>10\GeV$ is required.
\shat~ is reconstructed 1) as the invariant mass squared of all
particles belonging to the two jets, and 2) from the jet directions
(pseudorapidities $\eta_j$) measured in the hadronic CMS,
\begin{equation}
    \shat = W^2 \exp ( -\eta_1-\eta_2).
\end{equation}
It is required that the two
methods agree to $|\Delta \sqrt{\shat}| <  10$ GeV.

\begin{figure}[tbh]
   \centering
   \epsfig{file=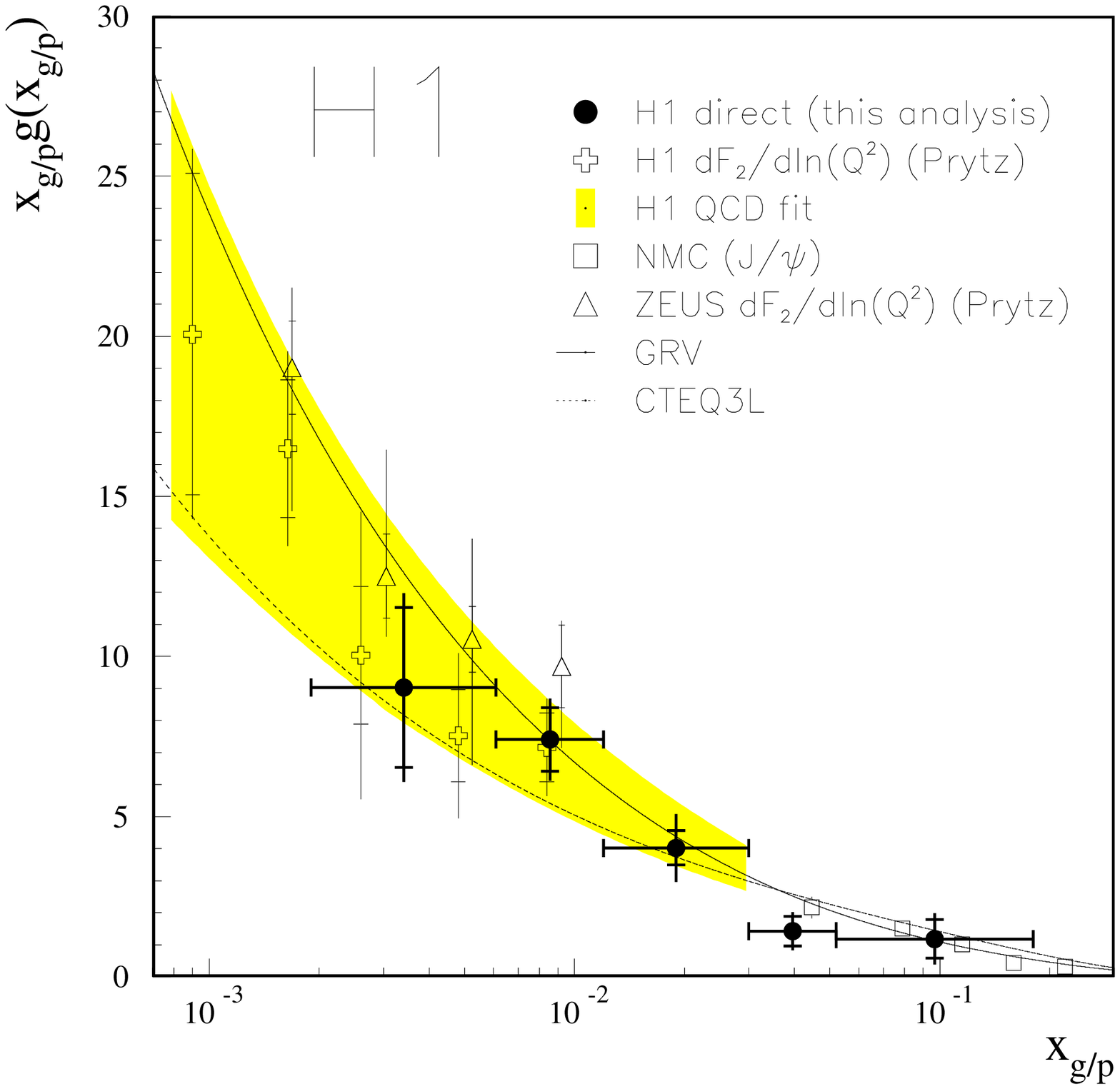,%
          width=10cm}
   \scaption{The unfolded LO gluon distribution as a function of
             fractional gluon momentum \xgluon~ at
             \av{\Qsq}=30\GeVsq \cite{h1:jetgluon}.
             The direct measurement from 2+1 jet events is compared
             to the indirect LO determinations from structure function
             measurements at \av{\Qsq}=20\GeVsq
             by H1 \cite{h1:f2gluon} and ZEUS \cite{z:f2gluon},
             and to data at large $x$ from NMC \cite{o:nmcgluon}
             obtained from $J/\psi$ production. The other data are
             evolved to the \av{\Qsq}=30\GeVsq of the jet analysis.
             Also shown are the LO gluon density parametrizations
             from GRV \cite{th:grv} and CTEQ \cite{th:cteq3}. }
   \label{logluon}
\end{figure}

The selected 2+1 jet event sample covers $0.002<\xi<0.2$ and
$0.0003<x<0.0015$. According to the Monte Carlo (LEPTO),
BGF and QCDC events are selected in a ratio 3:1.
By unfolding the observed $\xi$ distribution,
the (LO) gluon distribution
$\xgp \cdot g(\xgp)$ at $\av{Q^2}=30\GeVsq$ is obtained (fig. \ref{logluon}).
The data extend the region previously covered by NMC \cite{o:nmcgluon},
$\xgluon>0.04$ down to $\xgluon = 0.0019$.
This direct measurement of the gluon density confirms the
steep rise of the gluon density at small $x$ which had been
deduced from the structure function data (see fig.~\ref{logluon}).
A preliminary ZEUS analysis using a similar method gives consistent results
\cite{h1:grindhammer}, within large errors.
We note that
sizeable NLO corrections for $2+1$ jet production have been
calculated \cite{th:mirkes}.

Recent analyses focus on the direct determination of the gluon
density in NLO \cite{lowx:ringcarli}.
Using the measured partonic jet rate $R_{2+1}$ (JADE algorithm)
for $\Qsq>40\GeVsq$
from the H1
\as analysis \cite{h1:as2} as input (see section \ref{sn:jas}),
the extracted gluon density for $\xgp>0.01$ is consistent within errors
with results from structure function analyses \cite{h1:mellin}.
Due to the phase space restriction implied
by the JADE algorithm, not very small \xgp~ are probed, $\xgp>0.01$.
Smaller \xgp values can be accessed by tagging gluon initiated
BGF events with identified charm decays \cite{h1:charm95},
see section \ref{sn:charm}.

Recent attempts to extract the NLO gluon density at small \xgp~ from
dijet production at low \xb and \Qsq are facing a problem
\cite{z:mikunas,h1:lobo}.
There are about 20-30\% more
dijets in the data than expected from NLO calculations or from e.g. LEPTO
with
gluon densities as input that are compatible with the \ftwo data.
At face value, that would indicate a directly measured gluon density
that is incompatible with the indirect extraction from the \ftwo measurements.
It could also be that there are effects contributing to jet production
at small \xb and \Qsq that had not been taken properly into account.
To clarify the situation, data on jet production are essential that
do not contain any theoretical bias from the correction to the parton level.
Such measurements are presented in the next section.

  \section{Jet Rates at Low $Q^2$ \label{sn:jrates}} 

In a preliminary ZEUS analysis \cite{z:mikunas} that addresses
the gluon density,
the dijet cross
section
(cone algorithm in the laboratory frame with $R=1$)
has been determined and corrected for detector and
hadronization effects to the parton level using LEPTO 6.3.
In the laboratory
frame, $\eta_{\rm jet}<2$ was required to exclude the forward
region. A cut
$\ptjet>4\GeV$ was applied in the laboratory frame and in the CMS.

\begin{figure}[tbh]
   \centering
   \epsfig{file=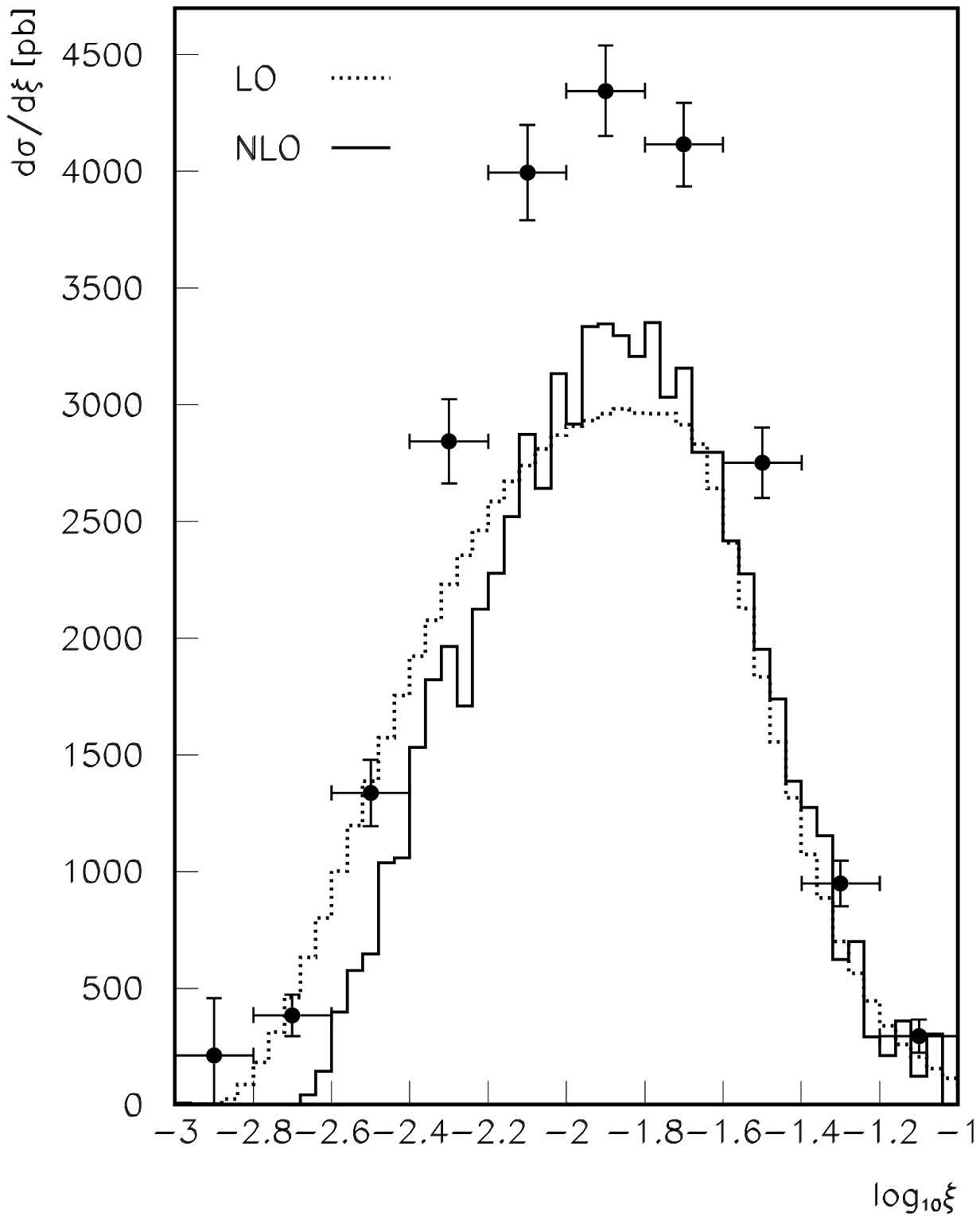,width=10cm,
        bbllx=66,bblly=71,bburx=421,bbury=520,clip=}
   \scaption{Differential dijet cross section (ZEUS prel. \cite{z:mikunas})
             as a function of
             $\xi$, corrected to the parton level (LEPTO 6.3).
             Shown are statistical errors only. The data are compared
             to QCD calculations in LO and NLO.}
   \label{zeusxi}
\end{figure}

The shapes of the measured cross sections as a function of \xb, \Qsq,
\ptjet~ and $\eta$ in the CMS,
and $\xi=x(1+\shat/\Qsq)$ are well described
by LEPTO and by NLO calculations \cite{mc:mepjet}.
However, the absolute dijet cross section is $\approx 34\%$ larger
than predicted by NLO QCD (see fig.~\ref{zeusxi}).
Systematic effects cannot account for such a
large discrepancy \cite{z:mikunas} .

The previous jet studies were aimed at the measurement of
observables defined in QCD - the strong coupling constant and
the gluon density. To that end
regions of phase space where
the data deviate from the theory - cast either into NLO calculations
or Monte Carlo generators - had been carefully avoided. These
phase space regions were found at low $x$ and low $Q^2$, and in
the forward region close to the remnant \cite{mk:paris,z:mikunas,h1:lobo}.
The forward region
at low $x$ is discussed in a special chapter on low $x$ physics,
section \ref{sn:fjets}.
Here we shall discuss jet production at low $Q^2$, with
\Qsq ranging from the DIS regime down to photoproduction,
$0<\Qsq<100\GeVsq$.
The object of these studies is a test of QCD and
to explore the validity of QCD models for jet production.
The data extend into
a new phase space region in DIS, where
the transverse momenta squared of the jets may
exceed $Q^2$, $p_T^2>Q^2$.
In such a situation it may be possible
to probe constituents of the virtual photon at a scale given
by the \pt of the jet \cite{th:klasen}.

\begin{figure}[p]
   \centering
   \epsfig{file=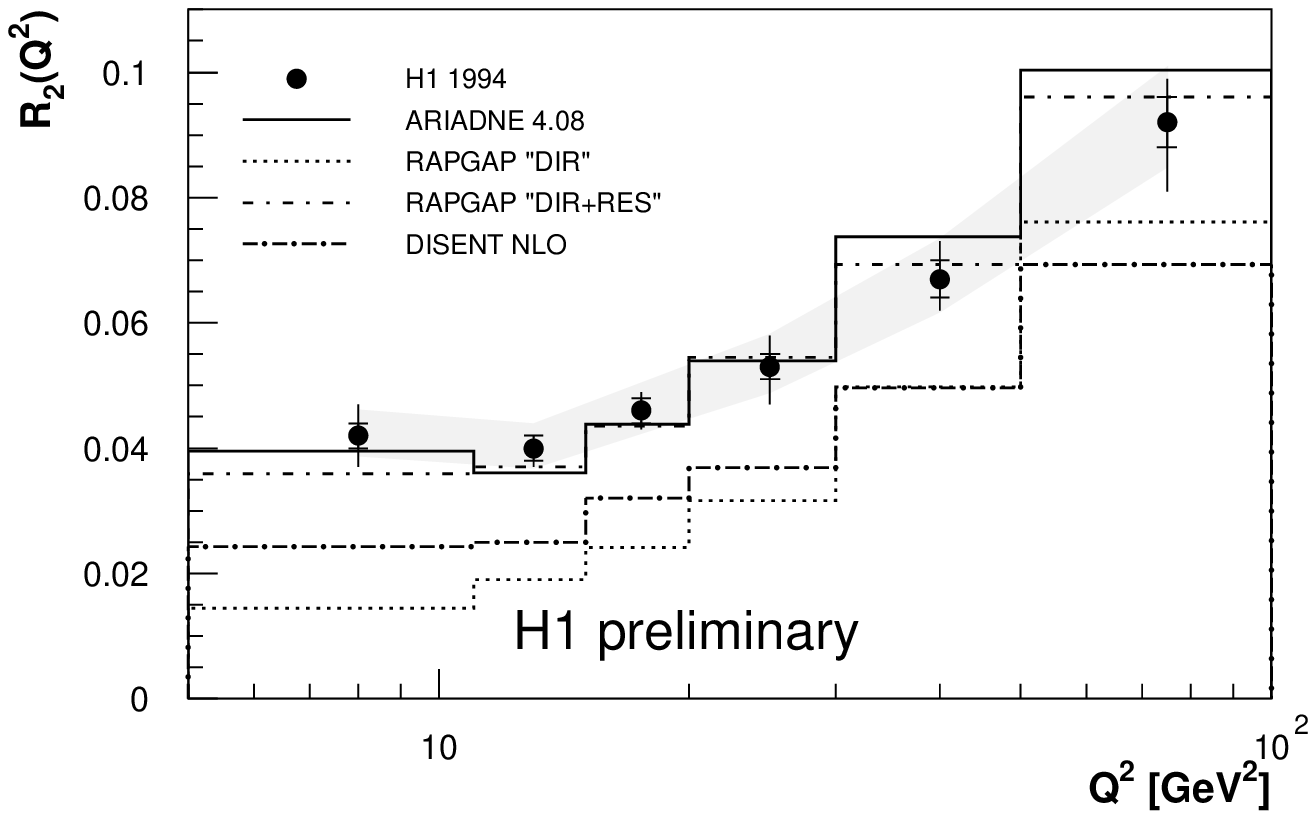,
   width=9cm}
   \epsfig{file=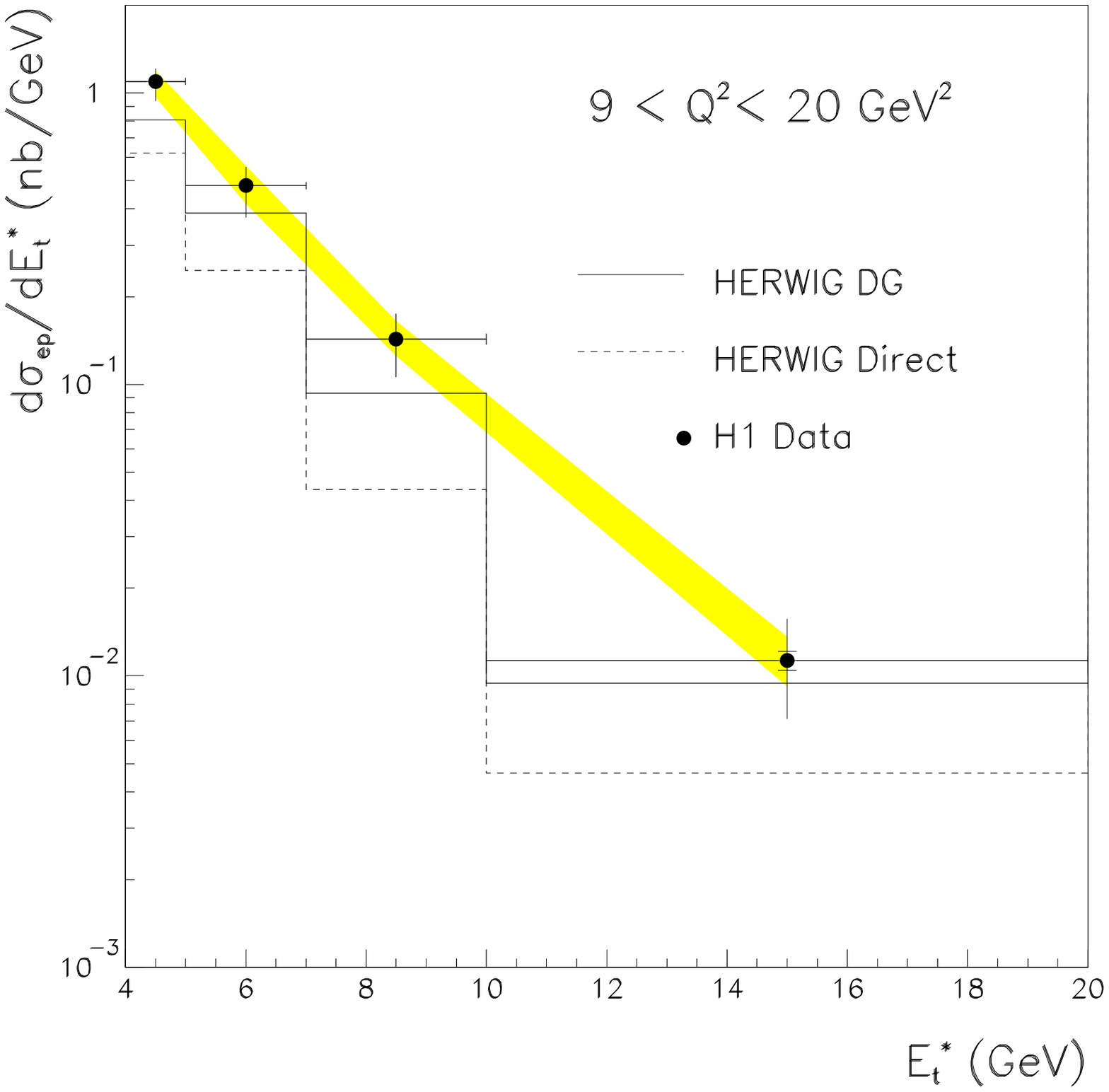,
   width=9cm,
   bbllx=9pt,bblly=142pt,bburx=532pt,bbury=659pt,clip=}
   \scaption{
             Top:
             the dijet rate $R_2$ as a function of \Qsq \cite{h1:dijet}.
             The data are corrected to the hadron level and
             compared to the models ARIADNE 4.08 \cite{mc:ariadne} and
             RAPGAP 2.06 \cite{mc:rapgap}
             with (DIR+RES) and without (DIR) a resolved
             component of the virtual photon.
             Also shown is the partonic prediction from the NLO
             program DISENT \cite{mc:disent}.
             Bottom:
             the single inclusive jet \et cross section, corrected to
             the hadron level,
             for jets
             with CMS $2.5<\eta^\ast<0.5$ (photon direction at $+z$)
             \cite{h1:jetlowq2}.
             The events are selected with
             $9<\Qsq<20\GeVsq$ and $0.3<y<0.6$.
             The inclusive DIS cross section in this kinematic range
             is $\approx 9.8\nb$.
             The data are compared to the HERWIG model with only
             direct interactions (Direct), and with resolved
             contributions added (DG, where the resolved photon structure
             is given by \cite{th:drees}).
             The shaded bands show an additional systematic error, mainly
             due to the uncertainty of the calorimeter calibration.
             }
   \label{dijets}
\end{figure}

In a preliminary H1 analysis \cite{h1:dijet}, low \Qsq DIS events
are selected with $y>0.05$, where the scattered electron angle
and energy are $156\degr<\theta_e<173\degr$ and
$E_e^\prime>11\GeV$. The data sample covers $5<\Qsq<100\GeVsq$ and
$10^{-4}<x<10^{-2}$.
Exactly two jets are required, reconstructed with the cone
algorithm ($\Delta R=1$) in the hadronic CMS
and with transverse momenta $\pt>5\GeV$.
The remnant is not counted.
The pseudorapidity difference $|\Delta \eta|$ between the two
jets has to be less than 2. In the CMS of the
scattering parton and the virtual photon, the jet angles
with respect to the photon axis
are given by
\begin{equation}
   \hat{\theta}_{1,2}=2 \arctan \exp(\pm\Delta\eta/2).
\end{equation}
The requirement $|\Delta \eta|<2$ thus excludes very forward jets,
because
$|\cos\hat{\theta}|<0.76$, or $z>0.12$.
Measured is the dijet rate $R_2$,
the fraction of DIS events that fulfil the dijet requirements.
The rate is corrected for
detector effects to the hadron level (fig.~\ref{dijets} top).
The dijet rate increases from 4 to 10\% for \Qsq between 10 and
100 \GeVsqx.

\begin{figure}[tbh]
   \centering
   \epsfig{file=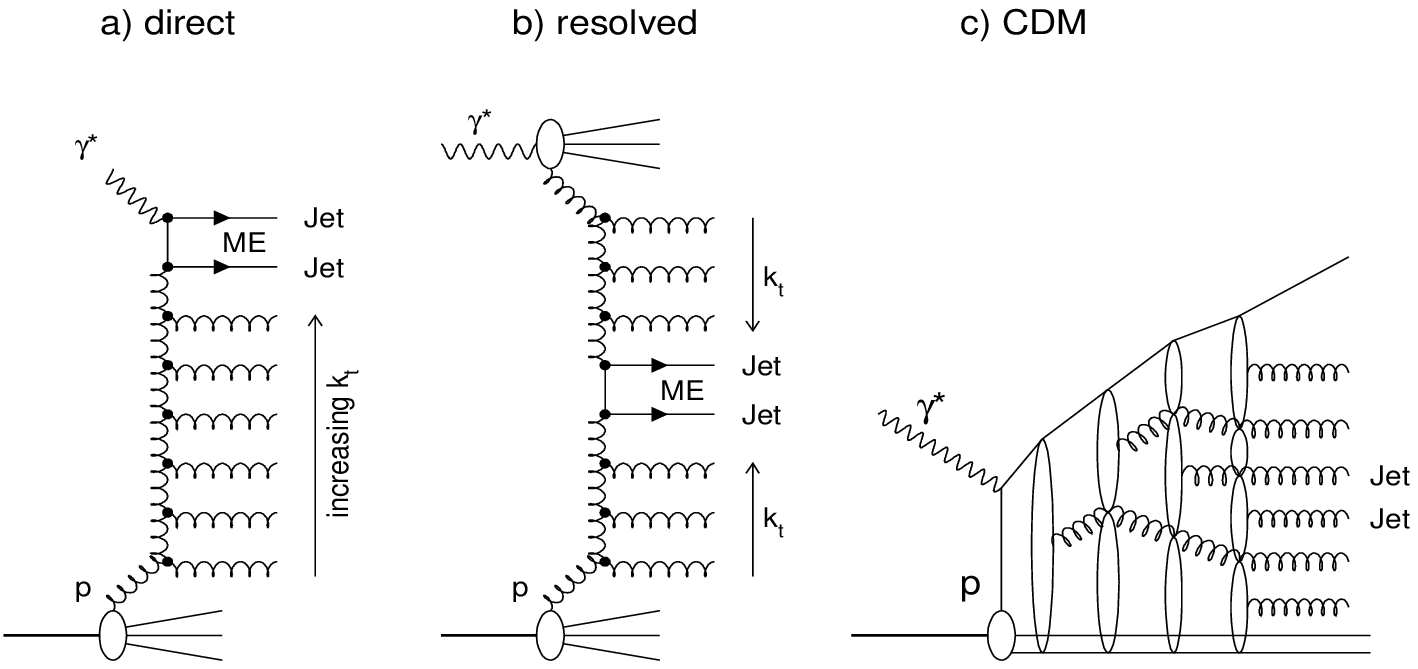,
           width=12cm}
   \scaption{Parton evolution scenarios.
             {\bf a)} ``Direct'': the hardest emission (with the largest \pt),
                given by the QCD matrix element, occurs at the top
                of the ladder. Further emissions down the ladder are
                ordered with decreasing \kt.
             {\bf b)} ``Resolved'': the hardest emission may occur anywhere
                in the ladder. Softer emissions decrease in \kt towards
                both ends.
             {\bf c)} CDM: gluons are emitted
                from succesively built colour dipoles. There is no
                restriction on \kt ordering.
                }
   \label{resdir}
\end{figure}

The data are compared to different QCD models, see fig.~\ref{resdir}.
In RAPGAP DIR (direct) \cite{mc:rapgap},
the virtual photon interacts directly with a proton constituent.
The hardest
interaction (with the highest $p_T$) takes place at the photon
vertex and is modelled with the LO QCD matrix element.
Softer emissions are generated with DGLAP parton showers.
In RAPGAP DIR+RES it is assumed that there exists an additional
contribution from so-called resolved processes, where the
structure of the virtual photon is resolved by a hard scale,
in this case the jet \pt with $p_T^2>Q^2$ \cite{lowx:jung}.
The hard interaction
takes place further down the ladder between (evolved)
partons from the photon and from the proton.
Virtual photon parton densities were taken from SaS-2D \cite{th:sas}.
The resolved contribution breaks the \kt ordering in the
parton evolution.

In the colour dipole model (CDM, program ARIADNE \cite{mc:ariadne}),
gluon emission stems from a chain of independenly radiating colour
dipoles.
The amount of radiation depends on the parameters that describe
the extendedness of the proton remnant colour charge, and on
the extendedness of the colour charge of the scattered quark,
which is determined by \Qsqx.
Emitted partons are not restricted to obey \kt ordering.
The radiation pattern thus shows similarity with BFKL evolution
\cite{mc:bfklcdm,lowx:rathsman}.

The data cannot be described by direct interactions only
(fig.~\ref{dijets} top), when conventional DGLAP models are applied.
At low \Qsq (and also small $x$),
the measured dijet rate is more than
a factor two above a model which contains direct interactions only.
At low \Qsq around 50\% of the dijet rate is ``missing'', that amounts
to $\approx 2\%$ of the total event rate.
Also a NLO QCD calculation for parton jets
with DISENT \cite{mc:disent} falls short
of the data. Hadronization is unlikely to account for the difference,
because
the difference between hadronic and partonic dijet rates
in the Monte Carlo was
found to be $<20\%$ for $\pt>5$ GeV.
The discrepancy diminishes with
increasing \Qsq (and also inreasing $x$).
The models where the hard interaction is
not tied to the photon vertex, implementing either a resolved
component of the virtual photon, or colour dipole radiation, provide
a good description of the data.

In a similar analysis the single inclusive jet cross section
(\kt algorithm with $\kcut=3\GeV$) has been measured
as a function of jet \et and pseudorapidity for events with
$0<\Qsq<49\GeVsq$ and $0.3<y<0.6$
\cite{h1:jetlowq2}.
The data are corrected to the hadron level.
It was found that when the transverse energy squared of the jet
exceeds $Q^2$,
conventional models (LEPTO 6.5 \cite{mc:lepto},
HERWIG 5.9 Direct \cite{mc:herwig})
with direct photon-parton interactions alone
undershoot the data (fig.~\ref{dijets} bottom) by a large amount.
A better description of the data is obtained
when contributions are taken into account where the hard
scale provided by the jet \et serves to resolve partons inside
the virtual photon (HERWIG 5.9 DG, with resolved photon component),
or by the colour dipole model (ARIADNE 4.08 \cite{mc:ariadne}).

In summary, the mechanism for jet production at small $x$ and \Qsq
is not yet very well understood. It appears that for regions of
phase space where the jet $p_T^2$ is larger than \Qsq the conventional
DIS framework, where the virtual photon resolves partons inside
the proton, ceases to be valid. Instead, the high \pt of the jet
could resolve both structures of the proton and of the virtual
photon.
We note that the linked dipole chain model \cite{th:ldc,mc:ldc}
should be able to treat both situations consistently where
the jet $p_T^2$ is smaller or larger than \Qsqx.
This field of research will continue with more detailed measurements.
It has to be seen also in connection with ``small-$x$'' physics, to
be discussed in the next chapter \ref{ch:lowx}.
Are these models alternatives to genuine ``small-$x$'' effects,
as expected from BFKL evolution, or are they just a different
language for the same effect?
It remains to be seen whether the competing models are actually
consistent with the large body of other hadronic final state data,
or whether they can describe just one facette of the hadronic final
state.

\chapter{Low-$x$ Physics \label{ch:lowx}}

  \section{Introduction \label{sn:lowxi}}          
Amongst the most interesting issues of HERA physics is QCD in the
newly accessible regime of small $x$, see section \ref{sn:interest}.
The observed rise of the
structure function \ftwo towards small $x$ suggests a strong
increase of the parton density in the proton,
but what is its dynamical origin?
Is DGLAP evolution sufficient to account for the data, or
is BFKL dynamics at work, at least to some extent?
The structure function measurements are probably too inclusive to
answer such questions unambiguously.
Complementary measurements of the hadronic final state provide more
detailed information on the reaction, which may help to uncover
the underlying dynamics.

In the simple quark parton model, a quark is scattered
out of the proton by the virtual boson
emitted from the scattering
lepton. QCD modifies this picture. Partons may be radiated
before and after the boson-quark vertex, and the boson may
also fuse with a gluon inside the proton by producing a
quark-antiquark pair.
In fact, the parton which is probed
by the boson may be the end point in
a whole cascade of parton branchings.
This parton shower materializes in the hadronic final state,
allowing experimental access to the dynamics governing the
cascade.

\begin{figure}[htb]
   \centering
   \vspace{-1cm}
   \epsfig{file=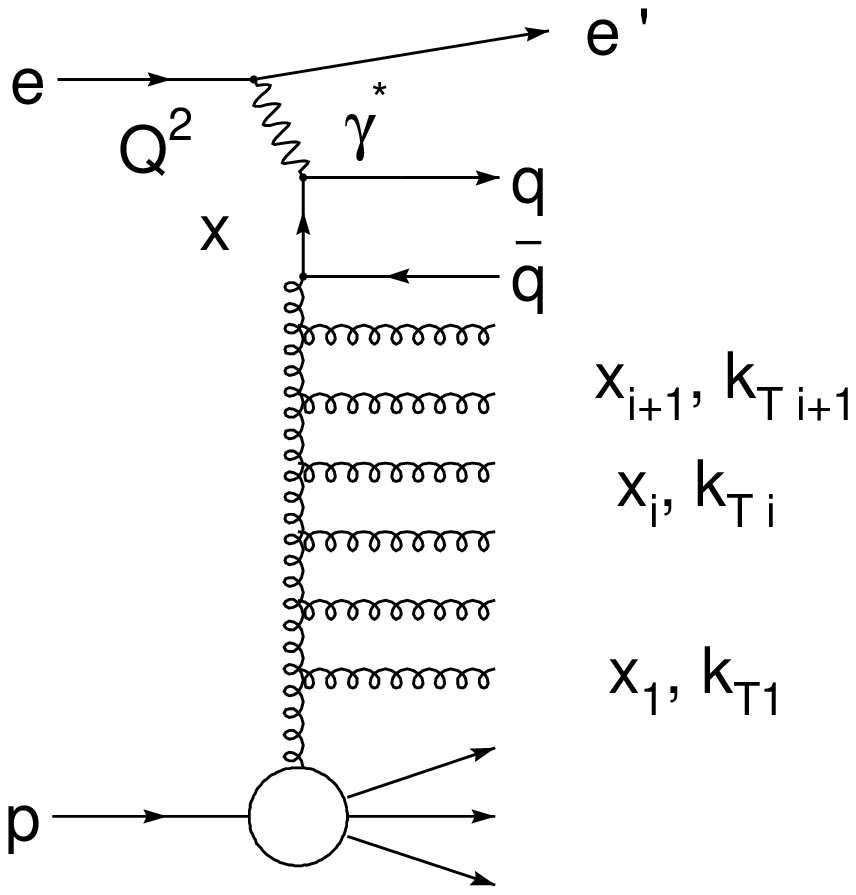,
    width=8cm}
   \scaption{Generic diagram for parton evolution at small $x$.
             A gluon ladder
             evolves between the quark box attached to the virtual photon
             and the proton. The gluon longitudinal momentum fractions
             and transverse momenta are labelled $x_i$ and $k_{Ti}$.
            }
   \label{ladder}
\end{figure}

Though there may be other
dynamical features to be discovered
that leave their footprint in the hadronic final state
(see for example section \ref{sn:jrates} on the virtual photon structure
or chapter \ref{ch:inst} on instantons),
most dedicated small $x$ measurements have concentrated on the predicted
signals for BFKL evolution.
The leading log
DGLAP resummation corresponds to a strong
ordering of the transverse momenta \kt (w.r.t. the proton beam)
in the parton cascade
($Q_0^2 \ll k_{T1}^2 \ll ... k_{Ti}^2 \ll ... Q^2$).
In the BFKL regime, the transverse momenta follow
a kind of random walk
($k_{Ti}^2 \approx k_{Ti+1}^2$)~\cite{lowx:ordering}.
The longitudinal momentum fractions are ordered according to
$x_i \gg x_{i+1}$ for BFKL and $x_i > x_{i+1}$ for DGLAP.

A radiated parton (or any other object) with CMS momentum fraction $x_i$
and longitudinal and transverse momentum components $p_{zi},p_{Ti}$
is
to be found at CMS rapidity
\begin{equation}
  y_i = \frac{1}{2} \ln \frac{E_i+p_{zi}}{E_i-p_{zi}}
      = \ln \sqrt{ \frac{E_i^2-p_{zi}^2}{(E_i-p_{zi})^2}}
      \approx \ln \frac{p_{Ti}}{x_i W}
      \approx \ln \frac{p_{Ti}}{x_i \sqrt{Q^2/x}}.
\end{equation}
For fixed $W$ and fixed $p_{Ti}$, the CMS rapidity $y_i$ is
determined by
$\ln x_i$.
Here the relations $E_i=x_iW/2$ and $p_{zi}\approx - x_i W/2$ in the CMS
have been used (with the proton direction along the $-z$ axis).
Finally,
\begin{equation}
   x_i  =  \sqrt{\frac{x}{Q^2}} \cdot p_{Ti} \exp (-y_i)  \hspace{2cm}
   y_i  =  - \frac{1}{2} \ln \frac{x_i^2 Q^2}{x p_{Ti}^2}.
\end{equation}

For DGLAP evolution the phase
space for parton radiation between the remnant
and the current is restricted by \kt ordering.
A generic signal for deviations from DGLAP evolution
is therefore enhanced activity in the central and forward
CMS rapidity region, between the current system and the proton remnant.
The observables which
have so far been exploited are the following:

\begin{itemize}
\item
{\bf Transverse energy flow:}
Increased parton activity should result in increased
transverse energy at central (CMS) rapidity \cite{lowx:et}.
\item
{\bf Forward jets:}
High energy jets with $\ptjet^2 \approx \Qsq$ (kinematically
bound to be measured in the forward calorimetric systems)
could tag events with BFKL evolution, because DGLAP
evolution is not allowed \cite{lowx:fwdjets,lowx:hotref}.
\item
{\bf Hadrons at large $p_T$:}
High \kt partons, disfavoured by the strong \kt ordering
in DGLAP, are signalled by measureable high \pt hadrons \cite{mk:method}.
\item
{\bf Other observables:}
Other observables in the DIS hadronic final state
like correlations and de-correlations \cite{lowx:askew,lowx:ejetcor},
multijet production \cite{lowx:jets,lowx:twojets} and
photon production \cite{lowx:pizerogam}
are being discussed.
\end{itemize}

Apart from a few theoretical calculations in the resummed
DGLAP and BFKL schemes, and calculations in fixed order perturbation
theory (mostly NLO), predictions for the final states observables
can be derived from Monte Carlo models only.
Theoretical calculations have the advantage that their theoretical
input is well defined.
However, most of them (an exception is
presented in section \ref{sn:ptspectra}) neglect hadronization,
making it difficult to compare them with data. On the other hand,
the draw back with Monte Carlo models is that often their complexity
and their flexibility to model hadronization makes it difficult to
pin down
which feature of their theoretical input is actually being tested
when comparing to data.

HERWIG \cite{mc:herwig} and LEPTO  \cite{mc:lepto}
which are based upon leading log DGLAP parton showers
are used as representatives for DGLAP evolution with strong \kt ordering.
Unfortunately a Monte Carlo program
based upon BFKL (or CCFM, which in the limits of  small and large $x$
approaches the BFKL and DGLAP evolutions)
dynamics is not yet available for
$ep$ reactions, though
there exist some promising developments in that direction
\cite{mc:smallx, mc:schmidt, lowx:ringsalam}.
Rather, the DGLAP model predictions are contrasted with the
colour dipole model ARIADNE \cite{mc:dipole,mc:ariadne}.
In certain aspects the CDM description
of gluon emission is similar to that of the BFKL evolution
\cite{mc:bfklcdm,lowx:rathsman}.
In particular
the gluons emitted by the dipoles
do not obey \kt ordering along rapidity (see section \ref{sn:meps}).

  \section{Energy Flows \label{sn:eflows}}      
\begin{figure}[tbh]
   \centering
\begin{picture}(0,0) \put(0,0){{\bf a)}} \end{picture}
   \epsfig{file=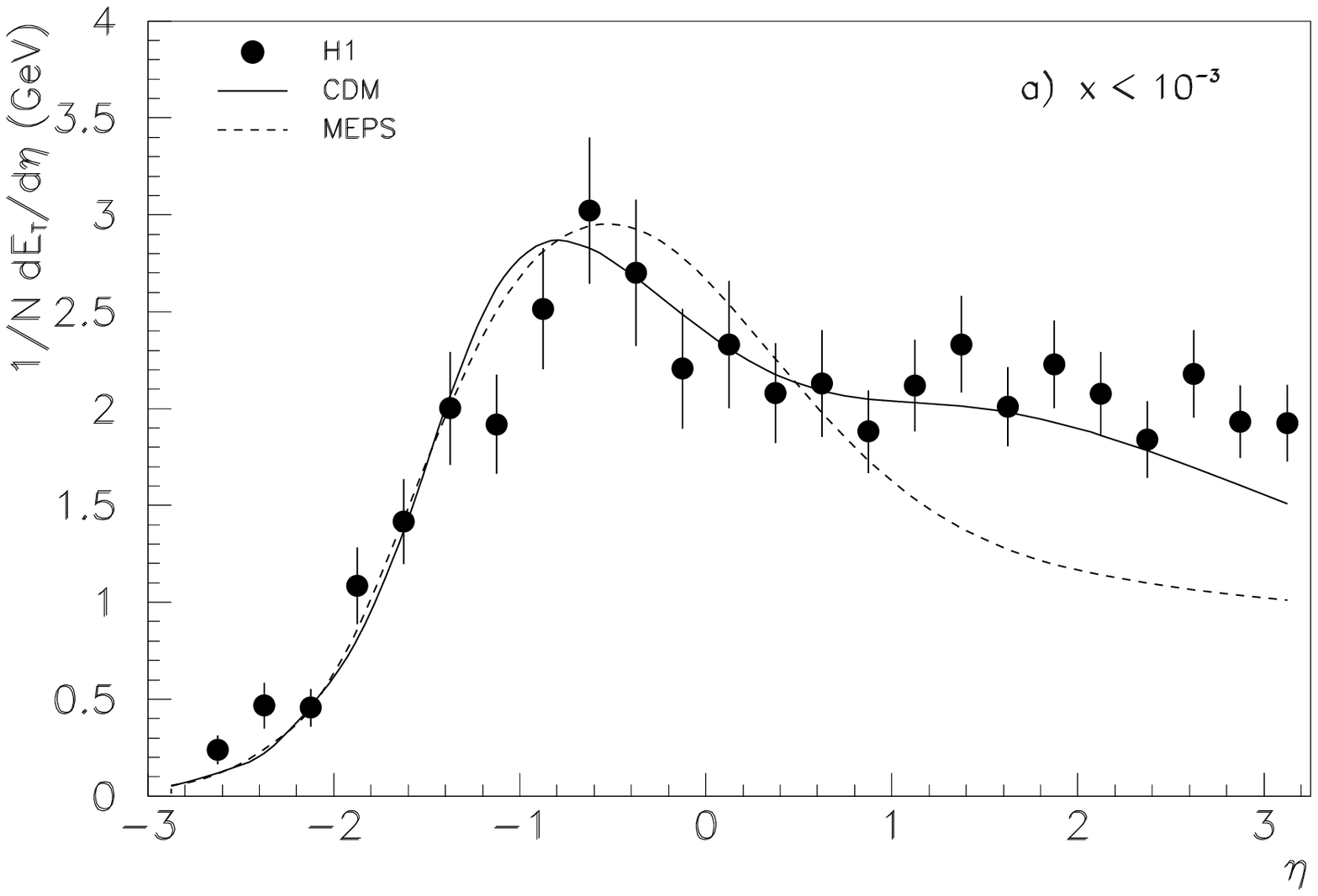,width=7.5cm,
           bbllx=1pt,bblly=357pt,bburx=491pt,bbury=691,clip=}
\begin{picture}(0,0) \put(0,0){{\bf b)}} \end{picture}
   \epsfig{file=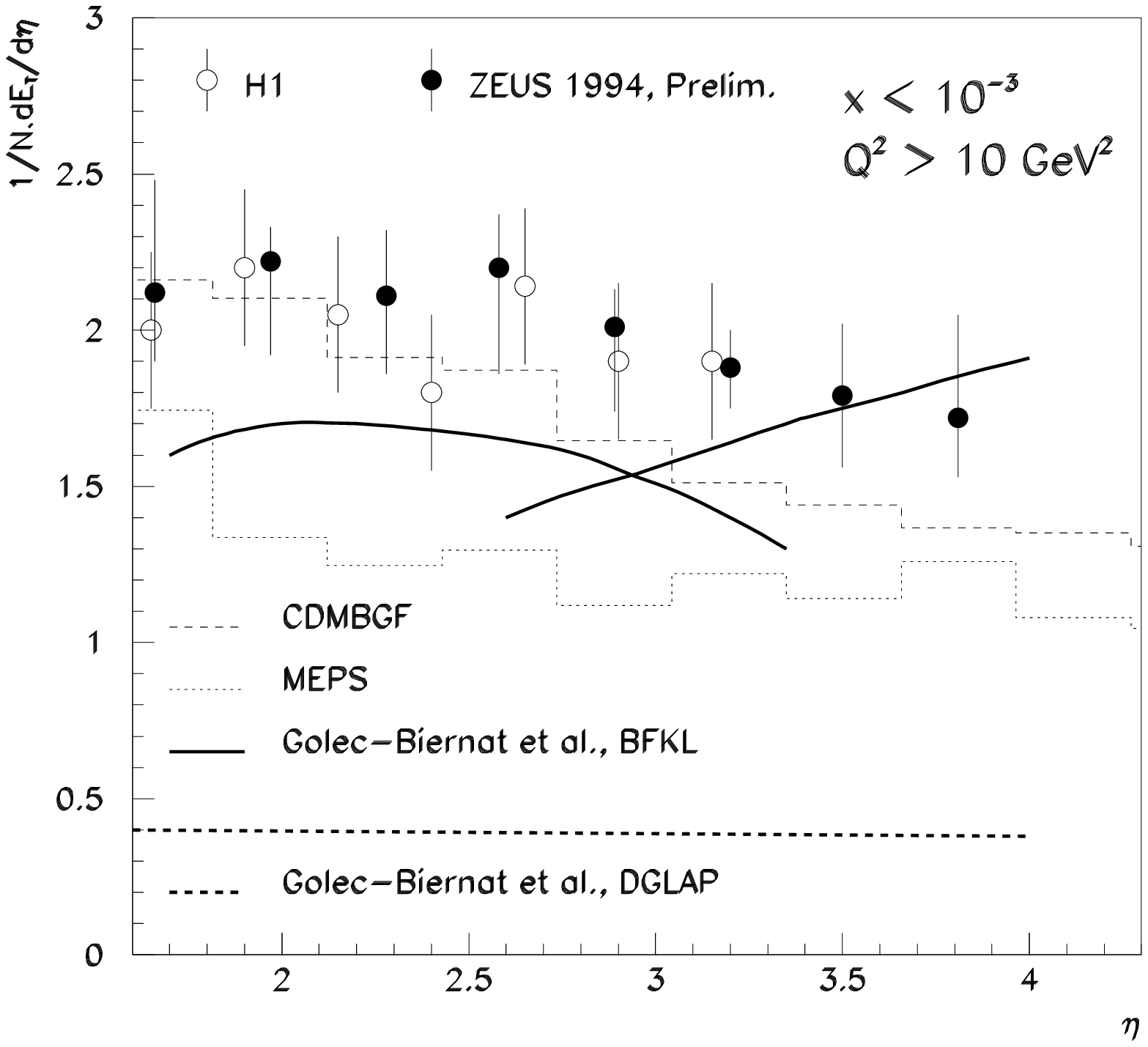,width=6.5cm}
   \scaption{The transverse energy flow \et as a function of
             pseudorapidity $\eta$ in the laboratory frame. The proton
             direction is to the right.
             {\bf a)} H1 data compared to the models MEPS (LEPTO 6.1)
             and CDM (ARIADNE 4.03) \cite{h1:flow2}.
             {\bf b)} The measured forward \et flow
             from H1 \cite{h1:flow2} and ZEUS \cite{z:ethep95}
             compared to the models MEPS (LEPTO 6.1) and
             CDM (ARIADNE 4.03, here labelled CDMBGF),
             and to BFKL and DGLAP based theoretical
             calculations (parton level) \cite{lowx:et}.
             The data are corrected for detector effects.}
   \label{etlab}
\end{figure}

The measurements of the transverse energy flow
\cite{h1:flow1,h1:flow2,h1:flow3,h1:flow4,h1:efldisgp,h1:eflcomp,
      z:flow1,z:ethep95,z:etdis96,z:eflcomp}
have had a great influence on our understanding and modelling of the
hadron production mechanism in DIS.
It had been noted already in
the early papers \cite{h1:flow2,z:flow1} that
some standard QCD models for the hadronic final state
failed to describe the measured energy flows, in particular at low $x$
(see also section \ref{sn:flow}).
The MEPS model (LEPTO 6.1), based upon the LO QCD matrix element with
DGLAP parton showers and string fragmentation,
undershoots the H1 data \cite{h1:flow2}
by a factor two in the ``forward region'', towards the proton remnant
(fig.~\ref{etlab} a).
In the following years, the energy flow and it's dependence on
kinematic conditions has been studied in great detail.

\begin{figure}[htb]
   \centering
   \epsfig{file=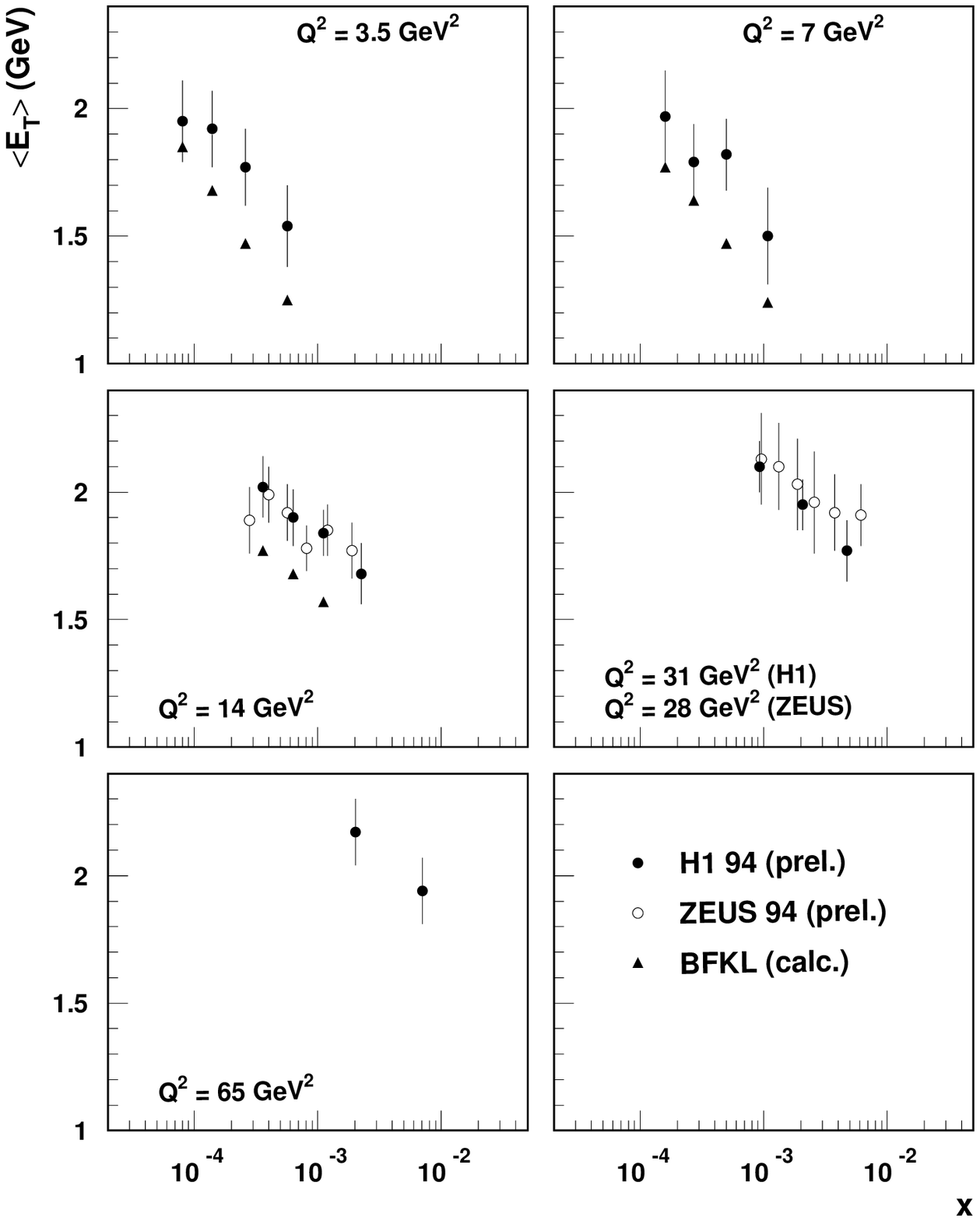,
           width=10cm}
   \scaption{The mean \et in the CMS pseudorapidity interval $-0.5<\eta<0.5$
             as a function of \xb for different \Qsq
             ranges \cite{mk:protvino}.
             Besides the preliminary data from H1 \cite{h1:flow4} and ZEUS
             \cite{z:etdis96} for hadrons,
             the BFKL expectation for partons is shown
             \cite{lowx:et}. Hadronization effects have to be taken
             into account when comparing this data with the partonic
             calculation.
             For the H1 data an additional 8\% overall normalization
             uncertainty is not shown.
             }
   \label{etx}
\end{figure}

The ``forward'' measurement close to
the beam hole at the edge of the calorimeter acceptance at
$\eta_{\rm lab}\approx 3.5$, corresponding to
$\theta_{\rm lab} \approx 4\degr$, is very difficult, as
particles are lost in the beam pipe, and backscatter from lower angle
particles hitting dead material
produce background in the forward
calorimeter region \cite{h1:lanius,h1:hess,z:et}.
An independent measurement is provided by ZEUS, where
the instrumentation and dead material distribution in the forward region
are quite different.
Fortunately, the data
from ZEUS \cite{z:ethep95} agree with the H1 data
(see figs.~\ref{etlab}b and \ref{etx}),
confirming the ``forward energy crisis'' \cite{mk:blois}.

The discrepancy between data and LEPTO,
considered as the reference model for hadron production,
triggered further investigations,
both theoretically
\cite{lowx:et,lowx:bartels}
and experimentally
\cite{h1:flow3,h1:flow4,z:ethep95,z:etdis96}
whether low $x$ effects,
for example BFKL dynamics, could be held responsible
for the excess.
Shown in fig.~\ref{etlab} b are also results of calculations \cite{lowx:et}
based upon either BFKL or DGLAP parton dynamics.
Qualitatively, much more \et is expected from BFKL than from
DGLAP evolution, roughly at the level of the data.
However,
these calculations do not include hadronization, and are therefore
not directly comparable to the data.

\begin{figure}[tbh]
   \centering
   \epsfig{file=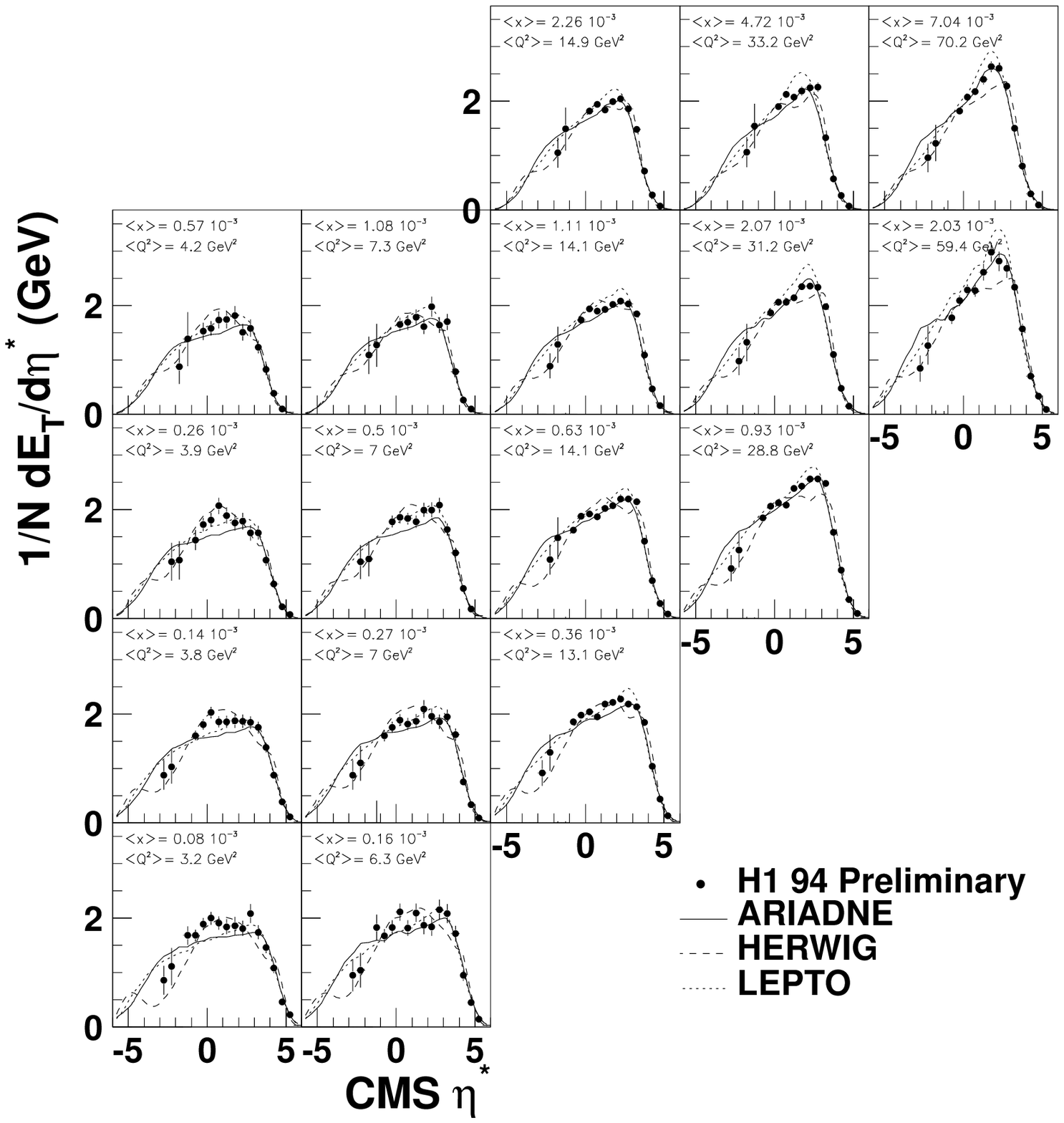,
    width=14.5cm,bbllx=39pt,bblly=159pt,bburx=513pt,bbury=665,clip=}
   \scaption{The transverse energy flow \et as a function of
             pseudorapidity $\eta$ in the hadronic CMS. The proton
             direction is to the left. The data \cite{h1:flow4}
             are compared to the models
             CDM (ARIADNE 4.08), MEPS (LEPTO 6.4), and HERWIG 5.8.
             Shown are statistical errors only, except for
             the two foremost data points from the plug.}
   \label{etplug}
\end{figure}

The \et flows have now been measured over a wide range of \xb and \Qsq
\cite{h1:flow3,h1:flow4,z:etdis96}, and are presented in the hadronic
CMS to eliminate the transverse boost introduced by the scattered electron.
The latest preliminary H1 data \cite{h1:flow4,mk:madrid}
(fig.~\ref{etplug})
also employ the plug calorimeter \cite{h1:panaro}
installed at small angle behind the liquid argon calorimeter,
extending the acceptance far into the target region.
The plug data have large errors due to the large amount
of dead material along the line of sight to the interaction vertex,
between 2 and 4 nuclear absorption lengths.

The \et flow is plateau-like ($\approx 2\GeV$ per unit $\eta$) at small
\xb and \Qsqx, and becomes more peaked in the current region at larger
\xb and \Qsqx. The average \et ($\av{E_T}$) measured centrally
($-0.5<\eta<0.5$) in the
CMS rises with
decreasing \xb (see fig.~\ref{etx})\footnote{The $x$ dependence for
fixed \Qsq implies also a $W$ dependence. In fig.~\ref{wdep} it is
shown that the average \et at central rapidity increases with
$W$ in DIS, and that this behaviour is in agreement with
data from hadron-hadron collisions.
A deeper understanding of \et production in DIS might also be benefitial
for the understanding of the hadron-hadron \et data, which so far could
be just parametrized phenomenologically.
}.
This behaviour
is predicted for the partonic $\av{E_T}$ from the BFKL calculation,
whereas for DGLAP evolution the opposite is expected \cite{lowx:et}.
CDM gives still the best overall description of the data.
The DGLAP based models, MEPS and
HERWIG, have developed and now give also a reasonable description
of the data (fig.~\ref{etplug}).

\begin{figure}[htb]
   \centering
  \begin{picture}(0,0) \put(0,0){{\bf a)}} \end{picture}
   \epsfig{file=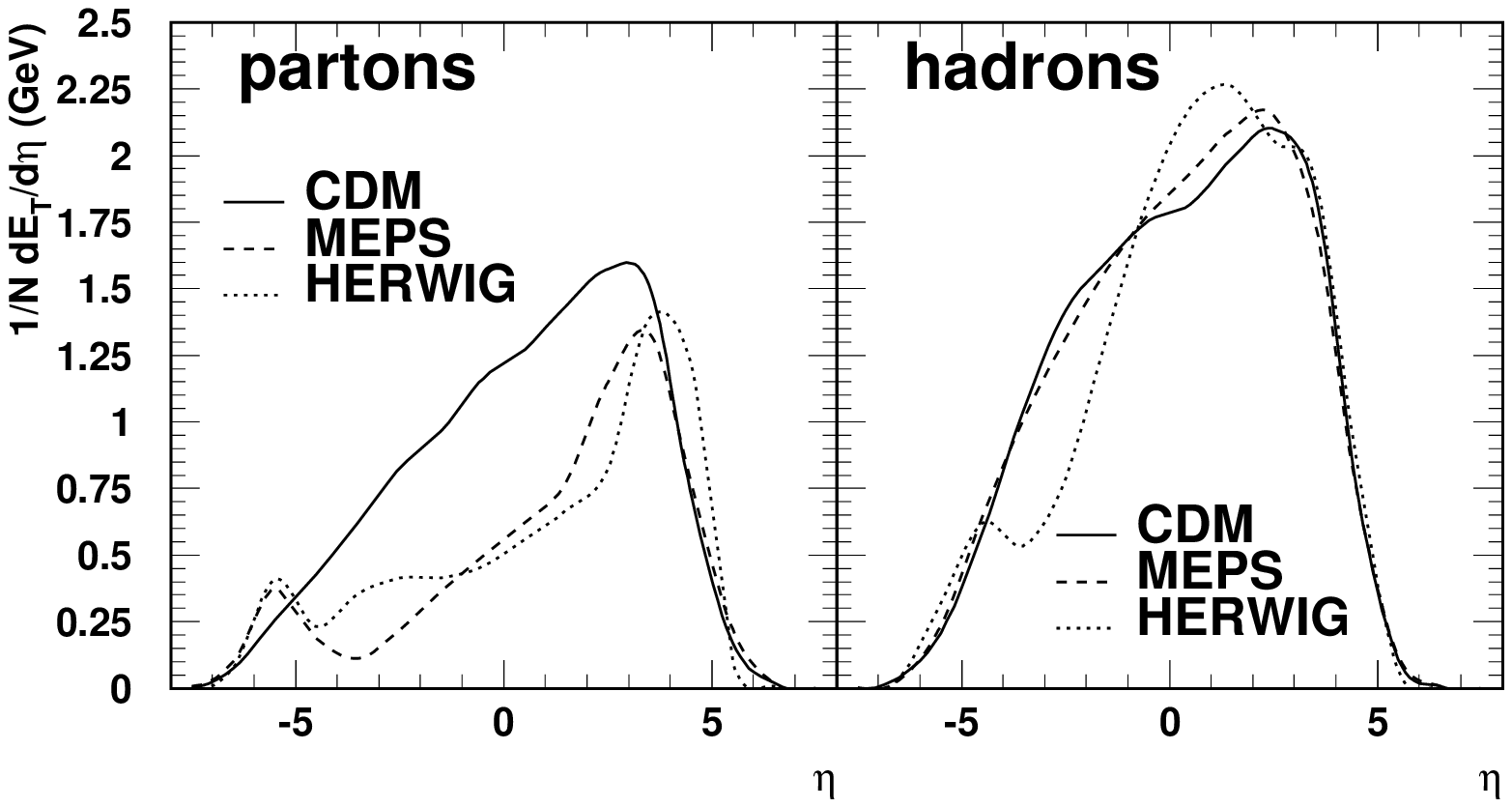,
           width=9.7cm,bbllx=2pt,bblly=276pt,bburx=518pt,bbury=521pt,clip=}
 \begin{picture}(0,0) \put(0,0){{\bf b)}} \end{picture}
   \epsfig{file=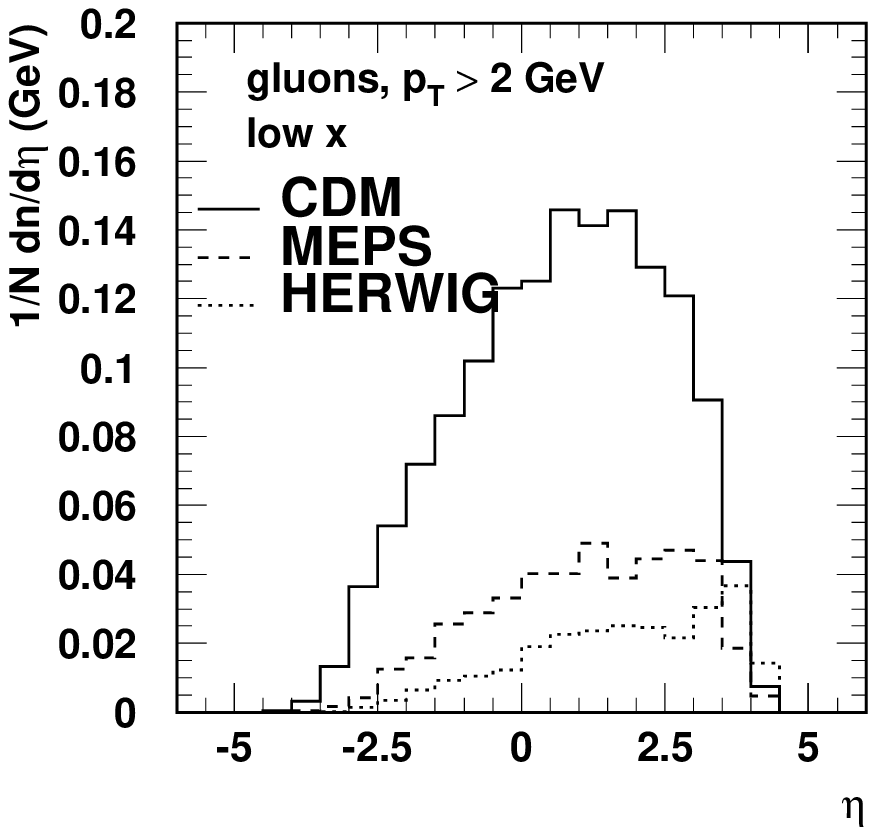,
           width=4.8cm,bbllx=4pt,bblly=275pt,bburx=257pt,bbury=516pt,clip=}
   \scaption{
      {\bf a)} Transverse energy flows
      as a function of the pseudorapidity $\eta$ for partons and
      for hadrons.
      {\bf b)} The multiplicity of hard
      gluons with $\pt>2\GeV$ vs. $\eta$.
      The events are generated with the models
      CDM (ARIADNE 4.08), MEPS (LEPTO 6.4) and HERWIG 5.8 in
      a ``low \xb'' kinematic bin
      with \av{\xb}=0.00037 and \av{\Qsq}=13.1 \GeVsq.
      The proton direction is to the left.}
   \label{etpar}
\end{figure}

In order to draw conclusions from comparing
the data with partonic QCD calculations,
hadronization effects need to be estimated from Monte Carlo models.
The partons produced in the CDM give about twice as much \et
in the central
rapidity region as
the ones emitted from the ordered cascades in
MEPS and HERWIG, see fig.~\ref{etpar}a \cite{mk:method}, which
can be traced to the fact that hard gluon radiation is much more
abundant in CDM, see fig.~\ref{etpar}b \cite{mk:heraws}.
However, the observable particles emerging
after hadronization give rise to very similar \et flows.
The \et flow predictions of the different models cannot be
well discriminated with the present data.
While hadronziation adds relatively
little ($\approx 0.5 \GeV/{\rm unit~rapidity}$)
to the partonic \et for CDM, most of the \et
($\approx 1.4 \GeV/{\rm unit~rapidity}$)
is generated by hadronization for MEPS and HERWIG.
It is noteworthy that CDM and MEPS use the same Lund string
fragmentation model for hadronization.
In addition,
the partonic and the hadronic \et are well correlated event-by-event
in CDM, as opposed to the other models \cite{mk:method}.

\begin{figure}[htb]
   \centering
   \epsfig{file=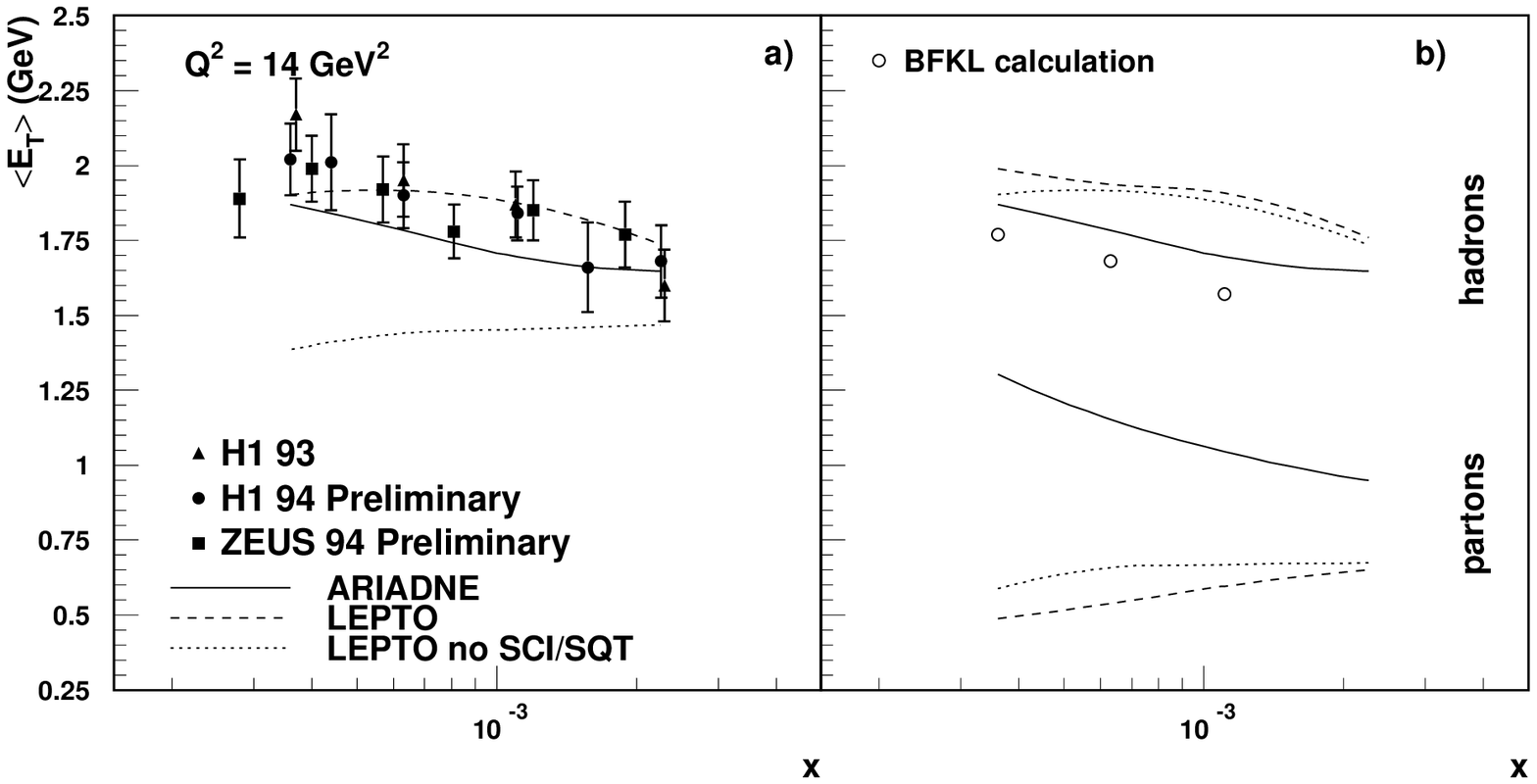,width=13cm}
   \scaption{
             The mean \et in the CMS pseudorapidity interval
              $-0.5<\eta<0.5$
             as a function of \xb for $\Qsq=14\GeVsq$.
             Besides the data from H1 \cite{h1:flow4}
             (which have an additional 8\% normalization uncertainty)
             and ZEUS
             \cite{z:etdis96}, the BFKL expectation for partons is shown
             \cite{lowx:et}, together with the results from
             the models CDM (ARIADNE 4.08), MEPS (LEPTO 6.4), and HERWIG 5.8
             for hadrons and for partons.
             The LEPTO predictions for hadrons without the features
             soft colour interaction (SCI) and  the new sea quark
             treatment (SQT) are also shown.}
   \label{etfix}
\end{figure}

With the \et data alone, the hadronization ambiguity cannot be
resolved. The situation is summarized in fig.~\ref{etfix}.
The $\av{\et}$ data rise with falling \xb and
agree with both the ordered and the unordered
Monte Carlo models for hadrons. BFKL predicts also a rise,
as is seen in the CDM partons, but at a somewhat lower level.
Apart from the normalization, the data are consistent with
a BFKL interpretation, assuming hadronization effects as
given by CDM.
The DGLAP model partons produce
much less \et than CDM, and have the opposite \xb behaviour,
as expected for DGLAP evolution \cite{lowx:et}.
However, the data are also consistent with DGLAP
evolution, when large hadronization effects are assumed.
How do they arise?

\begin{figure}[htb]
   \centering
   \epsfig{file=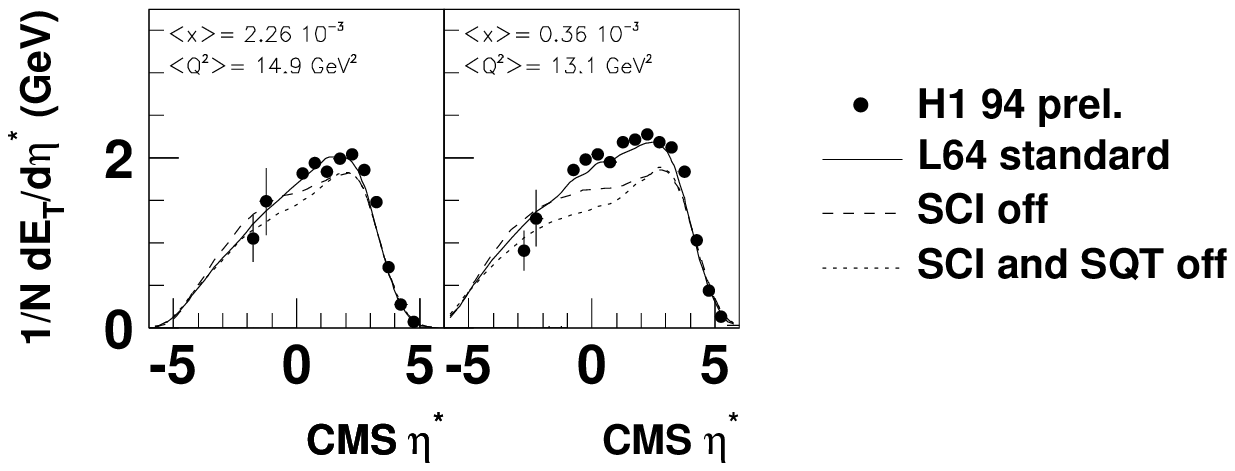,
           width=12cm,bbllx=44pt,bblly=527pt,bburx=370pt,bbury=672,clip=}
   \scaption{The \et flow from H1 \cite{h1:flow4} at
             $\av{x}\approx$ 0.002 (left) and 0.0004 (right)
             for $\av{Q^2}\approx 14$ \GeVsqx.
             The data (shown are statistical errors only) are
             compared to
             the MEPS model (LEPTO 6.4, structure function MRSH)
             in the standard version, without soft colour interactions (SCI),
             and with neither soft colour interactions nor the
             new sea quark treatment (SQT).}
   \label{scisqt}
\end{figure}

In LEPTO 6.4 changes to the non-perturbative part were introduced
to simultaneously describe the large \et flow seen in the data,
and to explain the rapidity gap events
\cite{h1:gap1,h1:gap2,z:gap1,z:gap2} without
invoking the explicit exchange of a Pomeron \cite{mc:sci}.
In the new concept of soft colour interactions (SCI),
colour quantum numbers may be exchanged between the outgoing partons
after the hard interaction, leading to a reconfiguration of
the fragmenting strings (see section \ref{sn:gaps}).
This may result in colour neutral subsystems,
separated by a rapidity gap from the rest of the event after
hadronization, or it may enhance the \et flow, when strings
are spanned back and forth (fig.~\ref{scistrings}).
Also the treatment of the remnant fragmentation has been changed
for events where a sea quark is hit (new sea quark treatment - SQT).
Instead of recombining the sea quark partner
(for example a $\ol{s}$ quark in case a sea quark $s$ was hit)
within the remnant to
form a hadron, a fragmenting string is now stretched between
the hit sea quark and the remnant.
At large $x$ these effects are minor,
but at small \xb they are significant, see fig.~\ref{scisqt}.
As long as the new SCI concept has not been ruled out by
other data --
from detailed studies of rapidity gap events, for example --
it has to be taken seriously.
The SCI does not only offer an explanation for the rapidity gap events
and the large \et flow, but has also been advocated
\cite{mc:scitev} to explain
the large production rate of heavy quarkonia in $p\ol{p}$ interactions
at the Tevatron \cite{coll:quarkonia}.
Presently the SCI model has difficulties describing the thrust
of the hadronic final state in rapidity gap events, but so do
the other models based upon Pomeron exchange \cite{h1:tgap}.
Further data will be necessary to discriminate the Pomeron exchange
and SCI models.

In the CDM hadronization effects are much smaller, because gluon
radiation is more abundant, and less energy remains for hadronization.
The large hadronization effects in HERWIG can possibly be explained
by its very simple cluster fragmentation scheme.
When there are few partons, the masses of
parton pairs (clusters) to be fragmented tend to be large.
In the cluster fragmentation model,
they undergo a two-body decay according to phase space,
leading to large transverse energies.
The same mechanism leads to a significant fraction of rapidity
gap events which are ``not exponentially suppressed''
in HERWIG without Pomeron exchange or soft colour
interactions.
In a later version (HERWIG 5.9) this problem is
being addressed by allowing large mass clusters to split
longitudinally before they decay.

  \section{Transverse Momentum Spectra
                              \label{sn:ptspectra}}
Not yet well understood hadronization effects,
however interesting they may be, precluded strong
conclusions on the underlying parton dynamics from
the transverse energy flow measurements.
It has been shown
that single particle
transverse momentum (\pt) spectra
represent a more direct measure of the partonic
activity, and corresponding measurements
have been suggested \cite{mk:rome,mk:method}.

\begin{figure}[tbh]
   \centering
   \vspace{-0.2cm}
   \epsfig{file=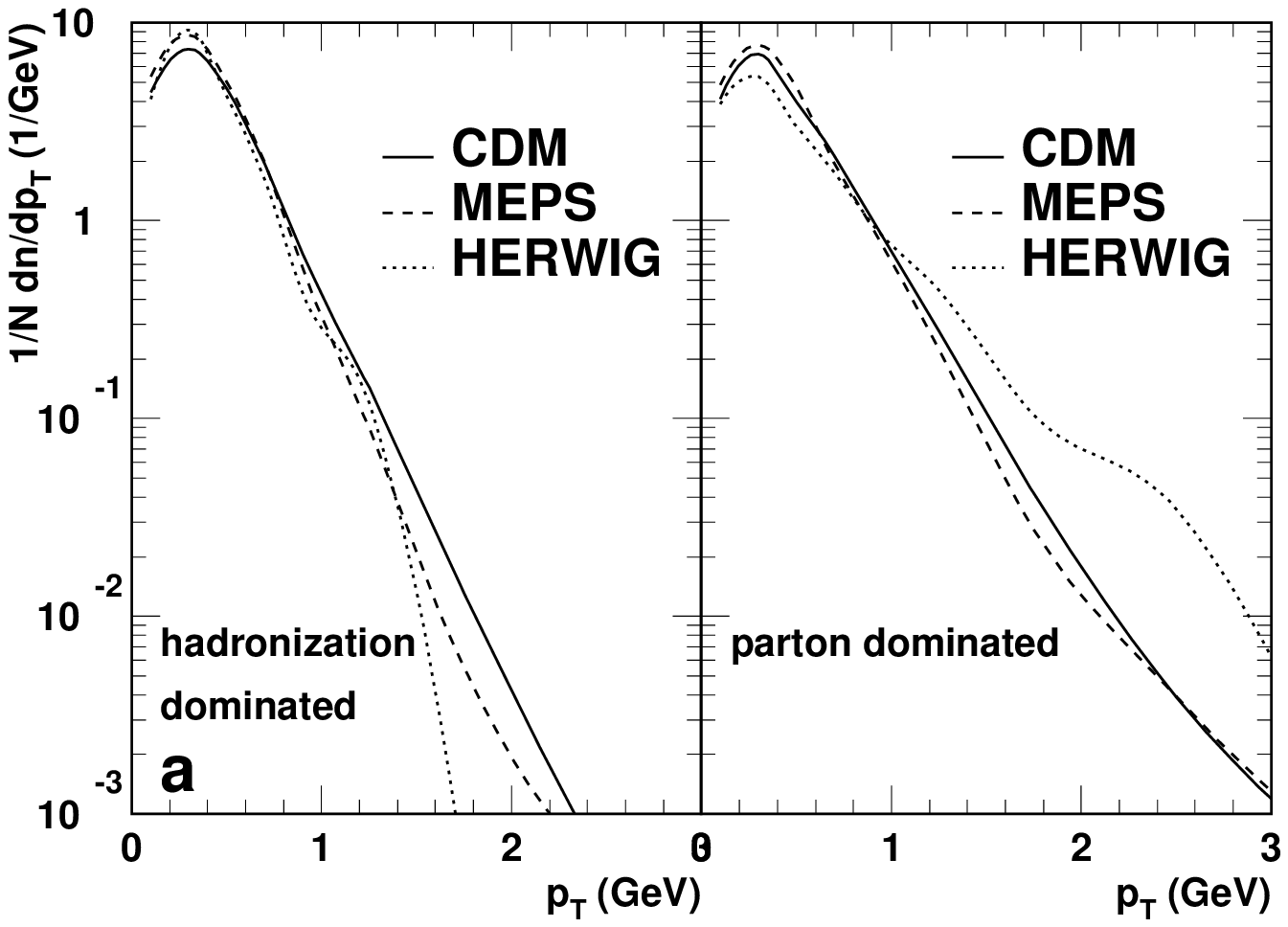,
    width=9cm}
   \epsfig{file=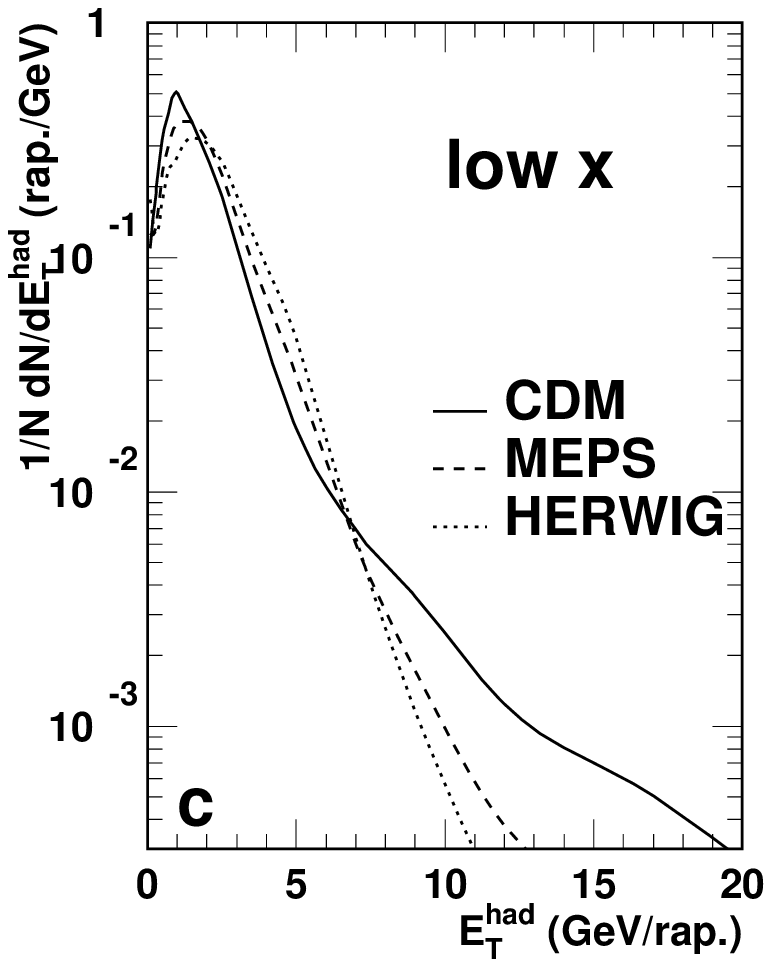,%
    width=4.5cm,bbllx=0pt,bblly=0pt,bburx=221pt,bbury=281pt,clip=}
   \scaption{
      {\bf a)} \pt spectra of charged particles with
      $0<\eta<2$ for events which
      are
      hadronization dominated (left, $\etpar < 0.2$ GeV/unit rapidity)
      or
      parton dominated (right, $\etpar/\ethad > 0.5$) \cite{mk:method}.
      The events are required to have
      \ethad between 1 and 2 GeV/unit rapidity.
      {\bf c)} The \et distribution (measured in $0<\eta<2$) at
         $\av{x}=0.00037$  \cite{mk:method}.
         The events are generated with CDM (ARIADNE 4.08),
         MEPS (LEPTO 6.4) and HERWIG 5.8.
      }
   \label{ptsens}
\end{figure}

To answer the question whether the \et observed in the data
is generated predominantly
by parton radiation or by hadronization,
inclusive \pt spectra are considered.
Hadronization
should produce typical spectra
which are limited in \pt, while
parton radiation should manifest itself in a hard tail of the \pt
distribution.
To test this idea particles from a ``central'' (in the CMS)
$\eta$  interval $0<\eta<2$ are examined.
Simulated events are compared which have similar hadronic
\et (\ethad) in that interval,
with either a large or small contribution \etpar from partons.
The parton dominated events indeed exhibit a harder \pt spectrum
than the hadronization dominated ones (s. Fig.~\ref{ptsens}a),
regardless of the mechanisms employed for parton evolution and
hadronization.
Different parton evolution scenarios which differ in the
production of hard gluons (for example BFKL vs. DGLAP, or
CDM vs. MEPS/HERWIG, see fig.~\ref{etlab}) can thus be
discriminated by the number of high-\pt hadrons which are produced.
In a similar fashion the distribution of the \et measured centrally
could be used for discrimination \cite{mk:method} (fig.~\ref{ptsens}c).
Preliminary H1 data \cite{h1:flow4} favour the CDM over the other
models, but the study of systematic errors has not yet been completed.

H1 has measured \cite{h1:pt}
the charged particle \pt spectra as much central
in the CMS
($0.5<\eta<1.5$)
as the detector acceptance
allows\footnote{The measurement
relies mostly on the central tracking device; when
systematic effects presently hampering the forward tracker are mastered,
the measurable rapidity region could be enlarged by one unit of rapidity
towards the remnant.}.
Measured is
\begin{equation}
  \frac{1}{N} \frac{\dd n}{\dd p_T} :=
  \int_{\Delta \eta}\dd \eta
                 \frac{ \dd^4 \sigma_h(ep\rightarrow e'Xh)}
                      { \dd \eta \dd p_T \dd x \dd Q^2 }
          \left/
         \frac{\dd^2 \sigma(ep\rightarrow e'X)}{ \dd x \dd Q^2} \right. ,
\end{equation}
where the differential
cross section to produce a hadron with transverse momentum
\pt in a pseudorapidity interval $\Delta \eta$ is normalized
to the differential event cross section for given $x,Q^2$.
In the measurement, the differential cross sections are averaged over
certain small $x,Q^2$ bins.

\begin{figure}[tbp]
   \centering
   \vspace{-1cm}
   \epsfig{file=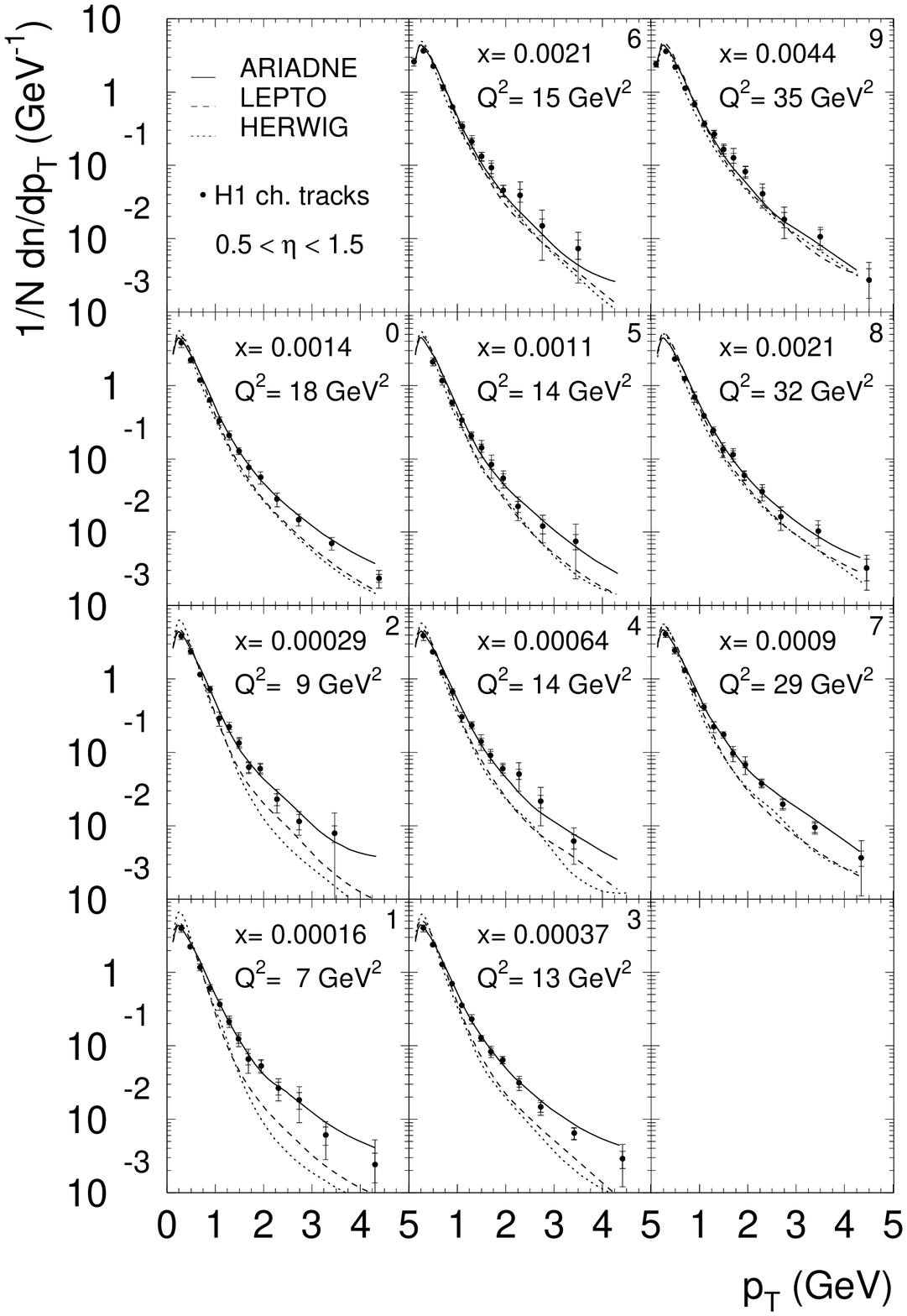,%
    width=14cm,bbllx=33pt,bblly=23pt,bburx=487pt,bbury=689pt,clip=}
   \scaption{
            The \pt spectra of charged particles,
            measured in the CMS in the
            pseudorapidity interval $0.5 < \eta < 1.5$ \cite{h1:pt}.
            Data are shown for nine different kinematic bins
            with the mean values of
            \xb and \Qsq as indicated,
            plus the combined sample (bin 0).
            The models
            ARIADNE 4.08, LEPTO 6.4 and
            HERWIG 5.8 are overlayed.}
   \label{ptcentral}
\end{figure}

The data are compared in fig. \ref{ptcentral} to
models with suppressed (LEPTO 6.4, including SCI and new SQT; HERWIG 5.8)
and unsuppressed (ARIADNE 4.08)
radiation patterns.
At ``large'' \xb and \Qsq
(here ``large'' means $Q^2\approx 35\,$GeV$^2$
and $x \approx 0.004$)
all models provide a good
description of the data. At smaller \xb and $Q^2$, LEPTO and HERWIG
fall significantly below the data for $p_T > 1\,$GeV. ARIADNE
gives a good description of the data over the full kinematic range.
The shortfall
of the models with suppressed gluon radiation indicates that
at small \xb there is more high $k_T$ parton radiation present than is
produced by the models based upon leading log DGLAP parton showers.
These models are closer to the data in the current region, or
when all charged particles are considered, including the soft ones
\cite{h1:pt}, see fig.~\ref{eta_charged}.


\begin{figure}[tbh]
   \centering
   \epsfig{file=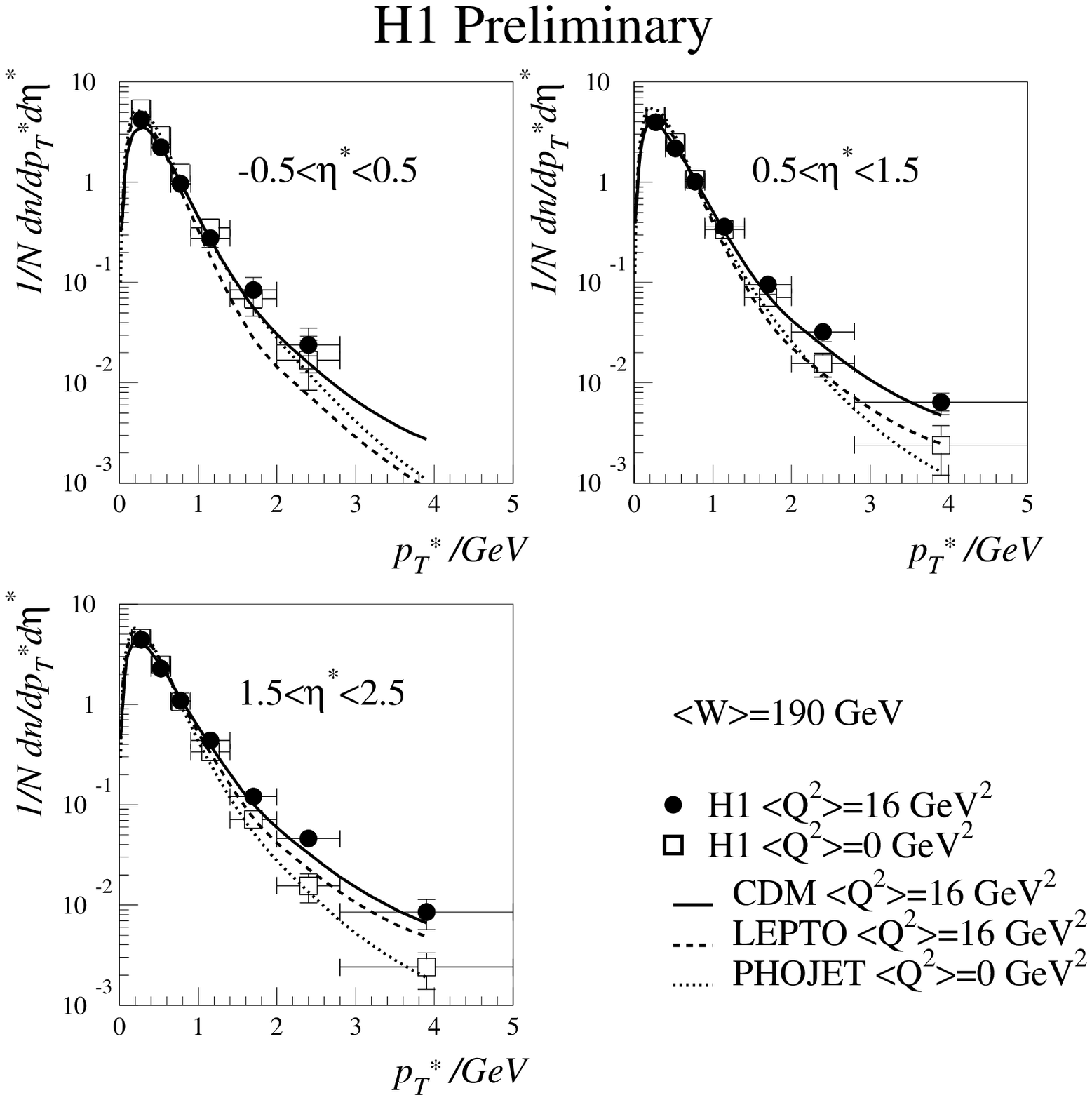,%
    width=12cm,bbllx=48pt,bblly=143pt,bburx=545pt,bbury=671pt,clip=}
   \scaption{
            The \pt spectra of charged particles for different
            CMS $\eta$ ranges (H1 preliminary, \cite{h1:ptq2}).
            DIS data with $\av{Q^2}=16\GeVsq$ are compared
            with photoproduction data with $\av{Q^2}=0$.
            Also shown are the DIS models CDM (ARIADNE 4.08)
            and LEPTO 6.5, and
            the photoproduction model PHOJET 1.3.
            The data samples are selected such that they have
            the same $\av{W}=190\GeV$.}
   \label{ptforw}
\end{figure}

In a preliminary analysis of the H1 forward tracker,
\pt data have been obtained also
in the ``truly central region'', $-0.5<\eta<0.5$.
In fig.~\ref{ptforw} the charged particle \pt spectra are shown
for three different $\eta$ ranges. At central rapidity,
$-0.5<\eta<0.5$, the spectra from photoproduction and DIS are
similar. Towards the photon fragmentation region, the DIS spectra
become harder than the photoproduction data.
Apparently, in that region, the
virtuality \Qsq influences the hardness of the \pt spectrum, while
there is little influence in the central region. A similar conlusion
has been arrived at from the study of the \et flow, see section
\ref{sn:flow}. The DIS data are well described by the CDM. LEPTO
becomes too soft away from the current region.
The prediction
by the photoproduction model
PHOJET \cite{mc:phojet} describes the photoproduction spectra well.

\begin{figure}[tbh]
   \centering
\begin{picture}(0,0) \put(0,0){{\bf a)}} \end{picture}
   \epsfig{file=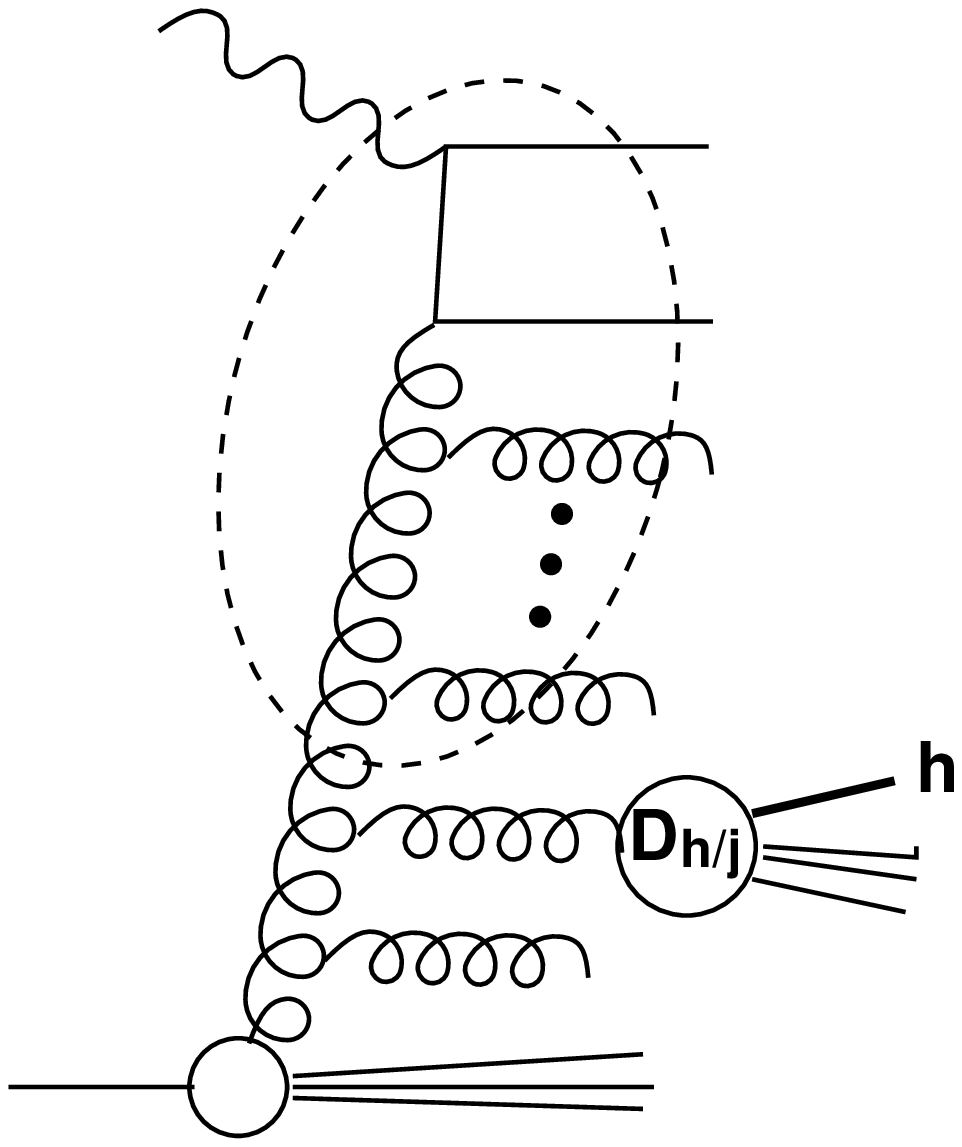,
    width=5cm,bbllx=146pt,bblly=219pt,bburx=451pt,bbury=562pt,clip=}
\begin{picture}(0,0) \put(0,0){{\bf b)}} \end{picture}
   \epsfig{file=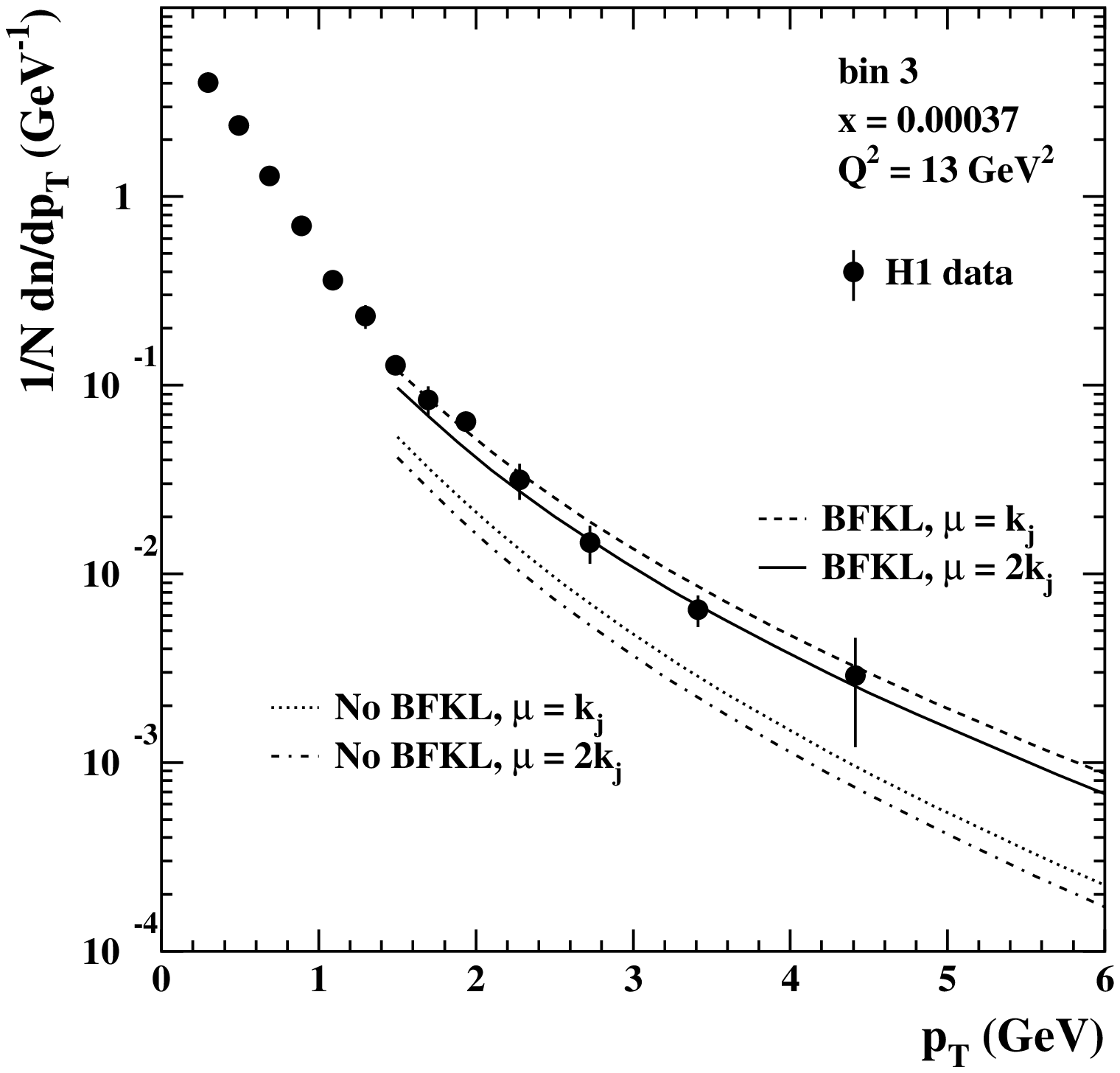,
    width=9cm,bbllx=49pt,bblly=210pt,bburx=490pt,bbury=645pt,clip=}
   \scaption{
             {\bf a)} The production of a hadron $h$ from a parton $j$ through
                a fragmentation function $D$.
                A gluon ladder connects the quark box attached to the
                virtual photon with the proton.
                The encircled part of the
                graph is calculated either with or without BFKL
                evolution.
             {\bf b)} The result of the calculation with and without
                BFKL calculation \cite{lowx:pt},
                compared to the H1 data at low $x$ \cite{h1:pt}.}
   \label{ptcalc}
\end{figure}

Theoretical calculations for the
charged particle \pt spectra have become
available \cite{lowx:pt},
where a cross section $\sigma_j$ to produce
a parton $j$ is folded with an
experimentally known fragmentation function $D_{h/j}(z,\mu^2)$
\cite{th:binnewies} to produce
a hadron $h$ with momentum fraction $z$ from the parton $j$,
see fig.~\ref{ptcalc}a.
Monte Carlo models for hadronization are thus avoided.
Symbolically,
\begin{equation}
   \sigma_h = \sigma_j \otimes D_{h/j}.
\end{equation}

The parton cross section $\sigma_j$ is divided into a conventional part
where gluon emissions in the ladder are ordered,
and a part with unordered emissions, which is calculated
with the BFKL equation (see fig~\ref{ptcalc}a).
The H1 \pt data at low $x$ are well described by this calculation,
see fig~\ref{ptcalc}b.
The absolute normalization for the BFKL part
is obtained by requiring that the calculated parton cross
section would match the measured H1 forward jets \cite{h1:fwdjet}
at the hadron level, see section \ref{sn:fjets}.
We note that this normalization neglects
the effects of the jet algorithm and hadronization effects, both
of which are not included in the single parton cross section calculation.
Without the BFKL part the calculation is by a factor $\approx 2$
below the data.
Both  predictions show little dependence on the
choice of the factorization scale
$\mu^2$ for the fragmentation function.

For the ``no BFKL'' calculation the LO cross section
for three partons was used, where the main contribution is
the diagram with the
\qqbar~ from the quark box plus one additional gluon,
see fig. \ref{ptcalc}a.
A complete NLO calculation is still missing.
DGLAP evolution was neglected,
because the kinematic conditions severely disfavour DGLAP evolution.
DGLAP evolution is not possible between the parton $j$ and the photon
if $k_{Tj} \approx Q$ due to strong \kt ordering. For the curves
shown in fig. ~\ref{ptcalc}b
this condition is approximately
fulfilled, $Q^2/2 < k_{Tj}^2 < 2 Q^2$, because a leading
hadron carries on average about half of the fragmenting parton's momentum,
$p_T \approx k_{Tj}/2$.

An even better signal is expected when the single particle detection
capability could be extended
further towards the proton remnant \cite{mk:heraws,lowx:pizerogam}.
Typically at HERA, the central trackers are limited to
$\thlab>20\degr$, and the forward trackers to $\thlab>7\degr$.
An upgrade with a forward silicon tracker could cover angles
$\thlab>3\degr$ \cite{mk:heraws}. Alternatively, one can
detect single photons or $\pi^0$s in the calorimeters, which
have acceptance for typically $\thlab>4\degr$.

\begin{figure}[tbh]
   \centering
   \epsfig{file=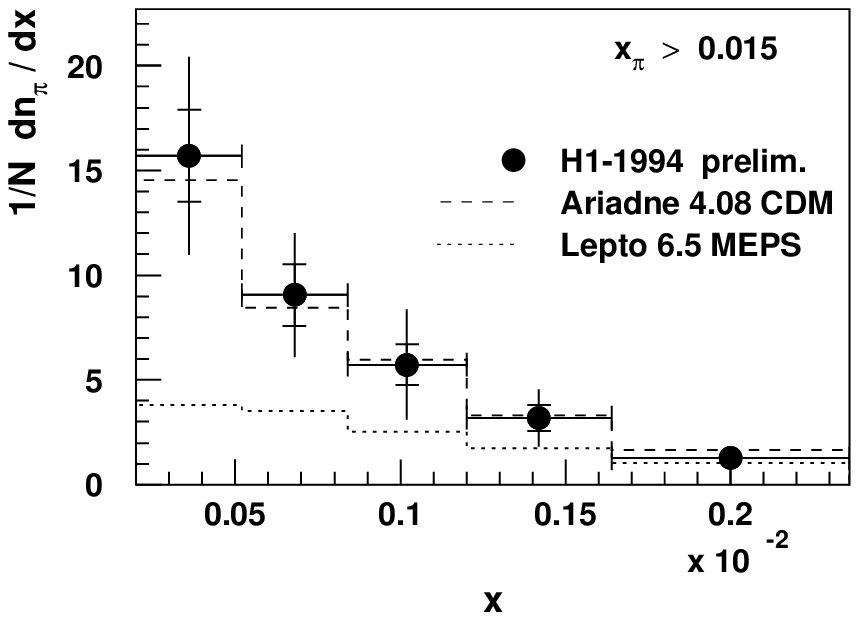,
    width=7cm,bbllx=27pt,bblly=493pt,bburx=288pt,bbury=690pt,clip=}
   \scaption{The number of ``forward'' $\pi^0$ mesons as a function
             of Bjorken $x$. $N$ is the number of events in the selected
             kinematic region.
             The preliminary H1 data \cite{h1:pizero} are compared
             to the MEPS model
             (LEPTO 6.5 \cite{mc:lepto}, including SCI and the new SQT)
             and the CDM
             (ARIADNE 4.08).
             }
   \label{pizero}
\end{figure}

H1 has measured ``forward'' $\pi^0$ production \cite{h1:pizero},
where
the $\pi^0$ meson is detected via the electromagnetic showers
of their decay photons in the H1 forward
liquid argon calorimeter \cite{h1:pizero}.
The $\pi^0$'s have been selected in
the laboratory frame
with $5\degr<\thlab<25\degr$, $E>8\GeV$, $\pt>1\GeV$ and
$x_\pi := E/E_p > 0.015$ (data with other $x_\pi$ cuts exist as well)
in order to enhance the phase space for BFKL evolution.
The number of such $\pi^0$'s increases with
decreasing Bjorken $x$, see fig.~\ref{pizero}.
The models studied, MEPS and CDM, agree at large \xb with the data.
The rise towards small \xb is only followed by CDM.
Calculations for forward $\pi^0$ production exist in principle
\cite{lowx:pizerogam,lowx:pt}, but need to be repeated for the
kinematic conditions of this H1 measurement.

\begin{figure}[tbh]
   \centering
   \epsfig{file=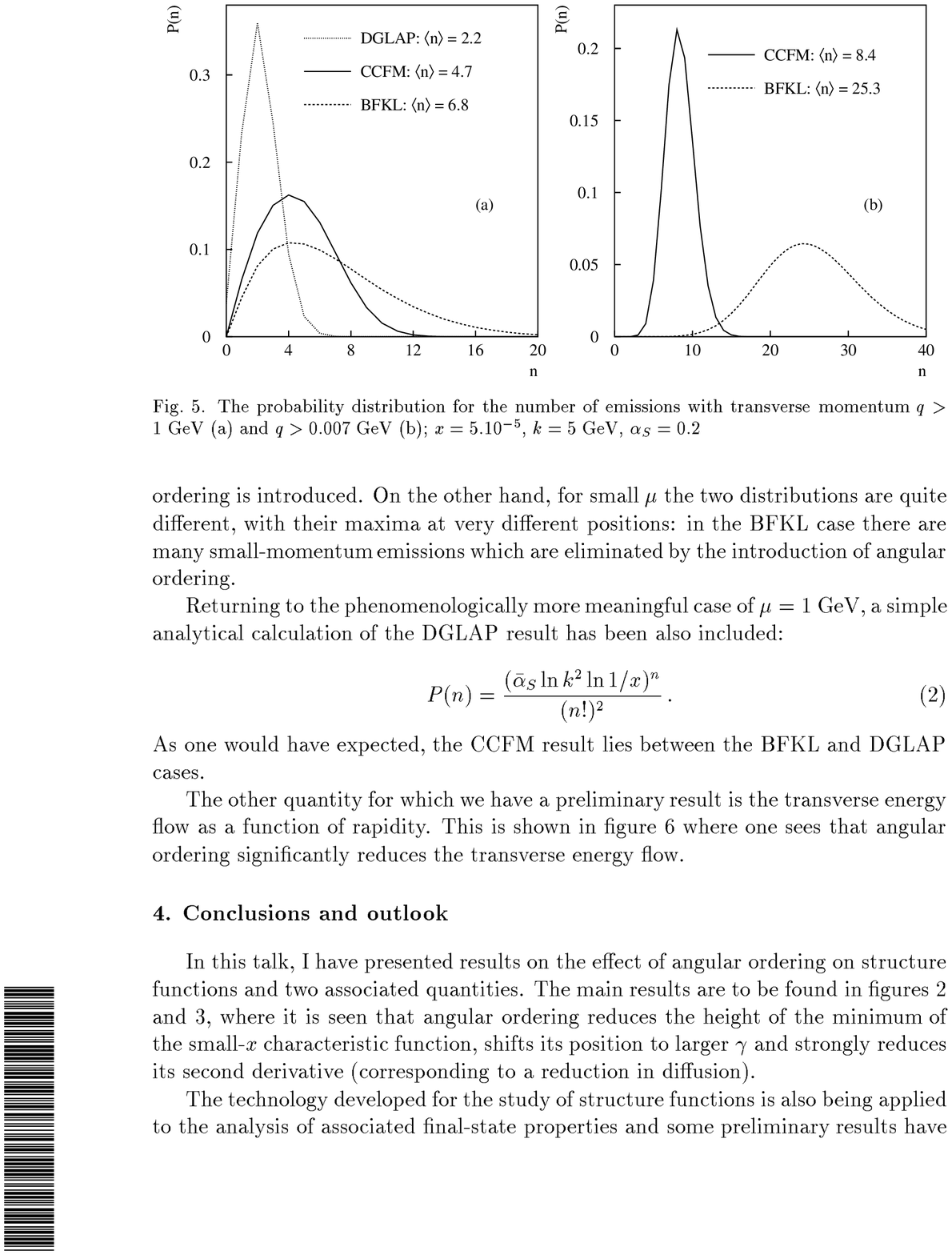,
    width=13cm,bbllx=92pt,bblly=487pt,bburx=523pt,bbury=706pt,clip=}
   \scaption{
                The probability distribution $P(n)$ for the number $n$
                of gluon emissions with transverse momentum
                {\bf a)} $k_T>1\GeV$ and {\bf b)} $k_T>0.007$ GeV.
                The calculations in the DGLAP, CCFM and BFKL
                schemes \cite{lowx:ringsalam} are for events
                with $x=5\cdot 10^{-5}$.}
   \label{ptsalam}
\end{figure}

Coherence effects, which can be taken into account
by angular ordering, are expected to modify the BFKL
behaviour. They are included in the
CCFM equation \cite{th:ccfm}.
Predictions for the hadronic final state at low $x$ based upon
the CCFM ansatz are now emerging \cite{lowx:ringsalam}.
In general, there is less \kt diffusion in CCFM than in BFKL.
The number of soft emissions is much reduced in CCFM when
compared to BFKL, due to angular ordering (fig.~\ref{ptsalam}).
For appreciable transverse momenta, the differences are not
so big, affecting mainly the high multiplicity tail.

The CCFM equation has been reformulated in the linked dipole chain model
(LDC) \cite{th:ldc}.
First results from a Monte Carlo generator implementing the LDC
in the ARIADNE framework are promising \cite{mc:ldc}.
The critical
distributions, \et flows and \pt spectra at small $x$, are much better
described than with LEPTO, though the agreement with the data is not
quite as good as with the CDM \cite{mc:ldc}.

  \section{Forward Jets  \label{sn:fjets}}         
The classic signature for BFKL evolution in the hadronic final
state are so-called ``forward jets'' \cite{lowx:fwdjets,lowx:hotref}
close to the remnant, that is ``forward'' for the HERA detectors.
The selection criteria are designed such as to suppress
the phase space for DGLAP evolution, and to maximize it for
BFKL evolution.
$\xjet=\ejet / E_p$, the ratio between the jet energy \ejet
and the proton beam energy $E_p$, is required to be
as large as possible, since BFKL evolution between the jet
and the quark box requires $\xjet \gg x$
to allow for strong ordering $x_{i+1}\gg x_i$ along the ladder
(see fig.~\ref{ladder}, where the forward jet is thought
to emerge from a radiated gluon close to remnant).
The
transverse momentum \kjet ~has to be close to
\Q ($\kjet \approx Q$).
The phase space for \kt ordered DGLAP evolution
($k_{Ti+1} \gg k_{Ti}$) is thus much reduced,
whereas it is still open for unordered BFKL evolution.
An enhanced rate of events with such jets is thus expected in the BFKL
scheme \cite{lowx:fwdjets}.
The experimental difficulty is to detect
jets close to the beam hole in the proton
direction in the forward calorimeter.

\begin{figure}[tbh]
   \centering
   \epsfig{file=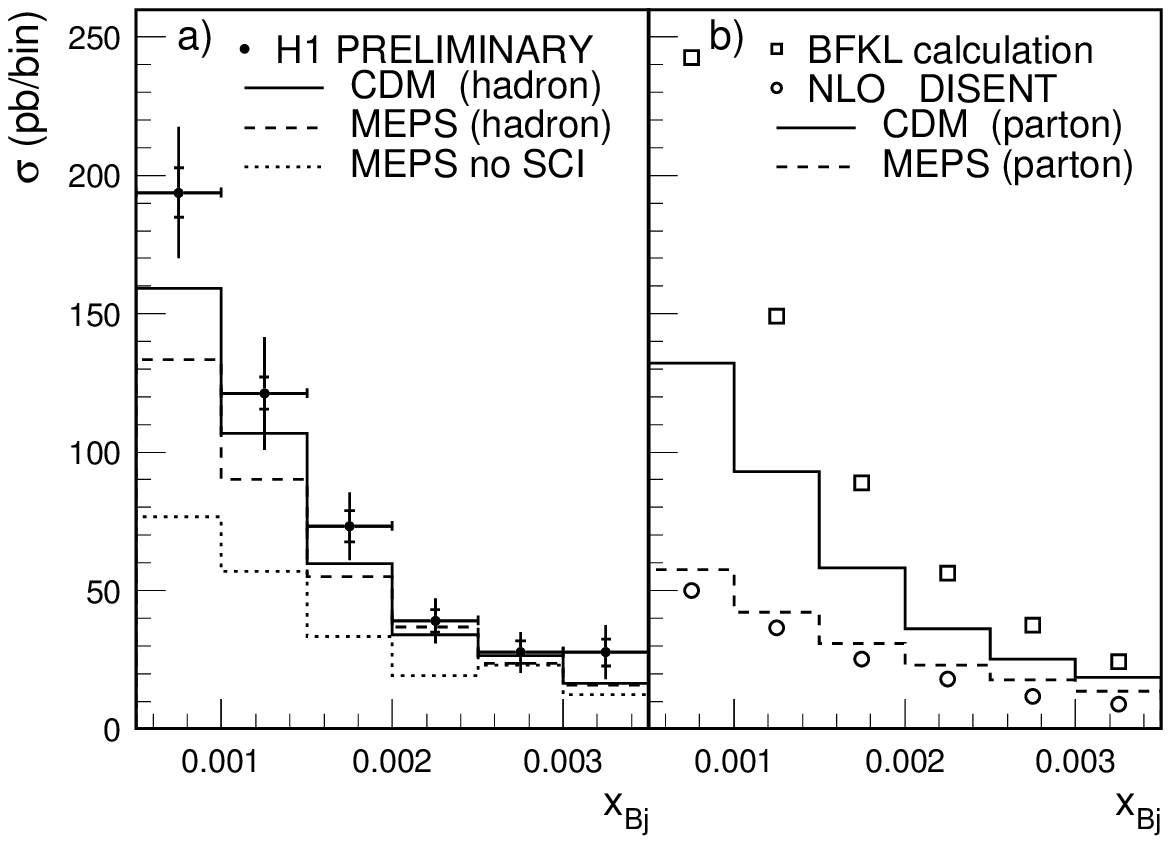,
           width=10cm}
   \begin{picture}(0,0) \put(0,0){{\bf c)}} \end{picture}
   \epsfig{file=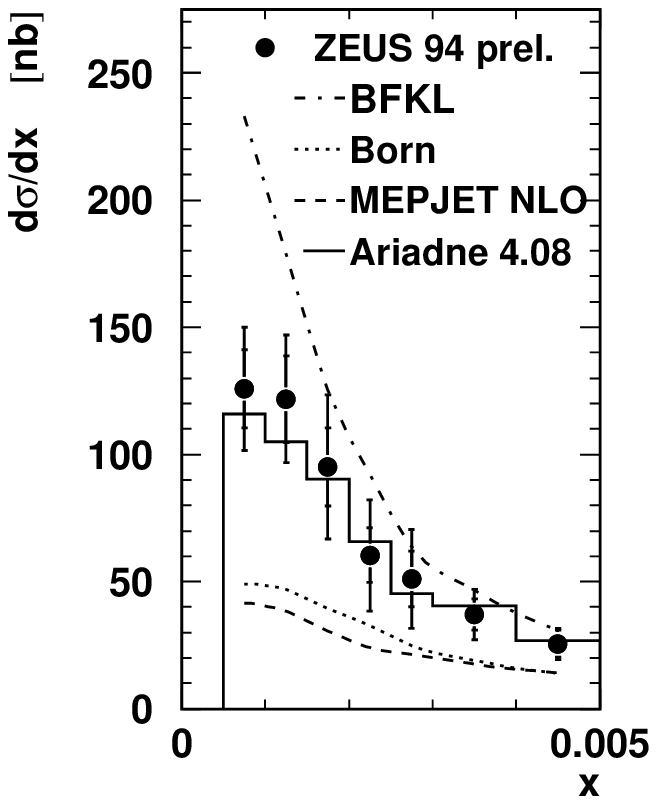,%
           width=4.2cm,bbllx=37pt,bblly=416pt,bburx=226pt,bbury=645,clip=}
   \scaption{
             {\bf a)} The forward jet cross section
             as a function of \xb from H1 \cite{h1:fwdjet} (1994 data).
             The data are corrected to the hadron level and
             compared to the models CDM (ARIADNE 4.08)
             and MEPS (LEPTO 6.4 with and without
             soft colour interactions).
             {\bf b)} The forward jet cross section from a NLO
             calculation \cite{mc:disent} and for parton jets
             found in CDM and MEPS, compared to a BFKL
             parton cross section calculation \cite{lowx:fwdcalc2}.
             {\bf c)}
             The ZEUS forward jet cross section \cite{z:fwdjet} (1994 data)
             vs. \xb, corrected to the parton level \cite{mk:madrid}.
             $\xjet>0.035$, $0.5 < \ptjet^2/Q^2 < 4$ and $\ptjet >$ 5 \GeV
             were required. The data are compared to a NLO jet calculation
             \cite{mc:mepjetz} and to parton jets from CDM (ARIADNE 4.08).
             Also shown are parton cross sections (no jet algorithm)
             with (``BFKL'') and without (``Born'') BFKL evolution
              \cite{lowx:fwdcalc2}. The systematic errors do not
             include uncertainties due to hadronization.
             }
   \label{fjeth1}
\end{figure}

Previous measurements by H1 \cite{h1:flow3} were
limited in statistics, but were consistent with BFKL calculations
\cite{lowx:fwdcalc1}.
In the preliminary analysis of the larger statistics sample from 1994
\cite{h1:fwdjet} forward jets with $\xjet>0.035$
are selected with the cone
algorithm (cone radius $R=\sqrt{\Delta \eta^2 + \Delta \phi^2}=1$)
in the HERA laboratory frame within $7\degr<\thjet<20\degr$.
The transverse jet momentum requirements are
$0.5 < \ptjet^2/Q^2 < 2$ and a lower
cut-off $\ptjet >$ 3.5 GeV.
The forward jet rate, corrected for detector effects to the hadron level,
increases sharply with falling $x$, see fig.~\ref{fjeth1}.
This is expected from BFKL
calculations, in contrast
to calculations without the BFKL ladder
\cite{lowx:fwdcalc2}.
The behaviour of the data is well described
by the CDM with the unsuppressed
gluon radiation pattern.
The DGLAP based model MEPS cannot describe the rise towards small $x$
when soft colour interactions are switched off.
When soft colour interactions are activated however,
without changing the parton dynamics,
MEPS comes
up rather close to the data.
A good description of the forward jet rate is also provided
by RAPGAP including a resolved (not point-like) component of
the virtual photon \cite{lowx:jung}. The physical significance
of this fact is still being debated.

Hadronization effects may play an important r\^{o}le for small
\ptjet, where it is possible to form a hadron jet
from hadronization fluctuations which is
uncorrelated with
any parton.
Only for $\ptjet>7\GeV$ hadronization effects were found to be
at the 10\% level \cite{lowx:haas}.
However, as one is leaving the BFKL condition $\ptjet \approx Q$ when
increasing the lower cut-off,
the BFKL signal diminishes. This could be compensated by
extending the measurements further into the forward region with
upgraded detectors \cite{lowx:fjetfut}.
The hadronization effect for jets depends
on the \et produced in hadronization.
In the CDM, where most of the \et is produced by parton radiation,
hadronization effects are much smaller.
The MEPS forward jet rate for partons agrees with a NLO calculation
with the program DISENT \cite{mc:disent} (see fig.~\ref{fjeth1}).
The BFKL calculation for partons shows an even larger increase
towards small $x$ than the CDM partonic forward jets. Similar
conclusions were reached in a preliminary ZEUS analysis
of 1994 data \cite{z:fwdjet},
see fig.~\ref{fjeth1}c.

\begin{figure}[tbh]
   \centering
   \epsfig{file=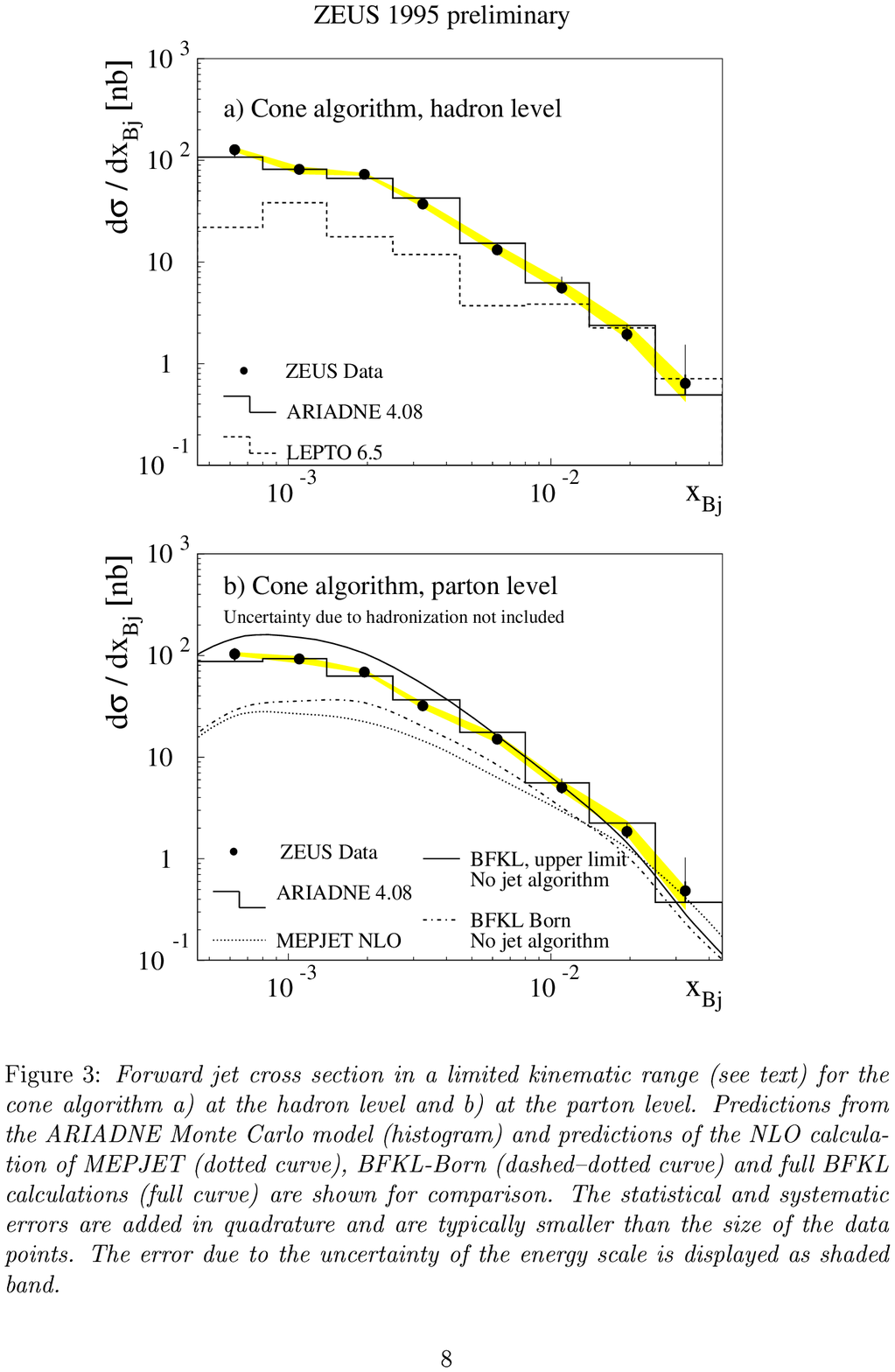,%
           width=8cm,bbllx=120pt,bblly=206pt,bburx=443pt,bbury=712,clip=}
   \scaption{The forward jet cross section as a function of \xb
             from ZEUS \cite{z:fjethep}. Statistical and systematic
             errors are added in quadrature.
             The shaded bands give the additional systematic error due
             to the uncertainty of the hadronic energy scale.
             {\bf a)} The data are corrected to the hadron level,
             and compared to the expectations from ARIADNE and LEPTO.
             {\bf b)} The data are corrected to the parton level.
                 Systematic errors due to the hadronziation correction
                 are not included.
                 Also shown are the
                 ARIADNE and the NLO predictions for jets \cite{mc:mepjet},
                 and the BFKL and Born graph calculations
                 for parton cross sections \cite{lowx:fwdcalc2}.
            }
   \label{fjethep}
\end{figure}

First results from a larger data sample from 1995
have been obtained, allowing for a larger \ptjet ~cut.
In the preliminary ZEUS analysis \cite{z:fjethep} events with
$0.00045<x<0.045$, $y>0.1$ and $E_e^\prime> 10\GeV$ are selected with
$\av{Q^2}\approx 15~\GeVsqx$.
The jets are required to lie in the target region of the Breit frame
and to satisfy $\ptjet>5~\GeVx$, $\xjet>0.036$ and
\mbox{$0.5 < \ptjet^2/Q^2 < 2$}. In the laboratory frame
$\thjet>8.5\degr$ is required.
Here we discuss the analysis with the cone algorithm (radius $R=1$).
An analysis with the \kt algorithm gives larger jet rates, but
one arrives at similar conclusions.
In a first step the data are corrected for detector effects.
At the hadron level (fig.~\ref{fjethep}a),
the increase towards small $x$ of the jet rate is well described
by ARIADNE 4.08. LEPTO 6.5 falls short of the data by a large amount
at small $x$.

In a second step, the data are corrected for hadronization effects
to the parton level
with ARIADNE, see fig.~\ref{fjethep}b.
Uncertainties due to this correction have not yet been
studied.
The data follow
the expectation for jets formed from the ARIADNE partons.
The partonic jet rate
at low $x$ is far above a NLO calculation.
The data are also compared to a calculation with and without BFKL
evolution, though these calculations do not include a jet algorithm.
However, the NLO result is close to the
Born graph calculation, which can be interpreted as a hint
that the differences are not so large.
The calculation without BFKL evolution
yields a low rate close to the
NLO calculation.
With BFKL effects included, a much larger rate at small $x$ is expected,
that even overshoots the data.
However, in view of the possibly large hadronization uncertainty
and the missing jet algorithm in the calculation, no firm
conclusion could be drawn from this comparison \cite{z:fjethep}.



  \section{Summary and Outlook \label{sn:lowxsum}}  

The hadronic final state has been searched for
an unsuppressed parton radiation pattern, which
could for example be expected from BFKL contributions,
in contrast to pure DGLAP evolution.
The high level of transverse energy is consistently
described when BFKL evolution is included, but
possibly large hadronization effects could also mimick
the effect. In order to describe the data with DGLAP evolution,
hadronization
phenomenology has introduced new concepts like soft
colour interactions in LEPTO, or it involuntarily produces
``not exponentially suppressed rapidity gaps''
by cluster fragmentation in HERWIG.

The forward jet measurements at low \pt are in rough agreement
with BFKL expectations,
but are also plagued
by hadronization uncertainties. With larger
statistics data sample, the measurements could be made
at higher \pt where a closer relation between hadron jets
and partons is expected \cite{lowx:haas}.
From the measured \et flows and forward
jets, a suppressed radiation scenario is presently only tenable
when large hadronization effects are assumed.

The measured single particle \pt spectra require
more parton radiation than is expected from pure DGLAP
evolution, regardless of the hadronziation model.
The spectra have  been calculated via known fragmentation functions
in the DGLAP and the BFKL scheme.
In the DGLAP scenario the calculation by far
underestimates the data.

Whereas pure DGLAP evolution fails,
BFKL evolution offers a plausible explanation
of the measurements. This does of course not
preclude other explanations. For example, it has been
suggested that contributions from resolved photons,
where the hard interaction happens not at the photon
vertex but further down the ladder, could be responsible for the additional
parton activity \cite{lowx:jung,lowx:klasen}.
We have seen possible signs for such effects
also in the inclusive jet rates, see section \ref{sn:jrates}.
More refined measurements
to unravel the details of parton dynamics at low $x$,
including correlations between jets and hadrons,
have been proposed
\cite{lowx:levin, lowx:ejetcor}.
It will also be important to check the consistency of the competing models
by confronting them with the entire data on the hadronic final state.

BFKL effects are also being looked for in
$p\ol{p}$ collisions at the Tevatron.
A jet-jet decorrelation is seen with increasing
rapidity difference, though the effect is larger
than expected from BFKL, and is described by
the HERWIG and PYTHIA models without BFKL dynamics \cite{coll:goussiou}.
Recently, BFKL effects have been brought forward to
explain the discrepancy between measured jet rates
and NLO calculations at small jet transverse energies
for different CM energies at the Tevatron \cite{th:kim}.

\chapter{Instantons \label{ch:inst}}

    \section{Introduction}                      
For a long time it has been recognized that
the standard model contains processes which cannot be
described by perturbation theory, and which violate
classical conservation laws like baryon number ($B$) and lepton number ($L$)
in the case of the electroweak interaction,
and chirality ($Q_5$) in the case of
the strong interaction \cite{inst:thooft}.
Such anomalous processes are induced by
so-called instantons \cite{inst:belavin}.
The name indicates that these are non-perturbative fluctuations
that are confined to ``an instant'' in space-time, with no
corresponding free particle solutions for $t\rightarrow \pm \infty$.
The interest in instantons
remained somewhat academic, as observable effects
were predicted to exist only at extremely
high energies of $\order{10^5~\TeVx}$,
until it was discovered that
their exponential suppression is much reduced
by the emission of gauge bosons \cite{inst:ringwald}.
In electroweak theory with massive gauge bosons
still rather high energies of $\order{\gtrsim 10~\TeVx}$ would be required,
but not so in QCD with massless gluons and strong coupling.
Instanton effects could play a r\^{o}le in QCD reactions
already at present day colliders.
Deep inelastic $ep$ scattering at HERA is particularly interesting,
because the virtuality of the photon probe \Qsq provides a
hard scale for the instanton subprocess,
which is needed for
theoretically sound
predictions \cite{inst:balitsky1,inst:vladimir,inst:yaroslavl}.
Instanton effects have not yet been observed in nature.
Their experimental discovery would be of fundamental
significance for particle physics.

Here a short
introduction to the basic theoretical ideas will be given
(a pedagogical treatment of instantons can be found in \cite{inst:abc}).
Instanton phenomenology in deep inelastic scattering (DIS) will
be discussed, covering cross sections and event topologies.
Finally, prospects for instanton searches and
first results from the analysis of HERA data
will be presented.

    \section{Instanton Theory}                  

Instantons originate from the non-trivial topological
structure of the vacuum in non-Abelian
gauge field theories, where
the vacuum is degenerate in
the Chern-Simons number $N_{\rm CS}$.
\ncs is defined as an integral over the gauge
fields $A^a_\mu$ with coupling $g$,
\begin{equation}
  \ncs := \frac{g^2}{16\pi^2} \int \dd^3x \epsilon_{ijk}
         \left( A_i^a \partial_j A^a_k -
               \frac{g}{3} \epsilon_{abc} A_i^a A_j^b A_k^c \right).
\end{equation}
\begin{wrapfigure}{r}{6cm}
\mbox{
   \epsfig{file=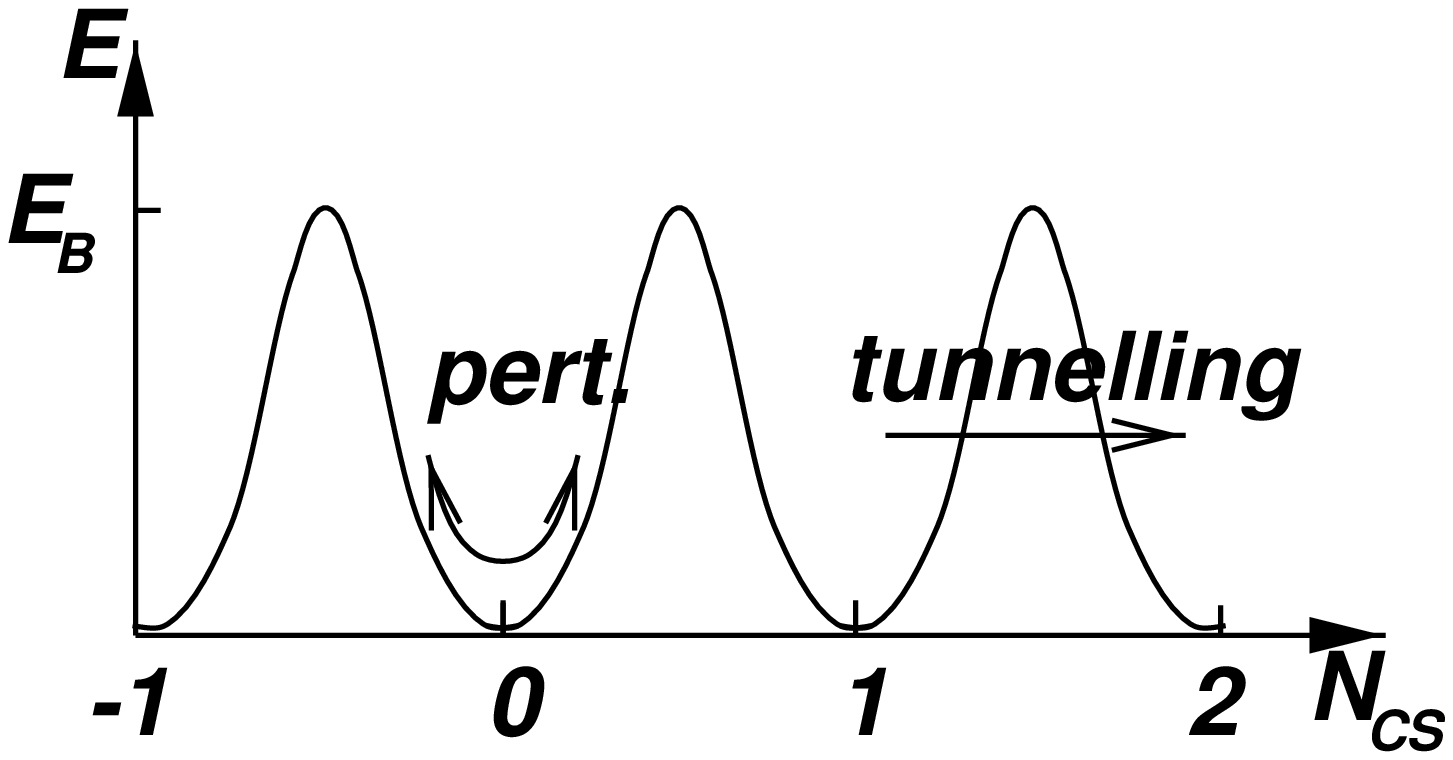,%
      width=5cm,bbllx=93pt,bblly=283pt,bburx=516pt,bbury=511,clip=}
     }
   \bild{5cm}{The structure of the vacuum.
                      Instanton solutions represent tunnelling
                      transitions between topological inequivalent
                      minima, which cannot be reached perturbatively.}
   \label{structure}
\end{wrapfigure}
Neighbouring vacua have the same (minimal) potential energy,
but differ in
their topological winding number \ncs, and
are separated by a potential barrier of height
$E_B$ (fig.~\ref{structure}).
The usual perturbative expansion of the scattering amplitudes
in the coupling constant $\alpha$
around {\em one} minimum (fig.~\ref{structure}),
conveniently represented by a series of Feynman graphs,
does not allow for transitions between neighbouring
minima.
They may however occur classically when the energy $E$ is large enough
$E>E_B$, or by quantum mechanical tunnelling when $E<E_B$,
corresponding to so-called instanton solutions
of the classical field equations.
The transition amplitude for the
instanton--induced tunnelling process
is exponentially suppressed
$\propto \exp(-4\pi/\alpha)$, a very small number.


\begin{figure}[htb]
   \centering
\begin{picture}(0,0) \put(0,0){{\bf a)}} \end{picture}
   \epsfig{file=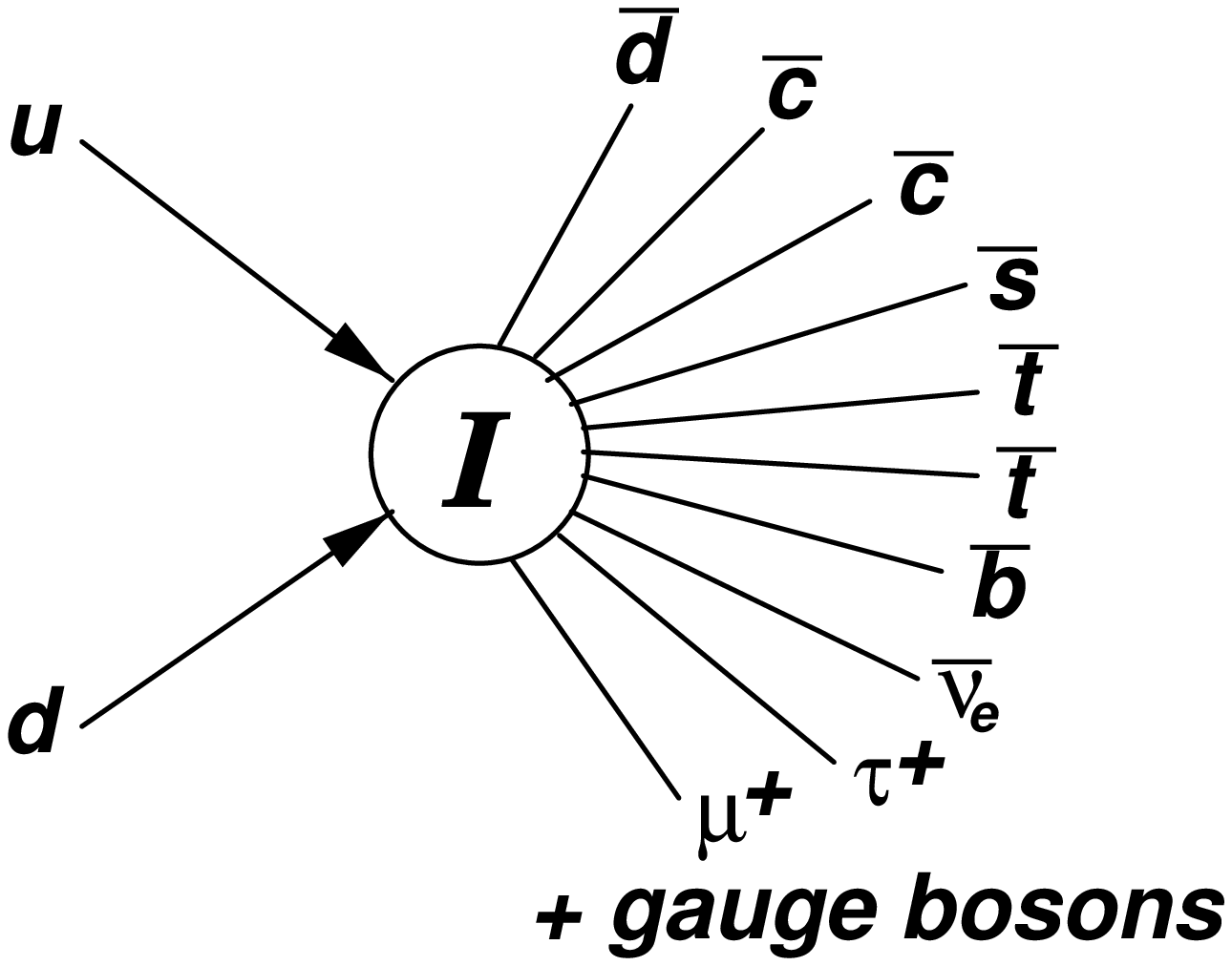,%
      width=6.0cm,bbllx=101pt,bblly=220pt,bburx=508pt,bbury=563,clip=}
   \hspace{1.5cm}
\begin{picture}(0,0) \put(0,0){{\bf b)}} \end{picture}
   \epsfig{file=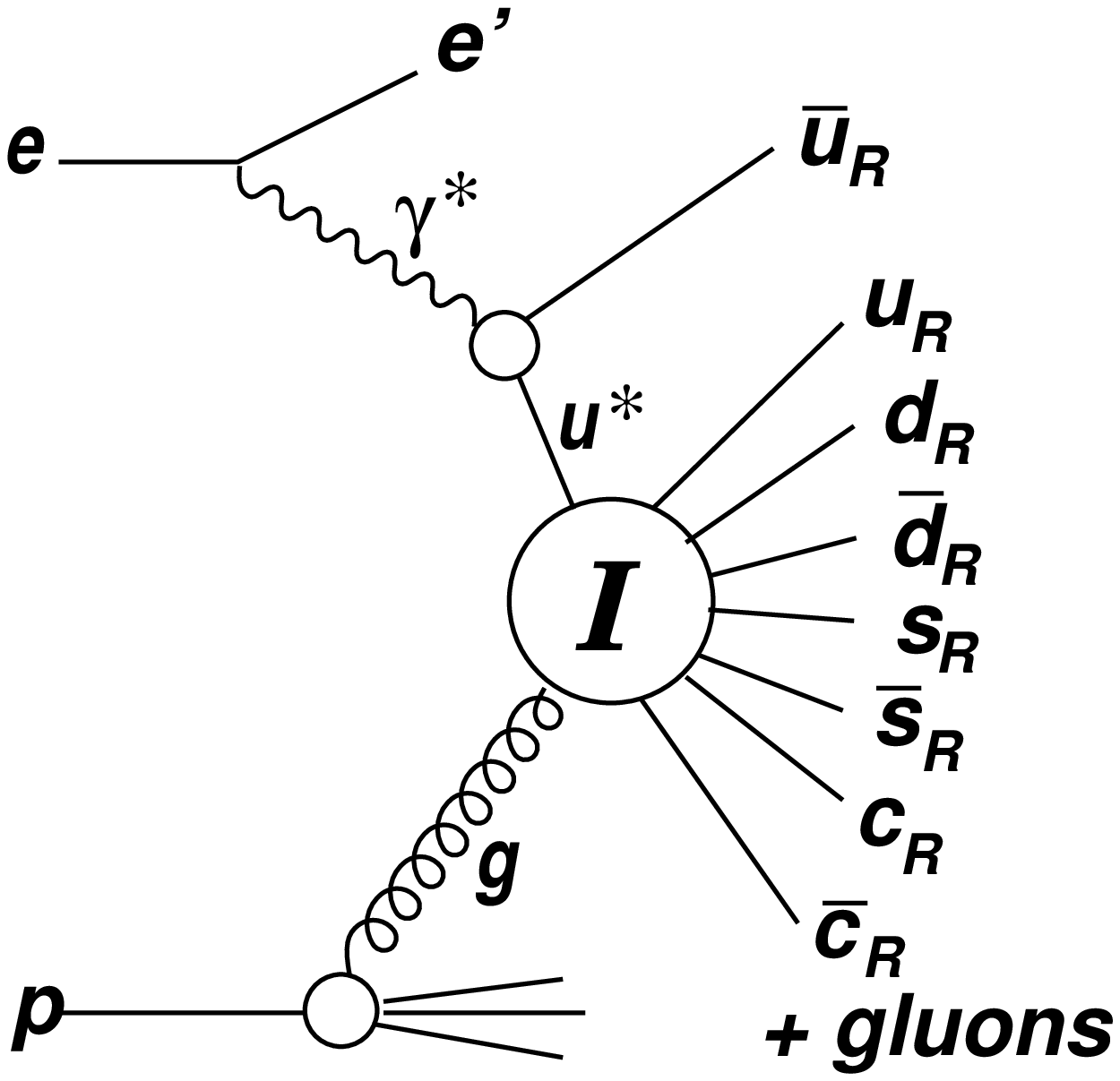,%
      width=6.0cm,bbllx=101pt,bblly=220pt,bburx=508pt,bbury=591,clip=}
\vspace{0.3cm}
   \scaption{
           {\bf a)} the electroweak interaction with $\Delta (B+L)=-6$ and in
             {\bf b)} the strong interaction with $\Delta Q_5 = 8$.}
   \label{instviol}
\end{figure}

In the electroweak theory, the minimal barrier height is
$E_B  \approx m_W/\alpha_w = \order{10 \TeVx}$.
Instanton transitions between vacua separated by $\Delta \ncs$
(see fig.~\ref{instviol}a for an example)
would violate baryon ($B$) and lepton numbers ($L=L_e+L_\mu+L_\tau$)
according to
\begin{equation}
 \Delta (B+L) = -2~ n_{\rm gener.} \cdot\Delta \ncs,
\end{equation}
but respect
\begin{equation}
  \Delta (B-L)=0 \hspace{2cm}
  \Delta L_e = \Delta L_\mu = \Delta L_\tau = \Delta B /3.
\end{equation}
 $n_{\rm gener.}=3$ is the number of fermion generations.

In instanton induced QCD reactions
(see fig.~\ref{instviol}b)
chirality is violated.
The chirality $Q_5$ is the difference between
the number of left- and right-handed fermions, $Q_5 = \#L -\#R$.
For $n_f$ active quark flavours, the selection rule is
\begin{equation}
 \Delta Q_5 = 2~ n_f \cdot \Delta \ncs.
\end{equation}
The minimal barrier height
is given by the hard scale of the process, e.g.
$E_B=\order{Q}$ for DIS \cite{inst:vladimir}.
The exponential suppression is less
severe than in the electroweak case, because $\alpha_s \gg \alpha_w$.

    \section{Instantons at HERA}                
\begin{figure}[h]
\begin{tabular}{ll}
\mbox{
 \epsfig{file=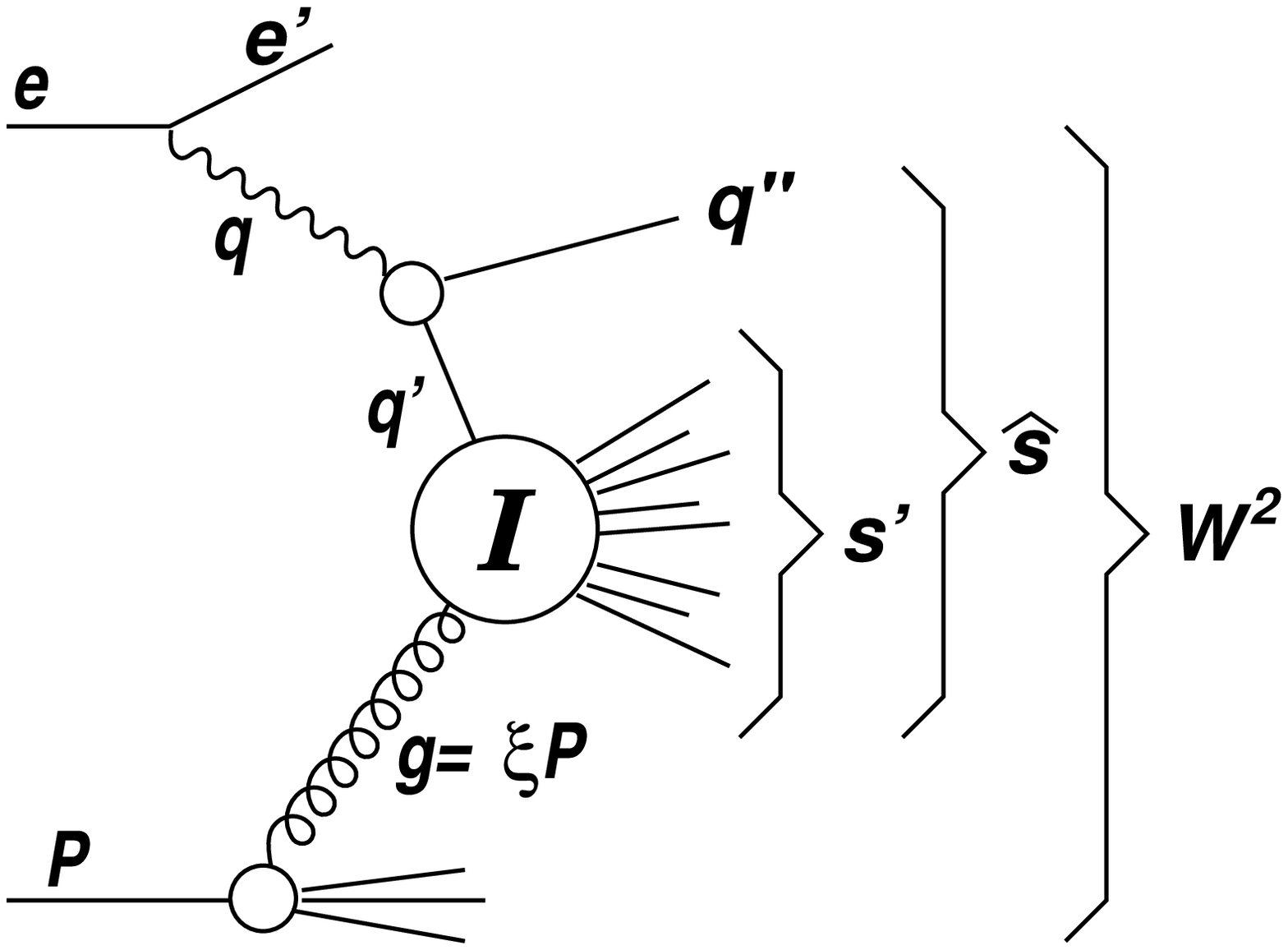,width=7.0cm,%
 bbllx=70pt,bblly=215pt,bburx=534pt,bbury=569,clip=}
}
&
\begin{tabular}{l}
 \vspace{-5.5cm} \\
 DIS variables: \\
   $\Qsq := - q^2 $ \\
   $ x := \Qsq / (2 P \cdot q) $ \\
   $ W^2 :=(q+P)^2 = \Qsq (1 - x)/x$ \\
   $ \hat{s} := (q+g)^2$ \\
   $ \xi = x (1+\hat{s}/Q^2)$ \\ \\
 Variables of instanton subprocess: \\
 $\qprimesq:=- q'^2 $ \\
 $x':= \qprimesq /(2 \; g \cdot q' ) $ \\
 $s':= (q'+g)^2 = \qprimesq ( 1 - \xprime )/ \xprime $
\end{tabular}
\end{tabular}
\vspace{0.3cm}
   \scaption{Kinematics of instanton induced processes in DIS.
    The labels denote the 4-vectors of the particles.
    A virtual photon $\gamma^\ast$ (4-momentum $q=e-e'$)
    emitted from the incoming electron fuses with
    a gluon (4-momentum $g$) from the proton (4-momentum $P$).
    The gluon carries a fraction $\xi$ of the proton momentum.
    The virtual quark $q^*$ entering
    the instanton subprocess has 4-momentum $q^\prime$, and the outgoing
    quark from the $\gamma^\ast\rightarrow q\ol{q}$ splitting
    has 4-momentum $q^{\prime\prime}$. The invariant masses squared of the
    $\gamma^\ast g$ and $q^*g$ systems are \shat~ and $s'$.
    $W$ is the
    invariant mass
    of the total hadronic system (the $\gamma^\ast p$ system).
    $0 \leq x \leq x/\xi \leq x' \leq 1$ holds.
    For completeness, we
    note $y:= (Pq)/(Pe) = Q^2/(sx)$, where $s=(e+P)^2$ is the $ep$
    invariant mass squared.
    }
   \label{fig:inst}
\end{figure}

In recent years, it has been
realized \cite{inst:vladimir,inst:yaroslavl,inst:moch,inst:schremppdis96,%
inst:ringwalddis97,inst:schremppdis97}
that quantitative
calculations are possible for processes induced
by QCD instantons in DIS due
to the presence of a hard
scale, $Q^2$.
In DIS, events due to QCD instantons $I$ (and anti-instantons \ol{I})
are predominantly produced
in a photon-gluon fusion
processes\footnote{Quark initiated processes have not yet been considered.
Due to the large gluon content of the proton
in the HERA domain at small $x$, they
are expected to be of minor importance.
In addition, they are expected to be suppressed by $\order{\alpha_s^2}$
with respect to the gluon initiated processes.}
~(Fig.~\ref{fig:inst})
\begin{equation}
  \gamma^\ast+g \stackrel{I}{\rightarrow}
            \sum_{n_f} (\ol{q}_R+q_R)+n_g g
 \hspace{1cm}
  \gamma^\ast+g \stackrel{\overline{I}}{\rightarrow}
            \sum_{n_f} (\ol{q}_L+q_L)+n_g g.
\end{equation}
In each event, quarks and antiquarks of all $n_f$ active flavours are found,
with $n_g$ gluons in addition.

The kinematics is depicted in fig.~\ref{fig:inst}.
The DIS variables Bjorken $x$ and $Q^2$
can be measured from the scattered electron, $q=e-e'$.
A measurement of the other variables is more challenging.
A measurement of the invariant
mass of the hadronic system, excluding the remnant, would determine \shat.
If the outgoing ``current jet'' could be identified and measured,
it's 4-momentum $q^{\prime\prime}$ would determine
$q'=q-q^{\prime\prime}$, and thus the
variables $x'$ and $Q'^2$ which characterise the instanton subprocess.
In practice, when not all of the five independent invariants
(for example ${x,Q^2,x',Q'^2,\hat{s}}$) can be measured, they are being
integrated over.

The instanton induced cross section is given by a convolution of the
probability to find a gluon in the proton $P_{g/p}$,
the probability that the virtual photon splits
into a quark-antiquark pair in the instanton background
$P^{(I)}_{q^*/\gamma^*}$, and
the cross section
$\sigma^{(I)}_{q^*g}(\xprimex,\qprimesqx)$ of the instanton
subprocess \cite{inst:yaroslavl,inst:moch}.
Multi-gluon emission enhances the cross section \cite{inst:ringwald}
\begin{equation}
\sigma_{q^*g;n_g}^{(I)} \propto \frac{1}{n_g!}
                 \left(\frac{1}{\alpha_s}\right)^{n_g}
                \exp(-4\pi/\alpha_s).
\end{equation}
The cross section of the instanton induced subprocess is
then \cite{inst:yaroslavl}:
\begin{equation}
 \sigma_{q^*g}^{(I)}(\xprimex,\qprimesqx)
 = \sum_{n_g=0}^{\infty}\sigma_{q^*g;n_g}^{(I)}
 \approx
 \frac{\Sigma(\xprimex)}{\qprimesqx}
{ \left(\frac{4 \pi}{\alpha_s(\mu(\qprimex))}\right)}^{\frac{21}{2}}
 {\rm exp} \left(\frac{- 4 \pi}{\alpha_s(\mu(\qprimex))} F(\xprimex) \right).
\label{eq:cross}
\end{equation}
It depends critically on the functions
$F(\xprimex)$ (called the ``holy grail'' function),
which modifies the exponent in the suppression factor
$\exp(-4 \pi /\alpha_s)$,
and on $\Sigma(\xprimex)$, which depends on $F(\xprimex)$.
There exists also a scale dependence due to the choice
of the renormalization scale $\mu(\qprimex)$.

\newpage

\begin{wrapfigure}{r}{8cm}
\mbox{
   \epsfig{file=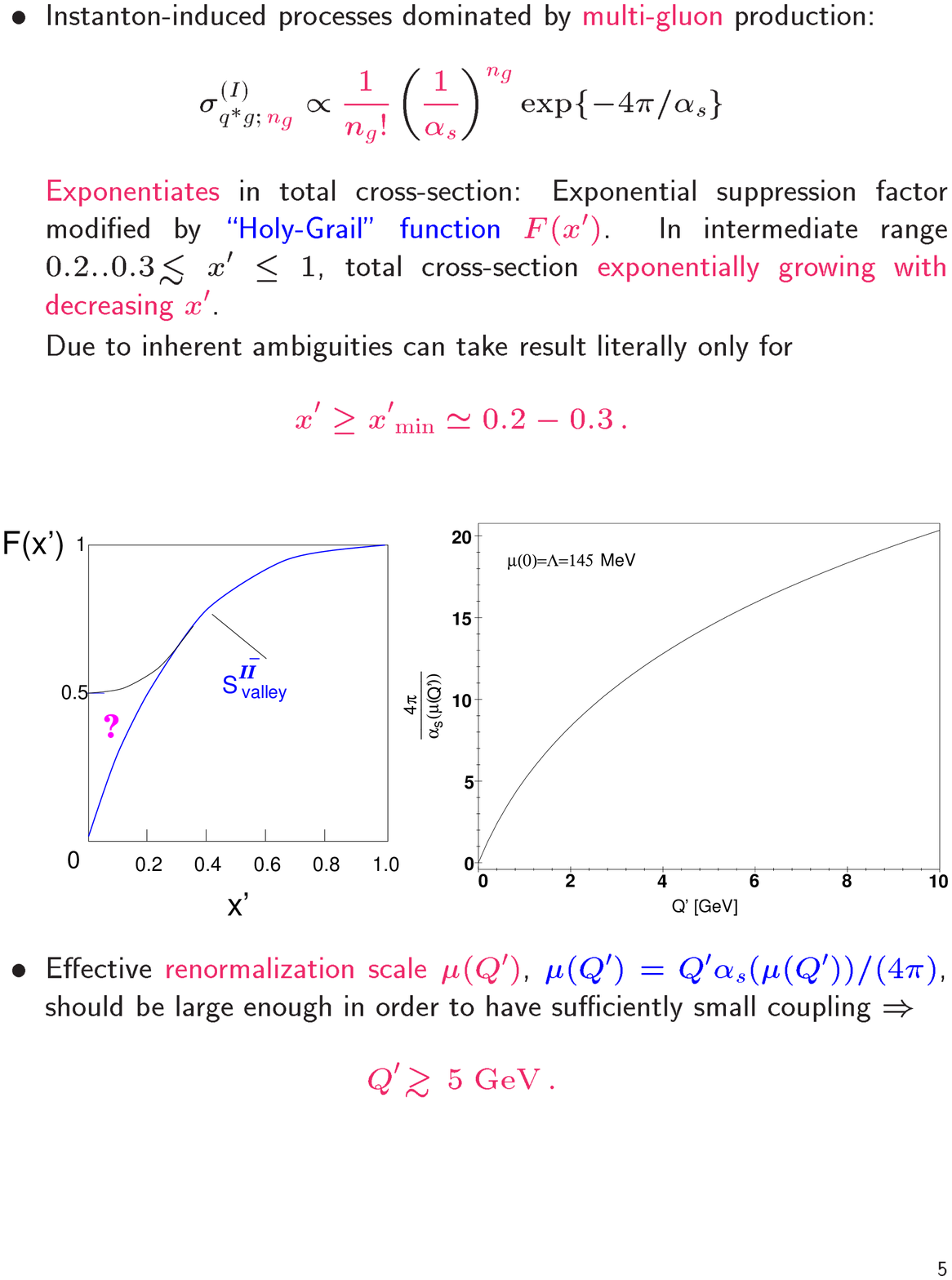,%
      width=5cm,bbllx=12pt,bblly=253pt,bburx=250pt,bbury=484,clip=}
     }
   \bild{8cm}{The holy grail function $F(x')$ \cite{inst:yaroslavl}.
              For small $s'$ ($x'\approx 1$), instanton perturbation
              theory is applied. The calculation with the valley
              method matches smoothly with the perturbative result.}
   \label{grail}
\end{wrapfigure}
$F(\xprimex)$
can be estimated reasonably well  (see fig.~\ref{grail})
for \xprime not
too small,
$\xprime \gtrsim 0.2$ \cite{inst:yaroslavl}.
The extrapolation to lower
values of \xprime is unreliable due to
inherent ambiguities.
In addition, multi-instanton effects
should be avoided
by limiting the instanton size $\rho_I$
(the spatial region occupied during the interaction)
to $\rho_I<2 {\rm ~GeV}^{-1}$ with
a cut-off
$\qprimesq \gtrsim 25 \GeVsq$ \cite{inst:yaroslavl,inst:moch}.
That requirement ensures also that $\alpha_s(\mu(\qprimex))$
stays small enough to apply instanton perturbation theory.

\begin{figure}[htb]
   \centering
 \begin{picture}(0,0) \put(0,0){{\bf a)}} \end{picture}
   \epsfig{file=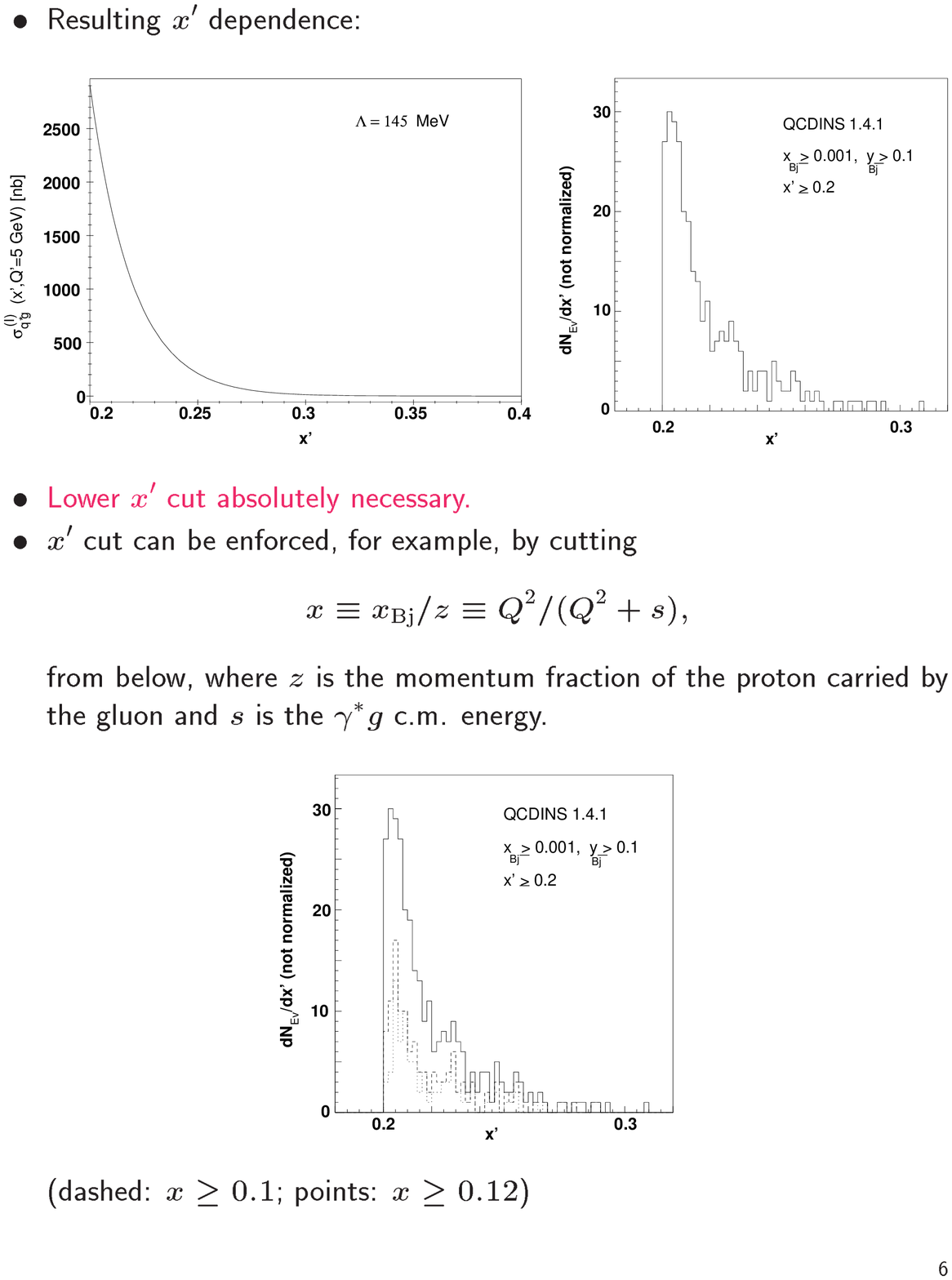,%
      width=7.3cm,bbllx=16pt,bblly=520pt,bburx=330pt,bbury=747,clip=}
 \begin{picture}(0,0) \put(0,0){{\bf b)}} \end{picture}
   \epsfig{file=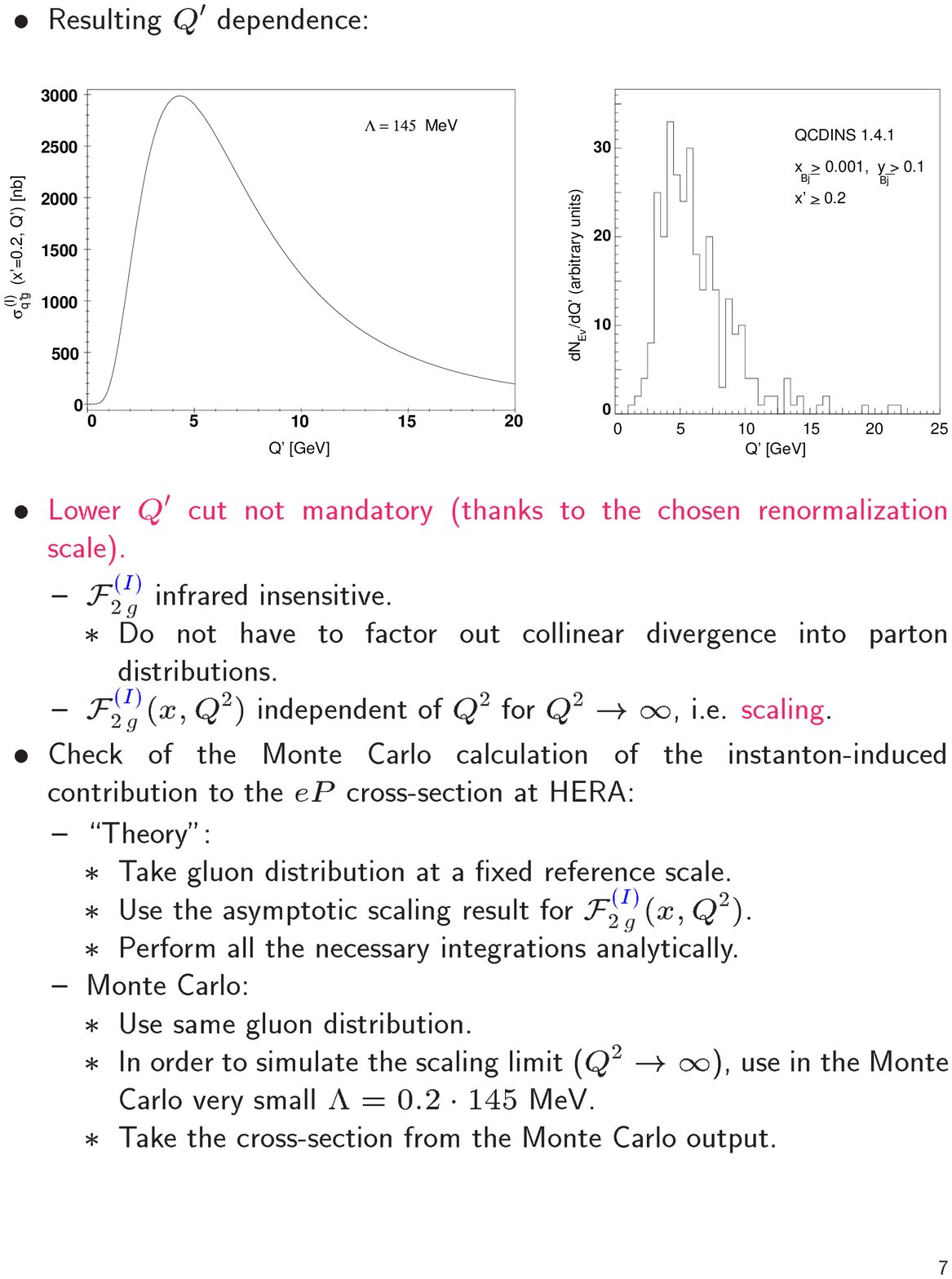,%
      width=7.3cm,bbllx=16pt,bblly=520pt,bburx=330pt,bbury=747,clip=}
   \scaption{
                The instanton subprocess
                cross section \cite{inst:ringwalddis97}
                for $q^\ast g \rightarrow$~hadrons
                as a function of
              {\bf a)} $x'$ for $Q'=5 \GeV$ and
              {\bf b)} $Q'$ for $x'=0.2$.}
   \label{sigsub}
\end{figure}

\begin{figure}[htb]
   \centering
   \epsfig{file=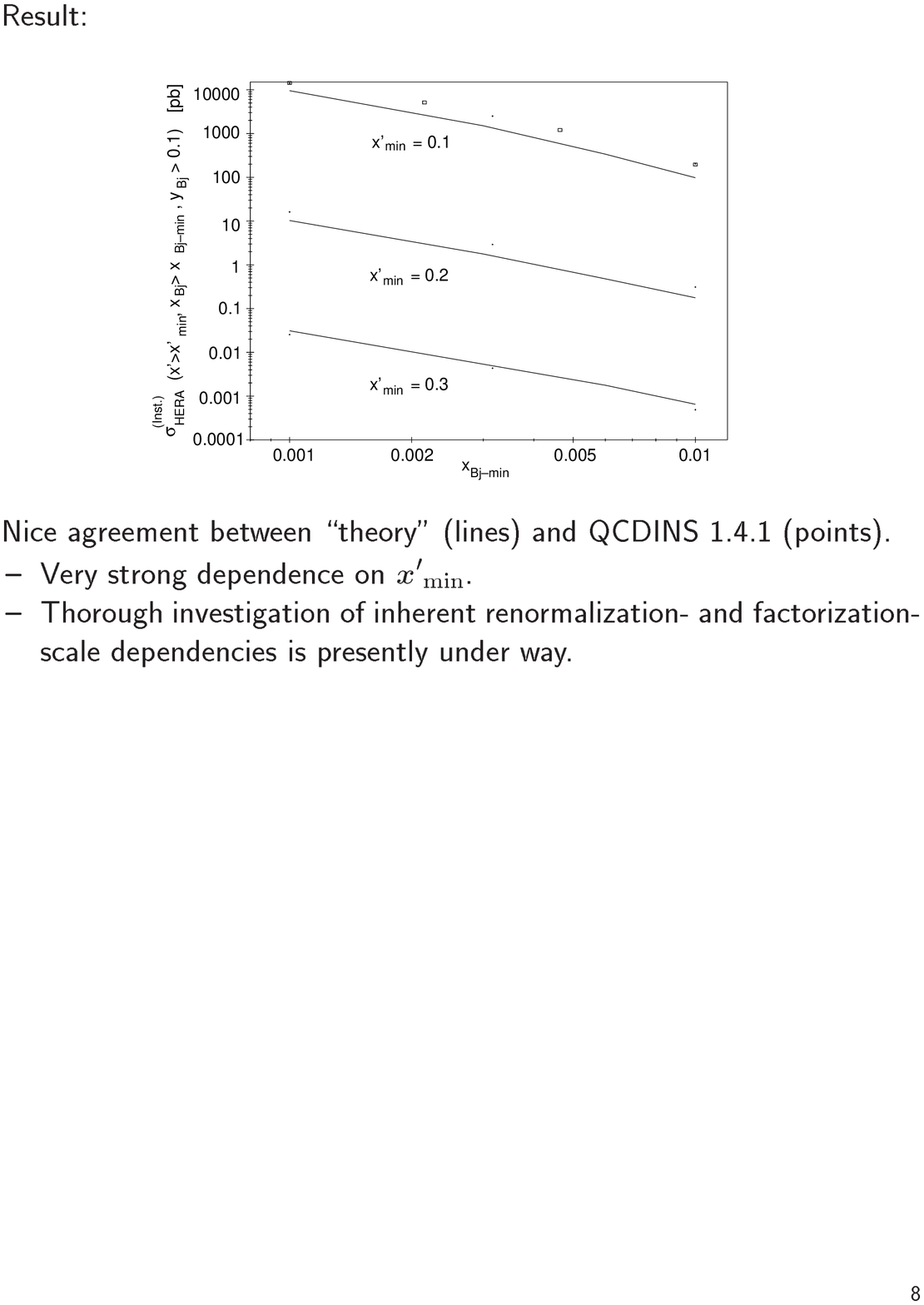,%
      width=10cm,bbllx=126pt,bblly=516pt,bburx=465pt,bbury=750,clip=}
\vspace{0.3cm}
   \scaption{
                The instanton induced DIS cross
                section \cite{inst:ringwalddis97}
                $ep \rightarrow e'X$ integrated
                over Bjorken $x>x_{\rm min}$, $y>0.1$, and over
                the regions $Q'>5\GeV$ and $x'>x'_{\rm min}$
                as indicated.}
   \label{sigi}
\end{figure}

The resulting instanton induced subprocess cross section
$\sigma_{q^*g}^{(I)}(\xprimex,\qprimesqx)$
(see fig.~\ref{sigsub})
is peaked at $Q'\approx 5 \GeV$
and exponentially grows with decreasing \xprimex.
The integrated instanton induced
$ep$ DIS cross section (see fig.~\ref{sigi})
is sizeable;
for $x>0.001$ and $x'>0.2$ it is of \order{10~\pbx}.
The cross section is approximately
scaling (depends only on $x$, not on \Qsq for large \Qsq) \cite{inst:moch}.
It grows towards small $x$, and increases dramatically
when the lower $x'$ cut-off is relaxed.
Eventually
higher order instanton effects have to dampen the growth of
the cross section.

Two kinematic regions have to be distinguished.
For $x'>0.2$ the predictions are relatively safe,
allowing the instanton theory to be tested. Either instantons are
discovered at the predicted level
-- including the substantial theoretical uncertainties, which still need
   to be quantified --,
or the theory has to be revised.
For $x'<0.2$ the cross section presumably continues to grow, but the
extrapolation is extremely uncertain.
For a discovery, this is the favourable region
due to the large cross section.
A negative result however cannot be turned against the theory,
it would rather restrict the unknown behaviour of $F(x')$ at small $x'$.
Most promising is the kinematic region of small Bjorken-$x$,
because both the total DIS cross section and the predicted
fraction of instanton induced events increase towards small $x$
(see fig.~\ref{dpdn}b).

    \section{Experimental Signatures}           
In the theoretically safe region, $x'>0.2$,
the expected fraction of instanton events in all DIS events is
of \order{10^{-3} - 10^{-4}} (compare fig.\ref{dpdn}b),
too small to be detected in
inclusive cross section measurements
(i.e. the structure function $F_2$).
Instead, dedicated searches for the characteristic features
of instanton events in the hadronic final state have to be performed.
A Monte Carlo generator (QCDINS \cite{inst:qcdins})
to simulate the hadronic final state of
instanton events in DIS is available.
In general, the event shape predictions are more stable than the
rate predictions, because poorly known factors cancel.
The instanton event properties can be contrasted with predictions
from event generators for normal DIS events
(ARIADNE \cite{mc:ariadne}, LEPTO \cite{mc:lepto} and HERWIG \cite{mc:herwig})
which give an overall satisfactory
description of the DIS final state properties \cite{mc:heratune}.

In the $q^\ast g$
rest frame
$2 n_f-1$ quark and antiquarks
and $n_g$ gluons are emitted isotropically
from the instanton subprocess.
$n_g$ is Poisson distributed with \cite{inst:moch,inst:ringwalddis97}
\begin{equation}
  \av{n_g} \approx \frac{2\pi}{\alpha_s} x' (1-x')
  \frac{\dd F(x')}{\dd x'}.
\end{equation}
After hadronization, this leads to a spherical system with a high
multiplicity of hadrons, depending mainly
on the available centre of mass energy
$\sqrt{s'} = Q'\sqrt{1/x'-1}$.
For a typical situation ($x'=0.2,Q'=5\GeV \Rightarrow \sqrt{s'}=10~\GeVx$),
$\av{n_g} = \order{2}$.
About $n_p=10$ partons
and $n=20$ hadrons are expected. The expected parton momentum spectrum
is semi-hard \cite{inst:vladimir} with transverse momentum
$\av{p_T} \approx (\pi/4)(\sqrt{s'}/\av{n_p})$.

Hadronic final state properties are conveniently
being studied in the centre of mass system (CMS)
of the incoming proton and the virtual boson, i.e. the CMS of
the hadronic final state.
Longitudinal and transverse quantities are calculated
with respect to the virtual boson direction (defining the $+z$ direction).
The pseudorapidity
$\eta$ is defined as $\eta= - \ln \tan (\theta/2) $, where
$\theta$ is the angle with respect to the virtual
photon direction.
When boosted to the CMS,
the hadrons emerging from the instanton subprocess occupy a
band in pseudorapidity of half width $\Delta \eta \approx 1$,
which is homogeneously populated in azimuth \cite{inst:vladimir}.

The characteristics of instanton events by which they
can be distinguished from normal DIS events are therefore:
high multiplicity with large transverse energy; spherical
event configuration (apart from the current jet);
and the presence of all flavours (twice!)
that are kinematically allowed in each event.
One would therefore look for events which in addition to
the other characteristics are rich in $K^0$s, charm decays,
secondary vertices,
muons etc..
In general, the strength of instanton signals in the hadronic
final state increases somewhat towards low $x'$ and large $Q'^2$
due to the increasing ``instanton mass'' $\sqrt{s'} = Q'\sqrt{1/x'-1}$.

\begin{figure}[h]
   \centering
 \begin{picture}(0,0) \put(0,0){{\bf a)}} \end{picture}
 \epsfig{%
    file=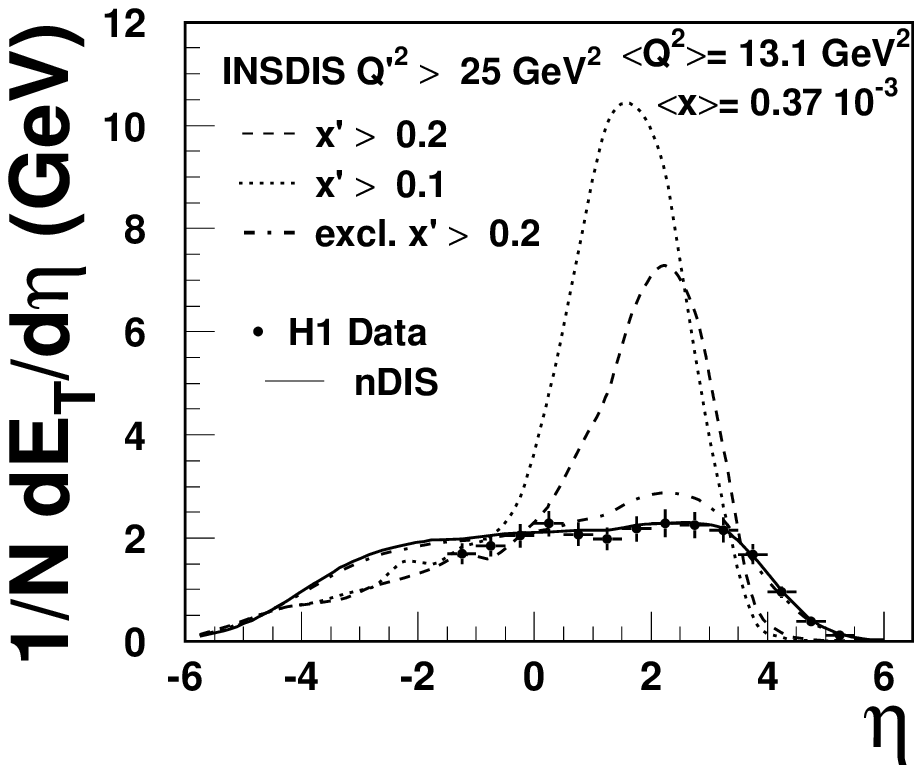,%
    width=7cm}%
 \begin{picture}(0,0) \put(0,0){{\bf b)}} \end{picture}
 \epsfig{file=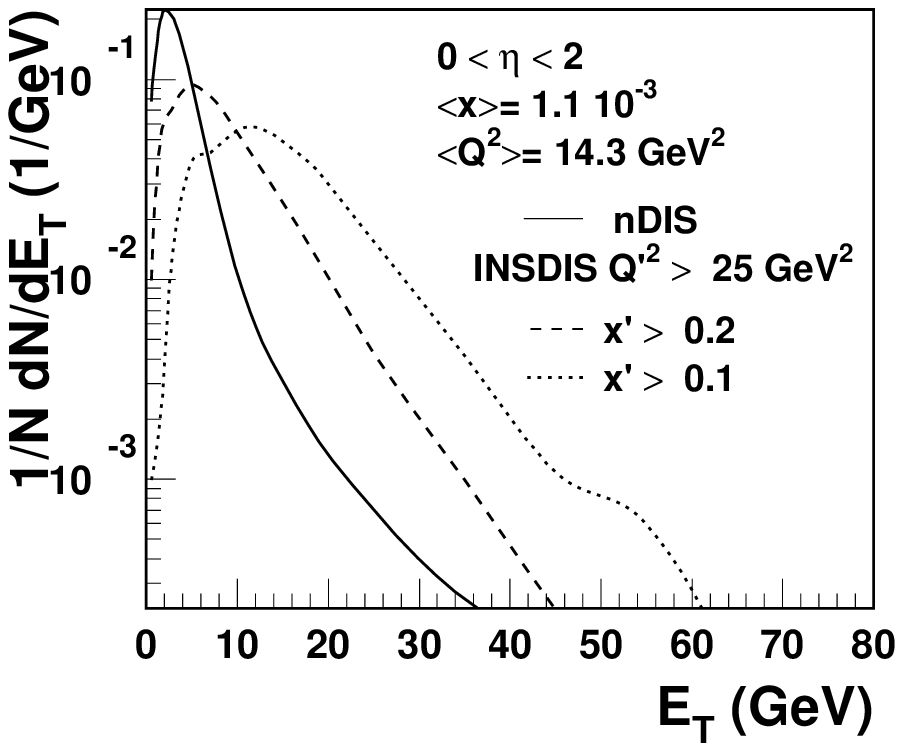,%
  width=7cm}%
\vspace{0.3cm}
   \scaption{Transverse energy \et in the hadronic CMS \cite{inst:bounds}.
             {\bf a)} The \et flow vs. $\eta$.
             The proton remnant direction is to the left.
             The standard QCD model (nDIS=ARIADNE) and
             different instanton scenarios are confronted with
             the H1 data \cite{h1:flow3}.
             The excluded scenario \cite{inst:bounds}
             with an instanton fraction $f_I> 11.8\%$ for
             $\xprime > 0.2$ is indicated.
             {\bf b)}
             The \et distribution, where the transverse energy
             is measured in the CMS rapidity bin $0<\eta<2$,
             for two instanton scenarios, and the standard QCD model
             (nDIS=ARIADNE).
             The plots are normalized to the
             total number of events $N$ that enter the distributions.
            }
   \label{flows}
\end{figure}

The ``instanton band'' shows up in
the flow of hadronic transverse energy \et as a function of
$\eta$
(fig.~\ref{flows}a).
It's height and position depends on \xprime and \qprimesq
(and also on \xb and \Qsq).
In normal DIS events on average an \et of $2\GeV$ per $\eta$ unit
is observed. In instanton induced events, that number may go up
to $10$~\GeV for low \xprimex.
A possible search strategy could involve
the \et distribution in a selected rapidity band (fig.~\ref{flows}b),
looking for high \et events in the
tail of the distribution \cite{inst:heraws96,inst:bounds}.

\begin{figure}[htb]
   \centering
   \epsfig{file=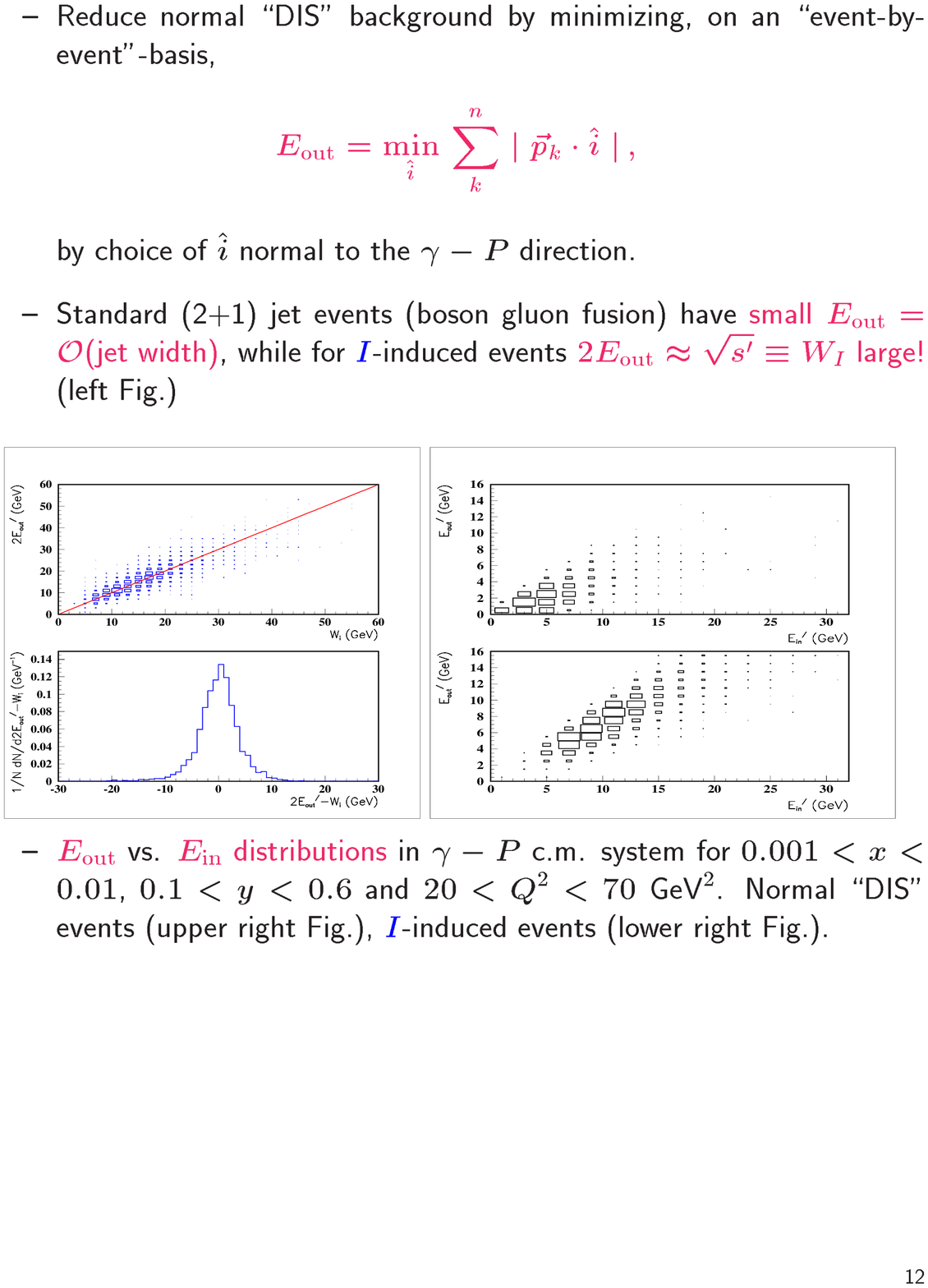,%
      width=14.5cm,bbllx=35pt,bblly=316pt,bburx=550pt,bbury=534,clip=}
   \scaption{
             {\bf a)}
                The correlation between $2\cdot\eoutp$ and
                the ``instanton mass'', $W_I=\sqrt{s'}$ (top),
                and the resolution for $\sqrt{s'}$ that can be achieved
                (bottom) \cite{inst:heraws96}.
                The primes indicate
                additional cuts in $\eta$ to minimize higher order
                QCD radiation which may wash out the relation
                between \eout and $\sqrt{s'}$.
             {\bf b)}~$\eoutp$ vs. $\einp$ for
                normal (top, HERWIG) and instanton
                induced events (bottom, QCDINS) \cite{inst:heraws96}.
                Both distributions are taken in the hadronic CMS
                for events with $0.001<x<0.01$, $0.1<y<0.6$ and
                $20 \GeVsq < Q^2 < 70 \GeVsq$.
             }
   \label{eti}
\end{figure}

Further discrimination can be obtained \cite{inst:heraws96}
from the fact that
for instanton events the \et should be distributed isotropically,
while normal DIS events are jet-like, in particular for large \et.
One defines
\begin{equation}
\eout  := \min \sum_{i} \left|\vec{p}_i \cdot \vec{n}\right| \hspace{2cm}
\ein   := \sum_{i} \left|\vec{p}_i \cdot \vec{n}' \right|.
\end{equation}
The sum runs over all final state hadrons $i$ with momentum $\vec{p}_i$.
$\vec{n}$ is the unit vector perpendicular to the virtual photon axis which
minimizes \eout and thus defines the event plane.
$\vec{n}'$ lies in the event plane and
is normal to both $\vec{n}$ and the virtual photon axis.
It is easy to show that for an ideal isotropic ``instanton decay'',
$\eout=\sqrt{s'}/2$ \cite{inst:heraws96}.
The ``instanton mass'' $\sqrt{s'}$ can thus be
reconstructed experimentally (fig.~\ref{eti}a).
Normal DIS events, either ``1+1'' or ``2+1'' jet events (the +1 refers
to the unobserved proton remnant) are contained in the event plane,
$\eout\ll\ein$, in contrast to instanton events with $\eout \approx \ein$
(see fig.~\ref{eti}b).

Instanton events are characterized by
a large particle density localized in rapidity. In normal
DIS events there are about 2 charged particles per unit of
pseudorapidity \cite{h1:pt}, rather uniformly distributed in $\eta$.
For a low \xprime cut-off, that number goes up to 10
in the peak of the instanton band \cite{inst:bounds}.
Very sensitive to instanton events is the charged particle
multiplicity distribution \cite{inst:bounds}, see fig.~{\ref{dpdn}a.
A significant fraction of the instanton events would
lead to charged multiplicities which are very
unlikely to be found in normal DIS events.
Furthermore, particle-particle correlation functions should
be influenced by instanton effects \cite{inst:kuvshinov}.

\begin{figure}[htb]
   \centering
\begin{picture}(0,0) \put(0,0){{\bf a)}} \end{picture}
\epsfig{
    file=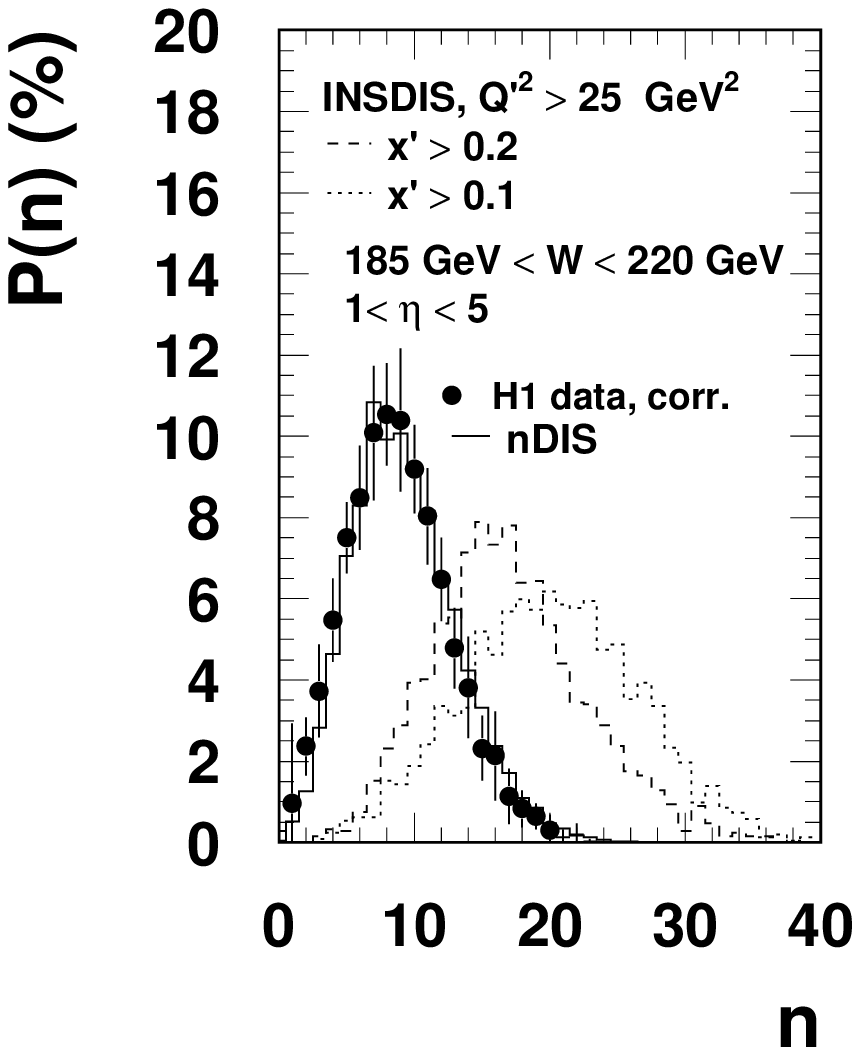,
    width=6cm,bbllx=62pt,bblly=436pt,bburx=310pt,bbury=742,clip=}
\begin{picture}(0,0) \put(0,0){{\bf b)}} \end{picture}
\epsfig{width=7.5cm,
   file=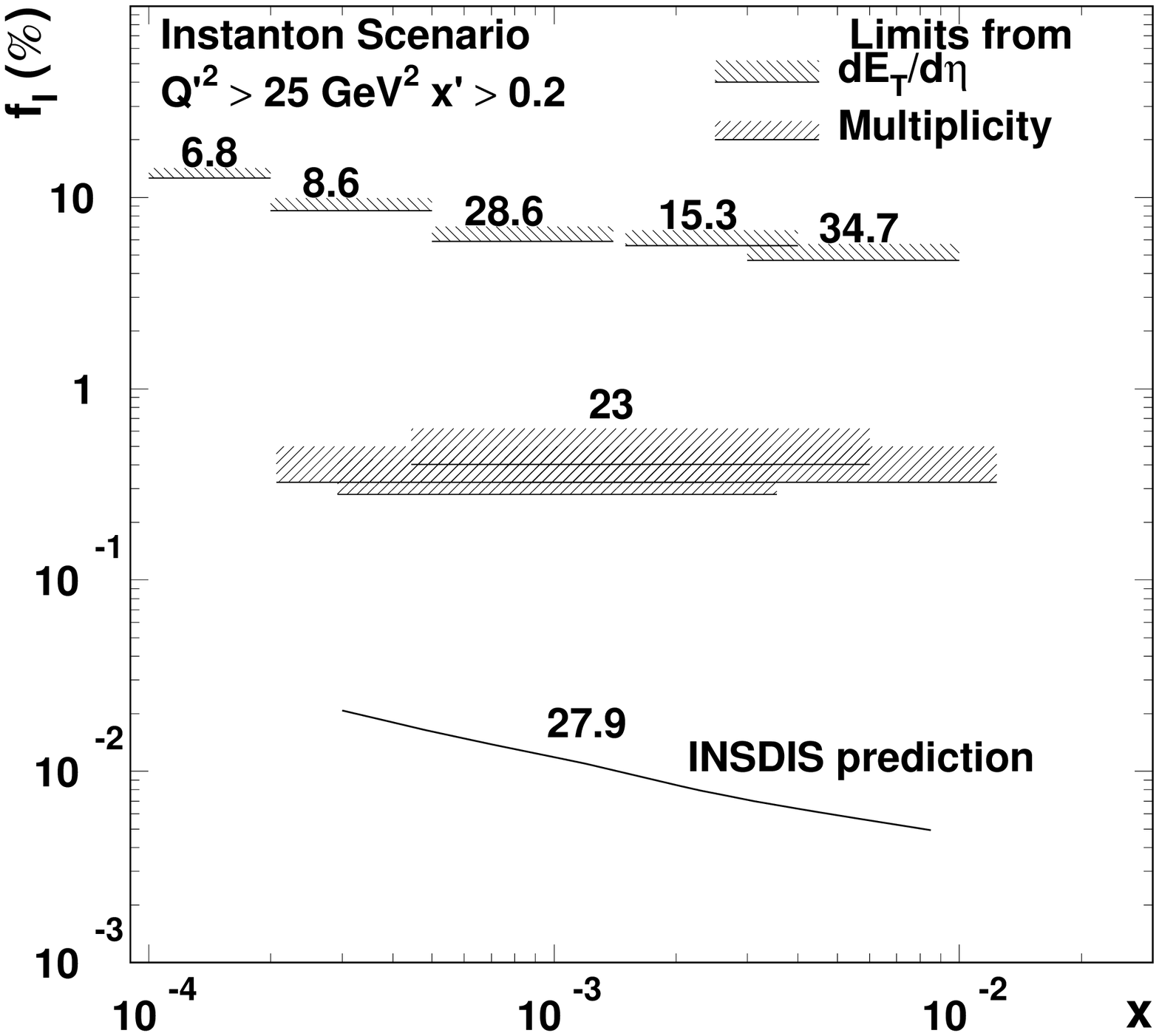,
   bbllx=0pt,bblly=-60pt,bburx=554pt,bbury=518,clip=}
\vspace{0.3cm}
   \scaption{
            {\bf a)}  The
              probability distribution $P(n)$ of the charged particle
              multiplicity $n$
              from the CMS pseudorapidity range $1<\eta<5$ for
              events with $185 \GeV < W < 220~\GeVx$.
              Shown are the unfolded H1 data \cite{h1:mult},
              the expectation from a standard
              DIS model (nDIS=ARIADNE)),
              and the predictions for instanton events with
              different cut-off scenarios \cite{inst:bounds}.
      {\bf b)}
      The maximally allowed fraction $f_{\rm lim}$
      of instanton induced events in DIS for
      $\qprimesq > 25$\GeVsq and $\xprime > 0.2$
      from transverse energy flows and the multiplicity distribution
      as function of $x$ \cite{inst:bounds}.
      Regions above the lines are excluded
      at 95\% C.L..
      The numbers give the average \Qsq values in \GeVsq for
      the $x$ bins.
      The theory prediction, calculated with QCDINS \cite{inst:qcdins},
      for $10 \GeVsq < \Qsq < 80$~\GeVsq
      is superimposed (full line, label INSDIS).
            }
   \label{dpdn}
\end{figure}

    \section{Searches for Instanton Processes}  
The fact that instanton events look very different from the expectation for
standard QCD events can be exploited to search for instanton signals
in the HERA data. One strategy is to compare the shape of
hadronic final state distributions to the expectation from
standard QCD events (nDIS) with an admixture of instanton events (INSDIS)
of fraction \finst. In case the measured distribution agrees with
the standard QCD expectation, a limit on the fraction of instanton
induced events in DIS $\finst<\flim$ can be set.
The caveat of this method is that one has to make an assumption
on what standard QCD looks like.
In particular at small $x$ that issue is
under debate
\cite{mk:madrid,lowx:ringcarli,lowx:ringgrindh},
see section \ref{sn:jrates} and chapter \ref{ch:lowx}.
There exists a danger that an instanton effect is tuned or explained
away by stretching the standard QCD predictions
by generator tuning, introducing BFKL effects etc..
A good understanding of standard QCD will therefore be crucial for the
positive identification of instanton effects.

In the first search for instanton events \cite{h1:k0} an anomalous
$K^0$ yield has been looked for.
For $x>10^{-3}$ about 0.12 $K^0$
mesons (including $\ol{K^0}$) have been measured per event and
unit pseudorapidity, with a relatively flat $\eta$ distribution.
For instanton events with $x'>0.2$,
a peaked distribution with
about 0.55 $K^0$ per event and unit $\eta$ is expected.
From the comparison with
standard QCD event generators \cite{mc:ariadne,mc:lepto,mc:herwig}
and the instanton generator \cite{inst:qcdins}
an upper limit of $f_I<\flim =6\%$ at 95\% C.L.
is obtained \cite{h1:k0}.

The charged particle multiplicity distribution $P(n)$ in high energy
reactions can often by described by a negative binomial distribution (NBD).
Also the DIS data are relatively well described by NBDs \cite{h1:mult}.
The multiplicity distribution
from the CMS interval $1<\eta<5$ for events with $W=80-115\GeV$
(corresponding to $x>0.0007$) can be parametrized
with an NBD of mean $\av{n}= 6.90 \pm 0.33$. Possible
deviations from an NBD allow for an
instanton fraction of at most \finst=2.7\% at 95\% C.L. \cite{h1:mult}.


Other measured event shapes have been systematically
analysed in terms of their sensitivity
to instanton events \cite{inst:bounds}, and their dependence on the
kinematic variables $x,Q^2,x',Q'^2$.
The most sensitive distributions were
the transverse energy flows \cite{h1:flow3},
the pseudorapidity distribution of charged particles and their
\pt spectra \cite{h1:pt}.
For example, the \et flow
has been measured over a wide range of $x$ and $Q^2$, allowing to extend
the search region down to $x=0.0001$.
From a shape analysis \cite{inst:bounds}
(see fig.~\ref{flows}),
instanton fractions \finst between 5 and 13 \%
can be excluded for $x'>0.2$ (see figs.~\ref{dpdn}b, \ref{limplane}).
For lower $x'$ the signal is more prominent, and somewhat
better limits are obtained.

\begin{figure}[htb]
 \centering
 \epsfig{%
   file=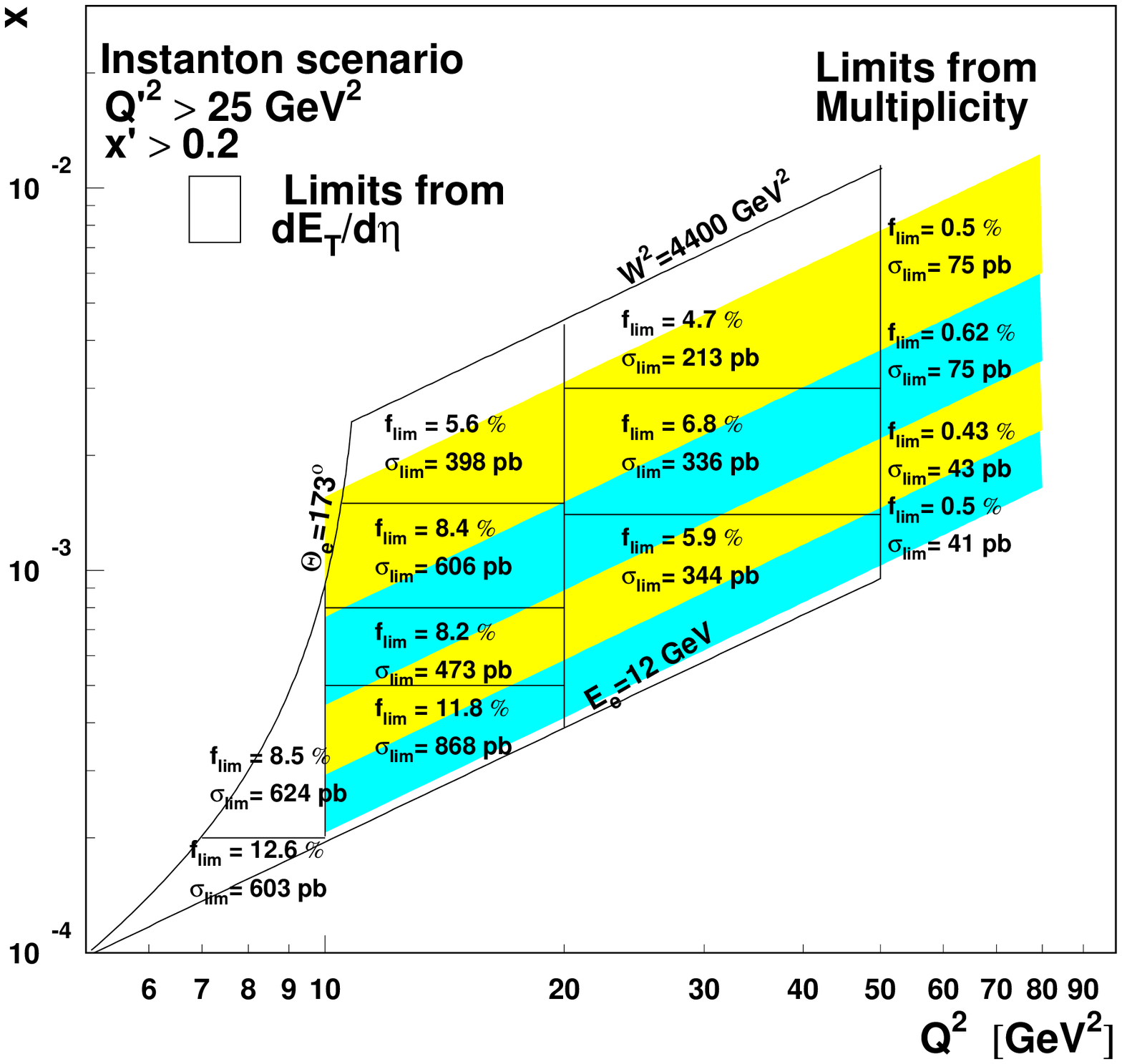,%
   width=12cm,bbllx=28pt,bblly=142pt,bburx=561pt,bbury=668,clip=}
 \scaption{
      95\% C.L. limits on instanton production with $\qprimesq > 25\GeVsq$ and
      $\xprime > 0.2$ \cite{inst:bounds}.
      The cross-section limits ($\sigma_{\rm lim}$)
      together with the maximally allowed instanton fraction
      $f_{\rm lim}$ are shown in the \xb--\Qsq plane.
      They are obtained from the \et flow analysis
      (open fields),
      and from the multiplicity analysis (shaded fields) with their
      numbers at the right edge.
      The boundaries implied by the analysis cuts of the energy flow
      analysis in the angle and energy of the scattered electron,
      $\theta_e < 173^\circ$
      and $E_e>12$ GeV,
      and by the requirement
      $W^2>4400 \GeVsq$ are indicated.}
 \label{limplane}
\end{figure}

The fact that H1 \cite{h1:mult} did not observe any events above a certain
multiplicity $\nmax$ has been exploited \cite{inst:bounds}
to place more stringent
limits on instanton production\footnote{The previous limits from the
H1 multiplicity analysis \cite{h1:mult}
were derived from the shape of the multiplicity
distribution for $n<\nmax$.}.
A significant fraction of instanton induced events would have
multiplicities $n>\nmax$ (compare fig.~\ref{dpdn}a).
Instanton fractions $\finst>0.4-0.6\%$
can therefore be excluded for $x'>0.2$
(see figs.~\ref{dpdn}b,~\ref{limplane}), and somewhat
lower \finst values for a lower cut-off
\mbox{$x'>0.1$ \cite{inst:bounds}.}
This search method has the advantage that, in contrast to the
previous shape comparisons, it does not rely on assumptions for standard
QCD event topologies, since no background needs to be subtracted.
Unavoidable of course is the dependence on the expected
instanton event shape, which may be even more uncertain than
the standard QCD event shapes.

\begin{footnotesize}
\begin{table}[tbh]
\label{tab:instlim}
\begin{tabular}{|c|c|c|c|c|c|c|}
\hline
analysis & \multicolumn{3}{c|}{DIS kinematics covered} &
           \multicolumn{2}{c|}{instanton scenario}     & limit \\
\hline
         & \Qsq (\GeVsqx) & $x$ & \W (GeV) & $Q'^2$(\GeVsqx) & $x'$ & \flim  \\
\hline
$K^0$ \cite{h1:k0} & 10 -- 70 & 0.001~ -- 0.01~ & 95 -- 230 &
               $\gtrsim 1$ &  $\gtrsim 0.2$ & 6 \% \\
multipl. \cite{h1:mult} & 10 -- 80 & 0.0007 -- 0.012 & 80 -- 115 &
                $\gtrsim 1$ &  $\gtrsim 0.2$ & 2.7 \% \\
\et flows \cite{inst:bounds} & ~5 -- 50 & 0.0001 -- 0.01~ & 65 -- 230 &
                   $>25$     & $>0.2$  & 5 -- 13 \% \\
multipl. \cite{inst:bounds} & 10 -- 80 & 0.0001 -- 0.01~ & 80 -- 220 &
                   $>25$     & $>0.2$  & 0.4 -- 0.6 \% \\
\hline
\end{tabular}
\scaption{Limits on QCD instantons in DIS. A fraction $\finst>\flim$
 of instanton induced events in DIS is excluded at 95\% C.L..}
\end{table}
\end{footnotesize}

The available bounds on instanton production are summarised in
tab.~\ref{tab:instlim}. The most stringent limits
for the theoretically ``safe'' scenario $x'>0.2$
are still a
factor 20 higher than what is predicted from the instanton theory,
see fig.~\ref{dpdn}b.
Limits for other scenarios can be found in \cite{inst:bounds}.
For $x'>0.1$ they are already below
the naive extrapolation into the
theoretically uncertain region, providing a constraint
for the theory and the holy grail function $F(x')$.

    \section{Conclusion}                        Instanton transitions,
a yet unexplored facette of non-abelian gauge field theories, have been
discussed. While in the electroweak theory the $B+L$ violating effects
induced by instantons
are expected only at energies $\gtrsim 10\TeV$, their chirality violating
pendant in QCD could lead to striking signatures already at present day
colliders. In DIS at HERA, these are a high particle multiplicity
with large transverse energy localized in rapidity,
and $s$, $c$ and possibly $b$ quarks
in the final state.

The expected contribution to DIS events from instantons
is of $\order{10^{-3}-10^{-4}}$,
with substantial theoretical uncertainties.
First analysis of HERA data taken in the years $\leq 1994$, corresponding
to an integrated luminosity of $\order{1.3\pbinv}$, are still a factor
$\approx 20$ above the prediction. With higher statistics data samples
($\order{40\pbinv}$ up to the end of 1997)
and improved search strategies, a fundamental discovery at HERA
appears to be in reach.

As the study of the hadronic
final state at HERA has shown, model predictions for normal DIS final states
are not unique.
It will probably be safe to take some uncertainty into account also for
the prediction of the instanton final state.
A good understanding of QCD in DIS will be vital for instanton searches.

It might be possible to exploit also other
reactions than DIS, such as photoproduction, where the hard scale needed for
reliable instanton calculations could be provided by the \pt of a jet.
Instantons have even be proposed \cite{inst:kochelev} as an explanation
of the possible excess of high \Qsq events at HERA
\cite{h1:highq2,z:highq2,h1:highq2update}.

\chapter{Summary}              

HERA has entered a new kinematic regime in deep inelastic scattering:
at small $x$, large $W$ and both small and large $Q^2$.
After five years of HERA operation,
the hadronic final state has been measured in great detail.
The large body of data has been confronted with QCD, the theory of the
strong interaction.

We have seen progress happening at different stages:
\begin{itemize}
\item QCD predictions are tested experimentally, and parameters
      of the theory are being extracted. An example are the
      measurements of jet rates and the extraction of \as or of the
      gluon density in the proton.

\item Where the theory is less well established,
      there is an interplay between experimental data and theory,
      in which new ideas are born and developed (or rejected).
      Examples are soft colour interactions as an explanation for
      rapidity gaps and large \et flows, or power corrections
      for hadronization effects.

\item The data are not yet understood. Either there exist no
      predictions, or there are conflicting explanations.
      An example are the measured forward jet rates, which
      could be explained by BFKL evolution, or by hadronziation
      effects, or by a resolved structure of the virtual photon.
\end{itemize}

The availability of as much data as possible
over a wide kinematic range
is very important
for the scientific process to work efficiently. A new idea which explains
one aspect of the data can be rejected early if it fails in another.
The data have severely constrained QCD models for the hadronic final
state which varied widely before the data had become available.

\subsubsection{General event properties in deep inelastic scattering at HERA}

In the hadronic CMS, the produced
hadrons cover about 5 units of rapidity in either hemisphere, depending
mainly on the available CM energy of the hadronic system, $W$.
There are about 2.5 charged particles per unit rapidity,
and
the transverse energy is on average 2 GeV per unit rapidity.
Multiplicity distributions exhibit the familiar KNO scaling,
and Bose-Einstein
correlations are measured similar to other reactions.
A large rapidity gap is found in about 10\% of the events.
Around 25\% of the events contain charm.
In the current hemisphere
about 0.9 $K^0$ mesons and 0.15 $\Lambda$ particles are measured per event.
In 5 to 10\% of the events
more than one jet is found (apart from the remnant).

The global event features are largely
determined by the available phase space given by the CM energy $W$
of the hadronic system.
The dependence on the photon virtuality \Qsq has also been studied.
In the current region the transverse activity
(\et flow, \pt spectra, jets) increases with $Q^2$.
In the central
rapidity region \Qsq dependencies are small.
Here the data are
similar to what has been measured in photoproduction (\Qsq=0) or
hadron-hadron collisions. The particle density per unit rapidity
increases slower
than in \epem~ annihilation with the available CM energy $W$.
When compared with DIS data at lower $W$, a faster than logarithmic
rise of the average multiplicity is observed.

\subsubsection{Comparisons with QCD}

Transverse energy and
momentum distributions provide evidence for hard QCD radiation beyond
the LO matrix element.
More \et has been measured than anticipated before HERA data taking.
It is not yet clear whether soft colour
interactions are required to describe the transverse energy flow,
or whether not yet well established perturbative QCD effects are
responsible for the large \et measured.
Soft colour interactions
are of interest also because they provide a mechanism for
producing rapidity gaps.
Alternatively, colour dipole radiation or BFKL evolution can be
employed to produce as much \et as seen in the data.

The scaled charged particle momentum spectra measured either in the
current hemispheres of the
hadronic
CMS with $W$ as scale or the Breit frame with $Q$ as scale
are in general well
described by QCD models in various approximations of
the perturbative shower evolution in conjunction with
non-perturbative hadronization models.
Where applicable, perturbative QCD calculations in NLO
employing fragmentation functions measured elsewhere are in agreement
with the data.
Scaling violations
of the spectra will provide a means to measure $\alpha_s$.

The growth of multiplicity with energy
is described by perturbative QCD calculations
in the MLLA approximation, assuming local parton hadron duality (LPHD).
The shape of the momentum spectra in the ``hump backed'' form is
approximately Gaussian. It's evolution with $Q$ is consistent with the
MLLA+LPHD expectation assuming soft colour coherence. Models without
soft colour coherence describe the data less well.
When allowance is made for the boson-gluon fusion process in DIS,
the particle spectra are similar to what has been
measured in \epem~ annihilation.
Evidence is found for the prediction
that the soft part of the
Lorentz invariant spectrum is independent of the energy
of the fragmenting parton.
The assumption of a constant coupling \as
is in stark conflict with the data.

Global event shape variables like thrust, jet mass and jet broadening
in the Breit frame current hemisphere show an increasing
event collimation
with energy $Q$.
``Power corrections'' supplementing the NLO calculations provide
a good description of the data.
The power corrections $\propto 1/Q$ parametrize
higher orders and hadronization effects and appear to be universal within
20\% -- the same
normalization parameter can be used for different shape variables
and $ep$ as well as \epem~ reactions.
The concept of ``power corrections''
may turn out to be very fruitful for the understanding of hadronization
effects. Assuming power corrections, the strong coupling \as has been
determined with competitive precision.

Of similar precision is the \as determination from the rate of 2+1 jet events.
The strong coupling \as has been measured as a function of \Qsq
from 2+1 jet events in a region where NLO QCD is able to describe
the data, namely where \Qsq is not too small.
The data are consistent with
the expected \Qsq dependence of the running coupling.
However,
the optimal strategy for \as measurements from jets has not yet been found.
A better understanding of the connection between measured hadron jets
and perturbative QCD predictions for parton jets appears to be needed.
Smaller theoretical uncertainties are anticipated with other jet
algorithms and increased statistics at higher \Qsq or jet $p_T$.

At small $x$ and \Qsq and in particular towards the proton remnant
the measured jet
rates are larger than expected from either NLO calculations
or conventional QCD models.
``Conventional'' stands for deep inelastic
scattering, where a virtual photon scatters on a parton in a proton, and
where the partons in the proton obey DGLAP evolution with $Q^2$ as scale.
A good description is obtained either by the colour dipole model, or
by a model in which the partonic content of the virtual photon is
resolved by the large jet \pt.

A strong growth of the gluon density towards small $x$ has been
inferred from DGLAP analyses of structure function measurements.
This is confirmed by
direct measurements from jet and charm production, by which
photon-gluon fusion processes can be tagged. The errors are still
large, however. It will be very important
to extend such direct measurements
towards smaller $x$ values.

Small $x$ physics has received considerable interest.
It has been speculated that the rise of $F_2$ with decreasing $x$
is a sign for BFKL evolution,
but DGLAP evolution was found to be able to account for the
data as well.
Traces of BFKL evolution have therefore
been searched for in less inclusive observables in the hadronic final
state, namely enhanced transverse activity in between the
current and the target fragmentation regions: transverse energy flow,
high \pt particles, ``forward jets''.
The data could not be described
by conventional DGLAP evolution.
Though BFKL calculations can explain
the measured enhanced activity, there exist competing explanations,
again making use of a virtual photon structure. It remains to be
seen though whether or not such an alternative model
would be in conflict with other
HERA data. In case both explanations work equally well, one might
speculate that they have more in common than is apparent.

Finally, the possibility exists to discover QCD instanton effects
at HERA. Such a discovery would open a new field of non-perturbative
QCD, and would imply also baryon and lepton number violation in
the electroweak sector already {\it within} the standard model.
Observable effects are predicted for the hadronic final state,
and first bounds on instanton production have been placed.
The predicted
cross sections are such that it appears likely that with
increased sensitivity instantons
can be discovered at HERA,
or the theory in its present form can be rejected.

\chapter*{Acknowledgements}    
This work has been made possible by a grant from the
Deutsche Forschungsgemeinschaft.
I would like to thank
G. Buschhorn from
the Max-Planck-Institut f\"ur Physik in Munich for
the hospitality I have experienced in his group
during the last two years.
I have enjoyed very much the close collaboration in the MPI group,
in particular
with F. Botterweck, T. Carli and G. Grindhammer.

I wish to thank all my
colleagues from the H1 and ZEUS experiments
on whose research this work is based
for their collaboration
and friendly competition at HERA.
I would like to thank J. Dainton, E. Lohrmann,
B. Naroska and A. Wagner
for their continuous encouragement
and their support to undertake the Habilitation
and to teach at the Universit\"at Hamburg.

I am grateful to
J. Bartels, E. De Wolf,
A. Martin,
G. Ingelman,
A. Ringwald and
F. Schrempp
for their patient explanations of sometimes heavy theory, and
for enlightening discussions, in which some very fruitful
ideas were born.
Helpful discussions with
B. Andersson,
Y. Dokshitzer,
D. Graudenz,
L. L\"onnblad,
M. Seymour,
S. Lang,
E. Mirkes and
B. Webber
are gratefully acknowledged.

Finally, I would like to thank
J. Bartels,
N. Brook,
T. Carli,
T. Doyle,
A. De Roeck,
E. De Wolf,
G. Grindhammer,
J. Hartmann,
G. Ingelman,
H. Jung,
A. Levy,
E. Lohrmann,
U. Martyn,
D. Milstead,
B. Naroska,
N. Pavel,
J. Repond,
A. Ringwald,
F. Schrempp,
F. Sefkow and
V. Shekelyan
for their comments on the manuscript, or parts of it. Their
critical reading has been invaluable.

\begin{footnotesize}
\newpage
%
%

\end{footnotesize}
\end{document}